\documentclass[prb,amsfonts,twocolumn,showpacs,amssymb,floatfix,bibnotes,superscriptaddress]{revtex4}
\usepackage{bm}
\usepackage{tabularx}
\usepackage{ascmac}
\usepackage{amsmath}
\usepackage{wrapfig}
 \usepackage{graphicx}
\usepackage{float}
\usepackage[FIGBOTCAP]{subfigure}
\usepackage{color}

\bmdefine{\bolds}{s}
\bmdefine{\boldi}{i}
\bmdefine{\boldj}{j}
\bmdefine{\boldzero}{0}
\bmdefine{\boldtau}{\tau}
\bmdefine{\boldsigma}{\sigma}
\bmdefine{\boldlambda}{\lambda}
\bmdefine{\boldPhi}{\Phi}
\bmdefine{\boldnabla}{\nabla}
\bmdefine{\boldx}{x}
\bmdefine{\boldX}{X}
\bmdefine{\boldk}{k}
\bmdefine{\boldv}{v}
\bmdefine{\boldK}{K}
\bmdefine{\boldH}{H}
\bmdefine{\boldq}{q}
\bmdefine{\boldQ}{Q}
\bmdefine{\boldA}{A}
\bmdefine{\boldr}{r}
\bmdefine{\boldj}{j}
\bmdefine{\boldJ}{J}

\begin{document}


\title{
Microscopic theory on charge transports 
of a correlated multiorbital system
}


\author{Naoya Arakawa}
\email{arakawa@hosi.phys.s.u-tokyo.ac.jp} 
\affiliation{RIKEN Center for Emergent Matter Science (CEMS), 
Wako, Saitama 351-0198, Japan}


\date{\today}

\begin{abstract}
Current vertex correction (CVC), 
the back-flow-like correction to the current, 
comes from conservation laws, 
and the CVC due to electron correlation contains 
information about many-body effects. 
However, 
it has been little understood how 
the CVC due to electron correlation affects the charge transports of 
a correlated multiorbital system. 
To improve this situation, 
I studied the inplane resistivity, $\rho_{ab}$, 
and the Hall coefficient in the weak-field limit, $R_{\textrm{H}}$, 
in addition to the magnetic properties and the electronic structure, 
for a $t_{2g}$-orbital Hubbard model on a square lattice 
in a paramagnetic state away from or near 
an antiferromagnetic (AF) quantum-critical point (QCP) 
in the fluctuation-exchange (FLEX) approximation 
with the CVCs arising from the self-energy ($\Sigma$), 
the Maki-Thompson (MT) irreducible four-point vertex function, 
and the main terms of the Aslamasov-Larkin (AL) one. 
Then, I found three main results about the CVCs. 
First, 
the main terms of the AL CVC does not qualitatively change 
the results obtained in the FLEX approximation 
with the $\Sigma$ CVC and the MT CVC. 
Second, 
$\rho_{ab}$ and $R_{\textrm{H}}$ near the AF QCP 
have 
high-temperature region, governed mainly by 
the $\Sigma$ CVC, 
and low-temperature region, governed mainly by 
the $\Sigma$ CVC and the MT CVC. 
Third, 
in case away from the AF QCP, 
the MT CVC leads to a considerable effect on only $R_{\textrm{H}}$ at low temperatures, 
although $R_{\textrm{H}}$ at high temperatures and $\rho_{ab}$ at all temperatures considered 
are sufficiently described by including only the $\Sigma$ CVC. 
Those findings reveal several aspects of many-body effects 
on the charge transports of a correlated multiorbital system. 
I also achieved the qualitative agreement with several experiments of 
Sr$_{2}$RuO$_{4}$ or Sr$_{2}$Ru$_{0.975}$Ti$_{0.025}$O$_{4}$. 
Moreover, I showed several better points of this theory than other theories. 
\end{abstract}


\maketitle

\section{Introduction} 
Many-body effects, 
effects of Coulomb interaction between itinerant electrons 
beyond the mean-field approximation, 
are important to discuss electronic properties~\cite{Fazekas,Moriya-review,Kon-review,MIT-review}. 
When the Coulomb interaction is very small compared with 
the bandwidth of itinerant electrons, which is of the order of magnitude $1$ eV, 
we can sufficiently describe its effects 
in the mean-field approximation 
as the static and effectively single-body potentials~\cite{Ashcroft-Mermin,Ziman}. 
However, 
in correlated electron systems such as transition metals or transition-metal oxides, 
the Coulomb interaction becomes moderately strong or strong, 
resulting in the derivations of  
their electronic properties from 
the single-body picture~\cite{Fazekas,Moriya-review,Kon-review,MIT-review}. 

For several correlated electron systems, 
many-body effects can be described in Landau's Fermi-liquid (FL) 
theory~\cite{Landau,Nozieres,AGD,NA-review}. 
This theory is based on two basic assumptions~\cite{NA-review}. 
One is the one-to-one correspondence between 
the noninteracting and the interacting systems. 
Because of this assumption, 
we can describe low-energy excitations of the interacting system 
in terms of quasiparticles (QPs) with the renormalized effective mass 
and the renormalized interactions described by the Landau parameters~\cite{NA-review}. 
The other assumption is lack of the temperature dependence of the Landau parameters. 
Because of this assumption, 
the temperature dependence of the electronic properties 
remains the same as that in the noninteracting system~\cite{NA-review}. 
Furthermore, 
as a result of those assumptions, 
many-body effects on the electronic properties are the changes of their coefficients 
due to the mass enhancement or the FL correction or both~\cite{NA-review}. 

Actually, 
Landau's FL theory well describes 
several electronic properties of Sr$_{2}$RuO$_{4}$ 
at low temperatures. 
First, 
this theory can explain 
the almost temperature-independent spin susceptibility~\cite{Maeno-RW} 
and the $T^{2}$ dependence of the inplane resistivity~\cite{resistivity-x2}. 
In addition, 
the importance of many-body effects 
has been suggested in the measurements 
of the de Haas-van Alphen (dHvA) effect~\cite{dHvA-x2,Mackenzie-review} 
and the Wilson ratio~\cite{Maeno-RW}: 
the effective mass of the $d_{xz/yz}$ or the $d_{xy}$ orbital 
measured in the dHvA 
becomes, respectively, $3-3.5$ or $5.5$ times 
as large as the mass obtained in the local-density approximation (LDA), 
a mean-field-type approximation; 
the Wilson ratio, the ratio of the spin susceptibility 
to the coefficient of the electronic specific heat, 
becomes $1.7-1.9$ times as large as the noninteracting value. 
Note that the enhancement of the Wilson ratio 
arises from the FL correction~\cite{Yamada-Yosida}. 

However,  
we observe non-FL-like behaviors, 
the deviations from the temperature dependence expected 
in Landau's FL theory, 
for correlated electron systems 
near a magnetic quantum-critical point (QCP)~\cite{Kon-review,MIT-review}. 
For example, 
Sr$_{2}$Ru$_{0.975}$Ti$_{0.025}$O$_{4}$, 
a paramagnetic (PM) ruthenate near an antiferromagnetic (AF) QCP, 
shows the Curie-Weiss-type temperature dependence of the spin susceptibility 
and the $T$-linear inplane resistivity~\cite{Ti214-nFL1,Ti214-nFL2}. 
Also, 
Ca$_{2-x}$Sr$_{x}$RuO$_{4}$ around $x=0.5$, 
a PM ruthenate near a ferromagnetic QCP, 
shows the similar non-FL-like behaviors~\cite{CSRO-nFL1,CSRO-nFL2}. 
Thus, 
those experimental results indicate the importance of many-body effects 
beyond Landau's FL theory near a magnetic QCP. 
Note, first, that 
the wave vector of the spin susceptibility enhanced most strongly 
in Sr$_{2}$Ru$_{0.975}$Ti$_{0.025}$O$_{4}$~\cite{Neutron-Ti214}
is the same for Sr$_{2}$RuO$_{4}$~\cite{Neutron-x2}, 
i.e. $\boldq\approx (\frac{2\pi}{3}, \frac{2\pi}{3})$; 
second, that 
Ti substitution does not cause any RuO$_{6}$ distortions~\cite{Ti214-nFL1}, 
while 
Ca substitution causes RuO$_{6}$ distortions 
such as the rotation and the tilting~\cite{xray-CSRO}, 
which drastically affect the electronic structure~\cite{Terakura,NA-GA}. 

Among correlated electron systems, 
the ruthenates are suitable to deduce general or characteristic aspects of many-body effects 
in correlated multiorbital systems 
because of the following three advantages. 
The first advantage is that 
the ruthenates show the FL or the non-FL-like behaviors, 
depending on the chemical composition or the crystal structure or 
both~\cite{Maeno-RW,resistivity-x2,Ti214-nFL1,Ti214-nFL2,CSRO-nFL1,CSRO-nFL2}. 
Due to this advantage, 
we can study 
how the FL state is realized and 
how the system changes from the FL state to the non-FL-like state, 
and we may obtain their general or characteristic properties. 
Then, the second advantage is that 
the ruthenates are the $t_{2g}$-orbital systems 
with moderately strong electron correlation~\cite{Mackenzie-review,X-ray10Dq}. 
This has been established for Sr$_{2}$RuO$_{4}$ by three facts: 
the Ru $t_{2g}$ orbitals are the main components of the density-of-states (DOS) 
near the Fermi level in the LDA~\cite{Oguchi,Mazin-LDA}; 
the LDA~\cite{Oguchi,Mazin-LDA} can reproduce 
the topology of the Fermi surface (FS) observed experimentally~\cite{dHvA-x2,ARPES-x2}; 
the experimentally estimated value of $U$, 
onsite intraorbital Coulomb interaction, 
is about $2$ eV~\cite{X-ray10Dq}, which is half of the bandwidth for the $t_{2g}$ orbitals 
in the LDA~\cite{Oguchi,Mazin-LDA}. 
In addition to the second advantage, 
the third advantage is 
the simple electronic structure~\cite{Oguchi,Mazin-LDA} compared with 
the other multiorbital systems. 
Due to the second and the third advantage, 
we can simply analyze many-body effects of a correlated multiorbital system, 
and 
that analysis may lead to a deep understanding of 
the general or characteristic aspects of the many-body effects. 

To describe many-body effects near a magnetic QCP, 
we need to use the theories that 
can satisfactorily take account of 
the effects of the critical electron-hole scattering 
arising from the characteristic spin fluctuation 
of that QCP. 
If the system approaches a magnetic QCP, 
we observe the enhancement of the spin fluctuation 
for the wave vector characteristic of that QCP~\cite{Kon-review,NA-review,Yanase-review}. 
That enhancement causes the strong temperature dependent 
critical electron-hole scattering mediated by the spin fluctuation. 
Then, 
that critical electron-hole scattering 
results in 
the emergence of both 
the hot spot of the QP damping and the Curie-Weiss-type temperature dependence 
of the reducible four-point vertex function 
for the momenta connected by the spin fluctuation. 
(Note that the reducible four-point vertex function 
describes the multiple electron-hole scattering~\cite{Nozieres}.) 
Thus, the emergence of the former violates 
the first basic assumption of Landau's FL theory 
since at the hot spot the QP lifetime is not so sufficiently long 
as to realize an approximate eigenstate as Landau's FL~\cite{NA-review}. 
Furthermore, 
the latter violates the second basic assumption 
because the temperature dependence of the reducible four-point vertex function 
and mass enhancement factor determines 
the temperature dependence of the Landau parameter~\cite{NA-review}. 
Thus, 
many-body effects near a magnetic QCP 
may be described by the theories beyond Landau's FL theory 
if the theories can satisfactorily treat 
the strongly enhanced temperature-dependent spin fluctuation.  

Actually, 
several non-FL-like behaviors near a magnetic QCP 
can be reproduced 
in fluctuation-exchange (FLEX) approximation~\cite{FLEX1,FLEX2,FLEX3,multi-FLEX1,multi-FLEX2} 
with the current vertex corrections (CVCs) 
arising from the self-energy ($\Sigma$) and 
the Maki-Thompson (MT) irreducible four-point vertex function~\cite{MT1,MT2} 
due to electron correlation~\cite{Kon-CVC,NA-CVC}. 
For example, 
this theory shows 
the Curie-Weiss-type temperature dependence of 
both the spin susceptibility and the Hall coefficient 
and the $T$-linear inplane resistivity 
for a single-orbital Hubbard model 
on a square lattice in a PM state near an AF QCP, 
where the spin fluctuation for $\boldq=(\pi,\pi)$ is enhanced~\cite{Kon-CVC}.  
Those results are consistent with 
the experiments of cuprates~\cite{cuprate-chiS,cuprate-rho,cuprate-RH}. 
Since the powerfulness of the FLEX approximation near a magnetic QCP 
arises from its satisfactory treatment of the momentum and temperature dependence 
of spin fluctuations~\cite{Yanase-review,NA-review}, 
the similar applicability will hold even for a multiorbital Hubbard model 
on a square lattice. 

Since the effects of the CVCs due to electron-electron interaction 
in a correlated multiorbital system had been unclear, 
I studied several electronic properties 
of an effective model of several ruthenates, 
a $t_{2g}$-orbital Hubbard model on a square lattice 
in a PM state away from or near an AF QCP, 
in the FLEX approximation with the $\Sigma$ CVC and the MT CVC~\cite{NA-review,NA-CVC},   
and then I obtained satisfactory agreement with several experiments
and three important aspects of many-body effects on the charge transports. 
First, 
the results away from the AF QCP 
qualitatively agree with five experimental results of Sr$_{2}$RuO$_{4}$, 
(i) the strongest enhancement of spin fluctuation~\cite{Neutron-x2} 
at $\boldq\approx (\frac{2\pi}{3}, \frac{2\pi}{3})$, 
(ii) the nearly temperature-independent spin susceptibility~\cite{Maeno-RW}, 
(iii) the larger mass enhancement~\cite{dHvA-x2,Mackenzie-review} of the $d_{xy}$ orbital 
than that of the $d_{xz/yz}$ orbital, 
(iv) the $T^{2}$ dependence of the inplane resistivity at low temperatures~\cite{resistivity-x2}, 
and (v) the non-monotonic temperature dependence of the Hall coefficient~\cite{Hall-x2}. 
Note that 
the Hall coefficient observed in Sr$_{2}$RuO$_{4}$ shows 
the following non-monotonic temperature dependence~\cite{Hall-x2}: 
at high temperatures above $130$K, 
the Hall coefficient is small and negative 
with a slight increase with decreasing temperature; 
after crossing over zero at $130$K, 
the Hall coefficient becomes positive with keeping an increase, 
and shows a peak at about $70$K; 
below about $70$K, 
the Hall coefficient monotonically decreases with decreasing temperature. 
Then, 
the results near the AF QCP 
can qualitatively explain three experimental results of Sr$_{2}$Ru$_{0.975}$Ti$_{0.025}$O$_{4}$, 
(i) the strongest enhancement of spin fluctuation~\cite{Neutron-Ti214} 
at $\boldq\approx (\frac{2\pi}{3}, \frac{2\pi}{3})$, 
(ii) the Curie-Weiss-type temperature dependence of the spin susceptibility~\cite{Ti214-nFL1}, 
and (iii) the $T$-linear inplane resistivity~\cite{Ti214-nFL2}. 
In this comparison, 
I assume that 
the main effect of Ti substitution is approaching the AF QCP compared with Sr$_{2}$RuO$_{4}$. 
Note that 
the measurement of the Hall coefficient in Sr$_{2}$Ru$_{0.975}$Ti$_{0.025}$O$_{4}$ 
has been restricted at very low temperature~\cite{Ti214-RH}, 
which is out of the region I considered. 
Moreover, 
I revealed the realization of the orbital-dependent transports, 
the emergence of a peak of the temperature dependence of the Hall coefficient, 
and the absence of the Curie-Weiss-type temperature dependence of the Hall coefficient 
near the AF QCP. 

However, 
the previous studies~\cite{NA-review,NA-CVC} contain two remaining issues.  
One is to clarify many-body effects of the Aslamasov-Larkin (AL) CVC, 
the CVC arising from the AL irreducible four-point vertex function~\cite{AL,Tewordt-AL}. 
In the previous studies~\cite{NA-review,NA-CVC}, 
I neglected the AL CVC in the FLEX approximation for simplicity 
since in a single-orbital Hubbard model~\cite{Kon-CVC} 
on a square lattice 
the AL CVC does not qualitatively change the results 
of the resistivity and Hall coefficient near an AF QCP 
and 
since the similar property 
would hold even in a multiorbital Hubbard model on a square lattice 
not far away from an AF QCP. 
However, it is necessary and important to analyze 
the effects of the AL CVC in that multiorbital Hubbard model 
since both the MT and the AL CVC are essential 
to hold conservation laws exactly~\cite{Yamada-Yosida}. 
In particular, 
that analysis is needed 
not only to check the validity neglecting the AL CVC 
for qualitative discussions 
but also to clarify many-body effects of the AL CVC. 
The other remaining issue is to give the comprehensive explanations about 
the formal derivations both of the transport coefficients 
in the extended \'{E}liashberg theory~\cite{Eliashberg-theory} to a multiorbital system 
and of the CVCs in the FLEX approximation for a multiorbital Hubbard model. 
My previous study~\cite{NA-CVC} reported 
a microscopic study about the effects of the CVCs 
due to electron correlation in a multiorbital system. 
In the previous studies~\cite{NA-review,NA-CVC}, however, 
I just gave brief explanations about those formal derivations. 
Thus, 
it is desirable to explain the detail of those formal derivations 
since those will be useful to adopt the same or similar method to 
the transport properties of other correlated electron systems. 

In this paper, 
after formulating the dc longitudinal and the dc transverse conductivities 
in the extended \'{E}liashberg theory to a multiorbital Hubbard model 
in the FLEX approximation with the CVCs, 
I study the effects of the main terms of the AL CVC 
on the in-plane resistivity and the Hall coefficient 
for the quasi-two-dimensional PM ruthenates near and away from the AF QCP. 
As the main results, 
I show the qualitative validity of the main results of 
the previous studies~\cite{NA-review,NA-CVC}, 
the existence of two almost distinct regions of 
the charge transports near the AF QCP as a function of temperature, 
and the different effects of the MT CVC 
on the low-temperature values of the in-plane resistivity and Hall coefficient 
away from the AF QCP 
in the presence of the $\Sigma$ CVC, the MT CVC, and 
the main terms of the AL CVC. 
I also present several results 
of the magnetic properties and the electronic structure, 
and show four main results about each of the magnetic properties 
and the electronic structure. 
Those are useful for deeper understanding 
than in the previous studies~\cite{NA-review,NA-CVC}. 

The remaining part of this paper is organized as follows. 
In Sec. II, 
I explain the method to calculate the electronic properties of 
some quasi-two-dimensional PM ruthenates without the RuO$_{6}$ distortions. 
In Sec. II A, 
I show the Hamiltonian of 
an effective model of some quasi-two-dimensional ruthenates, 
explain the parameter choice for the noninteracting Hamiltonian, 
and briefly remark on the spin-orbit interaction. 
In Secs. II B 1 and II B 2,  
I explain the extended \'{E}liashberg theory to 
the dc longitudinal and the dc transverse conductivities for a multiorbital system 
and give several theoretical remarks about their general properties. 
In Sec. II C, 
I explain several advantages of the FLEX approximation 
with the CVCs, 
formulate the FLEX approximation for a multiorbital Hubbard model, 
and derive the Bethe-Salpeter equation for the current 
with the $\Sigma$ CVC, the MT CVC, and the AL CVC 
in the FLEX approximation. 
Furthermore, 
I derive a simplified Bethe-Salpeter equation 
by approximating the AL CVC to its main terms. 
In Sec. III, 
I show the results of 
several electronic properties of the quasi-two-dimensional PM ruthenates 
near and away from the AF QCP in the FLEX approximation 
with the $\Sigma$ CVC, 
the MT CVC, and the main terms of the AL CVC; 
in addition to that case, 
I consider three other cases considered in Ref. \onlinecite{NA-review} 
for discussions about the transport properties 
in order to deduce the main effects of the AL CVC. 
After discussing the magnetic properties in Sec. III A 
and the electronic structure in Sec. III B, 
I discuss the main effects of the AL CVC 
on the inplane resistivity and the Hall coefficient in Sec. III C. 
Then, 
I compare the obtained results with 
several experiments of 
Sr$_{2}$RuO$_{4}$ or Sr$_{2}$Ru$_{0.975}$Ti$_{0.025}$O$_{4}$ in Sec. IV A, 
and other theories in Sec. IV B. 
In Sec. V, 
I summarize the obtained results and their conclusions, 
and show several remaining issues. 

\section{Method}
In this section, 
I explain an effective model of some quasi-two-dimensional ruthenates 
and a general theoretical method 
to analyze the resistivity and the Hall coefficient 
for a correlated multiorbital system in a PM state. 
In Sect. II A, 
we see the Hamiltonian of the effective model, 
determine the parameters of the noninteracting Hamiltonian, 
and remark on the spin-orbit interaction, neglected in the effective model. 
In Sect. II B 1, 
to analyze the resistivity, 
we explain the formal derivation of the dc longitudinal conductivity 
of a multiorbital Hubbard model in a PM state 
without an external magnetic field 
in the linear-response theory with 
the most-divergent-term approximation~\cite{Eliashberg-theory}, 
and show general properties of the derived longitudinal conductivity and 
their consequences for the resistivity.
In Sect. II B 2, 
we derive the dc transverse conductivity of a multiorbital Hubbard model in a PM state 
in the weak-field limit by using the linear-response theory 
with the most-divergent-term approximation, 
and see general properties of the derived transverse conductivity and the Hall coefficient 
in combination with the results for the longitudinal conductivity. 
The general formulations in Sects. II B 1 and II B 2 
are the extensions of the single-orbital cases for the resistivity~\cite{Eliashberg-theory} 
and the Hall coefficient~\cite{Fukuyama-RH,Kohno-Yamada}, respectively. 
In Sect. II C, 
we remark on several advantages of the FLEX approximation with the CVCs, 
formulate the FLEX approximation in Matsubara-frequency representation 
for a multiorbital Hubbard model in a PM state 
in the similar way for Refs. \onlinecite{multi-FLEX1} and \onlinecite{multi-FLEX2}, 
and derive the CVCs 
in the FLEX approximation by extending the formulation for a single-orbital case~\cite{Kon-CVC}. 

Hereafter, 
we use the following unit and notations: 
We set $\hbar=1$, $c=1$, $e=1$, $\mu_{\textrm{B}}=1$, and $k_{\textrm{B}}=1$. 
In the equations, 
the $d_{xz}$, $d_{yz}$, and $d_{xy}$ orbitals 
are labeled $1$, $2$, and $3$, respectively. 
In Matsubara-frequency representation of several quantities, 
we use the fermionic and the bosonic Matsubara frequency, 
$\epsilon_{m}=\pi T(2m+1)$ and $\Omega_{n}=2\pi T n$, respectively. 
In real-frequency representation, 
we use frequency variables such as $\epsilon$ and $\omega$ 
and abbreviations such as $k\equiv (\boldk,\epsilon)$ and $q\equiv (\boldq,\omega)$, 
with momenta $\boldk$ and $\boldq$. 
We use the abbreviations 
such as $\textstyle\sum_{\{a\}}\equiv \textstyle\sum_{a,b,c,d}$, 
$\textstyle\sum_{\{s_{1}\}}\equiv \textstyle\sum_{s_{1},s_{2},s_{3},s_{4}}$, 
and $\Gamma_{\{a\}}(\boldk,i\epsilon_{m},\boldk^{\prime},i\epsilon_{m^{\prime}};\boldzero,i\Omega_{n})
\equiv 
\Gamma_{abcd}(\boldk,i\epsilon_{m},\boldk^{\prime},i\epsilon_{m^{\prime}};\boldzero,i\Omega_{n})$.   

\subsection{Effective model }

In this section, 
I introduce the total Hamiltonian of an effective model 
for some quasi-two-dimensional ruthenates 
and explain how to choose the parameters in the noninteracting Hamiltonian. 
I also give a brief remark about the spin-orbit interaction. 

To describe the electronic properties of several 214-type ruthenates 
such as Sr$_{2}$RuO$_{4}$, 
I use a $t_{2g}$-orbital Hubbard model~\cite{NA-review,NA-CVC} on a square lattice 
because several 214-type ruthenates are categorized as 
quasi-two-dimensional $t_{2g}$-orbital correlated systems 
and Ru ions on a two-dimensional layer form a square lattice~\cite{Mackenzie-review}. 
The Hamiltonian of this model is 
\begin{align}
\hat{H}=\hat{H}_{0}+\hat{H}_{\textrm{int}},\label{eq:Htot}
\end{align}
where $\hat{H}_{0}$ and $\hat{H}_{\textrm{int}}$ 
are the noninteracting and the interacting Hamiltonian, respectively. 

\begin{figure}[tb]
\begin{center}
\includegraphics[width=86mm]{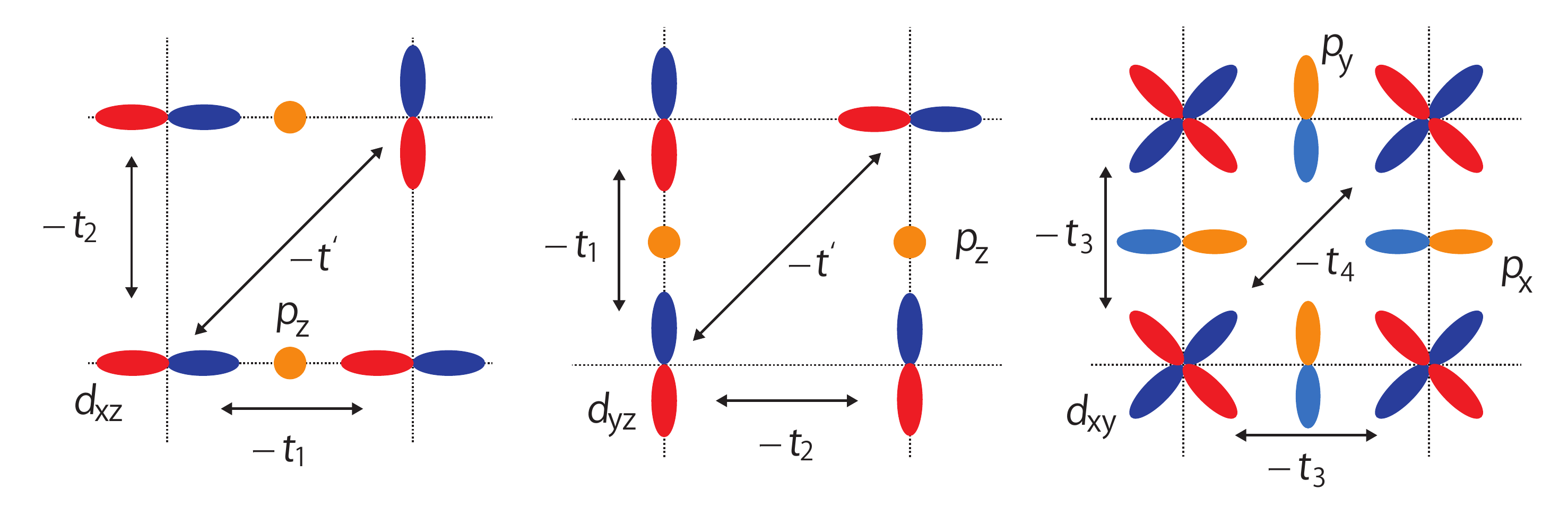}
\end{center}
\vspace{-20pt}
\caption{
Schematic pictures of the hopping processes of the $t_{2g}$ orbitals 
on a two-dimensional layer. 
The difference in the color of each orbital shows 
the difference in the sign of its wave function.}
\label{fig:Fig1}
\end{figure}

First, 
$\hat{H}_{0}$ is given by 
\begin{align}
\hat{H}_{0}
&=
\sum\limits_{\boldk}
\sum\limits_{a,b=1}^{3}
\sum\limits_{s=\uparrow,\downarrow}
\epsilon_{ab}(\boldk)
\hat{c}^{\dagger}_{\boldk a s} 
\hat{c}_{\boldk b s}.\label{eq:H0}
\end{align}
Here $\hat{c}_{\boldk a s}$ and $\hat{c}^{\dagger}_{\boldk a s}$ 
are the annihilation and the creation operator of an electron 
of momentum $\boldk$, orbital $a$, and spin $s$, 
and $\epsilon_{ab}(\boldk)$ is given by 
\begin{align}
\epsilon_{11}(\boldk)=&
-\frac{\Delta_{t_{2g}}}{3}-2 t_{1} \cos k_{x}-2 t_{2} \cos k_{y}-\mu,\label{eq:e11}\\ 
\epsilon_{12}(\boldk)=&\ \epsilon_{21}(\boldk)=\
 4 t^{\prime} \sin k_{x} \sin k_{y},\label{eq:e12}\\ 
\epsilon_{22}(\boldk)=&
-\frac{\Delta_{t_{2g}}}{3}-2 t_{2} \cos k_{x}-2 t_{1} \cos k_{y}-\mu,\label{eq:e22}\\ 
\epsilon_{33}(\boldk)=&\
\frac{2\Delta_{t_{2g}}}{3}-2t_{3}(\cos k_{x}+\cos k_{y})\notag\\
&-4t_{4}\cos k_{x} \cos k_{y}-\mu,\label{eq:e33}
\end{align}
and otherwise $\epsilon_{ab}(\boldk)=0$, 
where $\Delta_{t_{2g}}$ is the difference between the crystalline-electric-field energies 
of the $d_{xy}$ and the $d_{xz/yz}$ orbital, 
$\mu$ is the chemical potential determined so that 
the electron number per site, $n_{\textrm{e}}$, satisfies $n_{\textrm{e}}=4$, 
and $t_{1}$, $t_{2}$, $t_{3}$, $t_{4}$, and $t^{\prime}$ 
are the hopping integrals of the $t_{2g}$ orbitals, 
whose schematic pictures are shown in Fig. \ref{fig:Fig1}. 
Since 
I neglect the effects~\cite{Terakura,NA-GA} of the RuO$_{6}$ distortions 
on $\hat{H}_{0}$,  
the targets of this paper are 
the $214$-type ruthenates without the RuO$_{6}$ distortions. 

Assuming that 
the LDA~\cite{Oguchi,Mazin-LDA} for Sr$_{2}$RuO$_{4}$ gives 
a good starting point to include many-body effects 
in the $214$-type ruthenates without the RuO$_{6}$ distortions, 
I choose the parameters in $\epsilon_{ab}(\boldk)$ 
so as to reproduce the electronic structure 
obtained in the LDA~\cite{Oguchi,Mazin-LDA} for Sr$_{2}$RuO$_{4}$. 
Namely, 
I set $t_{1}=0.675$ eV, $t_{2}=0.09$ eV, $t_{3}=0.45$ eV, $t_{4}=0.18$ eV, 
$t^{\prime}=0.03$ eV, and $\Delta_{t_{2g}}=0.13$ eV. 
As explained in Ref. \onlinecite{NA-review}, 
the obtained electronic structure 
is consistent with the LDA~\cite{Oguchi,Mazin-LDA}: 
the bandwidth for the $t_{2g}$ orbitals is about $4$ eV; 
the quasi-one-dimensional $d_{xz}$ and $d_{yz}$ orbitals form 
the hole-like $\alpha$ and electron-like $\beta$ sheets, 
and the quasi-two-dimensional $d_{xy}$ orbital forms 
the electron-like $\gamma$ sheet; 
the van Hove singularity of the $d_{xy}$ orbital exists above the Fermi level; 
the occupation numbers of the $d_{xz/yz}$ and the $d_{xy}$ orbital 
are $n_{xz/yz}=1.38$ and $n_{xy}=1.25$. 

Then, 
$\hat{H}_{\textrm{int}}$ is given by 
\begin{align}
\hat{H}_{\textrm{int}}
=&\dfrac{1}{4}
\sum\limits_{\boldj}
\sum\limits_{\{a\}}
\sum\limits_{\{s_{1}\}}
U_{abcd}^{s_{1}s_{2}s_{3}s_{4}}
\hat{c}^{\dagger}_{\boldj a s_{1}}\hat{c}^{\dagger}_{\boldj d s_{4}}
\hat{c}_{\boldj c s_{3}}\hat{c}_{\boldj b s_{2}}\notag\\
=&\
 U 
\sum\limits_{\boldj}
\sum\limits_{a=1}^{3}
\hat{n}_{\boldj a \uparrow} \hat{n}_{\boldj a \downarrow}
+ U^{\prime}  
\sum\limits_{\boldj}
\sum\limits_{a=1}^{3}
\sum\limits_{b<a}
\hat{n}_{\boldj a} \hat{n}_{\boldj b}\notag\\
&- 
J_{\textrm{H}} 
\sum\limits_{\boldj}
\sum\limits_{a=1}^{3}
\sum\limits_{b<a}
( 
2 \hat{\bolds}_{\boldj a} \cdot 
\hat{\bolds}_{\boldj b} 
+ 
\frac{1}{2} \hat{n}_{\boldj a} \hat{n}_{\boldj b} 
)\notag\\
&+
J^{\prime} 
\sum\limits_{\boldj}
\sum\limits_{a=1}^{3}
\sum\limits_{b\neq a}
\hat{c}_{\boldj a \uparrow}^{\dagger} 
\hat{c}_{\boldj a \downarrow}^{\dagger} 
\hat{c}_{\boldj b \downarrow} 
\hat{c}_{\boldj b \uparrow}.\label{eq:Hint}
\end{align}
Here
$U_{abcd}^{s_{1}s_{2}s_{3}s_{4}}$ is a bare four-point vertex function, 
$U$ is intraobital Coulomb interaction, 
$U^{\prime}$ is interorbital Coulomb interaction, 
$J_{\textrm{H}}$ is Hund's rule coupling, 
$J^{\prime}$ is pair hopping term, 
$\hat{n}_{\boldj a}$ is 
$\hat{n}_{\boldj a}=\sum_{s}\hat{n}_{\boldj a s}=
\sum_{s}\hat{c}^{\dagger}_{\boldj a s}\hat{c}_{\boldj a s}$, 
and $\hat{\bolds}_{\boldj a}$ is $\hat{\bolds}_{\boldj a}=\frac{1}{2}
\sum_{s,s^{\prime}}\hat{c}^{\dagger}_{\boldj a s} 
\boldsigma_{s s^{\prime}} \hat{c}_{\boldj a s^{\prime}}$ 
with the Pauli matrices $\boldsigma_{s s^{\prime}}$. 
Among the terms of $U_{abcd}^{s_{1}s_{2}s_{3}s_{4}}$, 
it is sufficient for a PM state to use 
$U_{abcd}^{\uparrow \downarrow}$, $U_{abcd}^{\uparrow \uparrow}$, and $U_{abcd}^{\pm}$, 
which are, respectively,
$U^{\uparrow\downarrow}_{abcd}
\equiv 
U_{abcd}^{\uparrow \uparrow \downarrow \downarrow}
=U_{abcd}^{\downarrow \downarrow \uparrow \uparrow}$,
$U^{\uparrow\uparrow}_{abcd}
\equiv 
U_{abcd}^{\uparrow \uparrow \uparrow \uparrow}
=U_{abcd}^{\downarrow \downarrow \downarrow \downarrow}$, 
and 
$U^{\pm}_{abcd}
\equiv 
U_{abcd}^{\uparrow \downarrow \uparrow \downarrow}
=U_{abcd}^{\downarrow \uparrow \downarrow \uparrow}$. 
In addition, 
in the absence of the spin-orbit interaction, 
it is more useful to introduce bare four-point vertex functions 
in spin and charge sector, 
$U_{abcd}^{\textrm{C}}$ and $U_{abcd}^{\textrm{S}}$, 
defined as 
\begin{align}
U_{abcd}^{s_{1}s_{2}s_{3}s_{4}}
=
\frac{1}{2}U_{abcd}^{\textrm{C}}\sigma_{s_{1} s_{2}}^{0}\sigma_{s_{4} s_{3}}^{0}
-\frac{1}{2}U_{abcd}^{\textrm{S}}\boldsigma_{s_{1} s_{2}}\cdot \boldsigma_{s_{4} s_{3}}.
\label{eq:bareSC}
\end{align} 
Namely, 
$U_{abcd}^{\textrm{C}}$ is $U_{abcd}^{\textrm{C}}
=U^{\uparrow\downarrow}_{abcd}+U^{\uparrow\uparrow}_{abcd}$, 
and 
$U_{abcd}^{\textrm{S}}$ is 
$U_{abcd}^{\textrm{S}}= U^{\uparrow\downarrow}_{abcd}-U^{\uparrow\uparrow}_{abcd}=-U^{\pm}_{abcd}$. 

I will explain how to treat the effects of $\hat{H}_{\textrm{int}}$ 
in Sec. II C, 
and how to choose the values of $U$, $U^{\prime}$, $J_{\textrm{H}}$, and $J^{\prime}$ 
in Sec. III. 

In the effective model,
I neglect the spin-orbit interaction of the Ru $t_{2g}$ orbitals 
for simplicity. 
This treatment may be sufficient to discuss the electronic properties 
analyzed in this paper 
since 
the coupling constant estimated 
in local-spin-density approximation~\cite{Oguchi-LS} for Sr$_{2}$RuO$_{4}$ 
is $0.167$ eV, 
which is smaller than the main terms of $\hat{H}_{0}$ and $\hat{H}_{\textrm{int}}$, 
and since  
its effects will not qualitatively change the results shown in Sect. III.  
(The main terms of $\hat{H}_{\textrm{int}}$ are of the order of magnitude $1$ eV, 
as described in Sec. III.) 
For several expected roles of the spin-orbit interaction, 
see the remaining issues in Sec. V. 

\subsection{Extended  \'{E}liashberg theory to charge transports of a multiorbital system}

In this section, 
we derive 
the dc longitudinal conductivity 
without an external magnetic field 
and the dc transverse conductivity 
in a weak-field limit 
in the linear-response theory with 
the most-divergent-term approximation. 
In Sec. II B 1, 
we derive the dc longitudinal conductivity to analyze 
the resistivity of a correlated multiorbital system. 
After deriving the exact expression 
in terms of the four-point vertex function or the three-point vector vertex function, 
we derive its approximate expression 
in the most-divergent-term approximation~\cite{Eliashberg-theory}, 
which is appropriate for the metallic systems 
with long-lived QPs at (at least) several momenta. 
We also explain four general properties seen from 
the derived expression of the conductivity 
and show the properties of the resistivity about the dominant excitations, 
the dependence on the QP lifetime, and the main effects of the CVCs. 
In Sec. II B 2, to analyze the Hall coefficient of a correlated multiorbital system 
for a weak magnetic field, 
we derive the dc transverse conductivity in the weak-field limit. 
Due to difficulty deriving the exact expression, 
I derive only the approximate expression 
in the most-divergent-term approximation~\cite{Fukuyama-RH,Kohno-Yamada}. 
In addition, 
after explaining four general properties of the derived conductivity, 
we deduce the properties of the Hall coefficient in the weak-field limit 
about the similar things for the resistivity. 

Before the formal derivations, 
I remark on the meanings of taking the $\omega$ limit and 
holding $\omega\tau_{\textrm{trans}} \ll 1$ in these derivations~\cite{NA-review} 
with $\tau_{\textrm{trans}}$, the transport relaxation time~\cite{Eliashberg-theory,Fukuyama-RH} 
(of the order of magnitude the QP damping). 
First, the $\omega$ limit, 
i.e. $\textstyle\lim_{\omega\rightarrow 0}\textstyle\lim_{\boldq\rightarrow \boldzero}$, 
is vital to obtain the observable currents 
since the dynamic and uniform field causes 
the observable currents; 
on the other hand, 
the $\boldq$ limit, 
i.e. $\textstyle\lim_{\boldq\rightarrow \boldzero}\textstyle\lim_{\omega\rightarrow 0}$, 
does not cause any observable currents 
as a result of the screening induced by 
the modulations of the charge distribution~\cite{Yamada-text}. 
Then, 
in taking $\textstyle\lim_{\omega\rightarrow 0}$, 
the QP lifetime should hold 
$\omega\tau_{\textrm{trans}} \ll 1$ 
since the inequality characterizes the relaxation process of transports;  
in $\omega\tau_{\textrm{trans}} \ll 1$, 
local equilibrium is realized due to the rapid relaxation 
compared with $\omega^{-1}$, a typical time scale of the field, 
and then the QPs near the Fermi level 
mainly govern the electronic transports. 

\subsubsection{Resistivity}
For discussions about the resistivity of a correlated multiorbital system, 
I use the Kubo formula~\cite{Kubo-formula} 
for the longitudinal conductivity, 
$\sigma_{\nu\nu}$,  
in the $\omega$ limit and $\omega\tau_{\textrm{trans}} \ll 1$, 
\begin{align}
\sigma_{\nu\nu}=&
\ 2
\lim\limits_{\omega\rightarrow 0}
\lim\limits_{\boldq\rightarrow \boldzero}
\dfrac{\tilde{K}_{\nu \nu}^{(\textrm{R})}(\boldq,\omega)
-\tilde{K}_{\nu \nu}^{(\textrm{R})}(\boldq,0)}{i\omega}\notag\\
=&
\ 2
\lim\limits_{\omega\rightarrow 0}
\dfrac{\tilde{K}_{\nu \nu}^{(\textrm{R})}(\boldzero,\omega)
-\tilde{K}_{\nu \nu}^{(\textrm{R})}(\boldzero,0)}{i\omega},\label{eq:sigmaxx}
\end{align}
where 
$\tilde{K}_{\nu \nu}^{(\textrm{R})}(\boldzero,\omega)$ is determined 
by $\tilde{K}_{\nu \nu}^{(\textrm{R})}(\boldzero,\omega)=
\tilde{K}_{\nu \nu}(i\Omega_{n}\rightarrow \omega+i0+)$ 
with $\tilde{K}_{\nu \nu}(i\Omega_{n})$, being
\begin{align}
&\tilde{K}_{\nu \nu}(i\Omega_{n})\notag\\
=&\
\lim\limits_{\boldq\rightarrow \boldzero}
\dfrac{1}{N}
\int^{T^{-1}}_{0}d\tau e^{i\Omega_{n}\tau}
\langle \textrm{T}_{\tau}  
\hat{J}_{\boldq \nu}(\tau)
\hat{J}_{-\boldq \nu}(0)\rangle\notag\\
=&\ 
\frac{1}{N}
\sum\limits_{\boldk,\boldk^{\prime}}
\sum\limits_{\{a\}}
\int^{T^{-1}}_{0}d\tau e^{i\Omega_{n}\tau}
(v_{\boldk \nu})_{ba}
(v_{\boldk^{\prime} \nu})_{cd}\notag\\
&\times
\langle \textrm{T}_{\tau}  
\hat{c}_{\boldk b}^{\dagger}(\tau)
\hat{c}_{\boldk a}(\tau)
\hat{c}_{\boldk^{\prime} c}^{\dagger}
\hat{c}_{\boldk^{\prime} d} \rangle\notag\\
=&
-\frac{T}{N}
\sum\limits_{\boldk}
\sum\limits_{m}
\sum\limits_{\{a\}}
(v_{\boldk \nu})_{ba}
(v_{\boldk \nu})_{cd}
G_{ac}(\boldk,i\epsilon_{m+n})
G_{db}(\boldk,i\epsilon_{m})\notag\\
&
-\frac{T^{2}}{N^{2}}
\sum\limits_{\boldk,\boldk^{\prime}}
\sum\limits_{m,m^{\prime}}
\sum\limits_{\{a\}}
\sum\limits_{\{A\}}
(v_{\boldk \nu})_{ba}
(v_{\boldk^{\prime} \nu})_{cd}G_{aA}(\boldk,i\epsilon_{m+n})\notag\\
&\times 
G_{dD}(\boldk^{\prime},i\epsilon_{m^{\prime}})G_{Bb}(\boldk,i\epsilon_{m})
G_{Cc}(\boldk^{\prime},i\epsilon_{m^{\prime}+n})\notag\\
&\times 
\Gamma_{\{A\}}(\boldk,i\epsilon_{m},\boldk^{\prime},i\epsilon_{m^{\prime}};\boldzero,i\Omega_{n}).
\label{eq:Ktild}
\end{align}
In Eq. (\ref{eq:Ktild}), 
$(v_{\boldk \nu})_{ab}$ is the group velocity, 
\begin{align}
(v_{\boldk \nu})_{ab}=\frac{\partial \epsilon_{ab}(\boldk)}{\partial k_{\nu}},
\end{align} 
and $\Gamma_{\{A\}}(\boldk,i\epsilon_{m},\boldk^{\prime},i\epsilon_{m^{\prime}};\boldzero,i\Omega_{n})$ is 
the reducible four-point vertex function, 
which is connected with the irreducible one 
through the Bethe-Salpeter equation~\cite{Nozieres},
\begin{align}
&\Gamma_{\{A\}}(\boldk,i\epsilon_{m},\boldk^{\prime},i\epsilon_{m^{\prime}};\boldzero,i\Omega_{n})\notag\\
=&\ 
\Gamma_{\{A\}}^{(1)}(\boldk,i\epsilon_{m},\boldk^{\prime},i\epsilon_{m^{\prime}};\boldzero,i\Omega_{n})
\notag\\
&+\frac{T}{N}
\sum\limits_{\boldk^{\prime\prime}}
\sum\limits_{m^{\prime\prime}}
\sum\limits_{\{A^{\prime}\}}
\Gamma_{ABC^{\prime}D^{\prime}}(\boldk,i\epsilon_{m},\boldk^{\prime\prime},i\epsilon_{m^{\prime\prime}};
\boldzero,i\Omega_{n})\notag\\
&\ \ \times 
G_{C^{\prime}A^{\prime}}(\boldk^{\prime\prime},i\epsilon_{m^{\prime\prime}+n})
G_{B^{\prime}D^{\prime}}(\boldk^{\prime\prime},i\epsilon_{m^{\prime\prime}})\notag\\
&\ \ \times 
\Gamma_{A^{\prime}B^{\prime}CD}^{(1)}(\boldk^{\prime\prime},i\epsilon_{m^{\prime\prime}},
\boldk^{\prime},i\epsilon_{m^{\prime}};\boldzero,i\Omega_{n}).
\end{align} 
The irreducible four-point vertex function can be determined 
in the way explained in Sec. II C. 

To obtain an exact expression of $\sigma_{\nu\nu}$, 
we carry out the analytic continuations~\cite{AGD,Eliashberg-theory} 
of the first and second terms of Eq. (\ref{eq:Ktild}) 
by using the analytic properties of the single-particle Green's function 
and the four-point vertex function. 
As we will carry out those analytic continuations in Appendix A, 
we obtain 
\begin{align}
\tilde{K}_{\nu \nu}^{(\textrm{R})}(\boldzero,\omega)&
=-\frac{1}{N}
\sum\limits_{\boldk,\boldk^{\prime}}
\sum\limits_{\{a\}}
(v_{\boldk \nu})_{ba}
(v_{\boldk^{\prime} \nu})_{cd}
\int^{\infty}_{-\infty}\dfrac{d\epsilon}{4\pi i}\notag\\
&\times 
\Bigl[
\tanh \dfrac{\epsilon}{2T}
K_{1;\{a\}}^{(\textrm{R})}(\boldk,\boldk^{\prime}; \epsilon; \omega)\notag\\
&\ \ 
+\Bigl(\tanh \dfrac{\epsilon+\omega}{2T}
-\tanh \dfrac{\epsilon}{2T}\Bigr)
K_{2;\{a\}}^{(\textrm{R})}(\boldk,\boldk^{\prime}; \epsilon; \omega)\notag\\
&\ \
-\tanh \dfrac{\epsilon+\omega}{2T}
K_{3;\{a\}}^{(\textrm{R})}(\boldk,\boldk^{\prime}; \epsilon; \omega)
\Bigr],\label{eq:sum-analytic}
\end{align}
with $K_{l;\{a\}}^{(\textrm{R})}(\boldk,\boldk^{\prime}; \epsilon; \omega)$, being 
\begin{align}
K_{l;\{a\}}^{(\textrm{R})}(\boldk,\boldk^{\prime}; \epsilon; \omega)
&=\
g_{l;acdb}(k;\omega)\delta_{\boldk,\boldk^{\prime}}\notag\\
&+
\frac{1}{N}\int^{\infty}_{-\infty}\frac{d\epsilon^{\prime}}{4\pi i}
\sum\limits_{\{A\}}
\sum\limits_{l^{\prime}=1}^{3}
g_{l;aABb}(k;\omega)\notag\\
& \ \ \ \times 
\mathcal{J}_{ll^{\prime};\{A\}}(k,k^{\prime};\omega)
g_{l^{\prime};CcdD}(k^{\prime};\omega).\label{eq:Kpart-analytic}
\end{align}
Here $\mathcal{J}_{ll^{\prime};\{A\}}(k,k^{\prime};\omega)$ 
is connected with 
the reducible four-point vertex function in real-frequency representation 
and is determined by the Bethe-Salpeter equation, 
\begin{align}
&\mathcal{J}_{ll^{\prime};\{A\}}(k,k^{\prime};\omega)
=\mathcal{J}^{(1)}_{ll^{\prime};\{A\}}(k,k^{\prime};\omega)\notag\\ 
& \ \ \ \ \
+\sum\limits_{l^{\prime\prime}= 1}^{3}
\dfrac{1}{N}
\sum\limits_{\boldk^{\prime\prime}}
\sum\limits_{\{A^{\prime}\}}
\int^{\infty}_{-\infty}\frac{d\epsilon^{\prime\prime}}{4\pi i}
\mathcal{J}_{ll^{\prime\prime};ABC^{\prime}D^{\prime}}(k,k^{\prime\prime};\omega)\notag\\
&\ \ \ \ \ \ \ \times
g_{l^{\prime\prime};C^{\prime}A^{\prime}B^{\prime}D^{\prime}}(k^{\prime\prime};\omega)
\mathcal{J}^{(1)}_{l^{\prime\prime}l^{\prime};A^{\prime}B^{\prime}CD}(k^{\prime\prime},k^{\prime};\omega),
\label{eq:4VC-red-irred}
\end{align} 
with the similar connection between $\mathcal{J}^{(1)}_{ll^{\prime};\{A\}}(k,k^{\prime};\omega)$ 
and the irreducible four-point vertex function in real-frequency representation. 

We also rewrite $\tilde{K}_{\nu \nu}^{(\textrm{R})}(\boldzero,\omega)$ 
in a more compact form 
by using the three-point vertex function in real-frequency representation, 
$\Lambda_{\nu;l;ab}(k;\omega)\equiv \Lambda_{\nu;l;ab}(\boldk,\epsilon+\omega,\boldk,\epsilon)$ 
(for the detail see Appendix B): 
\begin{align}
&\tilde{K}_{\nu \nu}^{(\textrm{R})}(\boldzero,\omega)\notag\\
=&
-\frac{1}{N}
\sum\limits_{\boldk}
\sum\limits_{\{a\}}
(v_{\boldk \nu})_{ba}
\int^{\infty}_{-\infty}\frac{d\epsilon}{4\pi i}\notag\\
&\times \Bigl[
\tanh \frac{\epsilon}{2T}
g_{1;acdb}(k;\omega)\Lambda_{\nu;1;cd}(k;\omega)\notag\\
&\ \ \ +\Bigl(\tanh \frac{\epsilon+\omega}{2T}
-\tanh \frac{\epsilon}{2T}\Bigr)
g_{2;acdb}(k;\omega)\Lambda_{\nu;2;cd}(k;\omega)\notag\\
&\ \ \ -\tanh \frac{\epsilon+\omega}{2T}
g_{3;acdb}(k;\omega)\Lambda_{\nu;3;cd}(k;\omega)\Bigr].\label{eq:sum-analytic2}
\end{align} 

Because of the difficulty solving the exact expression of $\sigma_{\nu\nu}$, 
we use the most-divergent-term approximation, introduced 
by \'{E}liashberg~\cite{Eliashberg-theory}, 
in order to derive an approximate expression. 
In this approximation~\cite{Eliashberg-theory}, 
we consider only the most divergent terms with respect to the QP lifetime 
in $\gamma_{\alpha}^{\ast}(\boldk_{\textrm{F}})/T\rightarrow 0$ 
with $\gamma_{\alpha}^{\ast}(\boldk_{\textrm{F}})$, 
the QP damping for band $\alpha$ at Fermi momentum $\boldk_{\textrm{F}}$. 
This approximation is based on the limiting properties~\cite{AGD,Nozieres} of 
the pairs of two single-particle Green's functions 
with external momentum and frequency, $\boldq$ and $\omega$, in $q\rightarrow 0$ 
and $\gamma_{\alpha}^{\ast}(\boldk_{\textrm{F}})/T\rightarrow 0$. 
More precisely, 
utilizing the limiting properties, 
we can use the approximation that 
among the pairs of two single-particle Green's functions, 
only a retarded-advanced pair gives the leading dependence on 
the QP damping and the external momentum and frequency. 
Namely, 
we can approximate the leading dependence of 
$g_{1;acdb}(k;\omega)$, $g_{2;acdb}(k;\omega)$, and $g_{3;acdb}(k;\omega)$ to~\cite{NA-review} 
\begin{align}
g_{1;acdb}(k;\omega)
\sim &
\sum\limits_{\alpha,\beta}
(U_{\boldk})_{a\alpha}(U_{\boldk}^{\dagger})_{\alpha c}
(U_{\boldk})_{d\beta}(U_{\boldk}^{\dagger})_{\beta b}\notag\\
&\times 
\dfrac{z_{\alpha}(\boldk)z_{\beta}(\boldk)}
{[\epsilon-\xi_{\alpha}^{\ast}(\boldk)+i0+]
[\epsilon-\xi_{\beta}^{\ast}(\boldk)+i0+]},\label{eq:g1-FL}\\
g_{2;acdb}(k;\omega)
\sim &\
2\pi i
\sum\limits_{\alpha,\beta}
(U_{\boldk})_{a\alpha}(U_{\boldk}^{\dagger})_{\alpha c}
(U_{\boldk})_{d\beta}(U_{\boldk}^{\dagger})_{\beta b}\notag\\
&\times 
\dfrac{z_{\alpha}(\boldk)z_{\beta}(\boldk)\delta(\epsilon-\xi_{\alpha}^{\ast}(\boldk))}
{\omega-\xi_{\alpha}^{\ast}(\boldk)+\xi_{\beta}^{\ast}(\boldk)
+i[\gamma_{\alpha}^{\ast}(\boldk)+\gamma_{\beta}^{\ast}(\boldk)]},\label{eq:g2-FL}
\end{align}
and
\begin{align}
g_{3;acdb}(k;\omega)
\sim &
\sum\limits_{\alpha,\beta}
(U_{\boldk})_{a\alpha}(U_{\boldk}^{\dagger})_{\alpha c}
(U_{\boldk})_{d\beta}(U_{\boldk}^{\dagger})_{\beta b}\notag\\
&\times 
\dfrac{z_{\alpha}(\boldk)z_{\beta}(\boldk)}
{[\epsilon-\xi_{\alpha}^{\ast}(\boldk)-i0+]
[\epsilon-\xi_{\beta}^{\ast}(\boldk)-i0+]},\label{eq:g3-FL}
\end{align}
respectively. 
Here $\xi_{\alpha}^{\ast}(\boldk)$ is the QP energy, 
$z_{\alpha}(\boldk)$ is the mass enhancement factor, 
and $(U_{\boldk})_{a\alpha}$ is the unitary matrix to obtain the QP dispersions. 
Since this treatment remains reasonable for $\gamma_{\alpha}^{\ast}(\boldk_{\textrm{F}})/T < 1$, 
the most-divergent-term approximation is not only exact in the FL 
but also appropriate in the correlated metallic systems 
having the long-lived QPs at least for several momenta. 

To derive an approximate expression of $\sigma_{\nu\nu}$ 
in the most-divergent-term approximation~\cite{Eliashberg-theory}, 
we introduce two quantities, $\mathcal{J}^{(0)}_{ll^{\prime};\{a\}}(k,k^{\prime};\omega)$ and 
$\Lambda_{\nu;l;ab}^{(0)}(k;\omega)$, 
which are irreducible only about a retarded-advanced pair, 
and rewrite $\tilde{K}_{\nu \nu}^{(\textrm{R})}(\boldzero,\omega)$ by using 
the two quantities. 
First, we define $\mathcal{J}^{(0)}_{ll^{\prime};\{a\}}(k,k^{\prime};\omega)$ and 
$\Lambda_{\nu;l;ab}^{(0)}(k;\omega)$ as
\begin{align}
&\mathcal{J}^{(0)}_{ll^{\prime};\{a\}}(k,k^{\prime};\omega)
= \mathcal{J}^{(1)}_{ll^{\prime};\{a\}}(k,k^{\prime};\omega)\notag\\
& \ \ \ \ \ 
+
\dfrac{1}{N}
\sum\limits_{\boldk^{\prime\prime}}
\sum\limits_{\{A\}}
\int^{\infty}_{-\infty}\frac{d\epsilon^{\prime\prime}}{4\pi i}
\sum\limits_{l^{\prime\prime}=1,3}
\mathcal{J}^{(0)}_{ll^{\prime\prime};abCD}(k,k^{\prime\prime};\omega)\notag\\
&\ \ \ \ \ \ \ \times
g_{l^{\prime\prime};CABD}(k^{\prime\prime};\omega)
\mathcal{J}^{(1)}_{l^{\prime\prime}l^{\prime};ABcd}(k^{\prime\prime},k^{\prime};\omega),\label{eq:4VC-0}
\end{align}
and
\begin{align}
&\Lambda_{\nu;l;ab}^{(0)}(k;\omega)
= 
(v_{\boldk \nu})_{ab}\notag\\
& \ \ \ \ +
\sum\limits_{\{A\}}
\sum\limits_{l^{\prime}=1,3}
\dfrac{1}{N}\sum\limits_{\boldk^{\prime}}
\int^{\infty}_{-\infty}\frac{d\epsilon^{\prime}}{4\pi i}
\mathcal{J}^{(0)}_{ll^{\prime};abCD}(k,k^{\prime};\omega)\notag\\
&\ \ \ \ \times 
g_{l^{\prime};CABD}(k^{\prime};\omega)
(v_{\boldk^{\prime} \nu})_{AB},\label{eq:3VC-0}
\end{align}
respectively. 
We can also connect $\Lambda_{\nu;l;ab}^{(0)}(k;\omega)$ with 
$\Lambda_{\nu;l;ab}(k;\omega)$ as follows: 
\begin{align}
&\Lambda_{\nu;l;ab}(k;\omega)
= 
\Lambda_{\nu;l;ab}^{(0)}(k;\omega)\notag\\
& \ \ \ \ +
\sum\limits_{\{A\}}
\dfrac{1}{N}
\sum\limits_{\boldk^{\prime}}
\int^{\infty}_{-\infty}\dfrac{d\epsilon^{\prime}}{4\pi i}
\mathcal{J}^{(0)}_{l2;abCD}(k,k^{\prime};\omega)\notag\\
&\ \ \ \ \times 
g_{2;CABD}(k^{\prime};\omega)
\Lambda_{\nu;2;AB}(k^{\prime};\omega).\label{eq:3VC-0and3VC}
\end{align}
Then, 
substituting Eq. (\ref{eq:3VC-0and3VC}) into Eq. (\ref{eq:sum-analytic2}) 
and using two equalities, 
\begin{align}
&-\frac{1}{N}
\sum\limits_{\boldk}
\sum\limits_{\{a\}}
\int^{\infty}_{-\infty}\frac{d\epsilon}{4\pi i}
(v_{\boldk \nu})_{ba}
\tanh \frac{\epsilon}{2T}
g_{1;acdb}(k;\omega)\notag\\
&\times 
\frac{1}{N}
\sum\limits_{\boldk^{\prime}}
\sum\limits_{\{A\}}
\int^{\infty}_{-\infty}\frac{d\epsilon^{\prime}}{4\pi i}
\mathcal{J}_{12;cdCD}^{(0)}(k,k^{\prime};\omega)
g_{2;CABD}(k^{\prime};\omega)\notag\\
&\times \Lambda_{\nu;2;AB}(k^{\prime};\omega)\notag\\
=&
-\frac{1}{N}
\sum\limits_{\boldk}
\sum\limits_{\{a\}}
\int^{\infty}_{-\infty}\frac{d\epsilon}{4\pi i}
\Bigl[\frac{1}{N}
\sum\limits_{\boldk^{\prime}}
\sum\limits_{\{A\}}
\int^{\infty}_{-\infty}\frac{d\epsilon^{\prime}}{4\pi i}\notag\\
&\times 
\mathcal{J}_{21;baBA}^{(0)}(\boldk,\epsilon,\boldk,\epsilon+\omega,
\boldk^{\prime},\epsilon^{\prime},\boldk^{\prime},\epsilon^{\prime}+\omega)\notag\\
&\times 
g_{1;BDCA}(\boldk^{\prime},\epsilon^{\prime},\boldk^{\prime},\epsilon^{\prime}+\omega)
(v_{\boldk^{\prime} \nu})_{DC}\Bigr]
\notag\\
&\times 
\Bigl(\tanh \frac{\epsilon+\omega}{2T}
-\tanh \frac{\epsilon}{2T}\Bigr)
g_{2;acdb}(k;\omega)\Lambda_{\nu;2;cd}(k;\omega),\label{eq:equality1}
\end{align}
and 
\begin{align}
&\frac{1}{N}
\sum\limits_{\boldk}
\sum\limits_{\{a\}}
\int^{\infty}_{-\infty}\frac{d\epsilon}{4\pi i}
(v_{\boldk \nu})_{ba}
\tanh \frac{\epsilon+\omega}{2T}
g_{3;acdb}(k;\omega)\notag\\
&\times 
\frac{1}{N}
\sum\limits_{\boldk^{\prime}}
\sum\limits_{\{A\}}
\int^{\infty}_{-\infty}\frac{d\epsilon^{\prime}}{4\pi i}
\mathcal{J}_{32;cdCD}^{(0)}(k,k^{\prime};\omega)
g_{2;CABD}(k^{\prime};\omega)\notag\\
&\times \Lambda_{\nu;2;AB}(k^{\prime};\omega)\notag\\
=&
-\frac{1}{N}
\sum\limits_{\boldk}
\sum\limits_{\{a\}}
\int^{\infty}_{-\infty}\frac{d\epsilon}{4\pi i}
\Bigl[
\frac{1}{N}
\sum\limits_{\boldk^{\prime}}
\sum\limits_{\{A\}}
\int^{\infty}_{-\infty}\frac{d\epsilon^{\prime}}{4\pi i}\notag\\
&\times 
\mathcal{J}_{23;baBA}^{(0)}(\boldk,\epsilon,\boldk,\epsilon+\omega,
\boldk^{\prime},\epsilon^{\prime},\boldk^{\prime},\epsilon^{\prime}+\omega)\notag\\
&\times 
g_{3;BDCA}(\boldk^{\prime},\epsilon^{\prime},\boldk^{\prime},\epsilon^{\prime}+\omega)
(v_{\boldk^{\prime} \nu})_{DC}\Bigr]
\notag\\
&\times 
\Bigl(\tanh \frac{\epsilon+\omega}{2T}
-\tanh \frac{\epsilon}{2T}\Bigr)
g_{2;acdb}(k;\omega)\Lambda_{\nu;2;cd}(k;\omega),\label{eq:equality2}
\end{align}
we can express $\tilde{K}_{\nu \nu}^{(\textrm{R})}(\boldzero,\omega)$ as two parts, 
the part excluding a retarded-advanced pair and the other part: 
\begin{align}
\tilde{K}_{\nu \nu}^{(\textrm{R})}(\boldzero,\omega)
=&
-\frac{1}{N}
\sum\limits_{\boldk}
\sum\limits_{\{a\}}
(v_{\boldk \nu})_{ba}
\int^{\infty}_{-\infty}\dfrac{d\epsilon}{4\pi i}\notag\\
&\times
\Bigl[\tanh \dfrac{\epsilon}{2T}
g_{1;acdb}(k;\omega)\Lambda_{\nu;1;cd}^{(0)}(k;\omega)\notag\\
&-\tanh \dfrac{\epsilon+\omega}{2T}
g_{3;acdb}(k;\omega)\Lambda_{\nu;3;cd}^{(0)}(k;\omega)\Bigr]\notag\\
& -\frac{1}{N}
\sum\limits_{\boldk}
\sum\limits_{\{a\}}
\int^{\infty}_{-\infty}\dfrac{d\epsilon}{4\pi i}
\Lambda_{\nu;2;ba}^{(0)}(\boldk,\epsilon,\boldk,\epsilon+\omega)\notag\\
&\times
\Bigl(\tanh \dfrac{\epsilon+\omega}{2T}
-\tanh \dfrac{\epsilon}{2T}\Bigr)
g_{2;acdb}(k;\omega)\notag\\
&\times
\Lambda_{\nu;2;cd}(k;\omega).\label{eq:sum-analytic3}
\end{align}
This expression remains exact at this stage.
In Eqs. (\ref{eq:equality1}) and (\ref{eq:equality2}), 
we have used Eqs. (\ref{eq:4VC-12}), (\ref{eq:4VC-21}), (\ref{eq:4VC-23}), and (\ref{eq:4VC-32}) 
and the exchange symmetry~\cite{Kohno-Yamada} of the four-point vertex function 
about its variables. 

Adopting the most-divergent term approximation to Eq. (\ref{eq:sum-analytic3}), 
extracting the $\omega$-linear term, 
and using (\ref{eq:sigmaxx}), 
we obtain an approximate expression of $\sigma_{\nu\nu}$, 
\begin{align}
\sigma_{\nu\nu}=&
\dfrac{2}{N}
\sum\limits_{\boldk}
\sum\limits_{\{a \}=1}^{3}
\int^{\infty}_{-\infty}\dfrac{d\epsilon}{2\pi}
\Bigl(-\dfrac{\partial f(\epsilon)}{\partial \epsilon}\Bigr)\notag\\
&\times
\Lambda_{\nu;2;ba}^{(0)}(k;0)
g_{2;acdb}(k;0)
\Lambda_{\nu;2;cd}(k;0).\label{eq:sigmaxx-approx}
\end{align} 
Here we can regard $\Lambda_{\nu;2;ba}^{(0)}(k;0)$ and $\Lambda_{\nu;2;cd}(k;0)$ 
as, respectively, 
the current including the CVC arising from the self-energy  
and the current including the CVCs arising from 
the self-energy and the irreducible four-point vertex function. 
This is because 
Eq. (\ref{eq:3VC-0}) for $l=2$ at $\omega=0$ becomes 
\begin{align}
\Lambda_{\nu;2;ab}^{(0)}(k;0)=
(v_{\boldk \nu})_{ab}
+
\dfrac{\partial \textrm{Re}\Sigma_{ab}^{(\textrm{A})}(k)}{\partial k_{\nu}},\label{eq:Lamb0}
\end{align}
as a result of a Ward identity~\cite{Nozieres}, and 
because 
Eq. (\ref{eq:3VC-0and3VC}) for $l=2$ at $\omega=0$ becomes 
\begin{align}
&\Lambda_{\nu;2;cd}(k;0)
=
\Lambda_{\nu;2;cd}^{(0)}(k;0)\notag\\
& \ \ \ \ 
+\frac{1}{N}
\sum\limits_{\boldk^{\prime}}
\sum\limits_{\{A\}}
\int^{\infty}_{-\infty}\frac{d\epsilon^{\prime}}{4\pi i}
\mathcal{J}_{22;cdCD}^{(1)}(k,k^{\prime};0)\notag\\
&\ \ \ \ \times 
g_{2;CABD}(k^{\prime};0)\Lambda_{\nu;2;AB}(k^{\prime};0),\label{eq:Lamb}
\end{align}
as a result of the disappearance of the second term of Eq. (\ref{eq:4VC-0}), 
the higher-order term~\cite{Eliashberg-theory} about $\omega\tau_{\textrm{trans}}$ than 
the first term of Eq. (\ref{eq:4VC-0}). 
Note that 
the second term of Eq. (\ref{eq:Lamb}) plays a similar role for 
the backflow correction~\cite{Nozieres} in the FL theory 
since that term connects the currents at $\boldk$ and $\boldk^{\prime}$. 

From Eq. (\ref{eq:sigmaxx-approx}), 
we see four general properties for the dc longitudinal conductivity 
of a correlated electron system. 
(The following arguments for $\sigma_{xx}$ are qualitatively the same even for $\sigma_{yy}$.) 
First, due to the factor $(-\frac{\partial f(\epsilon)}{\partial \epsilon})$ 
in Eq. (\ref{eq:sigmaxx-approx}), 
the main excitations arise from the QPs near the Fermi level.  
This property indicates the importance of the coherent part 
of the single-particle Green's function in discussing $\rho_{ab}$. 
Such importance holds even if its incoherent part evolves, 
as shown in dynamical-mean-field theory (DMFT)~\cite{DMFT-trans-QP} 
for a single-orbital Hubbard model on a square lattice 
in a PM metallic state near a Mott transition. 
Second, 
Eq. (\ref{eq:sigmaxx-approx}) with the approximate form of $g_{2;acdb}(k;0)$ 
shows that 
the intraband excitations become dominant compared with 
the interband excitations. 
This is because the intraband components of Eq. (\ref{eq:g2-FL}) (i.e., $\alpha=\beta$) 
give larger finite contributions to $\sigma_{xx}$ 
than the interband components (i.e., $\alpha\neq \beta$) 
due to the factor $-\xi_{\alpha}^{\ast}(\boldk)+\xi_{\beta}^{\ast}(\boldk)$ 
in the denominator of Eq. (\ref{eq:g2-FL}) for $\omega=0$. 
Third, 
combining Eqs. (\ref{eq:sigmaxx-approx}) and (\ref{eq:g2-FL}) 
with the above second general property, 
we find that 
$\sigma_{xx}$ is inversely proportional to the QP damping. 
Note that 
the dependence of $\sigma_{xx}$ on the QP damping 
can be determined by the dependence of $g_{2;acdb}(k;0)$ 
since $\Lambda_{x;2;ba}^{(0)}(k;0)$ and $\Lambda_{x;2;cd}(k;0)$ 
are independent of the QP damping~\cite{Eliashberg-theory}. 
Fourth, 
due to the CVCs in $\Lambda_{x;2;ba}^{(0)}(k;0)$ and $\Lambda_{x;2;cd}(k;0)$, 
$\sigma_{xx}$ is affected both by the CVC arising from the self-energy 
and by the CVCs arising from the self-energy and irreducible four-point vertex function, 
and the dominant effect arise from the magnitude changes of the currents. 
This property can be deduced 
from the following arguments: 
Since $\Lambda_{x;2;ba}^{(0)}(k;0)$ includes 
the $\Sigma$ CVC [see Eq. (\ref{eq:Lamb0})], 
its effect is the renormalization of the group velocity, 
resulting in a magnitude change of the current~\cite{Eliashberg-theory}. 
On the other hand, 
the effects of the CVCs in $\Lambda_{x;2;cd}(k;0)$ 
are not only a magnitude change of the current 
but also an angle change 
since the CVC arising from the irreducible four-point vertex function 
connects the currents at $\boldk$ and $\boldk^{\prime}$, 
which are not always parallel or antiparallel~\cite{Kon-CVC}. 
Those effects on $\sigma_{xx}$ 
can be described by 
\begin{align}
&\Lambda_{x;2;ba}^{(0)}(k;0)\Lambda_{x;2;cd}(k;0)\notag\\
=&|\Lambda_{2;ba}^{(0)}(k)|\cos \varphi_{ba}^{(0)}(k)
|\Lambda_{2;cd}(k)|\cos \varphi_{cd}(k)\notag\\
\sim &
|\Lambda_{2;ba}^{(0)}(k)|\cos \varphi_{ba}^{(0)}(k)
|\Lambda_{2;cd}(k)|
\cos \varphi_{cd}^{(0)}(k)[1-\tfrac{\Delta \varphi_{cd}(k)^{2}}{2}],
\label{eq:sigmaxx-leadingPhi}
\end{align} 
where 
$|\Lambda_{2;ab}^{(0)}(k)|$ and $|\Lambda_{2;ab}(k)|$ represent the magnitudes 
of $\Lambda_{x;2;ba}^{(0)}(k;0)$ and $\Lambda_{x;2;cd}(k;0)$, respectively, 
and 
$\varphi_{ab}^{(0)}(k)$ and  
$\varphi_{ab}(k)=\varphi_{ab}^{(0)}(k)+\Delta \varphi_{ab}(k)$ 
represent the angles. 
Thus, 
even for the CVCs in $\Lambda_{x;2;cd}(k;0)$, 
the magnitude change is dominant for $\sigma_{xx}$. 

From those properties, 
we can deduce the properties of the resistivity 
about the dominant excitations, 
the dependence on the QP lifetime, 
and the main effects of the CVCs. 
Since the resistivity is the inverse of the longitudinal conductivity, 
the dominant excitations and the main effects of the CVCs are the same for $\sigma_{xx}$, 
and the resistivity is inversely proportional to the QP lifetime 
(in the same way for the relaxation-time approximation~\cite{Ziman}). 

\subsubsection{Hall coefficient}
For discussions of the usual Hall effect of a correlated multiorbital system 
for a weak external magnetic field, 
we consider a uniform static external magnetic field along the $z$-direction, 
which is so weak that the cyclotron frequency, 
$\omega_{\textrm{c}}$, satisfies $\omega_{\textrm{c}}\tau_{\textrm{trans}}\ll 1$, 
and derive an approximate expression of the Hall coefficient 
in the weak-field limit on the basis of the linear-response theory 
in the most-divergent-term approximation~\cite{Fukuyama-RH,Kohno-Yamada}. 
In this derivation, 
we assume that 
the system has the mirror symmetries about the $xz$- and the $yz$-plane 
and the equivalence between the $x$- and the $y$-direction~\cite{Kohno-Yamada}; 
these are valid for some 214-type ruthenates without the RuO$_{6}$ distortions. 
Because of the mirror symmetries and 
the Onsager reciprocal theorem~\cite{Onsager-thm1,Onsager-thm2}, 
we can treat the Hall coefficient, 
which is generally a third-rank axial tensor~\cite{Konno}, 
as a scalar. 
In addition, 
because of the equivalence between the $x$- and the $y$-direction, 
the Hall coefficient in the linear-response theory~\cite{Kubo-formula} 
in the weak-field limit becomes
\begin{align}
R_{\textrm{H}}=&
\dfrac{1}{\sigma_{xx}\sigma_{yy}}
\lim\limits_{H\rightarrow 0}
\dfrac{\sigma_{xy}}{H}\notag\\
=&
\dfrac{1}{\sigma_{xx}^{2}}
\lim\limits_{H\rightarrow 0}
\dfrac{\sigma_{xy}}{H}.\label{eq:RH}
\end{align}
Since we had derived $\sigma_{xx}$ in Sect. II B 1, 
we need to calculate $\lim\limits_{H\rightarrow 0}\frac{\sigma_{xy}}{H}$ 
in the linear-response theory with 
the most-divergent-term approximation~\cite{Fukuyama-RH,Kohno-Yamada} in this section. 

To calculate $\lim\limits_{H\rightarrow 0}\frac{\sigma_{xy}}{H}$, 
we need to derive the $H$-linear terms of $\sigma_{xy}$. 
For that purpose, 
we use the vector potential, $\boldA$, instead of $H$ itself, 
and derive the $\boldq$-linear and $\boldA$-linear terms. 
Thus, 
the Kubo formula for $\lim\limits_{H\rightarrow 0}\frac{\sigma_{xy}}{H}$ 
becomes~\cite{Fukuyama-RH,Kohno-Yamada}
\begin{align}
\lim\limits_{H\rightarrow 0}
\dfrac{\sigma_{xy}}{H}
=&
2\lim\limits_{\omega\rightarrow 0}
\lim\limits_{\boldq\rightarrow 0}
\dfrac{1}{i(q_{x}A_{y}(\boldq)-q_{y}A_{x}(\boldq))e^{i\boldq\cdot \boldr}}\notag\\
&\times
\dfrac{\Phi_{xy}^{(\textrm{R})}(\boldq,\omega)-\Phi_{xy}^{(\textrm{R})}(\boldq,0)}{i\omega},
\label{eq:SigmaXY-limit}
\end{align}
where $\Phi_{xy}^{(\textrm{R})}(\boldq,\omega)$ is obtained by 
$\Phi_{xy}^{(\textrm{R})}(\boldq,\omega)=
\Phi_{xy}(\boldq,i\Omega_{n}\rightarrow \omega+i0+)$ with 
\begin{align}
\Phi_{xy}(\boldq,i\Omega_{n})
=&\dfrac{T}{N}
\int^{T^{-1}}_{0}\hspace{-10pt}d\tau
\int^{T^{-1}}_{0}\hspace{-10pt}d\tau^{\prime} 
e^{i\Omega_{n}(\tau-\tau^{\prime})}\notag\\
&\times 
\langle \textrm{T}_{\tau}  
\hat{J}_{x}^{H}(\boldq,\tau)
\hat{J}_{y}^{H}(\boldzero,\tau^{\prime})\rangle_{H}\notag\\
=&\sum\limits_{\nu}
K_{xy\nu}(\boldq,i\Omega_{n})A_{\nu}(\boldq).\label{eq:phi-K}
\end{align}
Here $\hat{\boldJ}^{\textrm{H}}(\boldzero)$ within the linear response becomes
\begin{align}
\hat{\boldJ}^{\textrm{H}}(\boldzero)
-\hat{\boldJ}(\boldzero)
=
-
\sum\limits_{\boldk}
\sum\limits_{a,b}
\boldA(\boldq)\cdot \boldnabla_{\boldk}(\boldv_{\boldk})_{ba}
\hat{c}^{\dagger}_{\boldk b} 
\hat{c}_{\boldk a},
\end{align}
and $K_{xy\nu}^{(\textrm{R})}(\boldq,\omega)$ is obtained by 
$K_{xy\nu}^{(\textrm{R})}(\boldq,\omega)
=K_{xy\nu}(\boldq,i\Omega_{n}\rightarrow \omega+i0+)$, 
where $K_{xy\nu}(\boldq,i\Omega_{n})$ is given by 
\begin{align}
K_{xy\nu}(\boldq,i\Omega_{n})
&=
-\delta_{\nu,y}
\frac{T}{N}
\int^{T^{-1}}_{0}\hspace{-3pt}d\tau \int^{T^{-1}}_{0}\hspace{-3pt}d\tau^{\prime} 
e^{i\Omega_{n}(\tau-\tau^{\prime})}\notag\\
&\times 
\langle \textrm{T}_{\tau}  
\hat{J}_{x}(\boldq,\tau)
\sum\limits_{\boldk}
\sum\limits_{a,b}
\frac{\partial (v_{\boldk y})_{ba}}{\partial k_{\nu}}
\hat{c}^{\dagger}_{\boldk b}(\tau^{\prime}) 
\hat{c}_{\boldk a}(\tau^{\prime}) \rangle \notag\\
&+\frac{T}{N}
\int^{T^{-1}}_{0}\hspace{-3pt}d\tau 
\int^{T^{-1}}_{0}\hspace{-3pt}d\tau^{\prime} 
\int^{T^{-1}}_{0}\hspace{-3pt}d\tau^{\prime\prime} 
e^{i\Omega_{n}(\tau-\tau^{\prime})}\notag\\
&\times 
\langle \textrm{T}_{\tau}  
\hat{J}_{x}(\boldq,\tau)
\hat{J}_{y}(\boldzero,\tau^{\prime})
\hat{J}_{\nu}(-\boldq,\tau^{\prime\prime})\rangle .\label{eq:phi-start}
\end{align} 
\begin{figure*}[tb]
\begin{center}
\includegraphics[width=150mm]{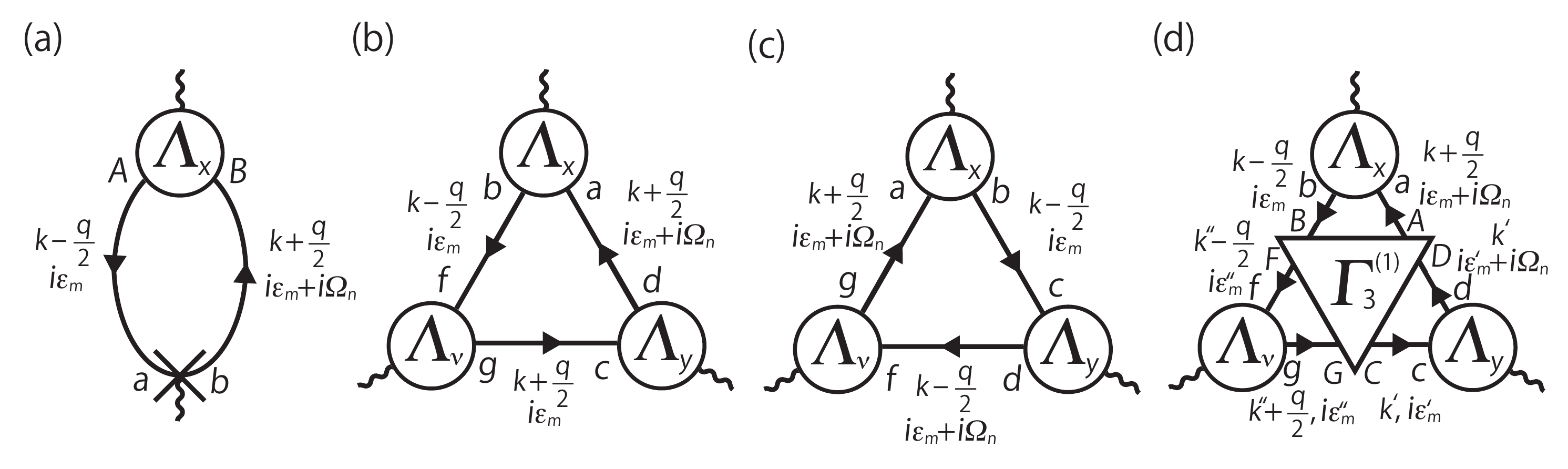}
\end{center}
\vspace{-20pt}
\caption{Diagrammatic representations of (a) the first, (b) the second, 
(c) the third, and (d) the fourth term of Eq. (\ref{eq:phi-next}). 
$\times$ in panel (a) represents the momentum derivative.  
}
\label{fig:Fig4}
\end{figure*}
Furthermore, 
using the three-point vector vertex function 
and introducing the irreducible six-point vertex function~\cite{Fukuyama-RH,Kohno-Yamada}, 
we can rewrite Eq. (\ref{eq:phi-start}) as
\begin{widetext}
\begin{align}
K_{xy\nu}(\boldq,i\Omega_{n})
=&\
\delta_{\nu,y}
\frac{T}{N}
\sum\limits_{\boldk}
\sum\limits_{m}
\sum\limits_{a,b,A,B}
\dfrac{\partial (v_{\boldk y})_{ba}}{\partial k_{\nu}}
G_{aA}(\boldk_{-},i\epsilon_{m})
\Lambda_{x;AB}(\boldk_{-},i\epsilon_{m},\boldk_{+},i\epsilon_{m+n})
G_{Bb}(\boldk_{+},i\epsilon_{m+n})\notag\\
&+\frac{T}{N}
\sum\limits_{\boldk}
\sum\limits_{m}
\sum\limits_{\{a\}}
\sum\limits_{f,g}
G_{fb}(\boldk_{-},i\epsilon_{m})
\Lambda_{x;ba}(\boldk_{-},i\epsilon_{m},\boldk_{+},i\epsilon_{m+n})
G_{ad}(\boldk_{+},i\epsilon_{m+n})\notag\\
&\ \ \times
\Lambda_{\nu;gf}(\boldk_{+},i\epsilon_{m},\boldk_{-},i\epsilon_{m})
G_{cg}(\boldk_{+},i\epsilon_{m})
\Lambda_{y;dc}(\boldk_{+},i\epsilon_{m+n},\boldk_{+},i\epsilon_{m})\notag\\
&+\frac{T}{N}
\sum\limits_{\boldk}
\sum\limits_{m}
\sum\limits_{\{a\}}
\sum\limits_{f,g}
G_{ag}(\boldk_{+},i\epsilon_{m+n})
\Lambda_{x;ba}(\boldk_{-},i\epsilon_{m},\boldk_{+},i\epsilon_{m+n})
G_{cb}(\boldk_{-},i\epsilon_{m})\notag\\
&\ \ \times
\Lambda_{\nu;gf}(\boldk_{+},i\epsilon_{m+n},\boldk_{-},i\epsilon_{m+n})
G_{fd}(\boldk_{-},i\epsilon_{m+n})
\Lambda_{y;dc}(\boldk_{-},i\epsilon_{m+n},\boldk_{-},i\epsilon_{m})\notag\\
&+\Bigl(\frac{T}{N}\Bigr)^{3}
\sum\limits_{\boldk,\boldk^{\prime},\boldk^{\prime\prime}}
\sum\limits_{m,m^{\prime},m^{\prime\prime}}
\sum\limits_{\{a\}}
\sum\limits_{\{A\}}
\sum\limits_{f,g,F,G}
G_{Bb}(\boldk_{-},i\epsilon_{m})
\Lambda_{x;ba}(\boldk_{-},i\epsilon_{m},\boldk_{+},i\epsilon_{m+n})
G_{aA}(\boldk_{+},i\epsilon_{m+n})\notag\\
&\ \ \times
G_{Gg}(\boldk^{\prime\prime}_{+},i\epsilon_{m^{\prime\prime}})
\Lambda_{\nu;gf}(\boldk^{\prime\prime}_{+},i\epsilon_{m^{\prime\prime}},
\boldk^{\prime\prime}_{-},i\epsilon_{m^{\prime\prime}})
G_{fF}(\boldk^{\prime\prime}_{-},i\epsilon_{m^{\prime\prime}})
G_{Dd}(\boldk^{\prime},i\epsilon_{m^{\prime}+n})
\Lambda_{y;dc}(\boldk^{\prime},i\epsilon_{m^{\prime}+n},\boldk^{\prime},i\epsilon_{m^{\prime}})
\notag\\
&\ \ \times 
G_{cC}(\boldk^{\prime},i\epsilon_{m^{\prime}}) \Gamma_{3; ABCDFG}^{(1)}
(\boldk_{+},i\epsilon_{m+n},\boldk_{-},i\epsilon_{m};
\boldk^{\prime},i\epsilon_{m^{\prime}},\boldk^{\prime},i\epsilon_{m^{\prime}+n};
\boldk^{\prime\prime}_{-},i\epsilon_{m^{\prime\prime}},\boldk^{\prime\prime}_{+},i\epsilon_{m^{\prime\prime}}),
\label{eq:phi-next}
\end{align}
\end{widetext}
with $\boldk_{\pm}\equiv \boldk\pm \frac{\boldq}{2}$.  
The terms of Eq. (\ref{eq:phi-next}) 
can be represented by the diagrams shown in Fig. \ref{fig:Fig4}. 
Thus, the remaining tasks 
are to derive the $\boldq$-linear terms of $K_{xy\nu}(\boldq,i\Omega_{n})$, 
to carry out its analytic continuation, 
and to combine the result with 
Eqs. (\ref{eq:SigmaXY-limit}) and (\ref{eq:phi-K}). 

\begin{figure*}[tb]
\begin{center}
\includegraphics[width=150mm]{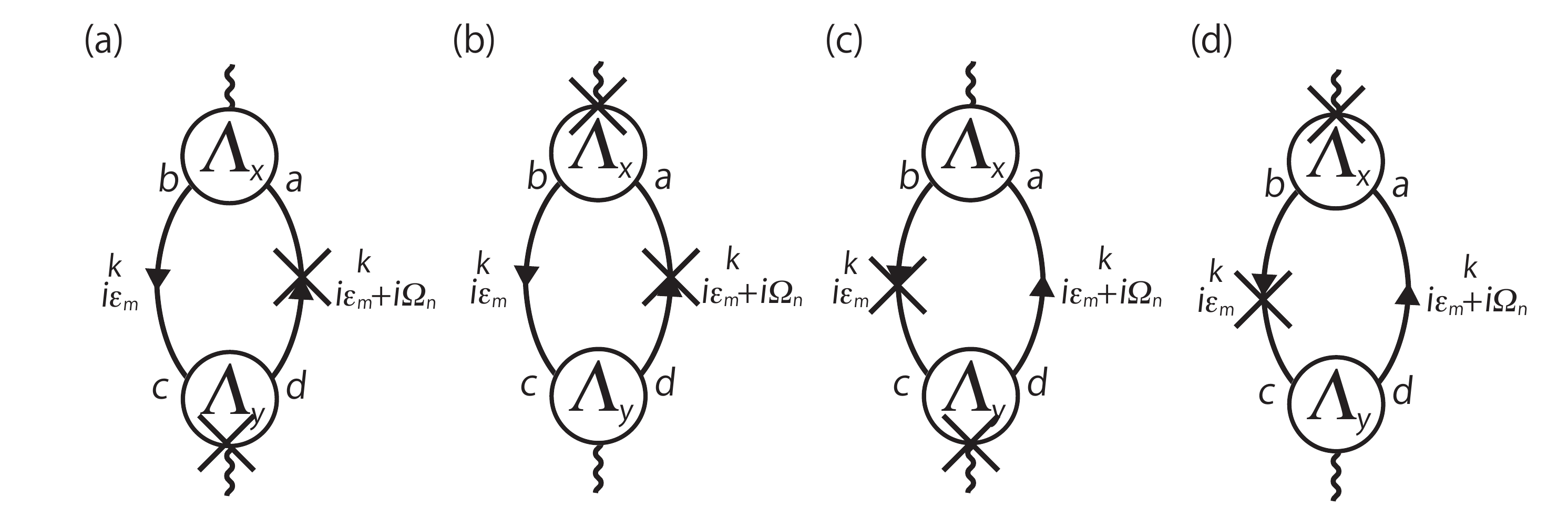}
\end{center}
\vspace{-20pt}
\caption{Diagrammatic representation of each term of Eq. (\ref{eq:phi-final}). 
The minus signs of the diagrams in panels (b) and (c) are not explicitly written. 
$\times$ represents the momentum derivative.  
}
\label{fig:Fig5}
\end{figure*}
We first derive the $\boldq$-linear terms of Eq. (\ref{eq:phi-next}) 
in the most-divergent-term approximation~\cite{Fukuyama-RH,Kohno-Yamada}. 
As I will explain in Appendix C in detail, 
the $\boldq$-linear terms are given by  
\begin{align}
&K_{xy\nu}(\boldq,i\Omega_{n})
=
\frac{1}{2}
(q_{x}\delta_{\nu,y}-q_{y}\delta_{\nu,x})
\frac{T}{N}
\sum\limits_{\boldk}
\sum\limits_{m}
\sum\limits_{\{a\}}\notag\\
&\times 
\Bigl[
\Lambda_{x;ba}(\boldk,i\epsilon_{m},\boldk,i\epsilon_{m+n})
\dfrac{\overleftrightarrow{\partial}}{\partial k_{y}}
\Lambda_{y;dc}(\boldk,i\epsilon_{m+n},\boldk,i\epsilon_{m})
\Bigr]\notag\\
&\times 
\Bigl[
G_{cb}(\boldk,i\epsilon_{m})
\dfrac{\overleftrightarrow{\partial}}{\partial k_{x}}
G_{ad}(\boldk,i\epsilon_{m+n})
\Bigr].\label{eq:phi-final}
\end{align} 
We can show the four terms of Eq. (\ref{eq:phi-final}) 
as the diagrams in Fig. \ref{fig:Fig5}. 

Then, 
we carry out the analytic continuation of Eq. (\ref{eq:phi-final}). 
This procedure can be done in the same way for $\sigma_{\nu\nu}$ in Sect. II B 1 
since the relevant parameter for analytic continuations about frequency 
is the frequency dependence 
and since the frequency dependence of Eq. (\ref{eq:phi-final}) 
is the same as $\tilde{K}_{\nu \nu}(i\Omega_{n})$ expressed 
in terms of the three-point vector vertex function, 
\begin{align}
\tilde{K}_{\nu \nu}(i\Omega_{n})
=&
-\frac{T}{N}
\sum\limits_{\boldk}
\sum\limits_{m}
\sum\limits_{\{a\}}
(v_{\boldk \nu})_{ba}
\Lambda_{\nu;cd}(\boldk,i\epsilon_{m};\boldzero,i\Omega_{n})\notag\\
&\times
G_{ac}(\boldk,i\epsilon_{m+n})
G_{db}(\boldk,i\epsilon_{m}).\label{eq:Ktild-3pVC}
\end{align}
Thus, we obtain 
$\Delta K_{xy\nu}^{(\textrm{R})}(\boldq,\omega)\equiv 
K_{xy\nu}^{(\textrm{R})}(\boldq,\omega)-K_{xy\nu}^{(\textrm{R})}(\boldq,0)$ 
in the most-divergent-term approximation~\cite{Fukuyama-RH,Kohno-Yamada} 
within the linear order of $\omega/T$: 
\begin{align}
\Delta K_{xy\nu}^{(\textrm{R})}(\boldq,\omega)
=&
-\frac{1}{2}
(q_{x}\delta_{\nu,y}-q_{y}\delta_{\nu,x})
\int^{\infty}_{-\infty}\frac{d\epsilon}{4\pi i}
2\omega  
\Bigl(-\dfrac{\partial f(\epsilon)}{\partial \epsilon}\Bigr)\notag\\
&\times 
\frac{1}{N}
\sum\limits_{\boldk}
\sum\limits_{\{a\}}
\Bigl[
\Lambda_{x;2;ba}(k;0)
\dfrac{\overleftrightarrow{\partial}}{\partial k_{y}}
\Lambda_{y;2;dc}(k;0)
\Bigr]\notag\\
&\times 
\Bigl[
G_{ad}^{(\textrm{R})}(k)
\dfrac{\overleftrightarrow{\partial}}{\partial k_{x}}
G_{cb}^{(\textrm{A})}(k)
\Bigr]. \label{eq:Phi-final}
\end{align} 
As described in Sect. II B 1, 
in the most-divergent-term approximation~\cite{Eliashberg-theory}, 
the contribution from a retarded-retarded or an advanced-advanced pair 
of two single-particle Green's functions 
is negligible compared with the contribution from a retarded-advanced pair. 

Combining Eq. (\ref{eq:Phi-final}) with Eqs. (\ref{eq:SigmaXY-limit}) and (\ref{eq:phi-K}), 
we finally obtain an approximate expression of the dc transverse conductivity 
in the weak-field limit within the most-divergent-term approximation:
\begin{align}
\lim\limits_{H\rightarrow 0}
\dfrac{\sigma_{xy}}{H}
=&\dfrac{1}{N}
\sum\limits_{\boldk}
\int^{\infty}_{-\infty}\frac{d\epsilon}{2\pi} 
\Bigl(-\dfrac{\partial f(\epsilon)}{\partial \epsilon}\Bigr)\notag\\
&\times
\sum\limits_{\{a\}}
\Bigl[
\Lambda_{x;2;ba}(k;0)
\dfrac{\overleftrightarrow{\partial}}{\partial k_{y}}
\Lambda_{y;2;dc}(k;0)
\Bigr]\notag\\
&\times
\textrm{Im}
\Bigl[
G_{ad}^{(\textrm{R})}(k)
\dfrac{\overleftrightarrow{\partial}}{\partial k_{x}}
G_{cb}^{(\textrm{A})}(k)
\Bigr]. \label{eq:sigmaXY-final}
\end{align}

Adopting the similar arguments for $\sigma_{xx}$ in Sect. II B 1 
to Eq. (\ref{eq:sigmaXY-final}), 
we see four general properties for $\lim\limits_{H\rightarrow 0}\frac{\sigma_{xy}}{H}$. 
First, 
the QPs near the Fermi level are dominant 
due to the factor $(-\frac{\partial f(\epsilon)}{\partial \epsilon})$. 
This is the same for $\sigma_{xx}$. 
Second, 
the dominance of the intraband excitations also holds 
because of the similar reason for $\sigma_{xx}$. 
Note that we can obtain the finite intraband components 
in $\lim\limits_{H\rightarrow 0}\frac{\sigma_{xy}}{H}$ 
since the quantities in the former square bracket in Eq. (\ref{eq:sigmaXY-final}) 
are odd about $k_{x}$ and even about $k_{y}$ 
due to the combination of the $k_{x}$ derivative in $\Lambda_{x;2;ba}(k;0)$, 
$\frac{\overleftrightarrow{\partial}}{\partial k_{y}}$, 
and the $k_{y}$ derivative in $\Lambda_{y;2;dc}(k;0)$, 
and since the quantities in the latter 
are odd about $k_{x}$ and even about $k_{y}$ 
due to the combination of $\frac{\overleftrightarrow{\partial}}{\partial k_{x}}$ 
and a product of the retarded and the advanced single-particle Green's function. 
Third, 
in contrast to $\sigma_{xx}$, 
$\lim\limits_{H\rightarrow 0}\frac{\sigma_{xy}}{H}$ is inversely proportional to 
the square of the QP damping. 
This is because the momentum derivative in a retarded-advanced pair 
leads to an additional factor of 
the inverse of the QP damping~\cite{Fukuyama-RH,Kohno-Yamada}. 
Fourth, 
the CVCs in $\Lambda_{x;2;ba}(k;0)$ and $\Lambda_{y;2;dc}(k;0)$ 
affect $\lim\limits_{H\rightarrow 0}\frac{\sigma_{xy}}{H}$, 
and the dominant effects are an angle change, which is 
different from the fourth property for $\sigma_{xx}$. 
This property arises from the dependence of the following quantity 
on the magnitude and angle changes of the currents: 
\begin{align}
&\Bigl[
\Lambda_{x;2;ba}(k;0)
\dfrac{\overleftrightarrow{\partial}}{\partial k_{y}}
\Lambda_{y;2;dc}(k;0)
\Bigr]\notag\\
=
&|\Lambda_{2;ba}(k)|\cos \varphi_{ba}(k)
|\Lambda_{2;dc}(k)|\cos \varphi_{dc}(k)
\dfrac{\partial \varphi_{dc}(k)}{\partial k_{y}}\notag\\
+&
|\Lambda_{2;ba}(k)|\sin \varphi_{ba}(k)
\dfrac{\partial \varphi_{ba}(k)}{\partial k_{y}}
|\Lambda_{2;dc}(k)|\sin \varphi_{dc}(k)\notag\\
\sim &
|\Lambda_{2;ba}(k)|\cos \varphi_{ba}^{(0)}(k)
|\Lambda_{2;dc}(k)|\cos \varphi_{dc}^{(0)}(k)
\dfrac{\partial \varphi_{dc}(k)}{\partial k_{y}}\notag\\
+&
|\Lambda_{2;ba}(k)|\sin \varphi_{ba}^{(0)}(k)
\dfrac{\partial \varphi_{ba}(k)}{\partial k_{y}}
|\Lambda_{2;dc}(k)|\sin \varphi_{dc}^{(0)}(k). \label{eq:sigmaxy-leadingPhi}
\end{align} 
Thus, due to the appearance of $\frac{\partial \varphi_{dc}(k)}{\partial k_{y}}$ 
or $\frac{\partial \varphi_{ba}(k)}{\partial k_{y}}$, 
the angle change of the current causes a more drastic effect on 
$\textstyle\lim_{H\rightarrow 0}\frac{\sigma_{xy}}{H}$ than $\sigma_{xx}$. 
Actually, 
the importance of such drastic effect has been obtained 
in a single-orbital Hubbard model on a square lattice~\cite{Kon-CVC}. 

Combining those properties with the four properties for $\sigma_{xx}$, 
we can deduce the properties of $R_{\textrm{H}}$ 
about the dominant excitations, 
the dependence on the QP lifetime, 
and the main effects of the CVCs. 
First, 
the dominant excitations are the intraband excitations near the Fermi level. 
Second, 
the dependence of the numerator and denominator of $R_{\textrm{H}}$ on the QP lifetime 
cancels each other out in the absence of the band dependence of the QP lifetime, 
while the cancellation is not perfect in the presence of the band dependence. 
This is because 
$\textstyle\lim_{H\rightarrow 0}\frac{\sigma_{xy}}{H}$ or $\sigma_{xx}$ 
consists of the sum of the corresponding intraband components, 
each of which has the dependence of the QP lifetime for the band. 
Note that 
the non-perfect cancellation is the origin of the temperature dependence 
of $R_{\textrm{H}}$ of a multiorbital system in the Fermi liquid. 
Third, 
the main effects of the CVCs on $R_{\textrm{H}}$ 
are the magnitude change of the current due to $\Lambda_{x;2;ba}^{(0)}(k;0)$ 
in the denominator of $R_{\textrm{H}}$ 
and the angle change of the current due to 
$\Lambda_{x;2;ba}(k;0)$ or $\Lambda_{y;2;dc}(k;0)$ in the numerator 
since there is the nearly perfect cancellation between 
the magnitude changes 
due to $\Lambda_{x;2;ba}(k;0)$ and $\Lambda_{y;2;dc}(k;0)$ in the numerator 
and due to the square of $\Lambda_{x;2;cd}(k;0)$ in the denominator. 

\subsection{FLEX approximation with the $\Sigma$ CVC, 
the MT CVC, and the AL CVC} 

In this section, 
after explaining several advantages 
of the FLEX approximation with the CVCs arising from 
the self-energy and irreducible four-point vertex function, 
I formulate the FLEX approximation in Matsubara-frequency representation 
for a multiorbital Hubbard model in a PM state 
and derive the CVCs arising from the irreducible four-point vertex function 
in the FLEX approximation. 
In the latter derivation, 
we first derive the irreducible four-point vertex function 
in Matsubara-frequency representation; 
second, 
we convert it into a real-frequency representation 
by using the analytic continuation; 
third, 
we calculate part of the kernel of the CVCs arising from 
the irreducible four-point vertex function; 
fourth, 
we derive the Bethe-Salpeter equation for the current including the CVCs. 
Furthermore, 
I introduce a simplified Bethe-Salpeter equation 
by approximating the AL CVC to its main terms. 

To describe the electronic properties near or away from a magnetic QCP, 
I use the FLEX approximation with the CVCs arising from 
the self-energy and irreducible four-point vertex function 
since its following three properties are the advantages 
in describing the electronic transports. 
One is that 
this approximation is 
one of the conserving approximations~\cite{FLEX1,FLEX2,FLEX3} 
that automatically satisfies conservation laws~\cite{Luttinger-Ward,Baym-Kadanoff}. 
This is powerful to describe transports 
since the treatment in keeping conservation laws is essential in transports~\cite{FLEX3}. 
Another advantage is that 
this approximation can take account of the many-body effects 
due to the self-energy itself 
and the CVCs arising from the self-energy and 
the irreducible four-point vertex function~\cite{Kon-review,NA-review}. 
In particular, 
this approximation can sufficiently treat the effects of 
spatial (i.e., momentum-dependent) correlation 
even near a magnetic QCP~\cite{Kon-review,NA-review,Yanase-review}. 
Due to this advantage, 
the FLEX approximation with the CVCs 
can analyze how those many-body effects influence 
the electronic properties 
beyond 
random-phase approximation (RPA), a mean-field-type approximation, 
and the relaxation-time approximation~\cite{Ziman}, 
where all the CVCs are neglected~\cite{Kon-review}, 
and improve several unrealistic results in the RPA; 
examples of the improvements are 
a reasonable value of $U$ for a magnetic transition 
and the Curie-Weiss-type temperature dependence of the spin susceptibility 
near an AF QCP~\cite{Yanase-review,NA-review}. 
(As described in Sect. II B, 
the CVCs are vital to satisfy conservation laws~\cite{Yamada-Yosida,Kon-review,NA-review}.) 
The other advantage is that
the FLEX approximation can sufficiently describe 
the coherent parts of the single-particle Green's function 
for a moderately strong electron correlation~\cite{FLEX3,Yanase-review,Kon-review,NA-review}. 
Actually, 
the FLEX approximation 
for a single-orbital Hubbard model on a square lattice 
at $U$ being a half of the bandwidth 
is in satisfactory agreement with 
the quantum Monte Carlo calculation 
about the imaginary-time dependence of the single-particle Green's function 
for several momenta~\cite{FLEX3}. 
Although it has been proposed in a diagrammatic Monte Carlo calculation~\cite{Georges-diag} 
for the same model that 
diagrammatic expansions based on the Luttinger-Ward functional~\cite{Luttinger-Ward} 
break down at a large $U$, 
I believe the above satisfactory agreement~\cite{FLEX3} remains valid 
since it has been shown~\cite{Eder} that 
this proposal results from an artifact of the technical pathological treatment 
of the noninteracting single-particle Green's function 
in the diagrammatic Monte Carlo calculation. 
This sufficient description of the coherent part 
is very useful to analyze the electronic dc transports  
since, as described in Sect. II B, the coherent parts 
almost dominate the electronic dc transports. 

We start to formulate the FLEX approximation for a multiorbital Hubbard model 
in a PM state in a similar way for 
Refs. \onlinecite{multi-FLEX1} and \onlinecite{multi-FLEX2}. 
A set of the equations in this approximation 
can be obtained by choosing the form of the Luttinger-Ward functional 
as the bubble and the ladder diagrams 
of the multiple electron-hole scattering and 
deriving the effective interaction and the Dyson equation. 
First, 
we can derive the effective interaction in the FLEX approximation 
by considering the bubble-type and the ladder-type multiple electron-hole scattering. 
Since we focus on a PM state, 
it is sufficient to consider 
the following three components: 
\begin{align}
V_{abcd}^{\uparrow \downarrow}(\boldq,i\Omega_{n})
=&\ \frac{1}{2}(U_{abcd}^{\textrm{S}}+U_{abcd}^{\textrm{C}})\notag\\
+&
\frac{1}{2}\sum\limits_{\{A\}}
U_{abAB}^{\textrm{S}}
\chi_{ABCD}^{\textrm{S}}(\boldq,i\Omega_{n})
U_{CDcd}^{\textrm{S}}\notag\\
-&
\frac{1}{2}\sum\limits_{\{A\}}
U_{abAB}^{\textrm{C}}
\chi_{ABCD}^{\textrm{C}}(\boldq,i\Omega_{n})
U_{CDcd}^{\textrm{C}},\label{eq:intFLEX-bubble-updown}\\
V_{abcd}^{\uparrow \uparrow}(\boldq,i\Omega_{n})
=&\ \frac{1}{2}(-U_{abcd}^{\textrm{S}}+U_{abcd}^{\textrm{C}})\notag\\
-&
\frac{1}{2}\sum\limits_{\{A\}}
U_{abAB}^{\textrm{S}}
\chi_{ABCD}^{\textrm{S}}(\boldq,i\Omega_{n})
U_{CDcd}^{\textrm{S}}\notag\\
-&
\frac{1}{2}\sum\limits_{\{A\}}
U_{abAB}^{\textrm{C}}
\chi_{ABCD}^{\textrm{C}}(\boldq,i\Omega_{n})
U_{CDcd}^{\textrm{C}},\label{eq:intFLEX-bubble-upup}
\end{align}
and 
\begin{align}
V_{abcd}^{\pm}(\boldq,i\Omega_{n})
=& -U_{abcd}^{\textrm{S}}\notag\\
-&\sum\limits_{\{A\}}
\sum\limits_{s^{\prime\prime}}
U_{abAB}^{\textrm{S}}
\chi_{ABCD}^{\textrm{S}}(\boldq,i\Omega_{n})
U_{CDcd}^{\textrm{S}},\label{eq:intFLEX-ladder}
\end{align}
with 
\begin{align}
\chi_{abcd}(\boldq,i\Omega_{n})
=&-
\frac{T}{N}
\sum\limits_{\boldk,m}
G_{ac}(\boldk+\boldq,i\epsilon_{m+n})
G_{db}(\boldk,i\epsilon_{m}),\label{eq:FLEX-1}\\
\chi_{abcd}^{\textrm{S}}(\boldq,i\Omega_{n})
&=\ 
\chi_{abcd}(\boldq,i\Omega_{n})\notag\\
&+\sum\limits_{\{A\}}
\chi_{abAB}(\boldq,i\Omega_{n})
U_{ABCD}^{\textrm{S}}
\chi_{CDcd}^{\textrm{S}}(\boldq,i\Omega_{n}),\label{eq:FLEX-2}
\end{align}
and 
\begin{align}
\chi_{abcd}^{\textrm{C}}(\boldq,i\Omega_{n})
&=\ 
\chi_{abcd}(\boldq,i\Omega_{n})\notag\\
&-\sum\limits_{\{A\}}
\chi_{abAB}(\boldq,i\Omega_{n})
U_{ABCD}^{\textrm{C}}
\chi_{CDcd}^{\textrm{C}}(\boldq,i\Omega_{n}).\label{eq:FLEX-3}
\end{align}
In deriving the effective interaction, 
we do not need to explicitly consider the ladder-type contributions in equal-spin-scattering case 
since those are included in part of Eq. (\ref{eq:intFLEX-bubble-upup}) 
as a result of the relation between 
the non-antisymmetrized and 
the antisymmetrized bare four-point vertex function~\cite{Hori-phD}. 
Combining the three components and 
using $\sigma_{s_{1}s_{2}}^{0}$, $\sigma_{s_{4}s_{3}}^{0}$, 
$\boldsigma_{s_{1} s_{2}}$, and $\boldsigma_{s_{4} s_{3}}$, 
we can express the effective interaction in the FLEX approximation 
as the following single equation: 
\begin{align}
&V_{abcd}^{s_{1}s_{2}s_{3}s_{4}}(\boldq,i\Omega_{n})\notag\\
=&\ 
\frac{1}{2}
\Bigl[
U_{abcd}^{\textrm{C}}
-\sum\limits_{\{A\}}
U_{abAB}^{\textrm{C}}\chi_{ABCD}^{\textrm{C}}(\boldq,i\Omega_{n})U_{CDcd}^{\textrm{C}}
\Bigr]\sigma_{s_{1} s_{2}}^{0}\sigma_{s_{4} s_{3}}^{0}\notag\\
-&\frac{1}{2}
\Bigl[
U_{abcd}^{\textrm{S}}
+\sum\limits_{\{A\}}
U_{abAB}^{\textrm{S}}\chi_{ABCD}^{\textrm{S}}(\boldq,i\Omega_{n})U_{CDcd}^{\textrm{S}}
\Bigr]\boldsigma_{s_{1} s_{2}}\cdot \boldsigma_{s_{4} s_{3}}.\label{eq:int-FLEX-total}
\end{align}
Then, using Eq. (\ref{eq:int-FLEX-total}) and excluding the double counting 
of the topologically equivalent term in the self-energy, 
we can derive the Dyson equation, 
\begin{align}
G_{ab}(\boldk,i\epsilon_{m})
&=\ G_{ab}^{0}(\boldk,i\epsilon_{m})\notag\\
&+\sum\limits_{A,B}
G_{aA}^{0}(\boldk,i\epsilon_{m})
\Sigma_{AB}(\boldk,i\epsilon_{m})
G_{Bb}(\boldk,i\epsilon_{m}),\label{eq:FLEX-4}
\end{align}
where $G_{ab}^{0}(\boldk,i\epsilon_{m})$ is the noninteracting single-particle Green's function, 
\begin{align}
G_{ab}^{0}(\boldk,i\epsilon_{m})=
\sum\limits_{\alpha}
(U_{\boldk}^{0})_{a\alpha}\dfrac{1}{i\epsilon_{m}-\epsilon_{\alpha}(\boldk)}
(U_{\boldk}^{0 \dagger})_{\alpha b},\label{eq:G0}
\end{align}  
and $\Sigma_{ab}(\boldk,i\epsilon_{m})$ 
is the self-energy in the FLEX approximation, 
\begin{align}
\Sigma_{ac}(\boldk,i\epsilon_{m})
=&\
\frac{T}{N}
\sum\limits_{\boldq,n}
\sum\limits_{b,d}
V_{abcd}(\boldq,i\Omega_{n})
G_{bd}(\boldk-\boldq,i\epsilon_{m-n}),\label{eq:FLEX-5}
\end{align}
with 
$(U_{\boldk}^{0})_{a\alpha}$, being the unitary matrix to diagonalize $\epsilon_{ab}(\boldk)$, 
and $V_{abcd}(\boldq,i\Omega_{n})$, being  
\begin{align}
&V_{abcd}(\boldq,i\Omega_{n})\notag\\
=&
-V_{abcd}^{\uparrow\uparrow\uparrow\uparrow}(\boldq,i\Omega_{n})
-V_{abcd}^{\uparrow\downarrow\uparrow\downarrow}(\boldq,i\Omega_{n})\notag\\
&-\sum\limits_{\{A\}}
U_{abAB}^{\textrm{S}}\chi_{ABCD}(\boldq,i\Omega_{n})U_{CDcd}^{\textrm{S}}\notag\\
=&\
\dfrac{3}{2}
\Bigl[
U_{abcd}^{\textrm{S}}
+\sum\limits_{\{A\}}
U_{abAB}^{\textrm{S}}
\chi_{ABCD}^{\textrm{S}}(\boldq,i\Omega_{n})
U_{CDcd}^{\textrm{S}}\Bigr]\notag\\
+&\dfrac{1}{2}
\Bigl[
-U_{abcd}^{\textrm{C}}
+\sum\limits_{\{A\}}
U_{abAB}^{\textrm{C}}
\chi_{ABCD}^{\textrm{C}}(\boldq,i\Omega_{n})
U_{CDcd}^{\textrm{C}}
\Bigr]\notag\\
-&\sum\limits_{\{A\}}
U_{aAbB}^{\uparrow\downarrow}
\chi_{ABCD}(\boldq,i\Omega_{n})
U_{CcDd}^{\uparrow\downarrow}.\label{eq:FLEX-6}
\end{align}
The reasons why the double counting term is the last term of Eq. (\ref{eq:FLEX-6}) 
are that 
the second-order terms in $V_{abcd}^{\uparrow\uparrow\uparrow\uparrow}(\boldq,i\Omega_{n})$ and 
$V_{abcd}^{\uparrow\downarrow\uparrow\downarrow}(\boldq,i\Omega_{n})$ lead to 
the topologically equivalent contributions to the self-energy, 
and that 
$V_{abcd}^{\uparrow\downarrow\uparrow\downarrow}(\boldq,i\Omega_{n})$ contains 
a relative $\frac{1}{2}$ factor arising from 
the coefficient $\sigma_{s_{1}s_{2}}^{+}\sigma_{s_{4}s_{3}}^{-}$ 
in $\boldsigma_{s_{1} s_{2}}\cdot \boldsigma_{s_{4} s_{3}}$. 
Solving a selfconsistent set of Eqs. (\ref{eq:FLEX-1}), (\ref{eq:FLEX-2}), 
(\ref{eq:FLEX-3}), (\ref{eq:FLEX-4}), (\ref{eq:FLEX-5}), and (\ref{eq:FLEX-6}) 
with Eq. (\ref{eq:G0}) and 
the equation to determine $\mu$, 
\begin{align}
n_{\textrm{e}}=& 
\dfrac{2}{N}
\sum\limits_{\boldk}
\sum\limits_{\alpha}
f(\epsilon_{\alpha}(\boldk))
+
\dfrac{2T}{N}
\sum\limits_{\boldk}
\sum\limits_{m}
\sum\limits_{a=1}^{3}
\Bigl[
G_{aa}(\boldk,i\epsilon_{m})\notag\\
&\ \ \ \ \ \ \ \ \ \ \ \ \ \ \ \ \ \ \ \ \ \ \ \ 
-G_{aa}^{0}(\boldk,i\epsilon_{m})
\Bigr],\label{eq:mu-int}
\end{align}
we can determine the single-particle or the two-particle quantities in the FLEX approximation. 
Its technical details for the numerical calculations will be described in Appendix D. 

We turn to the Bethe-Salpeter equation for the current 
with the CVCs in the FLEX approximation. 
The derivation consists of four steps. 
The four steps are 
to derive the irreducible four-point vertex function in the FLEX approximation 
in Matsubara-frequency representation, 
to convert it into in real-frequency representation 
by the analytic continuations, 
to calculate part of the kernel of the CVCs, 
and to combine the part and Eq. (\ref{eq:Lamb}). 

\begin{figure}[tb]
\begin{center}
\includegraphics[width=86mm]{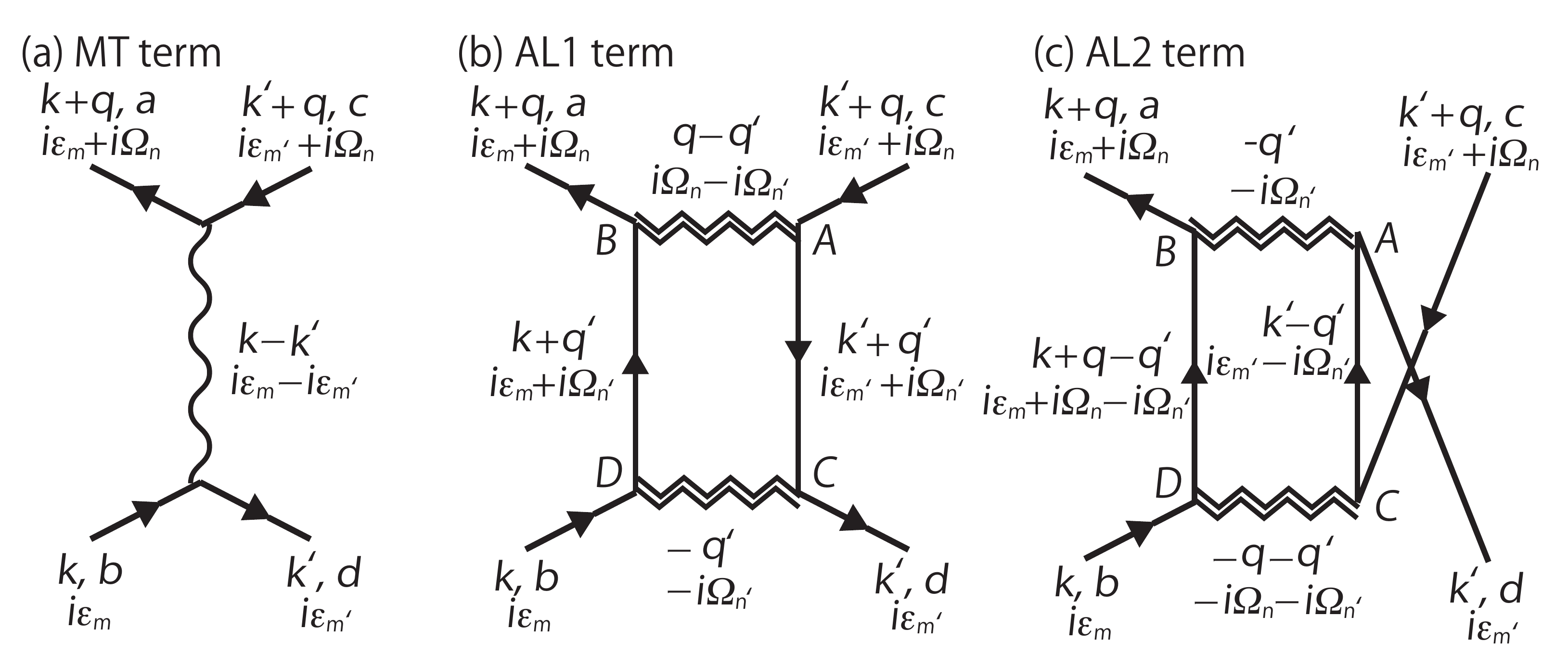}
\end{center}
\vspace{-20pt}
\caption{Diagrammatic representations of the irreducible four-point vertex functions 
in the FLEX approximation. 
Neglecting orbital indices and relabeling momentum and frequency variables, 
we can show that 
each diagram is equivalent to the corresponding diagram in Ref. \onlinecite{Kon-CVC}.  
}
\label{fig:Fig6}
\end{figure}

First, 
we derive the irreducible four-point vertex function 
in the FLEX approximation 
in Matsubara-frequency representation. 
Since the irreducible four-point vertex function is generally 
determined by~\cite{Baym-Kadanoff}
\begin{align}
\Gamma_{abcd}^{(1)}(\boldk,i\epsilon_{m},\boldk^{\prime},i\epsilon_{m^{\prime}}; \boldq,i\Omega_{n})
=\ 
\dfrac{\delta \Sigma_{ab}(\boldk,i\epsilon_{m})}
{\delta G_{cd}(\boldk^{\prime},i\epsilon_{m^{\prime}})},\label{eq:Gamma1-consv}
\end{align}
we adopt this equation to the self-energy in the FLEX approximation. 
For the actual calculation, 
we calculate the right hand side of Eq. (\ref{eq:Gamma1-consv}) 
at $\boldq=\boldzero$ and $\Omega_{n}=0$, 
and then we label $\boldq$ and $\Omega_{n}$ 
so as to represent the electron-hole scattering process 
among an electron of orbital $b$ with $(\boldk,i\epsilon_{m})$, 
a hole of orbital $d$ with $(\boldk^{\prime},i\epsilon_{m^{\prime}})$, 
an electron of orbital $a$ with $(\boldk+\boldq,i\epsilon_{m}+i\Omega_{n})$, 
and a hole of orbital $c$ with $(\boldk^{\prime}+\boldq,i\epsilon_{m^{\prime}}+i\Omega_{n})$. 
After the actual calculation explained in Appendix E, 
we obtain the irreducible four-point vertex function 
in Matsubara-frequency representation in the FLEX approximation: 
\begin{widetext}
\begin{align}
\Gamma_{abcd}^{(1)}(\boldk,i\epsilon_{m},\boldk^{\prime},i\epsilon_{m^{\prime}}; 
\boldq,i\Omega_{n})
=&\ \Gamma_{abcd}^{(1)\textrm{MT}}(\boldk,i\epsilon_{m},\boldk^{\prime},i\epsilon_{m^{\prime}}; 
\boldq,i\Omega_{n})+\Gamma_{abcd}^{(1)\textrm{AL}1}(\boldk,i\epsilon_{m},\boldk^{\prime},i\epsilon_{m^{\prime}}; 
\boldq,i\Omega_{n})\notag\\
&+\Gamma_{abcd}^{(1)\textrm{AL}2}(\boldk,i\epsilon_{m},\boldk^{\prime},i\epsilon_{m^{\prime}}; 
\boldq,i\Omega_{n}),\label{eq:Gamma-Matsu-sum}
\end{align}
with 
\begin{align}
\Gamma_{abcd}^{(1)\textrm{MT}}(\boldk,i\epsilon_{m},\boldk^{\prime},i\epsilon_{m^{\prime}}; 
\boldq,i\Omega_{n})
=& \ \delta_{\boldq,\boldzero}\delta_{n,0}
V_{acbd}(\boldk-\boldk^{\prime},i\epsilon_{m}-i\epsilon_{m^{\prime}}),\label{eq:MT-Matsu}\\
\Gamma_{abcd}^{(1)\textrm{AL}1}(\boldk,i\epsilon_{m},\boldk^{\prime},i\epsilon_{m^{\prime}}; 
\boldq,i\Omega_{n})
=& \
-\dfrac{T}{N}
\sum\limits_{\boldq^{\prime}}
\sum\limits_{n^{\prime}}
\sum\limits_{\{A\}}
W^{\textrm{AL}}_{aBcA;dCbD}(\boldq-\boldq^{\prime},i\Omega_{n-n^{\prime}};
-\boldq^{\prime},-i\Omega_{n^{\prime}})\notag\\
&\times 
G_{CA}(\boldk^{\prime}+\boldq^{\prime},i\epsilon_{m^{\prime}+n^{\prime}})
G_{BD}(\boldk+\boldq^{\prime},i\epsilon_{m+n^{\prime}}),\label{eq:AL1-Matsu}
\end{align}
and 
\begin{align}
\Gamma_{abcd}^{(1)\textrm{AL}2}(\boldk,i\epsilon_{m},\boldk^{\prime},i\epsilon_{m^{\prime}}; 
\boldq,i\Omega_{n})
=& \
-\dfrac{T}{N}
\sum\limits_{\boldq^{\prime}}
\sum\limits_{n^{\prime}}
\sum\limits_{\{A\}}
W^{\textrm{AL}}_{aBAd;CcbD}(-\boldq^{\prime},-i\Omega_{n^{\prime}};
-\boldq-\boldq^{\prime},-i\Omega_{n+n^{\prime}})\notag\\
&\times 
G_{AC}(\boldk^{\prime}-\boldq^{\prime},i\epsilon_{m^{\prime}-n^{\prime}})
G_{BD}(\boldk+\boldq+\boldq^{\prime},i\epsilon_{m+n+n^{\prime}}),\label{eq:AL2-Matsu}
\end{align}
\end{widetext}
where $W^{\textrm{AL}}_{abcd;ABCD}(\boldq_{1},i\Omega_{n_{1}};\boldq_{2},i\Omega_{n_{2}})$ is 
\begin{align}
&W^{\textrm{AL}}_{abcd;ABCD}(\boldq_{1},i\Omega_{n_{1}};\boldq_{2},i\Omega_{n_{2}})\notag\\
=&\dfrac{3}{2}
\tilde{N}_{abcd}^{\textrm{S}}(\boldq_{1},i\Omega_{n_{1}})
\tilde{N}_{ABCD}^{\textrm{S}}(\boldq_{2},i\Omega_{n_{2}})\notag\\
&+\dfrac{1}{2}
\tilde{N}_{abcd}^{\textrm{C}}(\boldq_{1},i\Omega_{n_{1}})
\tilde{N}_{ABCD}^{\textrm{C}}(\boldq_{2},i\Omega_{n_{2}})\notag\\
&-
U_{acbd}^{\uparrow\downarrow}
U_{ACBD}^{\uparrow\downarrow},\label{eq:W-Matsu}
\end{align}
with 
\begin{align}
\tilde{N}^{\textrm{S}}_{abcd}(\boldq^{\prime},i\Omega_{n^{\prime}})
=&\
U_{abcd}^{\textrm{S}}
+\sum\limits_{\{A\}}
U_{abCD}^{\textrm{S}}
\chi_{CDAB}^{\textrm{S}}(\boldq^{\prime},i\Omega_{n^{\prime}})
U_{ABcd}^{\textrm{S}},\label{eq:tildNS-Matsu}
\end{align}
and 
\begin{align}
\tilde{N}^{\textrm{C}}_{abcd}(\boldq^{\prime},i\Omega_{n^{\prime}})
=&\
U_{abcd}^{\textrm{C}}
-\sum\limits_{\{A\}}
U_{abCD}^{\textrm{C}}
\chi_{CDAB}^{\textrm{C}}(\boldq^{\prime},i\Omega_{n^{\prime}})
U_{ABcd}^{\textrm{C}}.\label{eq:tildNC-Matsu}
\end{align}
We can represent the terms of 
Eqs. (\ref{eq:MT-Matsu}), (\ref{eq:AL1-Matsu}), and (\ref{eq:AL2-Matsu}) 
as the diagrams of Figs. \ref{fig:Fig6}(a), \ref{fig:Fig6}(b), and \ref{fig:Fig6}(c),  
respectively. 

Second, 
we carry out the analytic continuations 
of Eqs. (\ref{eq:MT-Matsu}), (\ref{eq:AL1-Matsu}), and (\ref{eq:AL2-Matsu}) 
to convert these into real-frequency representation. 
This is because the irreducible four-point vertex functions 
in real-frequency representation 
are necessary to calculate part of the kernel of the CVCs, 
$\mathcal{J}_{22;cdCD}^{(1)}(k,k^{\prime};0)$ [see Eq. (\ref{eq:Lamb})]. 
Carrying out the analytic continuations, 
we obtain the MT, the AL1, and the AL2 terms 
for regions 22-II, 22-III, and 22-IV (see Appendix F). 

Third, 
using the MT, the AL1, and the AL2 terms 
in regions 22-II, 22-III, and 22-IV, 
we can calculate $\mathcal{J}_{22;cdCD}^{(1)}(k,k^{\prime};0)$ 
in the FLEX approximation. 
Since the irreducible four-point vertex function 
is the sum of the MT, the AL1, and the AL2 term, 
$\mathcal{J}_{22;cdCD}^{(1)}(k,k^{\prime};0)$ in the FLEX approximation 
is given by 
\begin{align}
\mathcal{J}_{22;abcd}^{(1)}(k,k^{\prime};0)=&
\mathcal{J}_{22;abcd}^{(1)\textrm{MT}}(k,k^{\prime};0)+\mathcal{J}_{22;abcd}^{(1)\textrm{AL}1}(k,k^{\prime};0)\notag\\
&+\mathcal{J}_{22;abcd}^{(1)\textrm{AL}2}(k,k^{\prime};0),\label{eq:4VC-tot} 
\end{align}
with 
\begin{align}
\mathcal{J}^{(1)\textrm{MT}}_{22;abcd}(k,k^{\prime};0)
=&\ 
F_{\textrm{ct}}^{\textrm{MT}}(\epsilon,\epsilon^{\prime};T)
2i \textrm{Im}V_{acbd}^{(\textrm{R})}(k-k^{\prime}),\label{eq:4VC-MT-real}\\
\mathcal{J}^{(1)\textrm{AL}1}_{22;abcd}(k,k^{\prime};0)
=&\
F_{\textrm{ct}}^{\textrm{AL}1}(\epsilon,\epsilon^{\prime};T) 
\Bigl(\frac{-i}{\pi}\Bigr) 
\frac{1}{N}
\sum\limits_{\boldq^{\prime}}
\sum\limits_{\{A \}}\notag\\
&\times
\int^{\infty}_{-\infty}d\omega^{\prime}
F_{\textrm{tt}}^{\textrm{AL}1}(\epsilon,\epsilon^{\prime},\omega^{\prime};T)\notag\\
&\times  
W^{\textrm{AL}(\textrm{RA})}_{aBcA;dCbD}(-q^{\prime};-q^{\prime})\notag\\
&\times
\textrm{Im}G_{CA}^{(\textrm{R})}(k^{\prime}+q^{\prime})
\textrm{Im}G_{BD}^{(\textrm{R})}(k+q^{\prime}),\label{eq:4VC-AL1-real}
\end{align}
and 
\begin{align}
\mathcal{J}^{(1)\textrm{AL}2}_{22;abcd}(k,k^{\prime};0)
=&\ 
F_{\textrm{ct}}^{\textrm{AL}2}(\epsilon,\epsilon^{\prime};T)
\Bigl(\frac{-i}{\pi}\Bigr) 
\frac{1}{N}
\sum\limits_{\boldq^{\prime}}
\sum\limits_{\{A \}}\notag\\
&\times
\int^{\infty}_{-\infty}d\omega^{\prime}
F_{\textrm{tt}}^{\textrm{AL}2}(\epsilon,\epsilon^{\prime},\omega^{\prime};T)\notag\\
&\times 
W^{\textrm{AL}(\textrm{RA})}_{aBAd;CcbD}(-q^{\prime};-q^{\prime})\notag\\
&\times
\textrm{Im}G_{AC}^{(\textrm{R})}(k^{\prime}-q^{\prime})
\textrm{Im}G_{BD}^{(\textrm{R})}(k+q^{\prime}),\label{eq:4VC-AL2-real}
\end{align}
where 
$F_{\textrm{ct}}^{\textrm{MT}}(\epsilon,\epsilon^{\prime};T)$, 
$F_{\textrm{ct}}^{\textrm{AL}1}(\epsilon,\epsilon^{\prime};T)$, 
$F_{\textrm{tt}}^{\textrm{AL}1}(\epsilon,\epsilon^{\prime},\omega^{\prime};T)$, 
$F_{\textrm{ct}}^{\textrm{AL}2}(\epsilon,\epsilon^{\prime};T)$, 
and 
$F_{\textrm{tt}}^{\textrm{AL}2}(\epsilon,\epsilon^{\prime},\omega^{\prime};T)$ are, 
respectively, 
\begin{align}
F_{\textrm{ct}}^{\textrm{MT}}(\epsilon,\epsilon^{\prime};T)
=&
\Bigl(\coth \frac{\epsilon-\epsilon^{\prime}}{2T}
+\tanh \frac{\epsilon^{\prime}}{2T}\Bigr),\label{eq:FctMT}\\
F_{\textrm{ct}}^{\textrm{AL}1}(\epsilon,\epsilon^{\prime};T)
=&
\Bigl(\coth \frac{\epsilon^{\prime}-\epsilon}{2T}
-\tanh \frac{\epsilon^{\prime}}{2T}\Bigr),\label{eq:FctAL1}\\ 
F_{\textrm{tt}}^{\textrm{AL}1}(\epsilon,\epsilon^{\prime},\omega^{\prime};T)
=&
\Bigl(\tanh \frac{\omega^{\prime}+\epsilon}{2T}
-\tanh \frac{\omega^{\prime}+\epsilon^{\prime}}{2T}\Bigr),\label{eq:FttAL1}\\
F_{\textrm{ct}}^{\textrm{AL}2}(\epsilon,\epsilon^{\prime};T)
=&
\Bigl(\coth \frac{\epsilon^{\prime}+\epsilon}{2T}
-\tanh \frac{\epsilon^{\prime}}{2T}\Bigr), \label{eq:FctAL2}
\end{align}
and 
\begin{align}
F_{\textrm{tt}}^{\textrm{AL}2}(\epsilon,\epsilon^{\prime},\omega^{\prime};T)
=
\Bigl(\tanh \frac{\omega^{\prime}+\epsilon}{2T}
-\tanh \frac{\omega^{\prime}-\epsilon^{\prime}}{2T}\Bigr).\label{eq:FttAL2}
\end{align}
In Eqs. (\ref{eq:4VC-MT-real}){--}(\ref{eq:4VC-AL2-real}),  
we have used the relations of the effective interaction 
and the single-particle Green's functions 
due to the time-reversal and the even-parity symmetry; 
in more general case, 
we should not use 
the relations such as 
$V_{acbd}^{(\textrm{R})}(k-k^{\prime})-V_{acbd}^{(\textrm{A})}(k-k^{\prime})
=2i \textrm{Im}V_{acbd}^{(\textrm{R})}(k-k^{\prime})$ 
and $G_{BD}^{(\textrm{R})}(k+q^{\prime})-G_{BD}^{(\textrm{A})}(k+q^{\prime})
=2i\textrm{Im}G_{BD}^{(\textrm{R})}(k+q^{\prime})$, 
and should retain the differences between the retarded and the advanced quantities. 

Fourth, 
substituting Eqs. (\ref{eq:4VC-tot}) with Eqs. (\ref{eq:4VC-MT-real}), 
(\ref{eq:4VC-AL1-real}), and (\ref{eq:4VC-AL2-real}) 
into Eq. (\ref{eq:Lamb}), 
we obtain the following Bethe-Salpeter equation 
with the CVCs in the FLEX approximation: 
\begin{align}
\Lambda_{\nu;2;cd}(k;0)
=&
\Lambda_{\nu;2;cd}^{(0)}(k;0)
+\Delta \Lambda_{\nu;2;cd}^{\textrm{MT}}(k;0)\notag\\
&+\Delta \Lambda_{\nu;2;cd}^{\textrm{AL}1}(k;0)
+\Delta \Lambda_{\nu;2;cd}^{\textrm{AL}2}(k;0),\label{eq:Lamb-FLEX-full}
\end{align}
where $\Delta \Lambda_{\nu;2;cd}^{\textrm{MT}}(k;0)$ is the MT CVC, 
\begin{align}
\Delta \Lambda_{\nu;2;cd}^{\textrm{MT}}(k;0)
=&\frac{1}{N}
\sum\limits_{\boldk^{\prime}}
\sum\limits_{\{A\}}
\int^{\infty}_{-\infty}\frac{d\epsilon^{\prime}}{2\pi}
F_{ct}^{\textrm{MT}}(\epsilon,\epsilon^{\prime};T)\notag\\
&\times 
\textrm{Im}V_{cCdD}^{(\textrm{R})}(k-k^{\prime})
\tilde{\Lambda}_{\nu;2;CD}(k^{\prime};0),\label{eq:Lamb-FLEX-MT}
\end{align}
$\Delta \Lambda_{\nu;2;cd}^{\textrm{AL}1}(k;0)$ is part of the AL CVC,
\begin{align}
\Delta \Lambda_{\nu;2;cd}^{\textrm{AL}1}(k;0)
=&
-
\frac{1}{4\pi^{2} N^{2}}
\sum\limits_{\boldk^{\prime},\boldq^{\prime}}
\sum\limits_{\{A\}}
\sum\limits_{\{A^{\prime\}}}
\int^{\infty}_{-\infty}d\epsilon^{\prime}
\int^{\infty}_{-\infty}d\omega^{\prime}\notag\\
&\times 
F_{\textrm{ct}}^{\textrm{AL}1}(\epsilon,\epsilon^{\prime};T) 
F_{\textrm{tt}}^{\textrm{AL}1}(\epsilon,\epsilon^{\prime},\omega^{\prime};T)\notag\\
&\times 
W^{\textrm{AL}(\textrm{RA})}_{cB^{\prime}CA^{\prime};DC^{\prime}dD^{\prime}}(-q^{\prime};-q^{\prime})\notag\\
&\times 
\textrm{Im}G_{C^{\prime}A^{\prime}}^{(\textrm{R})}(k^{\prime}+q^{\prime})
\textrm{Im}G_{B^{\prime}D^{\prime}}^{(\textrm{R})}(k+q^{\prime})\notag\\
&\times 
\tilde{\Lambda}_{\nu;2;CD}(k^{\prime};0),\label{eq:Lamb-FLEX-AL1}
\end{align}
and $\Delta \Lambda_{\nu;2;cd}^{\textrm{AL}2}(k;0)$ 
is the other part of the AL CVC, 
\begin{align}
\Delta \Lambda_{\nu;2;cd}^{\textrm{AL}2}(k;0)
=&
-
\frac{1}{4\pi^{2} N^{2}}
\sum\limits_{\boldk^{\prime},\boldq^{\prime}}
\sum\limits_{\{A\}}
\sum\limits_{\{A^{\prime\}}}
\int^{\infty}_{-\infty}d\epsilon^{\prime}
\int^{\infty}_{-\infty}d\omega^{\prime}\notag\\
&\times 
F_{\textrm{ct}}^{\textrm{AL}2}(\epsilon,\epsilon^{\prime};T)
F_{\textrm{tt}}^{\textrm{AL}2}(\epsilon,\epsilon^{\prime},\omega^{\prime};T)\notag\\
&\times 
W^{\textrm{AL}(\textrm{RA})}_{cB^{\prime}A^{\prime}D;C^{\prime}CdD^{\prime}}(-q^{\prime};-q^{\prime})\notag\\
&\times 
\textrm{Im}G_{A^{\prime}C^{\prime}}^{(\textrm{R})}(k^{\prime}-q^{\prime})
\textrm{Im}G_{B^{\prime}D^{\prime}}^{(\textrm{R})}(k+q^{\prime})\notag\\
&\times 
\tilde{\Lambda}_{\nu;2;CD}(k^{\prime};0),\label{eq:Lamb-FLEX-AL2}
\end{align}
with 
\begin{align}
\tilde{\Lambda}_{\nu;2;CD}(k^{\prime};0)
=&
\sum\limits_{A,B}
g_{2;CABD}(k^{\prime};0)\Lambda_{\nu;2;AB}(k^{\prime};0).\label{eq:tild-Lamb}
\end{align}
Equations (\ref{eq:Lamb-FLEX-MT}){--}(\ref{eq:Lamb-FLEX-AL2}) show that 
the MT and the AL CVC connect the currents at different momenta; 
for example, 
the MT CVC connects the current at $\boldk$ with the current at $\boldk^{\prime}$. 

In the actual numerical calculations, 
instead of the above Bethe-Salpeter equation, 
I use the simplified Bethe-Salpeter equation where 
the AL CVC is simplified by only its main terms. 
The main terms of the AL CVC can be determined 
by using the following two properties satisfied in the present model: 
The terms arising from $U$ are dominant compared with 
the terms arising from the other interactions 
in a realistic parameter set (i.e., $U>U^{\prime}$, $U>J_{\textrm{H}}$, and $U>J^{\prime}$); 
the intraorbital components of the current are larger than
the interorbital ones due to the large intraorbital hopping integrals compared with 
the interorbital hopping integrals (i.e., the larger intraorbital components 
of the group velocity). 
Namely, 
the main terms of the AL CVC are given by the sum of the following two quantities: 
\begin{align}
\Delta \Lambda_{\nu;2;cd}^{\textrm{AL}1}(k;0)
=&
-\delta_{c,d}
\frac{1}{4\pi^{2} N^{2}}
\sum\limits_{\boldk^{\prime},\boldq^{\prime}}
\int^{\infty}_{-\infty}d\epsilon^{\prime}
\int^{\infty}_{-\infty}d\omega^{\prime}\notag\\
&\times 
F_{\textrm{ct}}^{\textrm{AL}1}(\epsilon,\epsilon^{\prime};T) 
F_{\textrm{tt}}^{\textrm{AL}1}(\epsilon,\epsilon^{\prime},\omega^{\prime};T)\notag\\
&\times 
W^{\textrm{AL}(\textrm{RA})}_{c}(-q^{\prime};-q^{\prime})\notag\\
&\times 
\textrm{Im}G_{cc}^{(\textrm{R})}(k^{\prime}+q^{\prime})
\textrm{Im}G_{cc}^{(\textrm{R})}(k+q^{\prime})\notag\\
&\times 
\tilde{\Lambda}_{\nu;2;cc}(k^{\prime};0),\label{eq:Lamb-FLEX-AL1-simple}
\end{align}
and 
\begin{align}
\Delta \Lambda_{\nu;2;cd}^{\textrm{AL}2}(k;0)
=&
-\delta_{c,d}
\frac{1}{4\pi^{2} N^{2}}
\sum\limits_{\boldk^{\prime},\boldq^{\prime}}
\int^{\infty}_{-\infty}d\epsilon^{\prime}
\int^{\infty}_{-\infty}d\omega^{\prime}\notag\\
&\times 
F_{\textrm{ct}}^{\textrm{AL}2}(\epsilon,\epsilon^{\prime};T)
F_{\textrm{tt}}^{\textrm{AL}2}(\epsilon,\epsilon^{\prime},\omega^{\prime};T)\notag\\
&\times 
W^{\textrm{AL}(\textrm{RA})}_{c}(-q^{\prime};-q^{\prime})\notag\\
&\times 
\textrm{Im}G_{cc}^{(\textrm{R})}(k^{\prime}-q^{\prime})
\textrm{Im}G_{cc}^{(\textrm{R})}(k+q^{\prime})\notag\\
&\times 
\tilde{\Lambda}_{\nu;2;cc}(k^{\prime};0),\label{eq:Lamb-FLEX-AL2-simple}
\end{align}
where $W^{\textrm{AL}(\textrm{RA})}_{c}(-q^{\prime};-q^{\prime})$ is given by
\begin{align}
&W^{\textrm{AL}(\textrm{RA})}_{c}(-q^{\prime};-q^{\prime})
=
\dfrac{3}{2}
\tilde{N}_{cccc}^{\textrm{S}(\textrm{R})}(-q^{\prime})
\tilde{N}_{cccc}^{\textrm{S}(\textrm{A})}(-q^{\prime})\notag\\
&+\dfrac{1}{2}
\tilde{N}_{cccc}^{\textrm{C}(\textrm{R})}(-q^{\prime})
\tilde{N}_{cccc}^{\textrm{C}(\textrm{A})}(-q^{\prime})-U^{2},\label{eq:W-AL-simple}
\end{align}
with 
\begin{align}
\tilde{N}_{cccc}^{\textrm{S}(\textrm{R})}(-q^{\prime})
=&\ U+U^{2}\chi_{cccc}^{\textrm{S}(\textrm{R})}(-q^{\prime}),\label{eq:NS-AL-simple}
\end{align}
and 
\begin{align}
\tilde{N}_{cccc}^{\textrm{C}(\textrm{R})}(-q^{\prime})
=&\ U-U^{2}\chi_{cccc}^{\textrm{C}(\textrm{R})}(-q^{\prime}).\label{eq:NC-AL-simple}
\end{align}
More precisely, 
by using the former of the above two properties 
(corresponding to considering only the terms arising from $U$), 
we can replace 
$W^{\textrm{AL}(\textrm{RA})}_{cB^{\prime}CA^{\prime};DC^{\prime}dD^{\prime}}(-q^{\prime};-q^{\prime})$ 
of the AL$1$ term 
and $W^{\textrm{AL}(\textrm{RA})}_{cB^{\prime}A^{\prime}D;C^{\prime}CdD^{\prime}}(-q^{\prime};-q^{\prime})$ 
of the AL$2$ term by, respectively,  
$\delta_{B^{\prime},c}\delta_{C,c}\delta_{A^{\prime},c}
\delta_{D,d}\delta_{C^{\prime},d}\delta_{D^{\prime},d}
W^{\textrm{AL}(\textrm{RA})}_{cccc;dddd}(-q^{\prime};-q^{\prime})$ and 
$\delta_{B^{\prime},c}\delta_{A^{\prime},c}\delta_{D,c}
\delta_{C^{\prime},d}\delta_{C,d}\delta_{D^{\prime},d}
W^{\textrm{AL}(\textrm{RA})}_{cccc;dddd}(-q^{\prime};-q^{\prime})$; 
furthermore, 
using the latter property, 
we obtain Eqs. (\ref{eq:Lamb-FLEX-AL1-simple}) and (\ref{eq:Lamb-FLEX-AL2-simple}). 
Solving Eqs. (\ref{eq:Lamb-FLEX-full}), (\ref{eq:Lamb-FLEX-MT}), 
(\ref{eq:Lamb-FLEX-AL1-simple}), and (\ref{eq:Lamb-FLEX-AL2-simple}) 
with Eq. (\ref{eq:Lamb0}) self-consistently, 
we can determine the current including the CVCs arising from 
the self-energy and irreducible four-point vertex function 
in the FLEX approximation. 
I will describe the technical remarks to numerically solve those equations in Appendix G. 

\section{Results}
In this section, 
I show the results of the magnetic properties, 
the electronic structure, and the transport properties 
for a PM state of the multiorbital Hubbard model 
away from or near the AF QCP. 
In Sec. III A, 
I present the results of the magnetic properties in the FLEX approximation. 
From those results, 
we discuss the dominant fluctuations, 
the static and the dynamic properties of the spin susceptibility, 
the role of each $t_{2g}$ orbital, 
and the effects of the spin fluctuations 
on the imaginary part of the retarded effective interaction. 
In Sec. III B, 
to discuss the effects of the self-energy on the electronic structure, 
I show the results of the FS, the mass enhancement factor, 
the unrenormalized QP damping, and the QP damping in the FLEX approximation. 
In Sec. III C, 
we discuss the main effects of the AL CVC 
on the inplane resistivity, $\rho_{ab}$, 
and the Hall coefficient in the weak-field limit, $R_{\textrm{H}}$, 
in the FLEX approximation 
with the $\Sigma$ CVC, the MT CVC, and 
the main terms of the AL CVC 
and more simplified three cases. 
In addition to the temperature dependence of those transport coefficients, 
I show the orbital depedences of 
$\sigma_{xx}$ and $\lim\limits_{H\rightarrow 0}\frac{\sigma_{xy}}{H}$ 
in order to determine the role of each $t_{2g}$ orbital. 

I obtained the results of this section by the numerical calculations 
using the techniques explained in Appendixes D and G 
and converting the quantities obtained in the FLEX approximation 
in Matsubara-frequency representation 
to the corresponding quantities in real-frequency representation 
by the Pad\'{e} approximation~\cite{Pade-approx,Kusunose-Pade}. 
In the numerical calculations, 
I used $N=64\times 64$ meshes and $M=1024$ Matsubara frequencies, 
set  
$\Delta \epsilon_{j}=\Delta \epsilon_{j^{\prime}}^{\prime}
=\Delta \omega_{j^{\prime\prime}}^{\prime}=0.0025$ eV, 
$\epsilon_{\textrm{c}}=0.2$ eV, 
$J^{\prime}=J_{\textrm{H}}$, $U^{\prime}=U-2J_{\textrm{H}}$, 
and chose $J_{\textrm{H}}$, $U$, and $T$ as parameters.  
($\Delta \epsilon_{j}$, $\Delta \epsilon_{j^{\prime}}^{\prime}$, and 
$\Delta \omega_{j^{\prime\prime}}^{\prime}$ are 
the intervals of the discretized real-frequency integrals, 
and $\epsilon_{\textrm{c}}$ is 
the cut-off frequency in the discretized real-frequency integrals.) 
The parameters of $J_{\textrm{H}}$, $U$, and $T$ were chosen as follows: 
I put $J_{\textrm{H}}=\frac{U}{6}$ 
except the analysis of the dominant fluctuation, 
in which the value of $J_{\textrm{H}}$ was chosen 
in the range of $0\leq J_{\textrm{H}}\leq \frac{U}{5}$; 
I considered the case of $U=1.8$ or $2.1$ eV as, respectively, 
case~\cite{NA-review,NA-CVC} away from or near the AF QCP 
except the results in noninteracting case; 
I considered several values of $T$ 
in the range of $0.006 \textrm{eV}\leq T\leq 0.03 \textrm{eV}$. 
In addition, 
in the conversion by the Pad\'{e} approximation, 
I numerically solved its recursive procedure~\cite{Pade-approx,Kusunose-Pade}  
using the quantities at the lowest four Matsubara frequencies; 
for example, 
we obtained $\Sigma_{ab}^{(\textrm{R})}(\boldk,\epsilon)$ 
by adopting that recursive procedure to 
a set of $\Sigma_{ab}(\boldk,i\epsilon_{m})$ at $m=0,1,2,3$ in the FLEX approximation. 
Note that 
the advanced quantities are obtained by using the relations 
such as $\Sigma_{ab}^{(\textrm{A})}(\boldk,\epsilon)=\Sigma_{ab}^{(\textrm{R})}(\boldk,\epsilon)^{\ast}$ 
due to the time-reversal and the even-parity symmetries. 

\begin{figure*}[tb]
\begin{center}
\includegraphics[width=178mm]{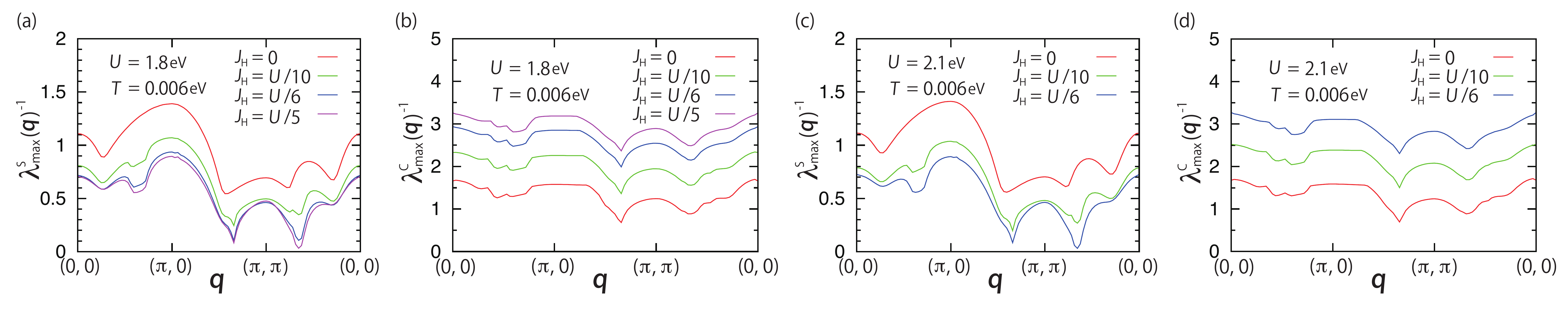}
\end{center}
\vspace{-20pt}
\caption{ 
Momentum dependence of $\lambda^{\textrm{S}}_{\textrm{max}}(\boldq)^{-1}$ 
and $\lambda^{\textrm{C}}_{\textrm{max}}(\boldq)^{-1}$ 
at $T=0.006$ eV and $U=1.8$ and $2.1$ eV for several values of $J_{\textrm{H}}$. 
}
\label{fig:Fig8}
\end{figure*}
\begin{figure*}[tb]
\begin{center}
\includegraphics[width=160mm]{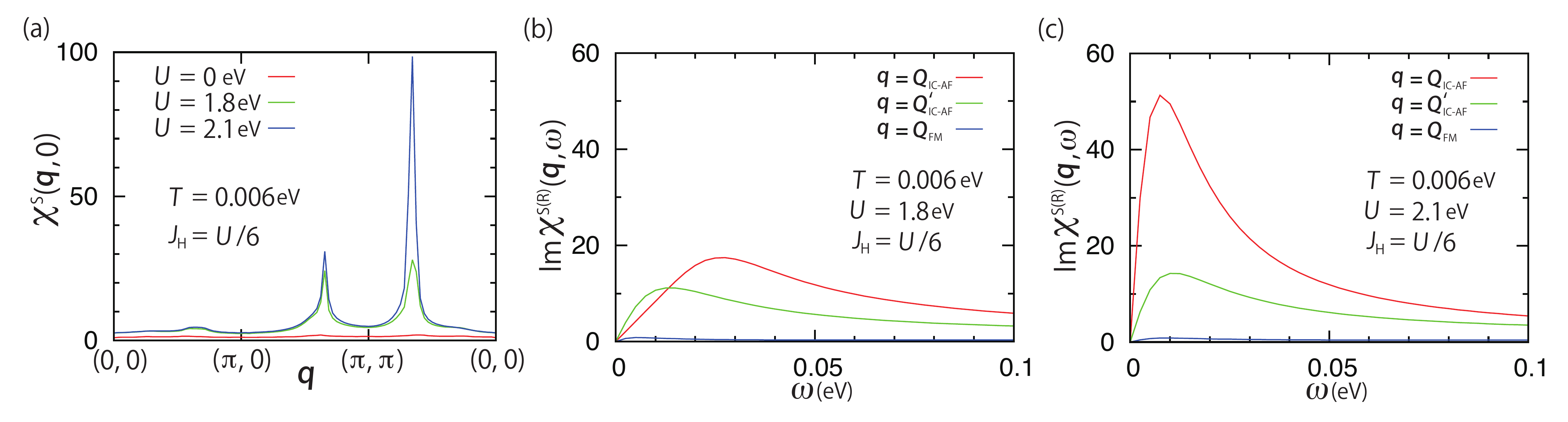}
\end{center}
\vspace{-20pt}
\caption{
(a) Momentum dependence of 
$\chi^{\textrm{S}}(\boldq,0)=\chi^{\textrm{S}(\textrm{R})}(\boldq,0)$ 
at $T=0.006$ eV, $U=0$, $1.8$, and $2.1$ eV and $J_{\textrm{H}}=\frac{U}{6}$, 
and frequency dependence of 
$\textrm{Im}\chi^{\textrm{S}(\textrm{R})}(\boldq,\omega)$ 
for $\boldq=\boldQ_{\textrm{IC-AF}}$, 
$\boldQ_{\textrm{IC-AF}}^{\prime}\equiv (\pi,\frac{21}{32}\pi)$, 
and $\boldQ_{\textrm{FM}}\equiv (0,0)$ 
at $T=0.006$ eV and $J_{\textrm{H}}=\frac{U}{6}$ 
for (b) $U=1.8$ eV and (c) $U=2.1$ eV. 
}
\label{fig:Fig9}
\end{figure*}

\subsection{Magnetic properties} 

In this section, 
I show four main results about the magnetic properties. 
First, 
the dominant fluctuations are the spin fluctuations. 
Second, 
an increase of electron correlation leads to 
the enhancement of low-energy spin fluctuation at 
$\boldq=\boldQ_{\textrm{IC-AF}}\equiv (\frac{21}{32}\pi,\frac{21}{32}\pi)$. 
Third, 
the diagonal and the nondiagonal component of 
$\chi_{aabb}^{\textrm{S}(\textrm{R})}(q)$ at $\boldq=\boldQ_{\textrm{IC-AF}}$ 
contribute to the enhancement 
of the spin fluctuation at $\boldq=\boldQ_{\textrm{IC-AF}}$, 
and the diagonal component of the $d_{xy}$ orbital is largest. 
Fourth, 
the orbital dependence of the effective interaction is determined by 
the orbital dependence of the spin fluctuation.

We first determine the dominant fluctuations in the present model. 
For that purpose, 
we analyze the effects of electron correlation 
on $\lambda^{\textrm{S}}_{\textrm{max}}(\boldq)^{-1}$ and $\lambda^{\textrm{C}}_{\textrm{max}}(\boldq)^{-1}$, 
the inverses of the maximum eigenvalues~\cite{Tsunetsugu-RPA,NA-RPA} of 
$\chi_{abcd}^{\textrm{S}}(\boldq,0)$ and $\chi_{abcd}^{\textrm{C}}(\boldq,0)$, 
respectively. 
This is because 
by analyzing the dependence of 
$\lambda^{\textrm{S}}_{\textrm{max}}(\boldq)^{-1}$ and $\lambda^{\textrm{C}}_{\textrm{max}}(\boldq)^{-1}$ 
on $U$ and $J_{\textrm{H}}$, 
we can determine the dominant fluctuations among four kinds of fluctuations, 
i.e. charge fluctuations, spin fluctuations, 
orbital fluctuations, and spin-orbital-combined fluctuations~\cite{Ueda-paramag,NA-paramag} 
(for more details see Appendix H). 
I show 
$\lambda^{\textrm{S}}_{\textrm{max}}(\boldq)^{-1}$ and $\lambda^{\textrm{C}}_{\textrm{max}}(\boldq)^{-1}$ 
at $T=0.006$ eV and $U=1.8$ eV for several values of $J_{\textrm{H}}$ 
in Figs. \ref{fig:Fig8}(a) and \ref{fig:Fig8}(b), 
respectively. 
We see that 
as $J_{\textrm{H}}$ increases, 
$\lambda^{\textrm{S}}_{\textrm{max}}(\boldq)^{-1}$ monotonically decreases, 
and $\lambda^{\textrm{C}}_{\textrm{max}}(\boldq)^{-1}$ monotonically increases. 
This behavior is characteristic of the enhancement of spin fluctuations 
and the suppression of the charge fluctuations~\cite{Ueda-paramag,NA-paramag} (see Appendix H). 
The similar results are obtained at $U=2.1$ eV, 
as shown in Figs. \ref{fig:Fig8}(c) and \ref{fig:Fig8}(d). 
Since approaching the inverse of the maximum eigenvalue towards zero 
characterizes the enhancement of the susceptibility, 
the results in Figs. \ref{fig:Fig8}(a){--}\ref{fig:Fig8}(d) show that 
spin fluctuations are dominant 
at $U=1.8$ and $2.1$ eV in the present model. 
This can be understood by considering the following three facts: 
the noninteracting susceptibility for $a=c$ and $b=d$, $\chi_{abab}^{(0)}(\boldq,i\Omega_{n})$, 
becomes very large in the present model 
since $G_{aa}^{(0)}(\boldk,i\omega_{m})$ 
is larger than $G_{ab}^{(0)}(\boldk,i\omega_{m})$ for $b\neq a$ 
due to the large intraorbital hopping integrals compared with 
the interorbital ones; 
the interactions between the different kinds of fluctuations 
may be generally very weak in the FLEX approximation 
due to lack of the vertex corrections of the susceptibilities; 
the terms arising from $U$ cause the strongest enhancement of the susceptibilities. 
Namely, 
due to those facts, 
the intraorbital components of $\chi_{abcd}^{\textrm{S}}(\boldq,i\Omega_{n})$, 
i.e. $\chi_{aaaa}^{\textrm{S}}(\boldq,i\Omega_{n})$, 
are strongly enhanced, 
resulting in the larger enhancement of spin fluctuations 
than the other fluctuations. 
Hereafter, we fix the value of $J_{\textrm{H}}$ at $J_{\textrm{H}}=\frac{U}{6}$.

\begin{figure}[tb]
\begin{center}
\includegraphics[width=86mm]{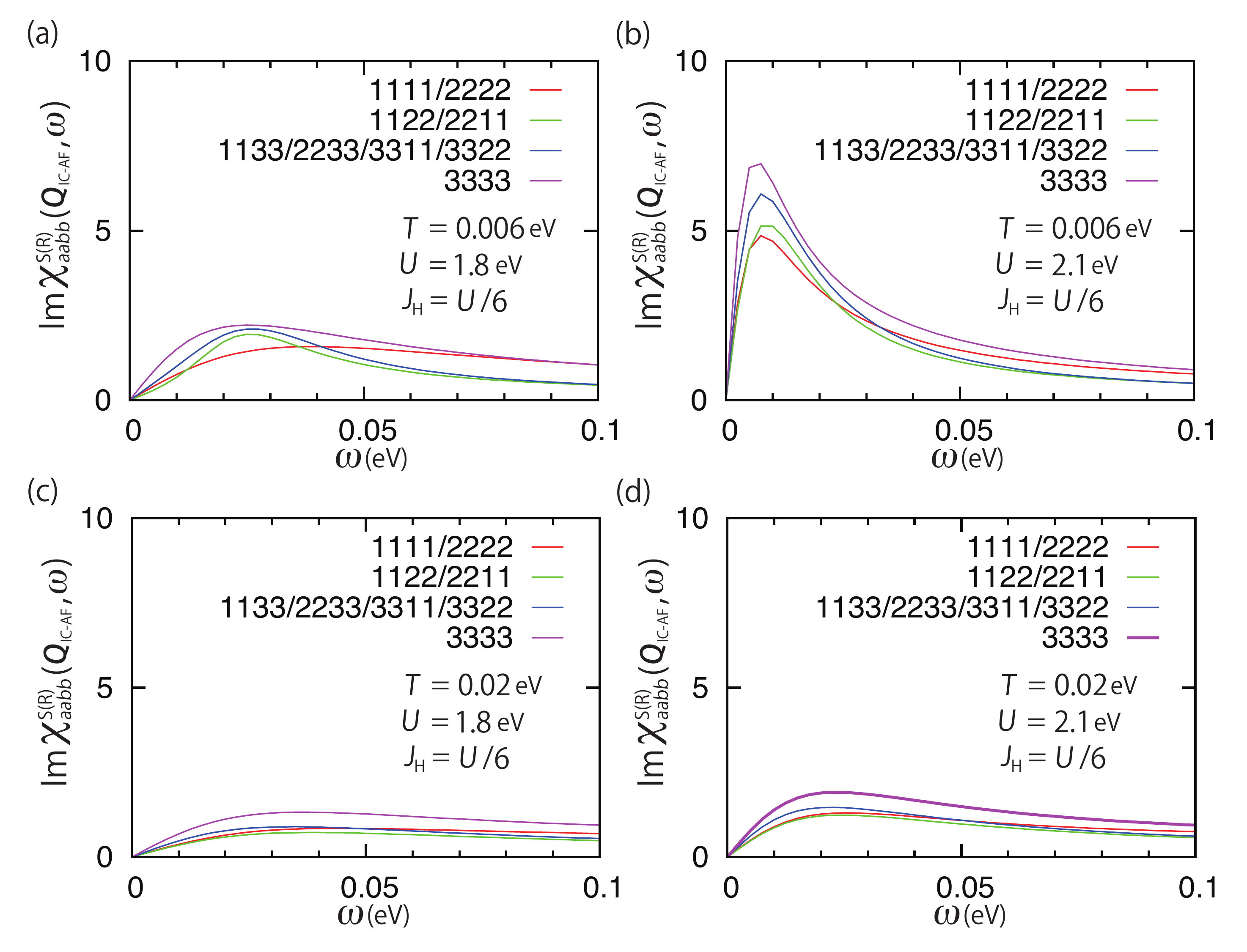}
\end{center}
\vspace{-20pt}
\caption{
Frequency dependence of 
$\textrm{Im}\chi^{\textrm{S}(\textrm{R})}_{aabb}(\boldQ_{\textrm{IC-AF}},\omega)$ 
at $J_{\textrm{H}}=\frac{U}{6}$ 
for $(T,U)=$(a)$(0.006,1.8)$, (b)$(0.006,2.1)$, (c)$(0.02,1.8)$, and (d)$(0.02,2.1)$ (eV). 
}
\label{fig:Fig10}
\end{figure}

Then, for a deeper understanding of spin fluctuations in the present model, 
I analyze the static and the dynamic properties of the spin susceptibility 
as a function of $\omega$, 
$\chi^{\textrm{S}(\textrm{R})}(\boldq,\omega)
=\sum_{a,b}\chi^{\textrm{S}(\textrm{R})}_{aabb}(\boldq,\omega)$. 
For the analysis of the static property, 
I show the momentum dependence of 
$\chi^{\textrm{S}}(\boldq,0)=\chi^{\textrm{S}(\textrm{R})}(\boldq,0)$ 
at $T=0.006$ eV for $U=1.8$ and $2.1$ eV in Fig. \ref{fig:Fig9}(a). 
The result shows that 
as $U$ increases, 
the spin fluctuation at $\boldq=\boldQ_{\textrm{IC-AF}}$ 
is most strongly enhanced 
and the enhancement at $\boldq=(0,0)$ is much weaker. 
That strongest enhancement can be understood as 
the combination of the merging of the nesting vectors 
of the $d_{xz/yz}$ and $d_{xy}$ orbitals around $\boldq=\boldQ_{\textrm{IC-AF}}$ 
due to the mode-mode coupling for the spin fluctuations around $\boldq=\boldQ_{\textrm{IC-AF}}$ 
and the nesting instability at $\boldq=\boldQ_{\textrm{IC-AF}}$ 
due to the RPA-type scattering process, 
as explained in Ref. \onlinecite{NA-CVC}. 
Next, 
for the analysis of the dynamic property, 
I show the frequency dependence of 
$\textrm{Im}\chi^{\textrm{S}(\textrm{R})}(\boldq,\omega)$ 
for several values of $\boldq$ at $T=0.006$ eV 
for $U=1.8$ and $2.1$ eV in Figs. \ref{fig:Fig9}(b) and \ref{fig:Fig9}(c). 
These figures show that 
low-energy spin fluctuation at $\boldq=\boldQ_{\textrm{IC-AF}}$ 
is dominant in the dynamic properties at $U=1.8$ and $2.1$ eV, 
and that 
the intensity at $\boldq=(0,0)$ is very small. 

\begin{figure}[tb]
\begin{center}
\includegraphics[width=86mm]{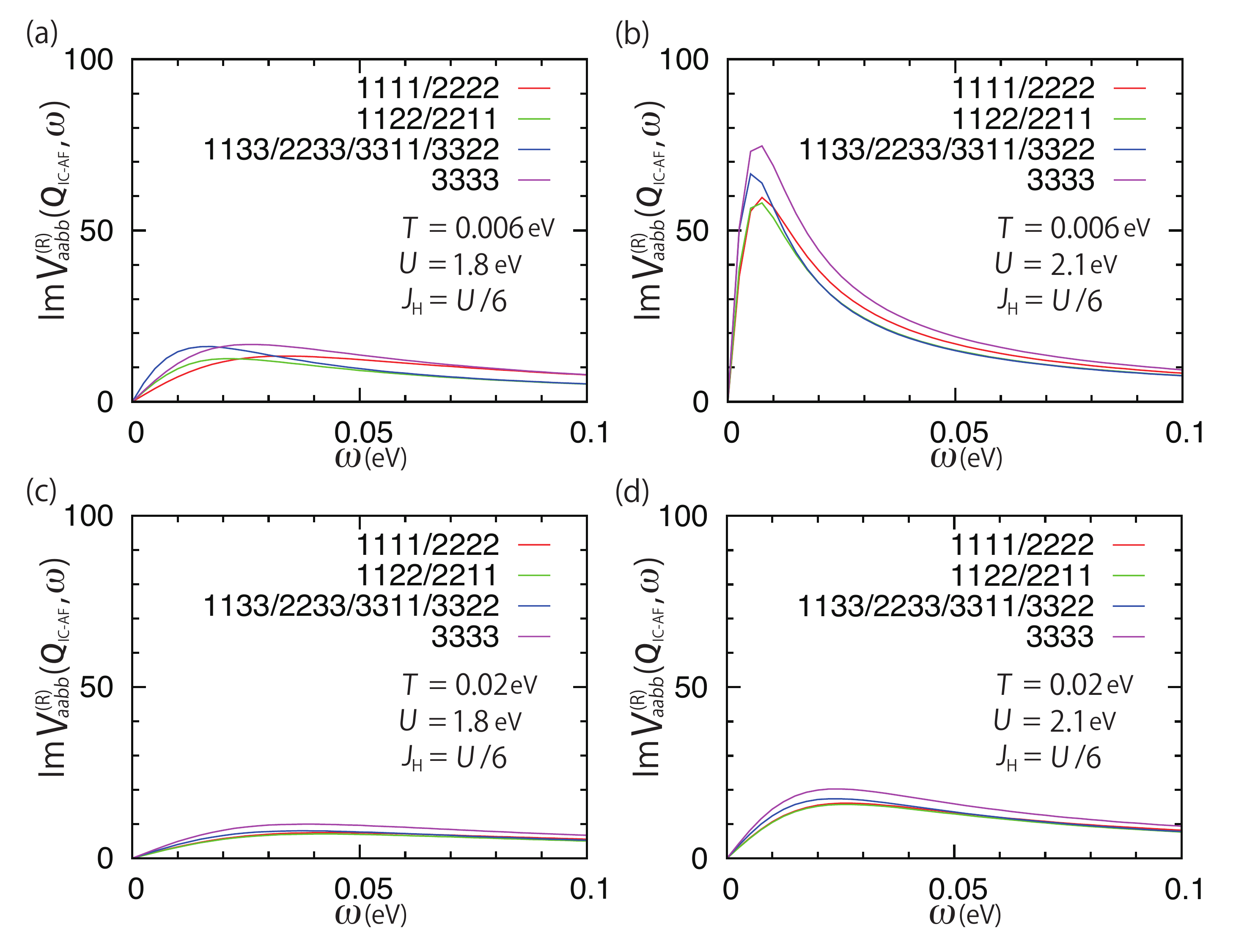}
\end{center}
\vspace{-20pt}
\caption{
Frequency dependence of 
$\textrm{Im}V^{(\textrm{R})}_{aabb}(\boldQ_{\textrm{IC-AF}},\omega)$ 
at $J_{\textrm{H}}=\frac{U}{6}$ 
for $(T,U)=$(a)$(0.006,1.8)$, (b)$(0.006,2.1)$, (c)$(0.02,1.8)$, and (d)$(0.02,2.1)$ (eV). 
}
\label{fig:Fig11}
\end{figure}

Moreover, 
I analyze the role of each $t_{2g}$ orbital in discussing the spin fluctuations. 
Figures \ref{fig:Fig10}(a){--}\ref{fig:Fig10}(d) show 
the frequency dependences of 
$\textrm{Im}\chi_{aabb}^{\textrm{S}(\textrm{R})}(\boldQ_{\textrm{IC-AF}},\omega)$ 
at $J_{\textrm{H}}=\frac{U}{6}$ 
for $(T,U)=(0.006,1.8)$, $(0.006,2.1)$, $(0.02,1.8)$, and $(0.02,2.1)$ (eV). 
We see that 
not only the diagonal but also the non-diagonal components are enhanced, 
and that 
the largest component is the diagonal one of the $d_{xy}$ orbital. 
First, 
the enhancement of the diagonal components arises from 
the combination of the large diagonal components of the noninteracting susceptibility 
of the $t_{2g}$ orbitals around $\boldq=\boldQ_{\textrm{IC-AF}}$, 
the merging of the nesting vectors of the $d_{xz/yz}$ and the $d_{xy}$ orbital 
around $\boldq=\boldQ_{\textrm{IC-AF}}$, 
and the larger enhancement due to the terms arising from $U$ 
than the other terms. 
Next, 
the nondiagonal components are enhanced due to the terms 
including $J_{\textrm{H}}$ and the diagonal components 
since $\chi_{aabb}^{\textrm{S}}(\boldq,i\Omega_{n})$ for $a\neq b$ 
are enhanced mainly through 
$\chi_{aaaa}(\boldq,i\Omega_{n})U\chi_{aabb}^{\textrm{S}}(\boldq,i\Omega_{n})
+\chi_{aaaa}(\boldq,i\Omega_{n})J_{\textrm{H}}\chi_{bbbb}^{\textrm{S}}(\boldq,i\Omega_{n})$ 
[see the second term of Eq. (\ref{eq:FLEX-2})]. 
Then, 
the diagonal component of the $d_{xy}$ orbital becomes largest 
due to the following three properties: 
the diagonal components of the noninteracting susceptibility are larger 
than the non-diagonal components 
due to the large intraorbital hopping integrals; 
the noninteracting susceptibility of the $d_{xy}$ orbital 
is larger than that of the $d_{xz/yz}$ orbital 
due to the larger DOS~\cite{NA-review} of the $d_{xy}$ orbital; 
the enhancement due to the terms arising from $U$ is largest 
in the terms arising from the Hubbard interaction terms. 

Finally, 
we see the effect of the spin fluctuations 
on the imaginary part of the retarded effective interaction of the FLEX approximation. 
The reason why that effect is analyzed is that 
its understanding is useful to 
understand the effect of the spin fluctuations 
on the MT CVC 
since the imaginary part of the retarded effective interaction 
is part of the kernel of the MT CVC [see Eq. (\ref{eq:Lamb-FLEX-MT})]. 
For that analysis, 
it is sufficient to present $\textrm{Im}V_{aabb}^{(\textrm{R})}(\boldq,\omega)$ 
since the other orbital components are much less important. 
This is due to the facts that 
the dominant fluctuations are the spin fluctuations 
and that 
their contributions to the effective interaction, $V_{acbd}(\boldq,i\Omega_{n})$, 
are given by 
$\textstyle\sum_{A,C}U_{acAA}^{\textrm{S}}
\chi_{AACC}^{\textrm{S}}(\boldq,i\Omega_{n})U_{CCbd}^{\textrm{S}}
=\delta_{a,c}\delta_{b,d}\textstyle\sum_{A,C}U_{aaAA}^{\textrm{S}}
\chi_{AACC}^{\textrm{S}}(\boldq,i\Omega_{n})U_{CCbb}^{\textrm{S}}$ 
[see Eq. (\ref{eq:FLEX-6})]. 
Figures \ref{fig:Fig11}(a){--}\ref{fig:Fig11}(d) show 
the frequency dependence of $\textrm{Im}V_{aabb}^{(\textrm{R})}(\boldQ_{\textrm{IC-AF}},\omega)$ 
for $(T,U)=(0.006,1.8)$, $(0.006,2.1)$, $(0.02,1.8)$, and $(0.02,2.1)$ (eV). 
The obtained orbital dependence is similar to 
that for $\textrm{Im}\chi_{aabb}^{\textrm{S}(\textrm{R})}(\boldq,\omega)$. 
Thus, 
the spin fluctuations lead to the main contributions to 
the MT CVC in the present model, 
and the orbital dependence of the MT CVC 
is determined by the orbital dependence of the spin fluctuations.

\subsection{Electronic structure}
\begin{figure}[tb]
\begin{center}
\includegraphics[width=75mm]{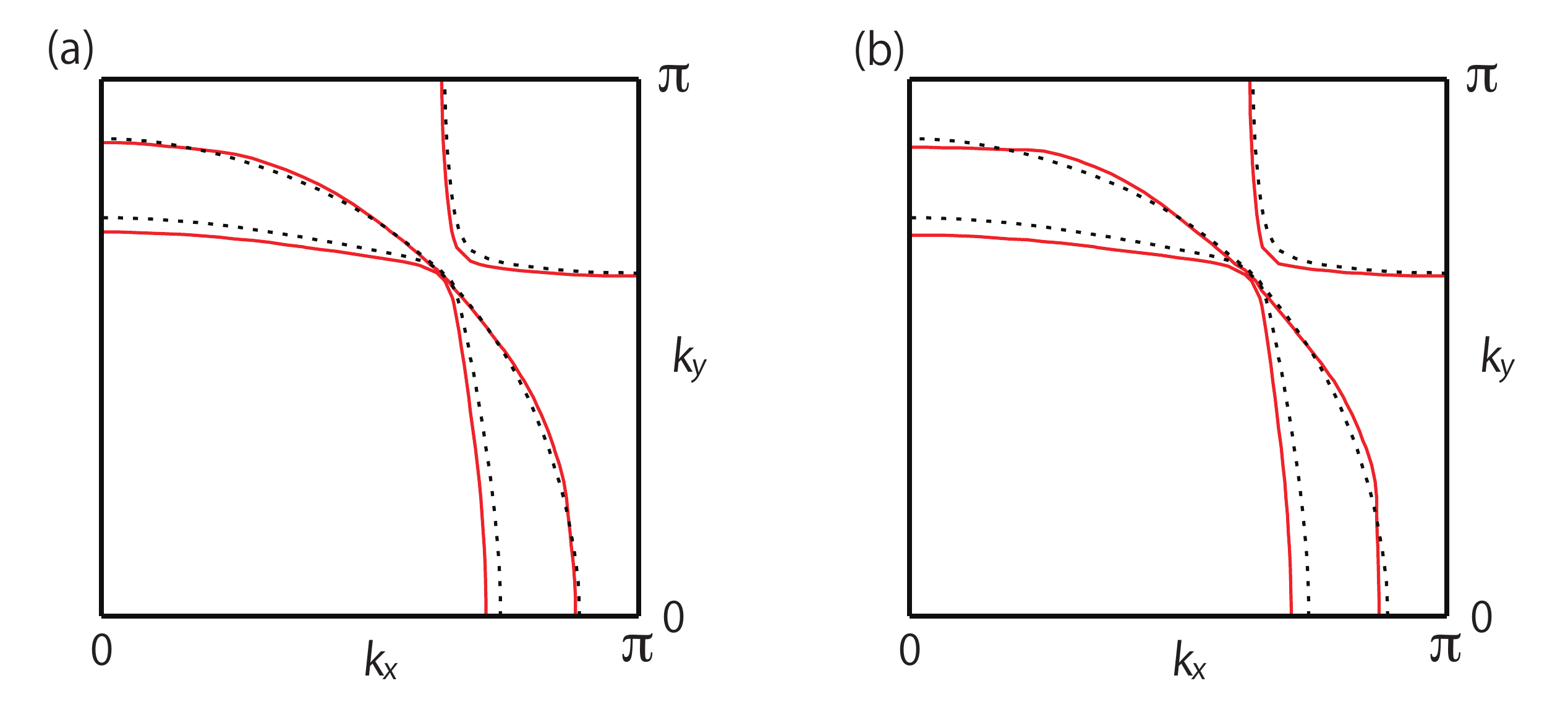}
\end{center}
\vspace{-20pt}
\caption{
FSs at $T=0.006$ eV and $J_{\textrm{H}}=\frac{U}{6}$ 
for (a) $U=1.8$ eV and (b) $U=2.1$ eV. 
In panels (a) and (b), 
the FS sheets at $U=0$ eV are shown by the dotted lines for comparison. 
}
\label{fig:Fig12}
\end{figure}
\begin{figure*}[tb]
\begin{center}
\includegraphics[width=178mm]{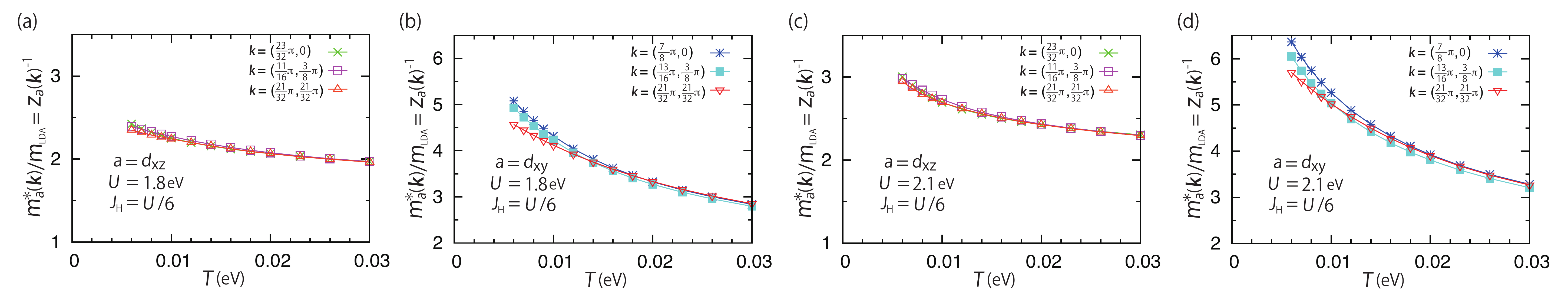}
\end{center}
\vspace{-20pt}
\caption{ 
Temperature dependence of $z_{a}(\boldk)^{-1}=
1-\frac{\partial \textrm{Re}\Sigma_{aa}^{(\textrm{R})}(\boldk,\omega)}
{\partial \omega}|_{\omega \rightarrow 0}$ for several $\boldk$ at $J_{\textrm{H}}=\frac{U}{6}$ 
for (a) $a=d_{xz}$ and $U=1.8$ eV, (b) $a=d_{xy}$ and $U=1.8$ eV, 
(c) $a=d_{xz}$ and $U=2.1$ eV, and (d) $a=d_{xy}$ and $U=2.1$ eV. 
$z_{d_{yz}}(\boldk)^{-1}$ is given by 
$z_{d_{yz}}(k_{x},k_{y})^{-1}=z_{d_{xz}}(k_{y},k_{x})^{-1}$. 
}
\label{fig:Fig13}
\end{figure*}

\begin{figure*}[tb]
\begin{center}
\includegraphics[width=178mm]{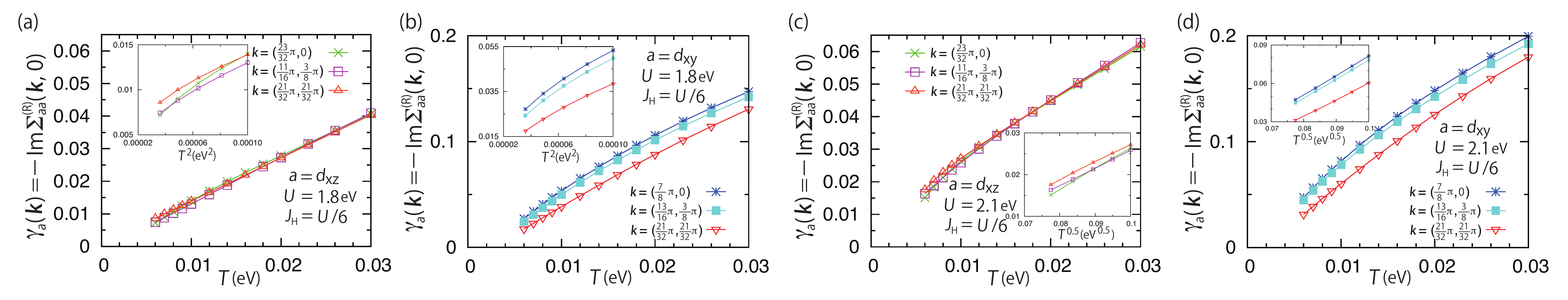}
\end{center}
\vspace{-20pt}
\caption{
Temperature dependence of $\gamma_{a}(\boldk)=-\textrm{Im}\Sigma_{aa}^{(\textrm{R})}(\boldk,0)$ 
for several $\boldk$ 
at $J_{\textrm{H}}=\frac{U}{6}$ 
for (a) $a=d_{xz}$ and $U=1.8$ eV, (b) $a=d_{xy}$ and $U=1.8$ eV, 
(c) $a=d_{xz}$ and $U=2.1$ eV, and (d) $a=d_{xy}$ and $U=2.1$ eV. 
The inset in panel (a) or (b) shows 
$\gamma_{a}(\boldk)$ against $T^{2}$ below $T=0.01$ eV, 
and the inset in panel (c) or (d) shows 
$\gamma_{a}(\boldk)$ against $T^{0.5}$ below $T=0.01$ eV. 
$\gamma_{d_{yz}}(\boldk)$ is given by 
$\gamma_{d_{yz}}(k_{x},k_{y})=\gamma_{d_{xz}}(k_{y},k_{x})$. 
}
\label{fig:Fig14}
\end{figure*}
\begin{figure*}[tb]
\begin{center}
\includegraphics[width=178mm]{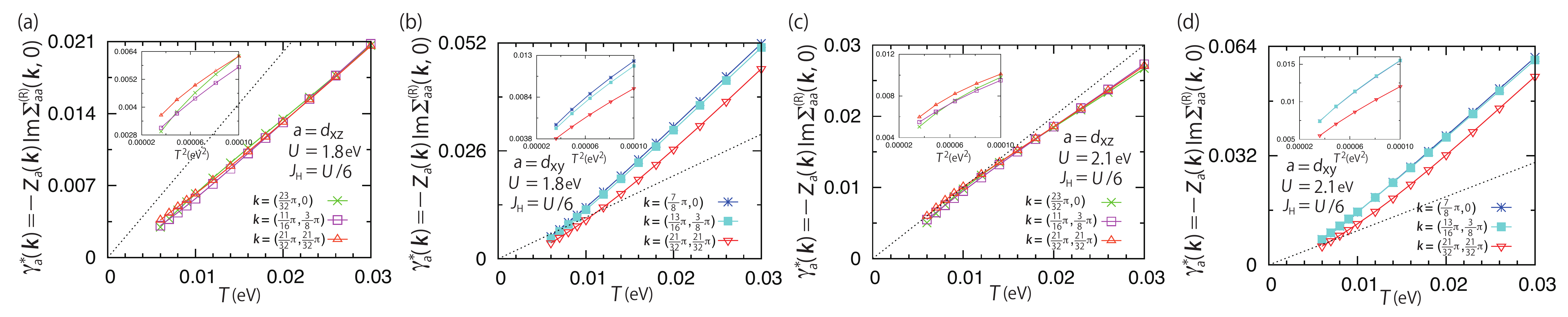}
\end{center}
\vspace{-20pt}
\caption{
Temperature dependence of $\gamma_{a}^{\ast}(\boldk)=z_{a}(\boldk)\gamma_{a}(\boldk)$ 
for several $\boldk$ 
at $J_{\textrm{H}}=\frac{U}{6}$ 
for (a) $a=d_{xz}$ and $U=1.8$ eV, (b) $a=d_{xy}$ and $U=1.8$ eV, 
(c) $a=d_{xz}$ and $U=2.1$ eV, and (d) $a=d_{xy}$ and $U=2.1$ eV. 
$\gamma_{a}^{\ast}(\boldk)=T$ are shown by the dotted lines 
to discuss whether the QP damping is cold-spot-type or hot-spot-type, 
and the insets show $\gamma_{a}^{\ast}(\boldk)$ against $T^{2}$ below $T=0.01$ eV. 
$\gamma_{d_{yz}}^{\ast}(\boldk)$ is given by 
$\gamma_{d_{yz}}^{\ast}(k_{x},k_{y})=\gamma_{d_{xz}}^{\ast}(k_{y},k_{x})$. 
}
\label{fig:Fig15}
\end{figure*}

In this section, 
I show four main results about the electronic structure. 
First, 
the topology of the FS remains the same 
as the noninteracting one 
even including the FS deformation due to the real part of the self-energy 
in the FLEX approximation. 
Second, 
the mass enhancement of the $d_{xy}$ orbital 
is larger than that of the $d_{xz/yz}$ orbital 
in a wide region of the parameter space in the present model. 
Third, 
the unrenormalized QP damping of the $d_{xy}$ orbital 
becomes larger than that of the $d_{xz/yz}$ orbital. 
Fourth, 
the orbital dependence of the QP damping is mainly determined 
by the orbital dependence of the unrenormalized QP damping. 

I begin with the effects of the real part of the self-energy in the FLEX approximation 
on the FS and the mass enhancement factor. 
I determine the FS 
by diagonalizing $[\epsilon_{ab}(\boldk)+\textrm{Re}\Sigma_{ab}^{(\textrm{R})}(\boldk,0)]$, 
where $\mu$ in $\epsilon_{ab}(\boldk)$ has been determined by Eq. (\ref{eq:mu-int}).

First, 
we see from Figs. \ref{fig:Fig12}(a) and \ref{fig:Fig12}(b) 
how the FS is modified with increasing $U$. 
Those figures show that the modification is slight. 
Thus, the real part of the self-energy in the FLEX approximation 
does not change the topology of the FS sheets 
(i.e., whether each sheet is electron-like or hole-like). 
This result can be understood by considering 
two facts that 
the occupation numbers of the $d_{xz/yz}$ and the $d_{xy}$ orbital 
do not become very close to integers, 
and that 
the van Hove singularity of the $d_{xy}$ orbital 
does not cross over the Fermi level. 
Note, first, that 
the occupation numbers of the $d_{xz/yz}$ and the $d_{xy}$ orbitals 
are $1.36$ and $1.28$, respectively, 
at $U=1.8$ and $2.1$ eV; 
second, that if the van Hove singularity crosses over the Fermi level, 
the $\gamma$ sheet touches the boundary of the Brillouin zone 
at $\boldk=(\pi,0)$ or $(0,\pi)$.

Next, 
we show the mass enhancement factor, $z_{a}(\boldk)^{-1}=
1-\frac{\partial \textrm{Re}\Sigma_{aa}^{(\textrm{R})}(\boldk,\omega)}
{\partial \omega}|_{\omega \rightarrow 0}$, at $U=1.8$ and $2.1$ eV in 
Figs. \ref{fig:Fig13}(a){--}\ref{fig:Fig13}(d). 
From those figures, 
we find three properties about the orbital, temperature, 
and momentum dependences of $z_{a}(\boldk)^{-1}$. 
The first property is that 
the mass enhancement of the $d_{xy}$ orbital is always larger than 
that of the $d_{xz/yz}$ orbital 
for all the temperatures considered. 
This arises from the stronger spin fluctuations of the $d_{xy}$ orbital 
than those of the $d_{xz/yz}$ orbital, 
as explained in Ref. \onlinecite{NA-CVC}. 
Combining this result with 
the similar orbital dependence~\cite{NA-CVC} of $z_{a}(\boldk)^{-1}$
as a function of $J_{\textrm{H}}$ (in $0\leq J_{\textrm{H}}\leq \frac{U}{5}$ 
at $T=0.006$ eV and $U=1.8$ eV), 
we deduce that 
the larger mass enhancement of the $d_{xy}$ orbital is realized 
in a wide region of the parameter space of the present model for a PM state 
in the FLEX approximation. 
It should be noted that 
although the spin fluctuations of the $d_{xy}$ orbital 
enhance $z_{a}(\boldk)^{-1}$ of not only the $d_{xy}$ orbital 
but also the $d_{xz/yz}$ orbital, 
the enhancement for the $d_{xy}$ is larger 
in a realistic set of the Hubbard interaction terms. 
This is because 
the spin fluctuations of an orbital cause the enhancement of $z_{a}(\boldk)^{-1}$ 
of the orbital proportional to the $U^{2}$ terms of 
$\frac{3}{2}\textstyle\sum_{A,B}U_{aaAA}^{\textrm{S}}
\chi_{AABB}^{\textrm{S}}(\boldq,i\Omega_{n})U_{BBaa}^{\textrm{S}}$ 
in $V_{aaaa}(\boldq,i\Omega_{n})$ in $\Sigma_{aa}(\boldk,i\epsilon_{m})$, 
and the enhancement of $z_{a}(\boldk)^{-1}$ of another orbital 
proportional to the $J_{\textrm{H}}^{2}$ terms. 
Then, 
the second property found in Figs. \ref{fig:Fig13}(a){--}\ref{fig:Fig13}(d) 
is that 
the temperature dependence is weak other than the case for the $d_{xy}$ orbital 
at $U=2.1$ eV. 
This results from the more significant enhancement of the spin fluctuations 
of the $d_{xy}$ orbital with decreasing temperature 
[see Figs. \ref{fig:Fig10}(a){--}\ref{fig:Fig10}(d)], 
and suggests that 
the mass enhancement of the $d_{xy}$ orbital may remain 
larger even at lower temperatures than the temperatures considered. 
The third property of Figs. \ref{fig:Fig13}(a){--}\ref{fig:Fig13}(d) 
is that 
the momentum dependence is negligible for the $d_{xz/yz}$ orbital, 
while the $d_{xy}$ orbital has the weak momentum dependence. 
This is due to the difference between 
the quasi-one-dimensionality of the $d_{xz/yz}$ orbital 
and the quasi-two-dimensionality of the $d_{xy}$ orbital: 
only the $d_{xy}$ orbital 
has the van Hove singularity due to the saddle points at 
$\boldk\approx (\frac{23}{32}\pi,0)$ and $(0,\frac{23}{32}\pi)$, 
resulting in a larger mass enhancement~\cite{vHs-cuprate}. 
Since this result shows that 
the momentum dependence of the mass enhancement factor 
is not important to discuss the magnitude difference of the mass enhancement, 
the present analysis is sufficient for that discussion. 

Then, 
we turn to the effects of the imaginary of the self-energy on 
the unrenormalized QP damping, $\gamma_{a}(\boldk)=-\textrm{Im}\Sigma_{aa}^{(\textrm{R})}(\boldk,0)$. 
From the results shown in Figs. \ref{fig:Fig14}(a){--}\ref{fig:Fig14}(d), 
we see three main features. 
The first one is about the orbital dependence: 
the magnitude for the $d_{xy}$ orbital is 
about three times as large as that for the $d_{xz/yz}$ orbital. 
This arises mainly from the larger DOS and stronger spin fluctuations of the $d_{xy}$ orbital. 
Note, first, that 
a ratio of the noninteracting DOSs of the $d_{xy}$ and the $d_{xz/yz}$ orbitals 
on the Fermi level is about $2.3$~\cite{NA-review}; 
second, that 
due to the similar reasons for $z_{a}(\boldk)^{-1}$, 
the spin fluctuations of the $d_{xy}$ orbital 
cause a larger enhancement of $\gamma_{a}(\boldk)$ of the $d_{xy}$ orbital 
in a realistic set of the Hubbard interaction terms. 
The second main feature is about the temperature dependence: 
the unrenormalized QP dampings of the $d_{xz/yz}$ orbital 
at $U=1.8$ eV show the $T^{2}$ dependence at low temperatures; 
the $T^{0.5}$ dependence of $\gamma_{a}(\boldk)$ for the $d_{xz/yz}$ orbital is realized 
for $\boldk=(\frac{21}{32}\pi,\frac{21}{32}\pi)$ at $U=2.1$ eV; 
the unrenormalized QP damping of the $d_{xy}$ orbital 
at $\boldk=(\frac{7}{8}\pi,0)$ is proportional to $T$ linear 
at $U=1.8$ and $2.1$ eV. 
The $T^{2}$ dependence is due to the formation of long-lived QPs~\cite{AGD}; 
the $T^{0.5}$ dependence results from the hot-spot structure~\cite{Hlubina-Rice} 
due to the enhanced AF spin fluctuation, as explained in Ref. \onlinecite{NA-review}; 
the $T$-linear behavior emerges as a result of 
the existence of the van Hove singularity~\cite{Hlubina-Rice}. 
The third main feature is about the momentum dependence: 
the unrenormalized QP damping of the $d_{xy}$ orbital depends weakly on momentum; 
the momentum dependence for the $d_{xz/yz}$ orbital is negligible. 
This arises from the considerable difference in the momentum dependence of 
the single-particle spectrum function 
due to the existence of the van Hove singularity only for the $d_{xy}$ orbital.   

Finally, 
we analyze the effects of the combination of the real and the imaginary part 
of the self-energy on the QP damping, $\gamma_{a}^{\ast}(\boldk)=z_{a}(\boldk)\gamma_{a}(\boldk)$. 
From the results shown in Figs. \ref{fig:Fig15}(a){--}\ref{fig:Fig15}(d), 
we see that 
even for the QP damping, 
the larger magnitude for the $d_{xy}$ orbital is realized. 
This is due to the larger difference in the unrenormalized QP damping 
compared with the difference in the mass enhancement factor, 
and suggests that 
the QPs of the $d_{xz/yz}$ orbital are more coherent than the QPs of the $d_{xy}$ orbital 
in the present model. 
In addition, 
we find the $T^{2}$ dependence for the $d_{xz/yz}$ orbital at low temperatures at $U=1.8$ eV, 
the deviation from the $T^{2}$ dependence for the $d_{xz/yz}$ orbital at $U=2.1$ eV, 
and the similar momentum dependence of the QP damping to that of the unrenormalized QP damping. 

\subsection{Transport properties}
In this section, 
I show three main results about the transport properties. 
First, 
the main results in the previous studies~\cite{NA-review,NA-CVC}  
remain qualitatively unchanged 
even including the main terms of the AL CVC. 
Second, 
the temperature dependences of $\rho_{ab}$ and $R_{\textrm{H}}$ 
near the AF QCP 
consist of two regions, 
high-temperature region, where 
only the $\Sigma$ CVC is sufficient, 
and low-temperature region, where 
only both the $\Sigma$ CVC and the MT CVC are sufficient. 
Third, 
in contrast to the case near the AF QCP, 
the effects of the MT CVC on $\rho_{ab}$ and $R_{\textrm{H}}$ at low temperatures 
are different in case away from the AF QCP: 
only for $R_{\textrm{H}}$, the effects are considerable.

To analyze the main effects of the AL CVC on $\rho_{ab}$ and $R_{\textrm{H}}$, 
we consider four cases, named 
MT$+$AL CVC case, MT CVC case, $\Sigma$ CVC case, and No CVC case. 
In the MT$+$AL CVC case, 
we take account of the $\Sigma$ CVC, the MT CVC, 
and the main terms of the AL CVC: 
$\Lambda_{\nu;2;cd}(k;0)$ in Eq. (\ref{eq:Lamb-FLEX-full}) 
includes those CVCs, 
and $\Lambda_{\nu;2;ab}^{(0)}(k;0)$ in Eq. (\ref{eq:Lamb0}) 
includes the $\Sigma$ CVC. 
In the MT CVC, 
we neglect only the AL CVC and take account of the other CVCs: 
the change from the MT$+$AL CVC case 
is neglecting the AL CVC in $\Lambda_{\nu;2;cd}(k;0)$. 
In the $\Sigma$ CVC case, 
we take account of only the $\Sigma$ CVC among the CVCs: 
$\Lambda_{\nu;2;cd}(k;0)$ becomes the same as $\Lambda_{\nu;2;ab}^{(0)}(k;0)$. 
In the No CVC case, 
we neglect all the CVCs: 
$\Lambda_{\nu;2;cd}(k;0)$ and $\Lambda_{\nu;2;ab}^{(0)}(k;0)$ 
are determined only by the noninteracting group velocity. 

\subsubsection{In-plane resistivity}
\begin{figure}[tb]
\begin{center}
\includegraphics[width=75mm]{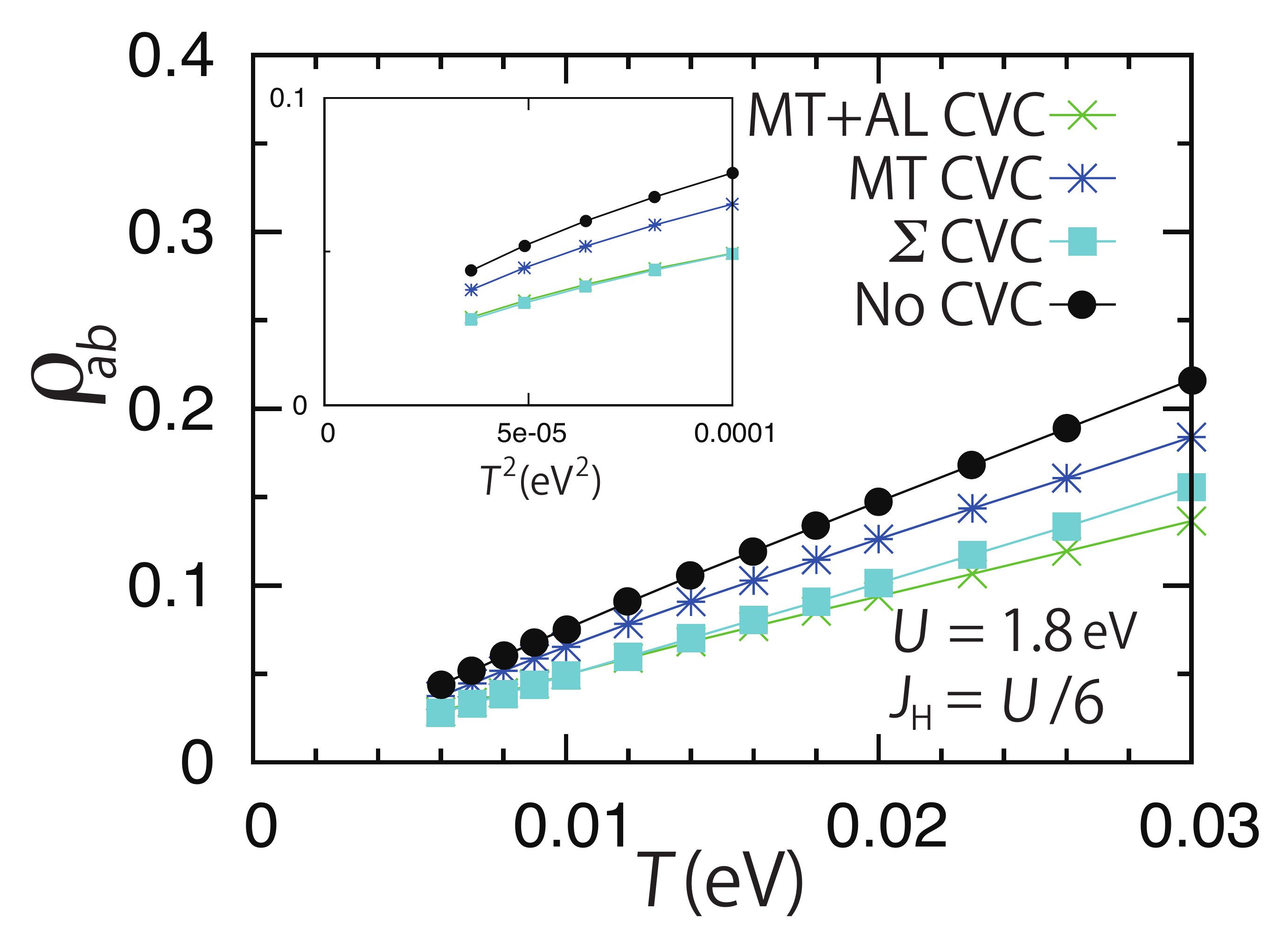}
\end{center}
\vspace{-20pt}
\caption{
Temperature dependence of $\rho_{ab}$ at $U=1.8$ eV and $J_{\textrm{H}}=\frac{U}{6}$ 
in the four cases. 
The inset shows $\rho_{ab}$ against $T^{2}$ below $T=0.01$ eV. 
}
\label{fig:Fig16}
\end{figure}
\begin{figure*}[tb]
\begin{center}
\includegraphics[width=174mm]{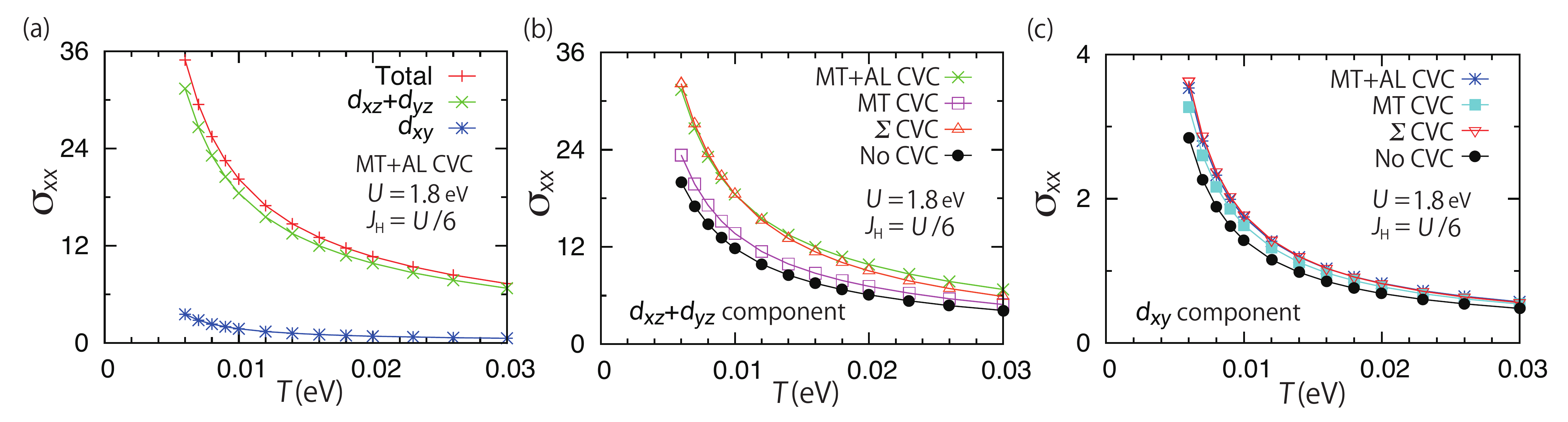}
\end{center}
\vspace{-20pt}
\caption{ 
Temperature dependence of (a) $\sigma_{xx}$ and orbital-decomposed components 
in the MT$+$AL CVC case at $U=1.8$ eV 
and the orbital-decomposed components of (b) the $d_{xz}$ and $d_{yz}$ orbitals 
and (c) $d_{xy}$ orbital in the four cases at $U=1.8$ eV. 
}
\label{fig:Fig18}
\end{figure*}
\begin{figure}[tb]
\begin{center}
\includegraphics[width=75mm]{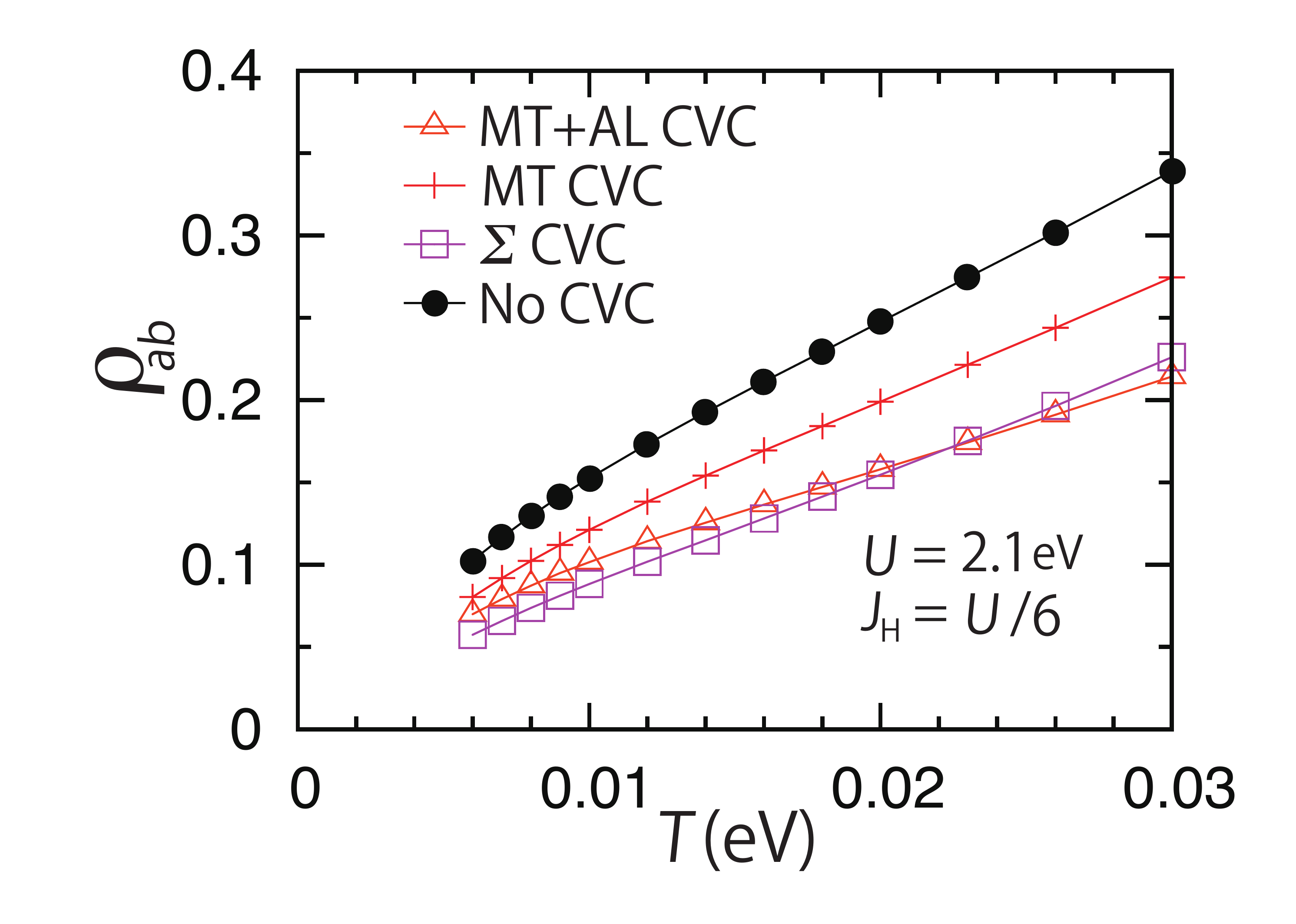}
\end{center}
\vspace{-20pt}
\caption{ 
Temperature dependence of $\rho_{ab}$ at $U=2.1$ eV and $J_{\textrm{H}}=\frac{U}{6}$ 
in the four cases. 
}
\label{fig:Fig19}
\end{figure}
\begin{figure*}[tb]
\begin{center}
\includegraphics[width=174mm]{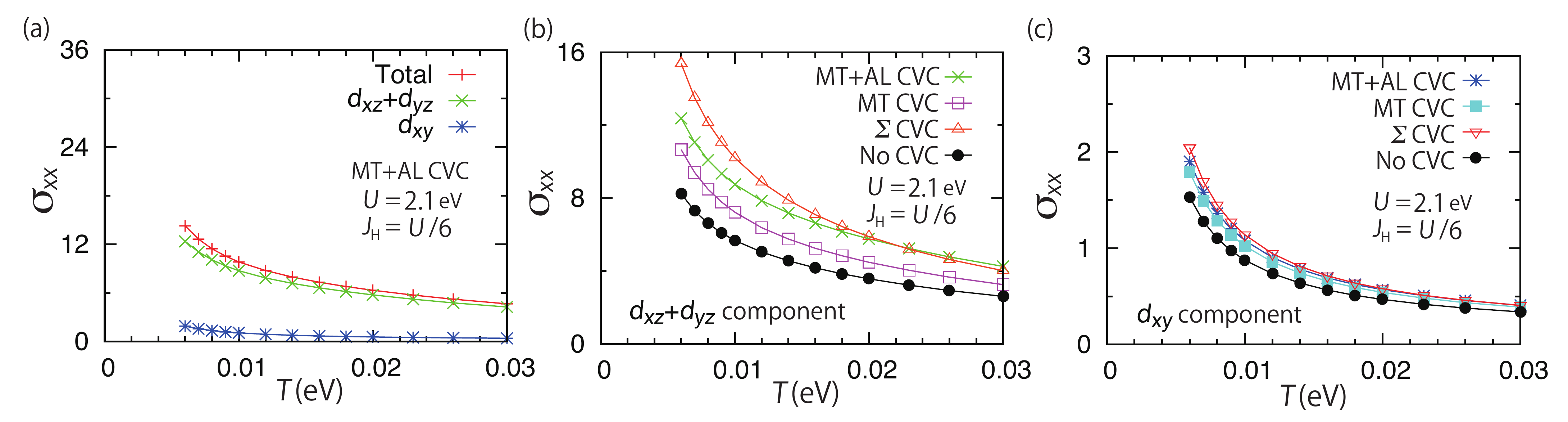}
\end{center}
\vspace{-20pt}
\caption{
Temperature dependence of (a) $\sigma_{xx}$ and orbital-decomposed components 
in the MT$+$AL CVC case at $U=2.1$ eV 
and the orbital-decomposed components of (b) the $d_{xz}$ and $d_{yz}$ orbitals 
and (c) $d_{xy}$ orbital in the four cases at $U=2.1$ eV. 
}
\label{fig:Fig20}
\end{figure*}

We begin with 
$\rho_{ab}(=\sigma_{xx}^{-1}=\sigma_{yy}^{-1})$ 
away from the AF QCP. 
We show the temperature dependence of $\rho_{ab}$ at $U=1.8$ eV in the four cases 
in Fig. \ref{fig:Fig16}, and find two main features. 
One is that 
the $T^{2}$ dependence below $T=0.008$ eV 
holds even in the MT$+$AL CVC case. 
This can be understood that 
the CVCs little affect the power of the temperature dependence of the resistivity. 
This is because the main effects of the CVCs on the resistivity 
arise from the magnitude changes of the current [see Sect. II B 1]
and because the magnitude changes appear in the equation of the resistivity 
as $\frac{1}{|\Lambda_{2;cd}^{(0)}(k)|+\Delta |\Lambda_{2;cd}(k)|}\sim 
\frac{1}{|\Lambda_{2;cd}^{(0)}(k)|}(1-\frac{\Delta |\Lambda_{2;cd}(k)|}{|\Lambda_{2;cd}^{(0)}(k)|})$, 
where $\frac{\Delta |\Lambda_{2;cd}(k)|}{|\Lambda_{2;cd}^{(0)}(k)|}$ is not large. 
The other main feature is that 
the value of $\rho_{ab}$ in the MT$+$AL CVC case 
becomes smaller than that in the MT CVC case 
and nearly the same as that in the $\Sigma$ CVC case. 
This is due to the small effects of the MT and the AL CVC; 
for high-temperature region, 
the small effects arise from the dominance of the QP damping 
compared with the spin susceptibility 
in determining the kernels of those CVCs; 
for low-temperature region, 
the small effects arise from 
the combination of the not large spin susceptibility 
and the partial cancellation between the effects of the MT and the AL CVC. 
The more detailed explanations about those are as follows: 
In discussing the effects of the MT and the AL CVC, 
the relative values of the spin susceptibility and the QP damping are important 
since the kernels of the MT and the AL CVC contain 
the spin susceptibility and the inverse of the QP damping 
[see Eqs. (\ref{eq:Lamb-FLEX-full}){--}(\ref{eq:Lamb-FLEX-AL2})]. 
Due to this property, 
at high temperatures, 
the kernels become small 
since the QP damping is large; 
thus, 
the effects of the MT and the AL CVC are small for high-temperature region. 
Furthermore, 
although the effects of the MT and the AL CVC are separately non-negligible 
at low temperatures since with decreasing temperature 
the QP damping decreases and the spin susceptibility remains almost unchanged, 
the effects of the AL CVC reduce 
the effects of the MT CVC 
as a result of the difference in the momentum dependence; 
due to this reduction, 
the total effects of the MT and the AL CVC are small. 
Such property due to the difference in the momentum dependence 
can be easily seen 
from a simple and sufficient case of the second-order perturbation theory 
for a single-orbital system 
since the momentum structure of each diagram of the MT, AL1, and AL2 terms 
remains the same as in the FLEX approximation: 
in this case, 
the MT CVC is given by $\textstyle\sum_{\boldk^{\prime},\boldq}\Delta_{0}(\boldk,\boldk^{\prime};
\boldk^{\prime}+\boldq,\boldk-\boldq)\boldPhi_{\boldk-\boldq}(\epsilon)$, 
and the AL1 and AL2 CVCs are 
$\textstyle\sum_{\boldk^{\prime},\boldq}\Delta_{0}(\boldk,\boldk^{\prime};
\boldk^{\prime}+\boldq,\boldk-\boldq)
[\boldPhi_{\boldk^{\prime}+\boldq}(\epsilon)-\boldPhi_{\boldk^{\prime}}(\epsilon)]$ 
(for more details, see Ref. \onlinecite{Yamada-Yosida}); 
since $\boldPhi_{\boldk}(\epsilon)$ is odd about momentum, 
the difference in the sign of $\boldq$ 
leads to the partial cancellation of the effects of the MT and the AL CVC. 

We next discuss the role of each $t_{2g}$ orbital 
in determining $\rho_{ab}$ away from the AF QCP. 
For that purpose, 
I show the orbital-decomposed components of $\sigma_{xx}$, 
the $d_{xz}+d_{yz}$ component and the $d_{xy}$ component; 
the former is obtained by replacing 
$\textstyle\sum_{\{a\}=1}^{3}$ in Eq. (\ref{eq:sigmaxx-approx}) 
by $\textstyle\sum_{\{a\}=1}^{2}$, 
and the latter is obtained by replacing 
$\textstyle\sum_{\{a\}=1}^{3}$ in Eq. (\ref{eq:sigmaxx-approx}) 
by $\textstyle\sum_{\{a\}=3}$. 
As explained in Ref. \onlinecite{NA-review}, 
only those components are sufficient in the present model 
since those (diagonal) components are larger than the non-diagonal ones 
due to the large intraorbital hopping integrals compared with 
the interorbital hopping integrals. 
First, 
we see from Fig. \ref{fig:Fig18}(a) that 
$\sigma_{xx/yy}$ is determined almost by the component of the $d_{xz/yz}$ orbital, 
and that the contributions from the component of the $d_{xy}$ orbital are very small. 
Namely, 
the $d_{xz}+d_{yz}$ component remains dominant 
even with the main terms of the AL CVC. 
We also see from Figs. \ref{fig:Fig18}(b) and \ref{fig:Fig18}(c) that 
the values in the MT$+$AL CVC case are nearly the same 
in the $\Sigma$ CVC case. 
Thus, 
the $\Sigma$ CVC is sufficient for discussions about 
the orbital dependence away from the AF QCP. 

From the results at $U=1.8$ eV, 
we deduce, first, that 
the main results obtained in the previous studies~\cite{NA-review,NA-CVC} away from the AF QCP, 
the $T^{2}$ dependence of $\rho_{ab}$ at low temperatures 
and the dominance of the $d_{xz/yz}$ orbital, 
remain qualitatively the same 
even with the main terms of the AL CVC; 
second, that 
the resistivity away from the AF QCP can be almost well described 
by taking account of only the $\Sigma$ CVC. 

Then, I turn to $\rho_{ab}$ near the AF QCP. 
From its temperature dependence shown in Fig. \ref{fig:Fig19}, 
we see three main features about $\rho_{ab}$ in the MT$+$AL CVC case. 
First, 
$\rho_{ab}$ in the MT$+$AL CVC case 
shows the $T$-linear dependence, 
which is similar for the other three cases. 
This origin is the same for the other three cases~\cite{NA-review,NA-CVC}, 
i.e. the $T^{0.5}$ dependence of the unrenormalized QP damping 
of the $d_{xz/yz}$ orbital around $\boldk=(\frac{21}{32}\pi,\frac{21}{32}\pi)$, 
since the CVCs little affect 
the power of the temperature dependence of $\rho_{ab}$ 
and since the $d_{xz}+d_{yz}$ component remains dominant 
even with the main terms of the AL CVC [see Fig. \ref{fig:Fig20}(a)]. 
Second, 
the values of $\rho_{ab}$ in the MT$+$AL CVC case 
at high temperatures are nearly the same as 
those in the $\Sigma$ CVC case at the corresponding temperatures. 
This is due to the same reason as that away from the AF QCP. 
Third, 
as temperature decreases, 
the value of $\rho_{ab}$ in the MT$+$AL CVC case 
approaches the value in the MT CVC case. 
This can be understood by combining two facts that 
the MT and the AL CVCs separately becomes non-negligible 
at low temperatures, 
and that 
the AL CVC near the AF QCP 
is negligible compared with the MT CVC 
in the presence of the even-parity symmetry and rotational symmetry.  
The mechanism of the former fact explained above, 
and the mechanism of the latter 
explained by the authors of Ref. \onlinecite{Kon-CVC}. 
The explanations about the latter are as follows: 
When the system approaches an AF QCP characterized by 
spin fluctuation at $\boldq=\boldQ_{\textrm{QCP}}$, 
that spin fluctuation gives the leading contributions 
to the MT, AL$1$, and AL$2$ CVCs 
through $\textrm{Im}V_{cCdD}^{(\textrm{R})}(k-k^{\prime})$ in Eq. (\ref{eq:Lamb-FLEX-MT}) 
and $W^{\textrm{AL}(\textrm{RA})}_{c}(-q^{\prime};-q^{\prime})$ in 
Eqs. (\ref{eq:Lamb-FLEX-AL1-simple}) and (\ref{eq:Lamb-FLEX-AL2-simple}), 
respectively. 
Then, 
although the MT CVC becomes more important near the AF QCP, 
the AL$1$ and AL$2$ CVCs become little important compared with 
the MT CVC near the AF QCP 
due to the cancellation between 
the contributions 
from $\boldk^{\prime}$ and $-\boldk^{\prime}$ 
arising from the spin fluctuation at $\boldq=\boldQ_{\textrm{QCP}}$. 
This cancellation 
is because in the terms of the AL$1$ or AL$2$ CVC 
only $\tilde{\Lambda}_{\nu;2;cc}(k^{\prime};0)$ is odd about momentum 
(i.e., the others are even) due to the even-parity symmetry 
and because 
the states at $-\boldk^{\prime}+\boldQ_{\textrm{QCP}}$ 
and $-\boldk^{\prime}-\boldQ_{\textrm{QCP}}$ 
are equivalent due to the rotational symmetry. 

Moreover, 
we determine the role of each $t_{2g}$ orbital 
in determining $\rho_{ab}$ near the AF QCP 
from the results of the orbital-decomposed components of $\sigma_{xx}$. 
Figure \ref{fig:Fig20}(a) shows that 
the main terms of the AL CVC keep the dominance of the $d_{xz}+d_{yz}$ component unchanged. 
Furthermore, 
from Figs. \ref{fig:Fig20}(b) and \ref{fig:Fig20}(c), 
we see a similar behavior for $\rho_{ab}$ at low temperatures, 
i.e. an approach of the value of the $d_{xz}+d_{yz}$ component 
or the $d_{xy}$ component in the MT$+$AL CVC case 
towards that in the MT CVC case with decreasing temperature. 

Combining the results at $U=2.1$ eV, 
we find that 
the $T$-linear $\rho_{ab}$ and the dominance of the $d_{xz/yz}$ orbital 
which are obtained in the MT CVC case are qualitatively unchanged 
even in the MT$+$AL CVC case, 
and that 
there are two almost distinct regions of the temperature dependence of $\rho_{ab}$. 
Those regions consist of 
high-temperature region, governed mainly by the $\Sigma$ CVC, 
and low-temperature region, governed mainly by 
the $\Sigma$ CVC and the MT CVC. 

\subsubsection{Hall coefficient}
\begin{figure}[tb]
\begin{center}
\includegraphics[width=75mm]{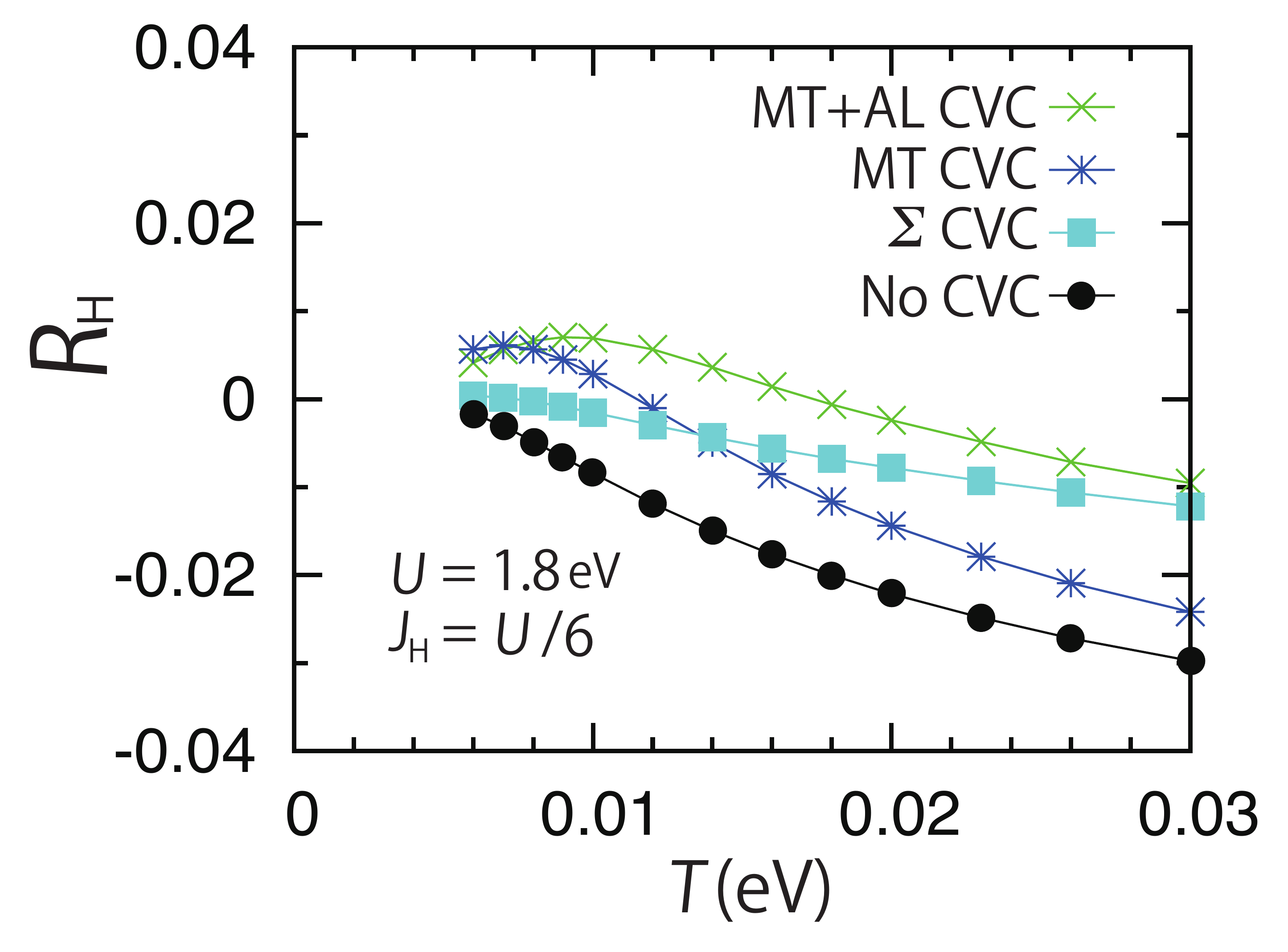}
\end{center}
\vspace{-20pt}
\caption{
Temperature dependence of $R_{\textrm{H}}$ at $U=1.8$ eV and $J_{\textrm{H}}=\frac{U}{6}$ 
in the four cases. 
}
\label{fig:Fig21}
\end{figure}
\begin{figure*}[tb]
\begin{center}
\includegraphics[width=174mm]{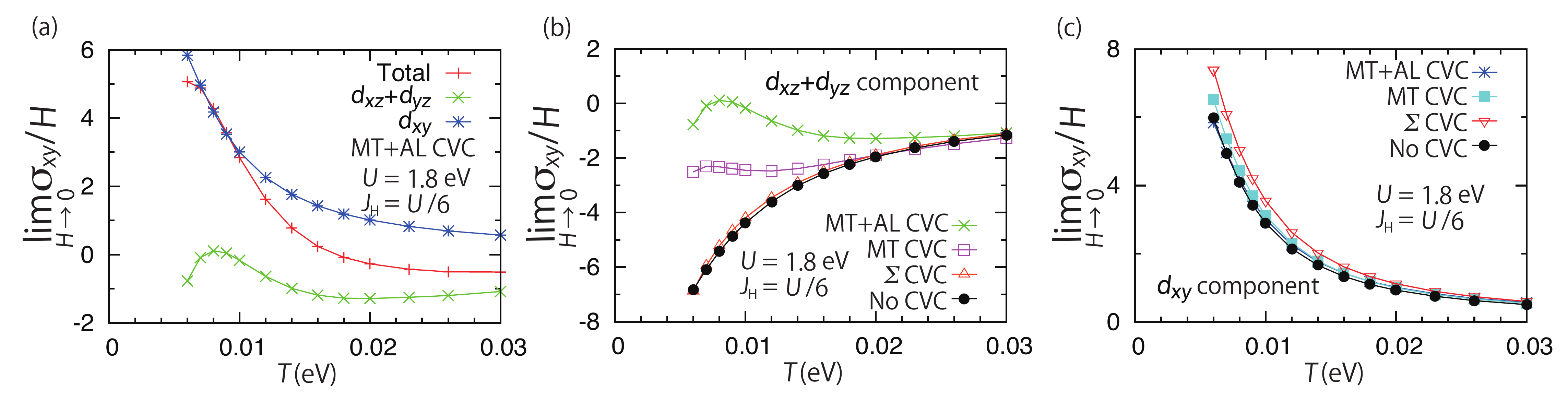}
\end{center}
\vspace{-20pt}
\caption{ 
Temperature dependence of $\textstyle\lim_{H\rightarrow 0}\frac{\sigma_{xy}}{H}$ and 
orbital-decomposed components in the MT$+$AL CVC case at $U=1.8$ eV 
and the orbital-decomposed components of (b) the $d_{xz}$ and $d_{yz}$ orbitals 
and (c) $d_{xy}$ orbital in the four cases at $U=1.8$ eV. 
}
\label{fig:Fig22}
\end{figure*}
\begin{figure}[tb]
\begin{center}
\includegraphics[width=75mm]{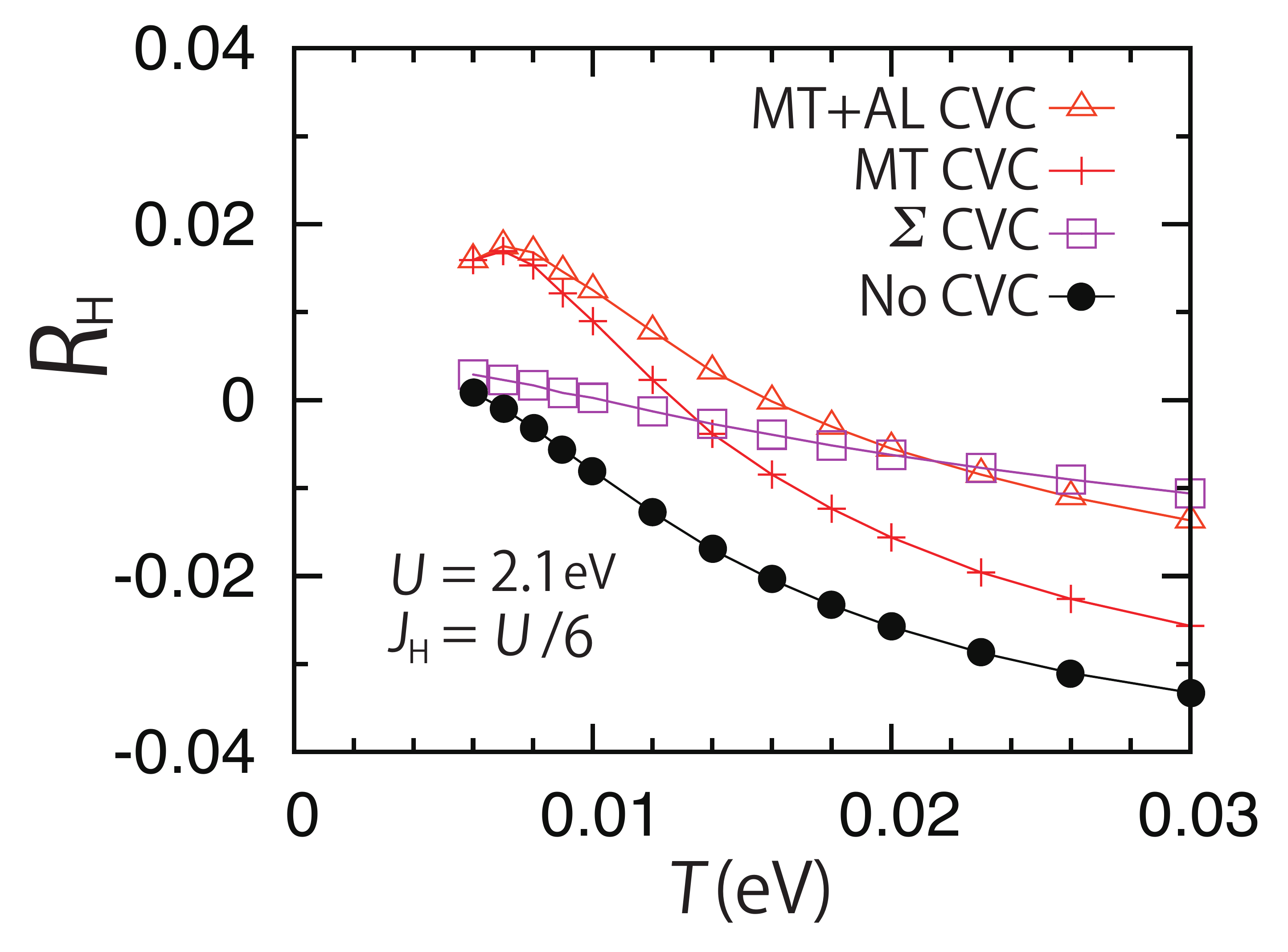}
\end{center}
\vspace{-20pt}
\caption{
Temperature dependence of $R_{\textrm{H}}$ at $U=2.1$ eV and $J_{\textrm{H}}=\frac{U}{6}$ 
in the four cases. 
}
\label{fig:Fig23}
\end{figure}
\begin{figure*}[tb]
\begin{center}
\includegraphics[width=174mm]{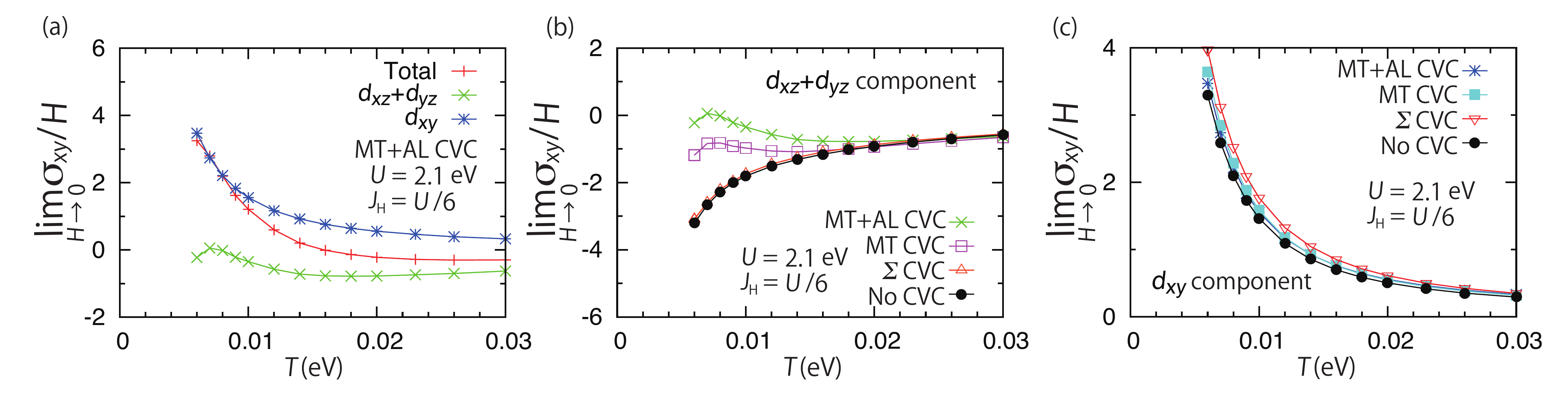}
\end{center}
\vspace{-20pt}
\caption{ 
Temperature dependence of $\textstyle\lim_{H\rightarrow 0}\frac{\sigma_{xy}}{H}$ 
and the orbital-decomposed components 
in the MT$+$AL CVC case at $U=2.1$ eV 
and the orbital-decomposed components of (b) the $d_{xz}$ and $d_{yz}$ orbitals 
and (c) $d_{xy}$ orbital in the four cases at $U=2.1$ eV. 
}
\label{fig:Fig24}
\end{figure*}

We start discussing  
$R_{\textrm{H}}(=\sigma_{xx}^{-2}\textstyle\lim_{H\rightarrow 0}\frac{\sigma_{xy}}{H})$  
away from the AF QCP. 
We show its temperature dependence in the four cases in Fig. \ref{fig:Fig21}, 
and see from that figure three main features in the MT$+$AL CVC case. 
First, 
at high temperatures, 
the values of $R_{\textrm{H}}$ in the MT$+$AL CVC case 
are close to the values in the $\Sigma$ CVC case. 
This origin is the same for $\rho_{ab}$ at high temperatures, 
i.e., the small effects of the MT and the AL CVCs 
due to the large QP damping. 
Second, 
when temperature is low, 
the value of $R_{\textrm{H}}$ in the MT$+$AL CVC case 
becomes almost the same as that in the MT CVC case. 
This result can be understood that 
the main effects of the MT CVC on $R_{\textrm{H}}$ at low temperatures 
remain leading even including the main terms of the AL CVC. 
Its mechanism is as follows: 
As shown in Ref. \onlinecite{NA-CVC}, 
the main effects of the MT CVC on $R_{\textrm{H}}$ 
at low temperatures 
are the decreases of the negative-sign contributions 
of the $d_{xz}+d_{yz}$ component of the transverse conductivity 
around $\boldk=(\frac{21}{32}\pi,\frac{21}{32}\pi)$ 
as a result of the magnitude changes of 
the currents of the $d_{xz/yz}$ orbital 
around $\boldk=(\frac{21}{32}\pi,\frac{21}{32}\pi)$ 
due to the MT CVC arising from 
the spin fluctuations of the $d_{xz/yz}$ orbital around $\boldq=\boldQ_{\textrm{IC-AF}}$. 
Although the currents of the $d_{xz/yz}$ orbital 
around $\boldk=(\frac{21}{32}\pi,\frac{21}{32}\pi)$ 
are affected by the AL1 and AL2 CVCs 
arising from the above spin fluctuations, 
the main effects of the MT CVC persist 
in the MT$+$AL CVC case 
due to the cancellation of those AL1 and AL2 CVCs 
in the presence of the even-parity and the rotational symmetry. 
(The mechanism of this cancellation~\cite{Kon-CVC} was explained in Sect. III C 1.) 
It should be noted that 
we can understand why only for $R_{\textrm{H}}$ 
the main effects of the MT CVC 
survive 
at low temperatures even with the main terms of the AL CVC 
by considering the difference between the important factors 
for $\sigma_{xx}$ and $\textstyle\lim_{H\rightarrow 0}\frac{\sigma_{xy}}{H}$: 
since the important factor for $\sigma_{xx}$ 
is the unrenormalized QP damping, 
the effects of the MT CVC on the currents of the $d_{xz/yz}$ orbital 
around $\boldk=(\frac{21}{32}\pi,\frac{21}{32}\pi)$ 
are little important for $\rho_{ab}$ away from the AF QCP 
due to the large unrenormalized QP damping; 
on the other hand, 
since not only the unrenormalized QP damping 
but also the momentum derivative of the angle of the current 
becomes important for $\textstyle\lim_{H\rightarrow 0}\frac{\sigma_{xy}}{H}$, 
the effects of the MT CVC on 
the currents of the $d_{xz/yz}$ orbital around $\boldk=(\frac{21}{32}\pi,\frac{21}{32}\pi)$ 
become considerable for $R_{\textrm{H}}$ 
due to the large momentum derivative. 
Third, 
three specific features of $R_{\textrm{H}}$ obtained in the MT CVC case 
survive even including the main terms of the AL CVC; 
the three specific features are 
emerging a peak, 
crossing over zero, 
and increasing monotonically in the high-temperature region. 
This can be understood that 
at high temperatures 
the effects of the AL CVC are small due to the large QP damping, 
and that 
the main effects of the MT CVC at low temperatures remain qualitatively unchanged. 

Next, 
we analyze the orbital-decomposed components of 
$\lim\limits_{H\rightarrow 0}\frac{\sigma_{xy}}{H}$ away from the AF QCP 
to determine the role of each $t_{2g}$ orbital. 
Due to the same reason for $\sigma_{xx}$, 
we consider only the $d_{xz}+d_{yz}$ component and $d_{xy}$ component 
of $\textstyle\lim_{H\rightarrow 0}\frac{\sigma_{xy}}{H}$; 
the former and the latter are defined 
as the equations that 
$\textstyle\sum_{\{a\}=1}^{3}$ in Eq. (\ref{eq:sigmaXY-final}) 
are replaced by, respectively, $\textstyle\sum_{\{a\}=1}^{2}$ and $\textstyle\sum_{\{a\}=3}$. 
From Figs. \ref{fig:Fig22}(a){--}\ref{fig:Fig22}(c), 
we find three main properties: 
at high temperatures, 
the effects of all the CVCs 
on those components are very small; 
in the low-temperature region, 
the temperature dependence 
in the MT$+$AL CVC case 
is qualitatively the same for the MT CVC case; 
a peak of the $d_{xz}+d_{yz}$ component survives even with the main terms of the AL CVC. 
Those results can be understood in the similar way for $R_{\textrm{H}}$. 
It should be noted that 
the difference in the important factor is the origin why 
the $d_{xy}$ orbital gives considerable contributions 
to only $\textstyle\lim_{H\rightarrow 0}\frac{\sigma_{xy}}{H}$, 
although its contributions to $\sigma_{xx}$ are negligible. 

Thus, 
we deduce from the results at $U=1.8$ eV 
that 
the qualitative behavior of $R_{\textrm{H}}$ away from the AF QCP 
can be captured by taking account of 
the $\Sigma$ CVC and the MT CVC.

Then, 
we go on to analyze the temperature dependence of $R_{\textrm{H}}$ near the AF QCP. 
The results at $U=2.1$ eV in the four cases are shown in Fig. \ref{fig:Fig23}. 
We see that 
even including the main terms of the AL CVC, 
$R_{\textrm{H}}$ shows three specific features 
(emerging a peak, crossing over zero, 
and increasing monotonically in high-temperature region), 
and that 
$R_{\textrm{H}}$ has two almost distinct regions 
as a function of temperature. 
The former result can be understood in the same way 
as the explanations about the third result of $R_{\textrm{H}}$ away from the AF QCP, 
and the latter can be understood in the same way 
for the similar property of $\rho_{ab}$ near the AF QCP (see Sect. III C 1). 

Moreover, 
we determine the orbital dependence of 
$\textstyle\lim_{H\rightarrow 0}\frac{\sigma_{xy}}{H}$ near the AF QCP 
by the analyses of the temperature dependence of 
the $d_{xz}+d_{yz}$ component and $d_{xy}$ component. 
From Figs. \ref{fig:Fig24}(a){--}\ref{fig:Fig24}(c), 
we see three features the same as those obtained away from the AF QCP. 
Furthermore, 
comparing the results away from and near the AF QCP, 
we find that 
the difference in the $d_{xz}+d_{yz}$ component or $d_{xy}$ component 
between the MT CVC case and the MT$+$AL CVC case 
becomes smaller near the AF QCP than away from the AF QCP. 
This results from the more importance of the MT CVC near the AF QCP. 

Thus, 
the results at $U=2.1$ eV 
show the validity of the qualitative behaviors of $R_{\textrm{H}}$ near the AF QCP 
in the MT CVC case 
and the existence of the two almost distinct regions of the temperature dependence 
of $R_{\textrm{H}}$ near the AF QCP, which are the same for $\rho_{ab}$ near the AF QCP. 

\section{Discussions}
In this section, 
I compare the results obtained away from or near the AF QCP 
with several experimental or theoretical results. 
The discussions in Sec. IV A are for the comparisons with several experiments, 
and the discussions in Sec. IV B are for the comparisons with other theories. 

\subsection{Comparisons with experiments}
In this section, 
we compare the results obtained away from or near the AF QCP 
with several experiments of Sr$_{2}$RuO$_{4}$ or Sr$_{2}$Ru$_{0.975}$Ti$_{0.025}$O$_{4}$, 
respectively, 
and see that 
the obtained results are qualitatively consistent with these experiments. 
In the comparisons with Sr$_{2}$Ru$_{0.975}$Ti$_{0.025}$O$_{4}$, 
I believe that 
the physical origins of several behaviors observed experimentally 
can be deduced by comparison with the results obtained near the AF QCP. 
This is because the main effect of the Ti-substitution 
may be the system approaching towards the AF QCP 
compared with Sr$_{2}$RuO$_{4}$; 
this main effect can be treated by increasing the value of $U$ 
in the model of Sr$_{2}$RuO$_{4}$. 
Although the microscopic mechanism why the Ti-substitution 
causes approaching towards the AF QCP is unclear, 
we can qualitatively understand several differences 
between Sr$_{2}$RuO$_{4}$ and Sr$_{2}$Ru$_{0.975}$Ti$_{0.025}$O$_{4}$ 
as a result of the difference in the distance from the AF QCP, 
as I will show below. 

We begin with the comparisons about the magnetic properties. 
The enhancement of the spin susceptibility at $\boldq=\boldQ_{\textrm{IC-AF}}$ 
away from the AF QCP agrees with the neutron~\cite{Neutron-x2} or 
the nuclear-magnetic-resonance (NMR)~\cite{Ishida-NMR-x2} measurement 
of Sr$_{2}$RuO$_{4}$, 
and the similar enhancement near the AF QCP 
is consistent with the neutron~\cite{Neutron-Ti214} or 
the NMR~\cite{Ishida-NMR-x05} measurement 
of Sr$_{2}$Ru$_{0.975}$Ti$_{0.025}$O$_{4}$. 
Also, 
no sizable commensurate ferromagnetic spin fluctuation 
obtained away from the AF QCP 
is in agreement with the neutron measurement~\cite{Neutron-x2} of Sr$_{2}$RuO$_{4}$. 
Thus, 
several magnetic properties can be well described in the FLEX approximation, 
as explained in Ref. \onlinecite{NA-CVC}. 
It should be noted that 
to discuss the anisotropy between the inplane and the out-of-plane 
spin susceptibilities, 
the spin-orbit interaction is necessary~\cite{Yanase-Ru}. 

Then, 
we turn to the comparisons about the electronic structure. 
As discussed in Ref. \onlinecite{NA-CVC}, 
the larger mass enhancement for the $d_{xy}$ orbital 
than for the $d_{xz/yz}$ orbital 
away from the AF QCP 
is consistent with an experiment~\cite{dHvA-x2,Mackenzie-review} in Sr$_{2}$RuO$_{4}$. 
In addition, 
the topology of the FS away from the AF QCP 
agrees with the measurement of the dHvA effect~\cite{dHvA-x2} 
or the angle-resolved photoemission spectroscopy~\cite{ARPES-x2}. 
However, 
there is a quantitative difference in the location of the FS sheet of the $d_{xy}$ orbital 
near $\boldk=(\frac{2}{3}\pi,\frac{2}{3}\pi)$: 
in my result, 
that FS sheet is very close to the inner FS sheet of the $d_{xz/yz}$ orbital 
[see Fig. \ref{fig:Fig12}(a)]; 
in the experiments~\cite{dHvA-x2,ARPES-x2}, 
that FS sheet is not very close to the inner FS sheet. 
That difference exists even in the LDA~\cite{Oguchi,Mazin-LDA}, 
as described in Ref. \onlinecite{NA-CVC}. 
To improve that difference, 
the spin-orbit interaction of the Ru ions 
will be necessary 
since the small spin-orbit interaction leads to the weak hybridization 
of the bands near $\boldk=(\frac{2}{3}\pi,\frac{2}{3}\pi)$~\cite{Oguchi-LS}. 
Actually, 
that difference is improved 
in the local-spin-density approximation~\cite{Oguchi-LS} 
including the spin-orbit interaction. 
Although that result indicates the importance of the spin-orbit interaction 
for quantitative discussions, 
the qualitative agreement of the FS is meaningful 
since that qualitative agreement suggests that 
the present theory can capture the aspects of many-body effects on the FS 
of Sr$_{2}$RuO$_{4}$ at least on a qualitative level. 
In addition, 
that qualitative agreement supports the suitability of the expectation 
that the electronic structure of Sr$_{2}$RuO$_{4}$ in the LDA~\cite{Oguchi,Mazin-LDA} 
may be regarded as a good starting point to include many-body effects. 
This is another meaningful aspect of that qualitative agreement 
since electron correlation sometimes modifies the FS drastically, 
resulting in the deviation of the FS from the experiment 
even on a qualitative level~\cite{Ikeda-FLEX}. 
From those comparisons, 
we deduce that 
the FLEX approximation 
can qualitatively well describe the electronic structure of Sr$_{2}$RuO$_{4}$. 

Finally, 
I compare the obtained results with the experimental results about the transport properties. 
As described in Ref. \onlinecite{NA-CVC}, 
the $T^{2}$ dependence of $\rho_{ab}$ at low temperatures 
and the non-monotonic temperature dependence of $R_{\textrm{H}}$ 
away from the AF QCP 
are qualitatively consistent with the experiments~\cite{resistivity-x2,Hall-x2} of 
$\rho_{ab}$ and $R_{\textrm{H}}$ of Sr$_{2}$RuO$_{4}$, 
and the $T$-linear $\rho_{ab}$ near the AF QCP can qualitatively explain 
the experimental result~\cite{Ti214-nFL2} in Sr$_{2}$Ru$_{0.975}$Ti$_{0.025}$O$_{4}$. 
(The detail of the temperature dependence of $R_{\textrm{H}}$ in Sr$_{2}$RuO$_{4}$ 
was described in Sec. I.) 
Although the spin-orbit interaction generally leads to 
an additional contribution to $R_{\textrm{H}}$ 
through the anomalous Hall effect~\cite{Kon-AHE}, 
it has been experimentally confirmed that 
such contribution is small~\cite{CSRO-nFL2}. 
Thus, neglecting the spin-orbit interaction will be sufficient 
for at least qualitative discussions about $R_{\textrm{H}}$. 
Combining those discussions, 
we find that 
the successful descriptions of the transport properties 
qualitatively hold 
in the FLEX approximation with 
the $\Sigma$ CVC, the MT CVC, and the main terms of the AL CVC.

\subsection{Comparisons with other theories}
In this section, 
we compare the results of this paper with other theoretical studies 
and show several better points of this theory. 
First, 
we focus on the comparisons with the DMFT~\cite{Haule-DMFT,Georges-DMFT-Ru} 
for a model of Sr$_{2}$RuO$_{4}$, 
and show that 
several electronic properties can be better described in the method I used 
due to a sufficient treatment of the spatially modulated spin fluctuations. 
Then, 
we compare my results with the transport properties of Sr$_{2}$RuO$_{4}$ 
obtained in the phenomenological Boltzmann theory 
within the relaxation-time approximation~\cite{Hall-theory-x2}, 
and point out the importance of the MT CVC arising from the spin fluctuations 
of the $d_{xz/yz}$ orbital in order to naturally obtain 
the non-monotonic temperature dependence of $R_{\textrm{H}}$ 
without any \textit{ad hoc} parameters of the unrenormalized QP damping. 
Finally, we discuss the similarities and differences between 
the main effects of the AL CVC on the charge transports 
in my case and the case~\cite{Kon-CVC} of the single-orbital Hubbard model near an AF QCP 
on a square lattice, 
and conclude that 
the existence of the two almost distinct regions 
of the charge transports near an AF QCP as a function of temperature 
is one of the important findings of this paper. 

We begin with the comparisons about the magnetic properties, 
the orbital dependence of the mass enhancement, 
and the modification of the FS sheets due to electron correlation 
in the DMFT~\cite{Haule-DMFT,Georges-DMFT-Ru} for a model of Sr$_{2}$RuO$_{4}$ 
in more detail than Ref. \onlinecite{NA-CVC}. 
First, in the DMFT, 
spatial correlations, 
the momentum dependent fluctuations, 
are completely neglected~\cite{Metzner}, 
and the effects of only local correlations 
are taken into account~\cite{Georges-review,Kotliar-review}. 
On the other hand, 
I have shown that 
not local but spatially modulated spin fluctuation with $\boldq=\boldQ_{\textrm{IC-AF}}$ 
is important to discuss the magnetic properties of Sr$_{2}$RuO$_{4}$. 
As explained in Sect. IV A, 
the neutron~\cite{Neutron-x2} and the NMR~\cite{Ishida-NMR-x2} measurement for Sr$_{2}$RuO$_{4}$ 
have observed the enhancement of that spatially modulated spin fluctuation. 
Thus, the magnetic properties can be better described in my case. 
Second, 
in the DMFT, 
the locations of the FS sheets are more drastically modified 
due to electron correlation than in my case: 
on $k_{x}=k_{y}$ line, 
the $\gamma$ sheet of the $d_{xy}$ orbital becomes the most inner sheet 
in the DMFT~\cite{Georges-DMFT-Ru}, 
while the most inner sheet in the LDA~\cite{Oguchi,Mazin-LDA} or my case 
is the $\beta$ sheet of the $d_{xz/yz}$ orbital. 
Since the experimental results~\cite{dHvA-x2,ARPES-x2} are consistent with the result 
in the LDA~\cite{Oguchi,Mazin-LDA} and my case, 
the agreement about the FS is better in my case than in the DMFT~\cite{Georges-DMFT-Ru}. 
Third, 
in the DMFT, 
the mass enhancement of the $d_{xy}$ orbital is larger than 
that of the $d_{xz/yz}$ orbital for finite values of $J_{\textrm{H}}$~\cite{Haule-DMFT}; 
e.g., 
the former and the latter become $3.2$ and $2.4$ 
at $U=2.3$ eV and $J_{\textrm{H}}=0.3$ eV ($\sim 0.13U$) 
or $4.5$ and $3.3$ at $U=2.3$ eV 
and $J_{\textrm{H}}=0.4$ eV ($\sim 0.174U$). 
Thus, 
the larger mass enhancement for the $d_{xy}$ orbital than for the $d_{xz/yz}$ orbital 
is obtained in both the DMFT~\cite{Haule-DMFT} and the FLEX approximation, 
although 
the large mass enhancement of the $d_{xy}$ orbital 
is realized in a wider region of the parameters space in my case 
than in the DMFT~\cite{Haule-DMFT}; 
in particular, 
that larger mass enhancement of the $d_{xy}$ orbital 
is obtained even at $J_{\textrm{H}}=0$ eV in my case~\cite{NA-CVC}. 
It should be noted that 
it is important and necessary to check 
whether spatial correlations, neglected in the DMFT~\cite{Metzner}, 
keep 
the orbital dependence of the mass enhancement qualitatively the same 
since the spatial correlations sometimes 
drastically change the results obtained in the DMFT 
(e.g., see case~\cite{cDMFT-Kotliar} of a single-orbital Hubbard model on a square lattice). 
Actually, even in a two-orbital Hubbard model~\cite{cDMFT-Arita} 
on a square lattice, 
spatial correlations included by considering a cluster 
cause the almost perfect disappearance of 
the unusual $J_{\textrm{H}}$ dependence of a critical value of $U$ for a Mott transition, 
which is obtained in the DMFT.  
Combining the discussions of this paragraph, 
I conclude that 
several electronic properties of Sr$_{2}$RuO$_{4}$ 
can be better described in my case 
than in the DMFT~\cite{Haule-DMFT,Georges-DMFT-Ru}. 
In particular, 
it is significant to find the importance of the MT CVC for obtaining 
the non-monotonic temperature dependence of $R_{\textrm{H}}$ of Sr$_{2}$RuO$_{4}$ 
in this paper 
since in the DMFT the CVCs are neglected 
due to lack of the momentum dependence of the self-energy~\cite{DMFT-CVC1,DMFT-CVC2}. 

I turn to the comparisons about the transport properties of Sr$_{2}$RuO$_{4}$ 
with the phenomenological Boltzmann theory 
in the relaxation-time approximation~\cite{Hall-theory-x2}, 
in which all the CVCs are neglected~\cite{Kon-review}. 
In the relaxation-time approximation for Sr$_{2}$RuO$_{4}$, 
the unrenormalized QP damping, $\tau_{a}^{-1}$, 
is given by $\tau_{a}^{-1}=\eta_{a}+\alpha_{a}T^{2}$ 
with the ad-hoc parameters $\eta_{a}$ and $\alpha_{a}$, 
which are chosen as 
$\eta_{d_{xz}}=\eta_{d_{yz}}=2.75$, $\eta_{d_{xy}}=3.25$, 
$\alpha_{d_{xz}}=0.035$, $\alpha_{d_{yz}}=0.04$, and $\alpha_{d_{xy}}=0.06$; 
this form of $\tau_{a}^{-1}$ as a function of temperature 
is typical of the FL~\cite{AGD,NA-review,Morel-Nozieres}. 
In addition, 
only for the calculation of the in-plane resistivity, 
the authors of Ref. \onlinecite{Hall-theory-x2} added $0.6T$ to $\tau_{d_{xy}}^{-1}$ 
since they assumed that 
Sr$_{2}$RuO$_{4}$ were close to a ferromagnetic instability. 
Although that assumption is experimentally incorrect~\cite{Neutron-x2,Ishida-NMR-x2}, 
their results at low temperatures will remain qualitatively unchanged 
since the contribution from the $d_{xz/yz}$ orbital 
is more important than that 
from the $d_{xy}$ orbital due to the smaller $\tau_{a}^{-1}$ of the $d_{xz/yz}$ orbital. 
Adopting the phenomenological Boltzmann theory in the relaxation-time approximation 
with those expressions of the unrenormalized QP damping to $\rho_{ab}$ and $R_{\textrm{H}}$, 
the authors of Ref. \onlinecite{Hall-theory-x2} 
obtained the $T^{2}$ dependence of $\rho_{ab}$ at low temperatures 
and the non-monotonic temperature dependence of $R_{\textrm{H}}$, 
which are consistent with the experiments~\cite{resistivity-x2,Hall-x2}. 
However, 
as they pointed out in Ref. \onlinecite{Hall-theory-x2}, 
the result of $R_{\textrm{H}}$ 
is very sensitive to the small relative variation of $\alpha_{d_{xz}}$ and $\alpha_{d_{yz}}$, 
and the sign change of $R_{\textrm{H}}$ disappears in some cases. 
In addition, 
$\alpha_{d_{xz}}$ and $\alpha_{d_{yz}}$ should be the same 
due to the tetragonal symmetry of the crystal. 
Thus, 
although the results obtained in the relaxation-time approximation~\cite{Hall-theory-x2} 
seem to be reasonable, 
the validity of the choice of the \textit{ad hoc} parameters is unclear. 
On the other hand, 
I have shown in Ref. \onlinecite{NA-CVC} without any \textit{ad hoc} parameters 
of the unrenormalized QP damping that 
the MT CVC arising from the spin fluctuations of the $d_{xz/yz}$ orbital 
is essential to obtain the non-monotonic temperature dependence of $R_{\textrm{H}}$. 
Furthermore, 
in this paper, I show that 
the importance of that MT CVC remains unchanged even if we consider 
the main terms of the AL CVC. 
From those arguments, 
I propose the importance of the MT CVC arising from the spin fluctuations 
of the $d_{xz/yz}$ orbital to understand the temperature dependence of $R_{\textrm{H}}$ 
of Sr$_{2}$RuO$_{4}$. 

I close this section with the comparisons about the main effects 
of the AL CVC with the case~\cite{Kon-CVC} of the single-orbital Hubbard model 
near an AF QCP on a square lattice. 
In that single-orbital case, 
the authors of Ref. \onlinecite{Kon-CVC} analytically or numerically 
studied the effects of the AL CVC on $\rho_{ab}$ and $R_{\textrm{H}}$ 
near the AF QCP where the spin susceptibility at $\boldq=(\pi,\pi)$ 
was most strongly enhanced. 
Then, 
their analytic study revealed the cancellation of the leading contributions 
in the AL$1$ or the AL$2$ CVC near the AF QCP, 
whose details were explained in Sect. III C 1, 
and their numerical study 
considering only the contributions of the MT CVC and the AL CVC from the states 
on the Fermi level revealed the qualitative validity of the results 
obtained without the AL CVC. 
However, 
due to neglecting the other contributions near the Fermi level, 
they did not obtain the existence of the two almost distinct regions 
of the transport properties near the AF QCP as a function of temperature, 
which is revealed in this paper. 
It should be noted that 
I obtained the similar results~\cite{Kon-CVC} to those they had obtained, 
a decrease of the value of the inplane resistivity due to the AL CVC 
and a larger decrease of that at high temperatures than at low temperatures 
(see Fig. \ref{fig:Fig19}). 
Thus, 
the aspects of the AL CVC in my case are qualitatively 
consistent with those in the single-orbital case~\cite{Kon-CVC}. 
In addition, 
it is one of the important findings 
of this paper to reveal the existence of the two almost distinct regions 
of the charge transports near an AF QCP. 

\section{Summary}
In summary, 
after explaining the formal derivations of $\rho_{ab}$, $R_{\textrm{H}}$, 
and the FLEX approximation with the $\Sigma$ CVC, the MT CVC, and the AL CVC, 
I studied $\rho_{ab}$ and $R_{\textrm{H}}$ for a $t_{2g}$-orbital Hubbard model 
in a PM state near or away from the AF QCP on a square lattice 
in the FLEX approximation with the $\Sigma$ CVC, the MT CVC, 
and the main terms of the AL CVC, 
and then found the three main results about many-body effects. 
The first main result is showing that 
the results of the previous studies~\cite{NA-CVC,NA-review} 
remain qualitatively unchanged even with the main terms of the AL CVC. 
This indicates the qualitative validity of the arguments 
in the previous studies~\cite{NA-CVC,NA-review}. 
The second main result is finding 
the two almost distinct regions of the charge transports near the AF QCP: 
$\rho_{ab}$ and $R_{\textrm{H}}$ in the high-temperature region 
are described by including only the $\Sigma$ CVC, 
while the $\Sigma$ CVC and the MT CVC are necessary 
for their descriptions in low-temperature region. 
The third main result is clarifying the difference of the effects of the MT CVC 
between $\rho_{ab}$ and $R_{\textrm{H}}$ away from the AF QCP at low temperatures: 
the MT CVC leads to the considerable effect only on the latter, 
although at high temperatures 
only the $\Sigma$ CVC affects $\rho_{ab}$ and $R_{\textrm{H}}$. 
Thus, 
the second and third main results highlight 
the important aspects of many-body effects on the charge transports. 
I also showed several results of the magnetic properties and the electronic structure 
for that model in the FLEX approximation. 
Those support a deeper understanding than in the previous studies~\cite{NA-CVC,NA-review}. 
Then, comparing the obtained results with several experiments, 
I achieved the qualitative agreement about 
the momentum dependence of the spin fluctuation 
in Sr$_{2}$RuO$_{4}$~\cite{Neutron-x2,Ishida-NMR-x2} 
or Sr$_{2}$Ru$_{0.975}$Ti$_{0.025}$O$_{4}$~\cite{Neutron-Ti214,Ishida-NMR-x05}, 
the orbital dependence of the mass enhancement in Sr$_{2}$RuO$_{4}$~\cite{dHvA-x2,Mackenzie-review}, 
the topology of the FS of Sr$_{2}$RuO$_{4}$~\cite{dHvA-x2,ARPES-x2}, 
the $T^{2}$ dependence~\cite{resistivity-x2} of $\rho_{ab}$ at low temperatures 
in Sr$_{2}$RuO$_{4}$, 
the non-monotonic temperature dependence~\cite{Hall-x2} of $R_{\textrm{H}}$ in Sr$_{2}$RuO$_{4}$, 
and the $T$-linear $\rho_{ab}$ in Sr$_{2}$Ru$_{0.975}$Ti$_{0.025}$O$_{4}$~\cite{Ti214-nFL2}. 
Furthermore, 
by the comparisons with other theories, 
I showed several stronger points 
to discuss the electronic properties of Sr$_{2}$RuO$_{4}$ 
than other theories~\cite{Haule-DMFT,Georges-DMFT-Ru,Hall-theory-x2}, 
and clarified the similarities and differences of the main effects of the AL CVC 
between the present multiorbital case and 
a single-orbital case~\cite{Kon-CVC}. 

Several important issues remain for further study. 
One is the extension of the present method 
to case with the spin-orbit interaction. 
In particular, 
it is desirable to study 
the anisotropy between the inplane and the out-of-plane susceptibility, 
the quantitative effects on the FS deformation including many-body effects, 
and the charge Hall effect including both the usual Hall effect 
and the anomalous Hall effect. 
It is also important to extend the present method to case~\cite{CSRO-nFL1,CSRO-nFL2} 
with the RuO$_{6}$ distortions near the ferromagnetic QCP 
since its results and the present results 
lead to deep understanding of the similarities and differences 
between many-body effects on the charge transports 
near the ferromagnetic and the AF QCPs. 
Furthermore, 
the study about the charge transports of the $3$D ruthenates~\cite{Ru113} 
using the extended method 
is useful to clarify the role of the dimensionality 
in the charge transports of a correlated multiorbital system. 
Then, 
another important issue is the extended study 
about the charge or heat transports of Sr$_{2}$RuO$_{4}$ 
in a superconducting phase~\cite{SC-trans1,SC-trans2} 
since it may clarify the role of each $t_{2g}$ orbital 
in the superconducting phase. 
Finally, 
it is challenging and important to study 
the charge transports of other transition-metal oxides~\cite{MIT-review} or 
the heavy-fermion materials such as CeCoIn$_{5}$~\cite{Ce115} and UPt$_{3}$~\cite{UPt3} 
or the organic conductors~\cite{organic-review} 
by extending the present method and 
adopting that to other multiorbital Hubbard models or the multiorbital Anderson models. 
Their achievements may provide deeper knowledge 
about ubiquitous or characteristic properties of many-body effects 
on the charge transports of a correlated multiorbital system.

\appendix
\section{Derivation of Eq. (\ref{eq:sum-analytic})}

In this appendix, 
we derive Eq. (\ref{eq:sum-analytic}) from Eq. (\ref{eq:Ktild}). 
In this derivation, 
we use the analytic properties of the single-particle Green's function 
and reducible four-point vertex function in terms of frequency variables. 
The former becomes singular for $\textrm{Im}\epsilon=0$ 
with $\epsilon$ being its frequency variable. 
The analytic property of the latter is the same for 
the two-particle Green's function~\cite{Eliashberg-theory}: 
it becomes singular 
when the frequency variables $\epsilon$, $\epsilon^{\prime}$, and $\omega$ satisfy 
$\textrm{Im}\epsilon = 0$ or  
$(\textrm{Im}\epsilon+\textrm{Im}\omega) = 0$ or 
$\textrm{Im}\epsilon^{\prime} = 0$ or 
$(\textrm{Im}\epsilon^{\prime}+\textrm{Im}\omega) = 0$ or 
$(\textrm{Im}\epsilon+\textrm{Im}\epsilon^{\prime}+\textrm{Im}\omega) = 0$ or 
$(\textrm{Im}\epsilon-\textrm{Im}\epsilon^{\prime}) = 0$, 
where 
we consider case for $\textrm{Im}\omega > 0$ 
to take $i\Omega_{n}\rightarrow \omega+i0+$. 
As a result, 
there are $16$ possibilities of the four-point vertex function 
in real-frequency representation, 
as shown in Table \ref{tab:4ptVF}.
\begin{table*}[tb]
\caption{Relation between the additional subscripts of the four-point vertex function 
and the inequalities of its frequency variables. 
Because of $\textrm{Im}\omega > 0$ and 
the analytic properties of the four-point vertex function, 
there are $16$ possibilities. 
}
\label{tab:4ptVF}
\vspace{5pt}
\begin{tabular}{ccccccc} \toprule\\[-10pt]
\multicolumn{1}{c}{ \ \ \ region \ \ \ } &
\multicolumn{1}{c}{ \ \ \ $\textrm{Im}\epsilon$ \ \ \ } & 
\multicolumn{1}{c}{ \ \ \ $\textrm{Im}\epsilon+\textrm{Im}\omega$ \ \ \ } &
\multicolumn{1}{c}{ \ \ \ $\textrm{Im}\epsilon^{\prime}$ \ \ \ } & 
\multicolumn{1}{c}{ \ \ \ $\textrm{Im}\epsilon^{\prime}+\textrm{Im}\omega$ \ \ \ } &  
\multicolumn{1}{c}{ \ \ \ $\textrm{Im}\epsilon+\textrm{Im}\epsilon^{\prime}+\textrm{Im}\omega$ \ \ \ } &
\multicolumn{1}{c}{ \ \ \ $\textrm{Im}\epsilon-\textrm{Im}\epsilon^{\prime}$ \ \ \ } 
\\[1pt] \hline\\[-8pt]
11-I&$>0$&$>0$&$>0$&$>0$&$>0$&$>0$\\[3pt]
11-II&$>0$&$>0$&$>0$&$>0$&$>0$&$<0$\\[3pt]
21&$<0$&$>0$&$>0$&$>0$&$>0$&$<0$\\[3pt]
31-II&$<0$&$<0$&$>0$&$>0$&$>0$&$<0$\\[3pt]
31-I&$<0$&$<0$&$>0$&$>0$&$<0$&$<0$\\[3pt]
32&$<0$&$<0$&$<0$&$>0$&$<0$&$<0$\\[3pt]
33-I&$<0$&$<0$&$<0$&$<0$&$<0$&$<0$\\[3pt]
33-II&$<0$&$<0$&$<0$&$<0$&$<0$&$>0$\\[3pt]
23&$<0$&$>0$&$<0$&$<0$&$<0$&$>0$\\[3pt]
13-II&$>0$&$>0$&$<0$&$<0$&$<0$&$>0$\\[3pt]
13-I&$>0$&$>0$&$<0$&$<0$&$>0$&$>0$\\[3pt]
12&$>0$&$>0$&$<0$&$>0$&$>0$&$>0$\\[3pt]
22-III&$<0$&$>0$&$<0$&$>0$&$>0$&$>0$\\[3pt]
22-II&$<0$&$>0$&$<0$&$>0$&$>0$&$<0$\\[3pt]
22-I&$<0$&$>0$&$<0$&$>0$&$<0$&$<0$\\[3pt]
22-IV&$<0$&$>0$&$<0$&$>0$&$<0$&$>0$\\[3pt]
 \toprule 
\end{tabular}
\end{table*}
\begin{figure}[tb]
\begin{center}
\includegraphics[width=86mm]{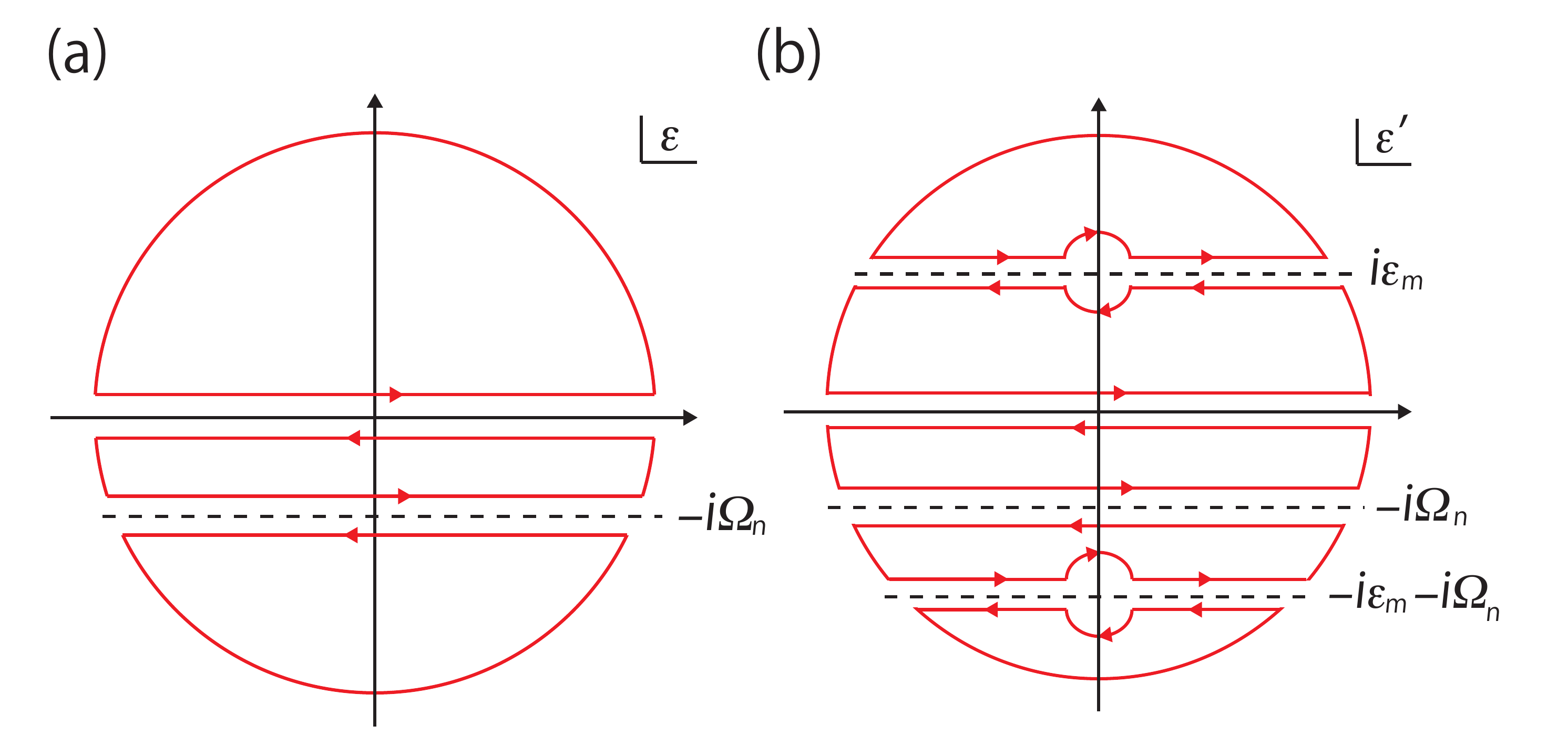}
\end{center}
\vspace{-20pt}
\caption{
Contours used for the analytic continuations of Eq. (\ref{eq:Ktild}). 
Contour shown in panel (b) is one of the possibilities.}
\label{fig:Fig2}
\end{figure}

Using those analytic properties, 
we can carry out the analytic continuations of 
the first and the second terms of Eq. (\ref{eq:Ktild}). 
This is because 
we can rewrite the summation about the Matsubara frequency 
as the corresponding contour integral~\cite{AGD}. 
Using the contour $C$ shown in Fig. \ref{fig:Fig2}(a), 
we can carry out the analytic continuation for the first term: 
\begin{align}
&-\frac{T}{N}
\sum\limits_{\boldk}
\sum\limits_{m}
\sum\limits_{\{a\}}
(v_{\boldk \nu})_{ba}
(v_{\boldk \nu})_{cd}
G_{ac}(\boldk,i\epsilon_{m+n})
G_{db}(\boldk,i\epsilon_{m})\notag\\
&=
-\frac{1}{N}
\sum\limits_{\boldk}
\int_{\textrm{C}}\frac{d\epsilon}{4\pi i}
\tanh \frac{\epsilon}{2T}
\sum\limits_{\{a\}}
(v_{\boldk \nu})_{ba}
(v_{\boldk \nu})_{cd}\notag\\
&\times 
G_{ac}(\boldk,\epsilon+i\Omega_{n})
G_{db}(\boldk,\epsilon)\notag\\
&\rightarrow 
-\frac{1}{N}
\sum\limits_{\boldk}
\int_{-\infty}^{\infty}\frac{d\epsilon}{4\pi i}
\sum\limits_{\{a\}}
(v_{\boldk \nu})_{ba}(v_{\boldk \nu})_{cd}\notag\\
&\times
\Bigl[
\tanh \frac{\epsilon}{2T}g_{1;acdb}(k;\omega)\notag\\
&+\Bigl(\tanh \dfrac{\epsilon+\omega}{2T}-\tanh \dfrac{\epsilon}{2T}\Bigr)
g_{2;acdb}(k;\omega)\notag\\
&-\tanh \dfrac{\epsilon+\omega}{2T}g_{3;acdb}(k;\omega)
\Bigr],\label{eq:1st-analytic}
\end{align}
where $\rightarrow$ represents the replacement of $i\Omega_{n}$ by $\omega+i0+$, 
and $g_{l;acdb}(k;\omega)$ are 
\begin{align}
g_{1;acdb}(k;\omega)=&\
G_{ac}^{(\textrm{R})}(\boldk,\epsilon+\omega)
G_{db}^{(\textrm{R})}(\boldk,\epsilon),\label{eq:g1}\\
g_{2;acdb}(k;\omega)=&\
G_{ac}^{(\textrm{R})}(\boldk,\epsilon+\omega)
G_{db}^{(\textrm{A})}(\boldk,\epsilon),\label{eq:g2}
\end{align}
and
\begin{align}
g_{3;acdb}(k;\omega)=&\
G_{ac}^{(\textrm{A})}(\boldk,\epsilon+\omega)
G_{db}^{(\textrm{A})}(\boldk,\epsilon).\label{eq:g3}
\end{align}
Next, 
by replacing $T\sum_{m}$ and $T\sum_{m^{\prime}}$
in the second term of Eq. (\ref{eq:Ktild}) 
by the contour integrals with 
the contour $C$ in Fig. \ref{fig:Fig2}(a) and 
the contour $C^{\prime}$ such as Fig. \ref{fig:Fig2}(b), respectively, 
we can similarly carry out the analytic continuation of the second term:
\begin{widetext} 
\begin{align}
&-\frac{T^{2}}{N^{2}}
\sum\limits_{\boldk,\boldk^{\prime}}
\sum\limits_{m,m^{\prime}}
\sum\limits_{\{a\}}
\sum\limits_{\{A\}}
(v_{\boldk \nu})_{ba}
(v_{\boldk^{\prime} \nu})_{cd}
G_{aA}(\boldk,i\epsilon_{m+n})
G_{dD}(\boldk^{\prime},i\epsilon_{m^{\prime}})
\Gamma_{\{A\}}(\boldk,i\epsilon_{m},\boldk^{\prime},i\epsilon_{m^{\prime}};\boldzero,i\Omega_{n})\notag\\
&\ \ \ \times 
G_{Bb}(\boldk,i\epsilon_{m})
G_{Cc}(\boldk^{\prime},i\epsilon_{m^{\prime}+n})\notag\\
=&
-\frac{1}{N^{2}}
\sum\limits_{\boldk,\boldk^{\prime}}
\int_{\textrm{C}}\frac{d\epsilon}{4\pi i}
\tanh \frac{\epsilon}{2T}
\sum\limits_{\{a\}}
\sum\limits_{\{A \}}
(v_{\boldk \nu})_{ba}
(v_{\boldk^{\prime} \nu})_{cd}
G_{aA}(\boldk,\epsilon+i\Omega_{n})G_{Bb}(\boldk,\epsilon)
\notag\\
&\  
\times \Bigl[\int_{\textrm{C}^{\prime}}\frac{d\epsilon^{\prime}}{4\pi i}
\tanh \frac{\epsilon^{\prime}}{2T}
G_{dD}(\boldk^{\prime},\epsilon^{\prime})
\Gamma_{\{A\}}(\boldk,\epsilon,\boldk^{\prime},\epsilon^{\prime}; \boldzero,i\Omega_{n})
G_{Cc}(\boldk^{\prime},\epsilon^{\prime}+i\Omega_{n})\notag\\
&\ \ \ \ 
+TG_{dD}(\boldk^{\prime},\epsilon)
\Gamma_{\{A\}}(\boldk,\epsilon,\boldk^{\prime},\epsilon;\boldzero,i\Omega_{n})
G_{Cc}(\boldk^{\prime},\epsilon+i\Omega_{n})
\notag\\
&\ \ \ \ 
+TG_{dD}(\boldk^{\prime},-\epsilon-i\Omega_{n})
\Gamma_{\{A\}}(\boldk,\epsilon,\boldk^{\prime},-\epsilon-i\Omega_{n};\boldzero,i\Omega_{n})
G_{Cc}(\boldk^{\prime},-\epsilon)
\Bigr]\notag\\
\rightarrow 
&-\frac{1}{N^{2}}
\sum\limits_{\boldk,\boldk^{\prime}}
\sum\limits_{\{a\}}
(v_{\boldk \nu})_{ba}
(v_{\boldk^{\prime} \nu})_{cd}
\sum\limits_{\{A \}}
\int^{\infty}_{-\infty}\frac{d\epsilon}{4\pi i}
\int^{\infty}_{-\infty}\frac{d\epsilon^{\prime}}{4\pi i}
\coth \dfrac{\epsilon^{\prime}-\epsilon}{2T}\notag\\
&\times 
\Bigl\{\Bigl(\tanh \dfrac{\epsilon+\omega}{2T}
-\tanh \dfrac{\epsilon}{2T}\Bigr)
g_{2;aABb}(k;\omega)g_{2;CcdD}(k^{\prime};\omega)
\Bigl[
\Gamma_{22\textrm{-II};\{A\}}(k,k^{\prime};\omega)-\Gamma_{22\textrm{-III};\{A\}}(k,k^{\prime};\omega)
\Bigr]\notag\\
&-\tanh \dfrac{\epsilon+\omega}{2T}
g_{3;aABb}(k;\omega)g_{3;CcdD}(k^{\prime};\omega)
\Bigl[
\Gamma_{33\textrm{-I};\{A\}}(k,k^{\prime};\omega)-\Gamma_{33\textrm{-II};\{A\}}(k,k^{\prime};\omega)
\Bigr]\notag\\
&+\tanh \dfrac{\epsilon}{2T}
g_{1;aABb}(k;\omega)g_{1;CcdD}(k^{\prime};\omega)
\Bigl[
\Gamma_{11\textrm{-II};\{A\}}(k,k^{\prime};\omega)-\Gamma_{11\textrm{-I};\{A\}}(k,k^{\prime};\omega)
\Bigr]\Bigr\}\notag\\
&-\frac{1}{N^{2}}
\sum\limits_{\boldk,\boldk^{\prime}}
\sum\limits_{\{a\}}
(v_{\boldk \nu})_{ba}
(v_{\boldk^{\prime} \nu})_{cd}
\sum\limits_{\{A \}}
\int^{\infty}_{-\infty}\dfrac{d\epsilon}{4\pi i}
\int^{\infty}_{-\infty}\dfrac{d\epsilon^{\prime}}{4\pi i}
\tanh \dfrac{\epsilon^{\prime}}{2T}\notag\\
&\times 
\Bigl\{\Bigl(\tanh \dfrac{\epsilon+\omega}{2T}
-\tanh \dfrac{\epsilon}{2T}\Bigr)
g_{2;aABb}(k;\omega)
\Bigl[
\Gamma_{21;\{A\}}(k,k^{\prime};\omega)g_{1;CcdD}(k^{\prime};\omega)
-\Gamma_{22\textrm{-II};\{A\}}(k,k^{\prime};\omega)g_{2;CcdD}(k^{\prime};\omega)
\Bigr]\notag\\
&-\tanh \dfrac{\epsilon+\omega}{2T}
g_{3;aABb}(k;\omega)
\Bigl[
\Gamma_{31\textrm{-I};\{A\}}(k,k^{\prime};\omega)g_{1;CcdD}(k^{\prime};\omega)
-\Gamma_{32;\{A\}}(k,k^{\prime};\omega)g_{2;CcdD}(k^{\prime};\omega)
\Bigr]\notag\\
&+\tanh \frac{\epsilon}{2T}
g_{1;aABb}(k;\omega)
\Bigl[
\Gamma_{11\textrm{-I};\{A\}}(k,k^{\prime};\omega)g_{1;CcdD}(k^{\prime};\omega)
-\Gamma_{12;\{A\}}(k,k^{\prime};\omega)g_{2;CcdD}(k^{\prime};\omega)
\Bigr] \Bigr\}\notag\\
&-\dfrac{1}{N^{2}}
\sum\limits_{\boldk,\boldk^{\prime}}
\sum\limits_{\{a\}}
(v_{\boldk \nu})_{ba}
(v_{\boldk^{\prime} \nu})_{cd}
\sum\limits_{\{A \}}
\int^{\infty}_{-\infty}\dfrac{d\epsilon}{4\pi i}
\int^{\infty}_{-\infty}\dfrac{d\epsilon^{\prime}}{4\pi i}
\tanh \dfrac{\epsilon^{\prime}+\omega}{2T}\notag\\
&\times 
\Bigl\{\Bigl(\tanh \dfrac{\epsilon+\omega}{2T}
-\tanh \dfrac{\epsilon}{2T}\Bigr)
g_{2;aABb}(k;\omega)
\Bigl[
\Gamma_{22\textrm{-IV};\{A\}}(k,k^{\prime};\omega)g_{2;CcdD}(k^{\prime};\omega)
-\Gamma_{23;\{A\}}(k,k^{\prime};\omega)g_{3;CcdD}(k^{\prime};\omega)
\Bigr]\notag\\
&-\tanh \dfrac{\epsilon+\omega}{2T}
g_{3;aABb}(k;\omega)
\Bigl[
\Gamma_{32;\{A\}}(k,k^{\prime};\omega)g_{2;CcdD}(k^{\prime};\omega)
-\Gamma_{33\textrm{-I};\{A\}}(k,k^{\prime};\omega)g_{3;CcdD}(k^{\prime};\omega)
\Bigr]\notag\\
&+\tanh \dfrac{\epsilon}{2T}
g_{1;aABb}(k;\omega)
\Bigl[
\Gamma_{12;\{A\}}(k,k^{\prime};\omega)g_{2;CcdD}(k^{\prime};\omega)
-\Gamma_{13\textrm{-I};\{A\}}(k,k^{\prime};\omega)g_{3;CcdD}(k^{\prime};\omega)
\Bigr] \Bigr\}\notag\\
&-\dfrac{1}{N^{2}}
\sum\limits_{\boldk,\boldk^{\prime}}
\sum\limits_{\{a\}}
(v_{\boldk \nu})_{ba}
(v_{\boldk^{\prime} \nu})_{cd}
\sum\limits_{\{A \}}
\int^{\infty}_{-\infty}\dfrac{d\epsilon}{4\pi i}
\int^{\infty}_{-\infty}\dfrac{d\epsilon^{\prime}}{4\pi i}
\coth \dfrac{\epsilon^{\prime}+\epsilon+\omega}{2T}\notag\\
&\times \Bigl\{\Bigl(\tanh \dfrac{\epsilon+\omega}{2T}
-\tanh \dfrac{\epsilon}{2T}\Bigr)
g_{2;aABb}(k;\omega)g_{2;CcdD}(k^{\prime};\omega)
\Bigl[
\Gamma_{22\textrm{-III};\{A\}}(k,k^{\prime};\omega)
-\Gamma_{22\textrm{-IV};\{A\}}(k,k^{\prime};\omega)
\Bigr]\notag\\
&-\tanh \dfrac{\epsilon+\omega}{2T}
g_{3;aABb}(k;\omega)
\Bigl[
\Gamma_{31\textrm{-II};\{A\}}(k,k^{\prime};\omega)-\Gamma_{31\textrm{-I};\{A\}}(k,k^{\prime};\omega)
\Bigr]
g_{1;CcdD}(k^{\prime};\omega)\notag\\
&+\tanh \dfrac{\epsilon}{2T}
g_{1;aABb}(k;\omega)g_{3;CcdD}(k^{\prime};\omega)
\Bigl[
\Gamma_{13\textrm{-I};\{A\}}(k,k^{\prime};\omega)-\Gamma_{13\textrm{-II};\{A\}}(k,k^{\prime};\omega)
\Bigr]\Bigr\}\notag\\
=&-\frac{1}{N^{2}}
\sum\limits_{\boldk,\boldk^{\prime}}
\int^{\infty}_{-\infty}\frac{d\epsilon}{4\pi i}
\sum\limits_{\{a\}}
\sum\limits_{\{A \}}
(v_{\boldk \nu})_{ba}
(v_{\boldk^{\prime} \nu})_{cd}
\Bigl[\tanh \frac{\epsilon}{2T}
g_{1;aABb}(k;\omega)
\int^{\infty}_{-\infty}\frac{d\epsilon^{\prime}}{4\pi i}
\sum\limits_{l=1}^{3}
\mathcal{J}_{1l;\{A\}}(k,k^{\prime};\omega)
g_{l;CcdD}(k^{\prime};\omega)\notag\\
&\ \ \ \ \ \ \ \ \ \ \ \ \ \
+\Bigl(\tanh \frac{\epsilon+\omega}{2T}
-\tanh \frac{\epsilon}{2T}\Bigr)
g_{2;aABb}(k;\omega)
\int^{\infty}_{-\infty}\frac{d\epsilon^{\prime}}{4\pi i}
\sum\limits_{l=1}^{3}
\mathcal{J}_{2l;\{A\}}(k,k^{\prime};\omega)
g_{l;CcdD}(k^{\prime};\omega)
\notag\\
&\ \ \ \ \ \ \ \ \ \ \ \ \ \ 
-\tanh \frac{\epsilon+\omega}{2T}
g_{3;aABb}(k;\omega)
\int^{\infty}_{-\infty}\frac{d\epsilon^{\prime}}{4\pi i}
\sum\limits_{l=1}^{3}
\mathcal{J}_{3l;\{A\}}(k,k^{\prime};\omega)
g_{l;CcdD}(k^{\prime};\omega)\Bigr],\label{eq:2nd-analytic}
\end{align}
where $\mathcal{J}_{ll^{\prime};\{A\}}(k,k^{\prime};\omega)$ are 
\begin{align}
\mathcal{J}_{11;\{A\}}(k,k^{\prime};\omega)
=&\
\tanh \frac{\epsilon^{\prime}}{2T} 
\Gamma_{11\textrm{-I};\{A\}}(k,k^{\prime};\omega)+\coth \frac{\epsilon^{\prime}-\epsilon}{2T} 
\Bigl[\Gamma_{11\textrm{-II};\{A\}}(k,k^{\prime};\omega)-
\Gamma_{11\textrm{-I};\{A\}}(k,k^{\prime};\omega)\Bigr],\label{eq:4VC-11}\\
\mathcal{J}_{12;\{A\}}(k,k^{\prime};\omega)
=&\
\Bigl(\tanh \frac{\epsilon^{\prime}+\omega}{2T} 
-\tanh \frac{\epsilon^{\prime}}{2T}\Bigr)
\Gamma_{12;\{A\}}(k,k^{\prime};\omega),\label{eq:4VC-12}\\
\mathcal{J}_{13;\{A\}}(k,k^{\prime};\omega)
=&
-\tanh \frac{\epsilon^{\prime}+\omega}{2T} 
\Gamma_{13\textrm{-I};\{A\}}(k,k^{\prime};\omega)-\coth \frac{\epsilon+\epsilon^{\prime}+\omega}{2T} 
\Bigl[\Gamma_{13\textrm{-II};\{A\}}(k,k^{\prime};\omega)-
\Gamma_{13\textrm{-I};\{A\}}(k,k^{\prime};\omega)\Bigr],\label{eq:4VC-13}\\
\mathcal{J}_{21;\{A\}}(k,k^{\prime};\omega)
=&\
\tanh \frac{\epsilon^{\prime}}{2T} 
\Gamma_{21;\{A\}}(k,k^{\prime};\omega),\label{eq:4VC-21}\\
\mathcal{J}_{22;\{A\}}(k,k^{\prime};\omega)
=&\
\Bigl(\coth \frac{\epsilon^{\prime}-\epsilon}{2T}
-\tanh \frac{\epsilon^{\prime}}{2T}\Bigr)
\Gamma_{22\textrm{-II};\{A\}}(k,k^{\prime};\omega)+
\Bigl(\coth \frac{\epsilon^{\prime}+\epsilon+\omega}{2T}
-\coth \frac{\epsilon^{\prime}-\epsilon}{2T}\Bigr)
\Gamma_{22\textrm{-III};\{A\}}(k,k^{\prime};\omega)\notag\\
&+\Bigl(\tanh \frac{\epsilon^{\prime}+\omega}{2T}
-\coth \frac{\epsilon^{\prime}+\epsilon+\omega}{2T}\Bigr)
\Gamma_{22\textrm{-IV};\{A\}}(k,k^{\prime};\omega),\label{eq:4VC-22}\\
\mathcal{J}_{23;\{A\}}(k,k^{\prime};\omega)
=&
-\tanh \frac{\epsilon^{\prime}+\omega}{2T} 
\Gamma_{23;\{A\}}(k,k^{\prime};\omega),\label{eq:4VC-23}\\
\mathcal{J}_{31;\{A\}}(k,k^{\prime};\omega)
=&\
\tanh \frac{\epsilon^{\prime}}{2T} 
\Gamma_{31\textrm{-I};\{A\}}(k,k^{\prime};\omega)
+\coth \frac{\epsilon+\epsilon^{\prime}+\omega}{2T} 
\Bigl[\Gamma_{31\textrm{-II};\{A\}}(k,k^{\prime};\omega)-
\Gamma_{31\textrm{-I};\{A\}}(k,k^{\prime};\omega)\Bigr],\label{eq:4VC-31}\\
\mathcal{J}_{32;\{A\}}(k,k^{\prime};\omega)
=&\
\Bigl(\tanh \frac{\epsilon^{\prime}+\omega}{2T} 
-\tanh \frac{\epsilon^{\prime}}{2T}\Bigr)
\Gamma_{32;\{A\}}(k,k^{\prime};\omega),\label{eq:4VC-32}
\end{align}
and 
\begin{align}
\hspace{-18pt}
\mathcal{J}_{33;\{A\}}(k,k^{\prime};\omega)
=&
-\tanh \frac{\epsilon^{\prime}+\omega}{2T} 
\Gamma_{33\textrm{-I};\{A\}}(k,k^{\prime};\omega)
-\coth \frac{\epsilon^{\prime}-\epsilon}{2T} 
\Bigl[\Gamma_{33\textrm{-II};\{A\}}(k,k^{\prime};\omega)-
\Gamma_{33\textrm{-I};\{A\}}(k,k^{\prime};\omega)\Bigr].\label{eq:4VC-33}
\end{align}
\end{widetext}
In Eq. (\ref{eq:2nd-analytic}), I have not explicitly written 
whether the integral is the principal integral 
(containing a hyperbolic cotangent of frequency). 

Combining Eqs. (\ref{eq:1st-analytic}) and (\ref{eq:2nd-analytic}), 
we finally obtain Eq. (\ref{eq:Kpart-analytic}).  

\section{Derivation of Eq. (\ref{eq:sum-analytic2})}
\begin{figure*}[tb]
\begin{center}
\includegraphics[width=150mm]{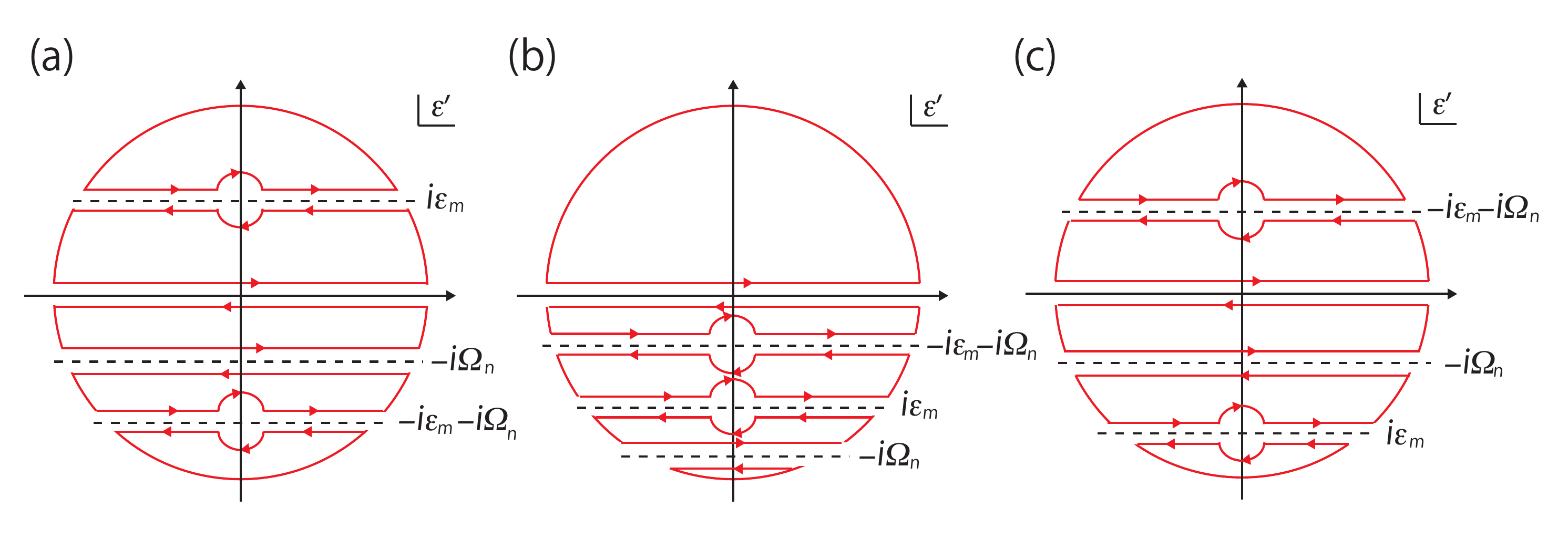}
\end{center}
\vspace{-20pt}
\caption{
Contours used for the analytic continuations of the three-point vector vertex function. 
$\textrm{Im}\epsilon > 0$ and $\textrm{Im}\epsilon+\textrm{Im}\omega > 0$ are satisfied 
in panel (a), 
$\textrm{Im}\epsilon < 0$ and $\textrm{Im}\epsilon+\textrm{Im}\omega > 0$ are satisfied 
in panel (b), 
and $\textrm{Im}\epsilon < 0$ and $\textrm{Im}\epsilon+\textrm{Im}\omega < 0$ 
are satisfied in panel (c).
}
\label{fig:Fig3}
\end{figure*}

In this appendix, 
we derive Eq. (\ref{eq:sum-analytic2}) 
after defining the three-point vector vertex function in Matsubara-frequency representation 
and carrying out its analytic continuation. 

We can express Eq. (\ref{eq:sum-analytic}) in a more compact form 
since $\mathcal{J}_{ll^{\prime};\{A\}}(k,k^{\prime};\omega)$ are connected with 
the three-point vector vertex function in real-frequency representation. 
We begin with the definition~\cite{Takada-text} in Matsubara-frequency representation, 
$\Lambda_{\nu;AB}(\boldk,i\epsilon_{m};\boldq,i\Omega_{n})\equiv 
\Lambda_{\nu;AB}(\boldk+\boldq,i\epsilon_{m+n}, \boldk,i\epsilon_{m})$: 
\begin{align}
&\textstyle\sum\limits_{A,B}
G_{aA}(\boldk+\boldq,i\epsilon_{m+n})
\Lambda_{\nu;AB}(\boldk,i\epsilon_{m};\boldq,i\Omega_{n})
G_{Bb}(\boldk,i\epsilon_{m})\notag\\
=&
\int^{T^{-1}}_{0}\hspace{-10pt}d\tau e^{i\epsilon_{m+n}\tau}
\int^{T^{-1}}_{0}\hspace{-10pt}d\tau^{\prime} e^{-i\Omega_{n}\tau^{\prime}}
\langle \textrm{T}_{\tau}  
\hat{c}_{\boldk+\boldq a}(\tau)
\hat{J}_{-\boldq \nu}(\tau^{\prime})
\hat{c}_{\boldk b}^{\dagger}
\rangle.\label{eq:Def-3VF-Matsu}
\end{align}
After some algebra for Eq. (\ref{eq:Def-3VF-Matsu}), 
we obtain the Bethe-Salpeter equation to determine 
$\Lambda_{\nu;AB}(\boldk,i\epsilon_{m};\boldq,i\Omega_{n})$,
\begin{align}
&\Lambda_{\nu;ab}(\boldk,i\epsilon_{m};\boldq,i\Omega_{n})
=
(v_{\boldk \nu})_{ab}\notag\\
&+\frac{T}{N}
\sum\limits_{\boldk^{\prime}}
\sum\limits_{m^{\prime}}
\sum\limits_{\{A \}}
\Gamma_{abCD}(\boldk,i\epsilon_{m},\boldk^{\prime}i\epsilon_{m^{\prime}};\boldq,i\Omega_{n})\notag\\
&\times 
G_{CA}(\boldk^{\prime}+\boldq,i\epsilon_{m^{\prime}+n})
G_{BD}(\boldk^{\prime},i\epsilon_{m^{\prime}})
(v_{\boldk^{\prime} \nu})_{AB}.\label{eq:3-point-VC-Matsu}
\end{align}
Since the second term of Eq. (\ref{eq:3-point-VC-Matsu}) 
has the similar analytic property for the second term of Eq. (\ref{eq:Ktild}), 
we can similarly carry out the analytic continuation of the former as follows: 
\begin{widetext}
\begin{align}
&\frac{T}{N}
\sum\limits_{\boldk^{\prime}}
\sum\limits_{m^{\prime}}
\sum\limits_{\{A \}}
\Gamma_{abCD}(\boldk,i\epsilon_{m},\boldk^{\prime}i\epsilon_{m^{\prime}};\boldq,i\Omega_{n})
G_{CA}(\boldk^{\prime}+\boldq,i\epsilon_{m^{\prime}+n})
G_{BD}(\boldk^{\prime},i\epsilon_{m^{\prime}})
(v_{\boldk^{\prime} \nu})_{AB}\notag\\
=\
&
\frac{1}{N}
\sum\limits_{\boldk^{\prime}}
\sum\limits_{\{A \}}
\Bigl[
\int_{C^{\prime}}\frac{d\epsilon^{\prime}}{4\pi i}
\tanh \frac{\epsilon^{\prime}}{2T}
\Gamma_{abCD}(\boldk,i\epsilon_{m},\boldk^{\prime},\epsilon^{\prime};\boldq,i\Omega_{n})
G_{CA}(\boldk^{\prime}+\boldq,\epsilon^{\prime}+i\Omega_{n})
G_{BD}(\boldk^{\prime},\epsilon^{\prime})\notag\\
&\ \ \ \ \ \ \ \ \ \ \ \
+T\Gamma_{abCD}(\boldk,i\epsilon_{m},\boldk^{\prime},i\epsilon_{m};\boldq,i\Omega_{n})
G_{CA}(\boldk^{\prime}+\boldq,i\epsilon_{m}+i\Omega_{n})
G_{BD}(\boldk^{\prime},i\epsilon_{m})\notag\\
&\ \ \ \ \ \ \ \ \ \ \ \
+T\Gamma_{abCD}(\boldk,i\epsilon_{m},\boldk^{\prime},-i\epsilon_{m}-i\Omega_{n};\boldq,i\Omega_{n})
G_{CA}(\boldk^{\prime}+\boldq,-i\epsilon_{m})
G_{BD}(\boldk^{\prime},-i\epsilon_{m}-i\Omega_{n})
\Bigr] (v_{\boldk^{\prime} \nu})_{AB},\label{eq:3-point-VC-before}
\end{align}
\end{widetext}
where the contour $C^{\prime}$ is the contour shown in 
Fig. \ref{fig:Fig3}(a) or \ref{fig:Fig3}(b) or \ref{fig:Fig3}(c), 
depending on the values of $\textrm{Im}\epsilon$ 
and $\textrm{Im}\epsilon+\textrm{Im}\omega$ 
for the second term of Eq. (\ref{eq:3-point-VC-Matsu}). 
Thus, the three-point vector vertex function in real-frequency representation 
is given by 
\begin{align}
&\Lambda_{\nu;l;ab}(k;q)
=\
(v_{\boldk \nu})_{ab}\notag\\
&+\frac{1}{N}
\sum\limits_{\boldk^{\prime}}
\sum\limits_{\{A\}}
\sum\limits_{l^{\prime}=1}^{3}
\int^{\infty}_{-\infty}\frac{d\epsilon^{\prime}}{4\pi i}
\mathcal{J}_{ll^{\prime};abCD}(k,k^{\prime};q)\notag\\
&\times 
g_{l^{\prime};CABD}(k^{\prime};q)
(v_{\boldk^{\prime} \nu})_{AB},\label{eq:3-point-VC-real}
\end{align} 
where case for $l=1$ or $2$ or $3$ corresponds to 
case of Fig. \ref{fig:Fig3}(a) or \ref{fig:Fig3}(b) or \ref{fig:Fig3}(c), 
respectively. 
Using Eqs. (\ref{eq:sum-analytic}), (\ref{eq:Kpart-analytic}), and (\ref{eq:3-point-VC-real}), 
we can rewrite Eq. (\ref{eq:sum-analytic}) as Eq. (\ref{eq:sum-analytic2}).  

\section{Derivation of Eq. (\ref{eq:phi-final})}

In this appendix, 
I explain the detail of the derivation of the $\boldq$-linear terms 
of $K_{xy\nu}(\boldq,i\Omega_{n})$ 
in the most-divergent-term approximation~\cite{Fukuyama-RH,Kohno-Yamada}. 

We start to show the leading-order $\boldq$ dependence of each quantity 
appearing in Eq. (\ref{eq:phi-next}). 
First, 
$G_{aA}(\boldk_{-},i\epsilon_{m})G_{Bb}(\boldk_{+},i\epsilon_{m+n})$ 
is approximated within the linear order of $\boldq$ as 
\begin{align}
&G_{aA}(\boldk_{-},i\epsilon_{m})
G_{Bb}(\boldk_{+},i\epsilon_{m+n})\notag\\
\sim &\ G_{aA}(\boldk,i\epsilon_{m})
G_{Bb}(\boldk,i\epsilon_{m+n})\notag\\
&+
\sum\limits_{\eta}\dfrac{q_{\eta}}{2}
\Bigl[G_{aA}(\boldk,i\epsilon_{m})
\dfrac{\overleftrightarrow{\partial}}{\partial k_{\eta}}
G_{Bb}(\boldk,i\epsilon_{m+n})\Bigr],\label{eq:qlinear-GG}
\end{align}
where we introduce a quantity, 
\begin{align}
\Bigl[g(x)
\dfrac{\overleftrightarrow{\partial}}{\partial x}
h(x)\Bigr]
\equiv 
g(x)\dfrac{\partial h(x)}{\partial x}
-
\dfrac{\partial h(x)}{\partial x}h(x).
\end{align}
Second, 
$\Lambda_{y;dc}(\boldk_{\pm},i\epsilon_{m+n},\boldk_{\pm},i\epsilon_{m})$ 
is approximated as
\begin{align}
&\Lambda_{y;dc}(\boldk_{\pm},i\epsilon_{m+n},\boldk_{\pm},i\epsilon_{m})
\sim \ 
\Lambda_{y;dc}(\boldk,i\epsilon_{m+n},\boldk,i\epsilon_{m})\notag\\
&\pm 
\sum\limits_{\eta}\dfrac{q_{\eta}}{2}
\dfrac{\partial}{\partial k_{\eta}}
\Lambda_{y;dc}(\boldk,i\epsilon_{m+n},\boldk,i\epsilon_{m}).\label{eq:qliner-3pointVCy}
\end{align}
Third, the leading-order $\boldq$ dependence of 
$\Lambda_{x;AB}(\boldk_{-},i\epsilon_{m},\boldk_{+},i\epsilon_{m+n})$ 
is given by
\begin{align}
&\Lambda_{x;AB}(\boldk_{-},i\epsilon_{m},\boldk_{+},i\epsilon_{m+n})
\sim \ \Lambda_{x;AB}(\boldk,i\epsilon_{m},\boldk,i\epsilon_{m+n})\notag\\
&+\Delta \Lambda_{x;AB}(\boldk,i\epsilon_{m},\boldk,i\epsilon_{m+n}),\label{eq:qlinear-3pointVCx}
\end{align}
with the equation derived from Eq. (\ref{eq:3-point-VC-Matsu}),
\begin{align}
&\Delta \Lambda_{x;AB}(\boldk,i\epsilon_{m},\boldk,i\epsilon_{m+n})\notag\\
=& 
\frac{T}{N}
\sum\limits_{\boldk^{\prime}}
\sum\limits_{m^{\prime}}
\sum\limits_{\{a \}}
\Gamma_{ABcd}^{(1)}(\boldk,i\epsilon_{m},\boldk,i\epsilon_{m+n}; 
\boldk^{\prime},i\epsilon_{m^{\prime}},\boldk^{\prime},i\epsilon_{m^{\prime}+n})\notag\\
&\times 
\sum\limits_{\eta}\dfrac{q_{\eta}}{2}
\Bigl[
G_{ca}(\boldk^{\prime},i\epsilon_{m^{\prime}})
\dfrac{\overleftrightarrow{\partial}}{\partial k_{\eta}}
G_{bd}(\boldk^{\prime},i\epsilon_{m^{\prime}+n})
\Bigr]\notag\\
&\times 
\Lambda_{x;ab}(\boldk^{\prime},i\epsilon_{m^{\prime}},\boldk^{\prime},i\epsilon_{m^{\prime}+n})\notag\\
&+\frac{T}{N}
\sum\limits_{\boldk^{\prime}}
\sum\limits_{m^{\prime}}
\sum\limits_{\{a \}}
\Gamma_{ABcd}^{(1)}(\boldk,i\epsilon_{m},\boldk,i\epsilon_{m+n}; 
\boldk^{\prime},i\epsilon_{m^{\prime}},\boldk^{\prime},i\epsilon_{m^{\prime}+n})\notag\\
&\times 
G_{bd}(\boldk^{\prime},i\epsilon_{m^{\prime}+n})
G_{ca}(\boldk^{\prime},i\epsilon_{m^{\prime}})\notag\\
&\times 
\Delta \Lambda_{x;ab}(\boldk^{\prime},i\epsilon_{m^{\prime}},\boldk^{\prime},i\epsilon_{m^{\prime}+n}).
\label{eq:BSeq-qlinear}
\end{align}
In Eq. (\ref{eq:BSeq-qlinear}), 
we have neglected 
the $\boldq$-linear term arising from the irreducible four-point vertex function 
because that is negligible compared with the main terms 
in the most-divergent-term approximation~\cite{Fukuyama-RH,Kohno-Yamada}. 
Fourth, 
due to the same reason, 
we can neglect the $\boldq$-linear term arising from 
the irreducible six-point vertex function~\cite{Fukuyama-RH,Kohno-Yamada}. 
Fifth, 
the $\boldq$ dependence of 
$\sum_{f,g}
G_{fb}(\boldk_{-},i\epsilon_{m})
\Lambda_{\nu;gf}(\boldk_{+},i\epsilon_{m},\boldk_{-},i\epsilon_{m})
G_{cg}(\boldk_{+},i\epsilon_{m})$ 
is negligible in the most-divergent-term approximation~\cite{Fukuyama-RH,Kohno-Yamada} 
since we can neglect the $\boldq$-linear term arising from 
a pair of two single-particle Green's functions whose 
frequencies are the same 
in the most-divergent-term approximation~\cite{Fukuyama-RH,Kohno-Yamada}. 
More precisely, 
$\sum_{f,g}
G_{fb}(\boldk_{-},i\epsilon_{m})
\Lambda_{\nu;gf}(\boldk_{+},i\epsilon_{m},\boldk_{-},i\epsilon_{m})
G_{cg}(\boldk_{+},i\epsilon_{m})$  is approximated as 
\begin{align}
&\sum\limits_{f,g}
G_{fb}(\boldk_{-},i\epsilon_{m})
\Lambda_{\alpha;gf}(\boldk_{+},i\epsilon_{m},\boldk_{-},i\epsilon_{m})
G_{cg}(\boldk_{+},i\epsilon_{m})\notag\\
\sim &\ \textstyle\sum\limits_{f,g}
G_{fb}(\boldk,i\epsilon_{m})
\Lambda_{\alpha;gf}(\boldk,i\epsilon_{m},\boldk,i\epsilon_{m})
G_{cg}(\boldk,i\epsilon_{m})\notag\\
=&\ \dfrac{\partial G_{cb}(\boldk,i\epsilon_{m})}{\partial k_{\alpha}}.
\label{eq:qlinear-3pointVC-alp}
\end{align}
In the final line, we use one of the Ward identities~\cite{Fukuyama-RH,Kohno-Yamada}. 

Using the $\boldq$ dependences of the quantities appearing in Eq. (\ref{eq:phi-next}), 
we obtain the $\boldq$-linear terms of $K_{xy\nu}(\boldq,i\Omega_{n})$ 
in the most-divergent-term approximation~\cite{Fukuyama-RH,Kohno-Yamada}:  
\begin{widetext}
\begin{align}
K_{xy\nu}(\boldq,i\Omega_{n})
=&\
\delta_{\nu,y}
\frac{T}{N}
\sum\limits_{\boldk}
\sum\limits_{m}
\sum\limits_{a,b,A,B}
\dfrac{\partial (v_{\boldk y})_{ba}}{\partial k_{\nu}}
G_{aA}(\boldk,i\epsilon_{m})
\Delta \Lambda_{x;AB}(\boldk,i\epsilon_{m},\boldk,i\epsilon_{m+n})
G_{Bb}(\boldk,i\epsilon_{m+n})\notag\\
&+
\delta_{\nu,y}
\frac{T}{N}
\sum\limits_{\boldk}
\sum\limits_{m}
\sum\limits_{a,b,A,B}
\dfrac{\partial (v_{\boldk y})_{ba}}{\partial k_{\nu}}
\Lambda_{x;AB}(\boldk,i\epsilon_{m},\boldk,i\epsilon_{m+n})
\sum\limits_{\eta}\dfrac{q_{\eta}}{2}
\Bigl[G_{aA}(\boldk,i\epsilon_{m})
\dfrac{\overleftrightarrow{\partial}}{\partial k_{\eta}}
G_{Bb}(\boldk,i\epsilon_{m+n})\Bigr]\notag\\
&+\frac{T}{N}
\sum\limits_{\boldk}
\sum\limits_{m}
\sum\limits_{\{a\}}
\Delta \Lambda_{x;ba}(\boldk,i\epsilon_{m},\boldk,i\epsilon_{m+n})
G_{ad}(\boldk,i\epsilon_{m+n})
\dfrac{\partial G_{cb}(\boldk,i\epsilon_{m})}{\partial k_{\nu}}
\Lambda_{y;dc}(\boldk,i\epsilon_{m+n},\boldk,i\epsilon_{m})\notag\\
&+\frac{T}{N}
\sum\limits_{\boldk}
\sum\limits_{m}
\sum\limits_{\{a\}}
\Lambda_{x;ba}(\boldk,i\epsilon_{m},\boldk,i\epsilon_{m+n})
\sum\limits_{\eta}\dfrac{q_{\eta}}{2}
\dfrac{\partial G_{ad}(\boldk,i\epsilon_{m+n})}{\partial k_{\eta}}
\dfrac{\partial G_{cb}(\boldk,i\epsilon_{m})}{\partial k_{\nu}}
\Lambda_{y;dc}(\boldk,i\epsilon_{m+n},\boldk,i\epsilon_{m})\notag\\
&+\frac{T}{N}
\sum\limits_{\boldk}
\sum\limits_{m}
\sum\limits_{\{a\}}
\Lambda_{x;ba}(\boldk,i\epsilon_{m},\boldk,i\epsilon_{m+n})
G_{ad}(\boldk,i\epsilon_{m+n})
\dfrac{\partial G_{cb}(\boldk,i\epsilon_{m})}{\partial k_{\nu}}
\sum\limits_{\eta}\dfrac{q_{\eta}}{2}
\dfrac{\partial \Lambda_{y;dc}(\boldk,i\epsilon_{m+n},\boldk,i\epsilon_{m})}{\partial k_{\eta}}\notag\\
&+\frac{T}{N}
\sum\limits_{\boldk}
\sum\limits_{m}
\sum\limits_{\{a\}}
\Delta \Lambda_{x;ba}(\boldk,i\epsilon_{m},\boldk,i\epsilon_{m+n})
G_{cb}(\boldk,i\epsilon_{m})
\dfrac{\partial G_{ad}(\boldk,i\epsilon_{m+n})}{\partial k_{\nu}}
\Lambda_{y;dc}(\boldk,i\epsilon_{m+n},\boldk,i\epsilon_{m})\notag\\
&-\frac{T}{N}
\sum\limits_{\boldk}
\sum\limits_{m}
\sum\limits_{\{a\}}
\Lambda_{x;ba}(\boldk,i\epsilon_{m},\boldk,i\epsilon_{m+n})
\sum\limits_{\eta}\dfrac{q_{\eta}}{2}
\dfrac{\partial G_{cb}(\boldk,i\epsilon_{m})}{\partial k_{\eta}}
\dfrac{\partial G_{ad}(\boldk,i\epsilon_{m+n})}{\partial k_{\nu}}
\Lambda_{y;dc}(\boldk,i\epsilon_{m+n},\boldk,i\epsilon_{m})\notag\\
&-\frac{T}{N}
\sum\limits_{\boldk}
\sum\limits_{m}
\sum\limits_{\{a\}}
\Lambda_{x;ba}(\boldk,i\epsilon_{m},\boldk,i\epsilon_{m+n})
G_{cb}(\boldk,i\epsilon_{m})
\dfrac{\partial G_{ad}(\boldk,i\epsilon_{m+n})}{\partial k_{\nu}}
\sum\limits_{\eta}\dfrac{q_{\eta}}{2}
\dfrac{\partial \Lambda_{y;dc}(\boldk,i\epsilon_{m+n},\boldk,i\epsilon_{m})}
{\partial k_{\eta}}\notag\\
&+\Bigl(\frac{T}{N}\Bigr)^{3}
\sum\limits_{\boldk,\boldk^{\prime},\boldk^{\prime\prime}}
\sum\limits_{m,m^{\prime},m^{\prime\prime}}
\sum\limits_{\{a\}}
\sum\limits_{\{A\}}
\sum\limits_{F,G}
G_{Bb}(\boldk,i\epsilon_{m})
\Delta \Lambda_{x;ba}(\boldk,i\epsilon_{m},\boldk,i\epsilon_{m+n})
G_{aA}(\boldk,i\epsilon_{m+n})\notag\\
&\ \ \times 
\dfrac{\partial G_{GF}(\boldk^{\prime\prime},i\epsilon_{m^{\prime\prime}})}{\partial k^{\prime\prime}_{\nu}}
G_{Dd}(\boldk^{\prime},i\epsilon_{m^{\prime}+n})
\Lambda_{y;dc}(\boldk^{\prime},i\epsilon_{m^{\prime}+n},\boldk^{\prime},i\epsilon_{m^{\prime}})
G_{cC}(\boldk^{\prime},i\epsilon_{m^{\prime}})\notag\\
&\ \ \times \Gamma_{3; ABCDFG}^{(1)}
(\boldk,i\epsilon_{m+n},\boldk,i\epsilon_{m};
\boldk^{\prime},i\epsilon_{m^{\prime}},\boldk^{\prime},i\epsilon_{m^{\prime}+n};
\boldk^{\prime\prime},i\epsilon_{m^{\prime\prime}},\boldk^{\prime\prime},i\epsilon_{m^{\prime\prime}})\notag\\
&+\Bigl(\frac{T}{N}\Bigr)^{3}
\sum\limits_{\boldk,\boldk^{\prime},\boldk^{\prime\prime}}
\sum\limits_{m,m^{\prime},m^{\prime\prime}}
\sum\limits_{\{a\}}
\sum\limits_{\{A\}}
\sum\limits_{F,G}
\Lambda_{x;ba}(\boldk,i\epsilon_{m},\boldk,i\epsilon_{m+n})
\sum\limits_{\eta}\dfrac{q_{\eta}}{2}
\Bigl[
G_{Bb}(\boldk,i\epsilon_{m})
\dfrac{\overleftrightarrow{\partial}}{\partial k_{\eta}}
G_{aA}(\boldk,i\epsilon_{m+n})\Bigr]\notag\\
&\ \ \times 
\dfrac{\partial G_{GF}(\boldk^{\prime\prime},i\epsilon_{m^{\prime\prime}})}{\partial k^{\prime\prime}_{\nu}}
G_{Dd}(\boldk^{\prime},i\epsilon_{m^{\prime}+n})
\Lambda_{y;dc}(\boldk^{\prime},i\epsilon_{m^{\prime}+n},\boldk^{\prime},i\epsilon_{m^{\prime}})
G_{cC}(\boldk^{\prime},i\epsilon_{m^{\prime}})\notag\\
&\ \ \times \Gamma_{3; ABCDFG}^{(1)}
(\boldk,i\epsilon_{m+n},\boldk,i\epsilon_{m};
\boldk^{\prime},i\epsilon_{m^{\prime}},\boldk^{\prime},i\epsilon_{m^{\prime}+n};
\boldk^{\prime\prime},i\epsilon_{m^{\prime\prime}},\boldk^{\prime\prime},i\epsilon_{m^{\prime\prime}})
.\label{eq:phi-third}
\end{align}
\end{widetext}
The above ninth and tenth terms can be rewritten in a simpler form 
by using Eqs. (\ref{eq:3-point-VC-Matsu}) and (\ref{eq:BSeq-qlinear}), 
the exchange symmetry~\cite{Kohno-Yamada} of the four-point vertex function, 
and the relation~\cite{Kohno-Yamada} 
between the irreducible four-point and the irreducible six-point vertex function,
\begin{widetext} 
\begin{align}
&\Bigl(\dfrac{\partial}{\partial k_{\nu}}+\dfrac{\partial}{\partial k^{\prime}_{\nu}}\Bigr)
\Gamma_{ABDC}^{(1)}(\boldk,i\epsilon_{m+n},\boldk,i\epsilon_{m};
\boldk^{\prime},i\epsilon_{m^{\prime}+n},\boldk^{\prime},i\epsilon_{m^{\prime}})\notag\\
=&\ 
\frac{T}{N}
\sum\limits_{\boldk^{\prime\prime}}
\sum\limits_{m^{\prime\prime}}
\sum\limits_{F,G}
\Gamma_{3; ABCDFG}^{(1)}
(\boldk,i\epsilon_{m+n},\boldk,i\epsilon_{m};
\boldk^{\prime},i\epsilon_{m^{\prime}},\boldk^{\prime},i\epsilon_{m^{\prime}+n};
\boldk^{\prime\prime},i\epsilon_{m^{\prime\prime}},\boldk^{\prime\prime},i\epsilon_{m^{\prime\prime}})
\dfrac{\partial G_{GF}(\boldk^{\prime\prime},i\epsilon_{m^{\prime\prime}})}
{\partial k^{\prime\prime}_{\nu}}.\label{eq:Ward-6point}
\end{align}
\end{widetext}
Namely, 
the ninth and tenth terms become
\begin{widetext}
\begin{align}
&\ \ \Bigl(\frac{T}{N}\Bigr)^{2}
\sum\limits_{\boldk,\boldk^{\prime}}
\sum\limits_{m,m^{\prime}}
\sum\limits_{\{a\}}
\sum\limits_{\{A\}}
G_{Bb}(\boldk,i\epsilon_{m})
\Delta \Lambda_{x;ba}(\boldk,i\epsilon_{m},\boldk,i\epsilon_{m+n})
G_{aA}(\boldk,i\epsilon_{m+n})
G_{Dd}(\boldk^{\prime},i\epsilon_{m^{\prime}+n})\notag\\
&\ \ \times 
\Lambda_{y;dc}(\boldk^{\prime},i\epsilon_{m^{\prime}+n},\boldk^{\prime},i\epsilon_{m^{\prime}})
G_{cC}(\boldk^{\prime},i\epsilon_{m^{\prime}})
\Bigl(\dfrac{\partial}{\partial k_{\nu}}+\dfrac{\partial}{\partial k^{\prime}_{\nu}}\Bigr)
\Gamma_{ABDC}^{(1)}(\boldk,i\epsilon_{m+n},\boldk,i\epsilon_{m};
\boldk^{\prime},i\epsilon_{m^{\prime}+n},\boldk^{\prime},i\epsilon_{m^{\prime}})\notag\\
&+\Bigl(\frac{T}{N}\Bigr)^{2}
\sum\limits_{\boldk,\boldk^{\prime}}
\sum\limits_{m,m^{\prime}}
\sum\limits_{\{a\}}
\sum\limits_{\{A\}}
\Lambda_{x;ba}(\boldk,i\epsilon_{m},\boldk,i\epsilon_{m+n})
\sum\limits_{\eta}\dfrac{q_{\eta}}{2}
\Bigl[
G_{Bb}(\boldk,i\epsilon_{m})
\dfrac{\overleftrightarrow{\partial}}{\partial k_{\eta}}
G_{aA}(\boldk,i\epsilon_{m+n})\Bigr]
G_{Dd}(\boldk^{\prime},i\epsilon_{m^{\prime}+n})\notag\\
&\ \ \times 
\Lambda_{y;dc}(\boldk^{\prime},i\epsilon_{m^{\prime}+n},\boldk^{\prime},i\epsilon_{m^{\prime}})
G_{cC}(\boldk^{\prime},i\epsilon_{m^{\prime}})
\Bigl(\dfrac{\partial}{\partial k_{\nu}}+\dfrac{\partial}{\partial k^{\prime}_{\nu}}\Bigr)
\Gamma_{ABDC}^{(1)}(\boldk,i\epsilon_{m+n},\boldk,i\epsilon_{m};
\boldk^{\prime},i\epsilon_{m^{\prime}+n},\boldk^{\prime},i\epsilon_{m^{\prime}})\notag\\
=\ 
&\frac{T}{N}
\sum\limits_{\boldk}
\sum\limits_{m}
\sum\limits_{a,b,A,B}
G_{Bb}(\boldk,i\epsilon_{m})
\Delta \Lambda_{x;ba}(\boldk,i\epsilon_{m},\boldk,i\epsilon_{m+n})
G_{aA}(\boldk,i\epsilon_{m+n})
\dfrac{\partial}{\partial k_{\nu}}
\Bigl[
\Lambda_{y;AB}(\boldk,i\epsilon_{m+n},\boldk,i\epsilon_{m})
-(v_{\boldk y})_{AB}\Bigr]\notag\\
-&
\Bigl(\frac{T}{N}\Bigr)^{2}
\sum\limits_{\boldk,\boldk^{\prime}}
\sum\limits_{m,m^{\prime}}
\sum\limits_{\{a\}}
\sum\limits_{\{A\}}
G_{Bb}(\boldk,i\epsilon_{m})
\Delta \Lambda_{x;ba}(\boldk,i\epsilon_{m},\boldk,i\epsilon_{m+n})
G_{aA}(\boldk,i\epsilon_{m+n})\notag\\
&\times 
\dfrac{\partial}{\partial k^{\prime}_{\nu}}
\Bigl[
G_{Dd}(\boldk^{\prime},i\epsilon_{m^{\prime}+n})
\Lambda_{y;dc}(\boldk^{\prime},i\epsilon_{m^{\prime}+n},\boldk^{\prime},i\epsilon_{m^{\prime}})
G_{cC}(\boldk^{\prime},i\epsilon_{m^{\prime}})\Bigr]
\Gamma_{ABDC}^{(1)}(\boldk,i\epsilon_{m+n},\boldk,i\epsilon_{m};
\boldk^{\prime},i\epsilon_{m^{\prime}+n},\boldk^{\prime},i\epsilon_{m^{\prime}})\notag\\
+&\frac{T}{N}
\sum\limits_{\boldk}
\sum\limits_{m}
\sum\limits_{a,b,A,B}
\Lambda_{x;ba}(\boldk,i\epsilon_{m},\boldk,i\epsilon_{m+n})
\sum\limits_{\eta}\dfrac{q_{\eta}}{2}
\Bigl[
G_{Bb}(\boldk,i\epsilon_{m})
\dfrac{\overleftrightarrow{\partial}}{\partial k_{\eta}}
G_{aA}(\boldk,i\epsilon_{m+n})\Bigr]\notag\\
&\times 
\dfrac{\partial}{\partial k_{\nu}}
\Bigl[
\Lambda_{y;AB}(\boldk,i\epsilon_{m+n},\boldk,i\epsilon_{m})
-(v_{\boldk y})_{AB}
\Bigr]\notag\\
-&\Bigl(\frac{T}{N}\Bigr)^{2}
\sum\limits_{\boldk,\boldk^{\prime}}
\sum\limits_{m,m^{\prime}}
\sum\limits_{\{a\}}
\sum\limits_{\{A\}}
\Lambda_{x;ba}(\boldk,i\epsilon_{m},\boldk,i\epsilon_{m+n})
\sum\limits_{\eta}\dfrac{q_{\eta}}{2}
\Bigl[
G_{Bb}(\boldk,i\epsilon_{m})
\dfrac{\overleftrightarrow{\partial}}{\partial k_{\eta}}
G_{aA}(\boldk,i\epsilon_{m+n})\Bigr]\notag\\
&\times 
\dfrac{\partial}{\partial k^{\prime}_{\nu}}
\Bigl[
G_{Dd}(\boldk^{\prime},i\epsilon_{m^{\prime}+n})
\Lambda_{y;dc}(\boldk^{\prime},i\epsilon_{m^{\prime}+n},\boldk^{\prime},i\epsilon_{m^{\prime}})
G_{cC}(\boldk^{\prime},i\epsilon_{m^{\prime}})
\Bigr] 
\Gamma_{ABDC}^{(1)}(\boldk,i\epsilon_{m+n},\boldk,i\epsilon_{m};
\boldk^{\prime},i\epsilon_{m^{\prime}+n},\boldk^{\prime},i\epsilon_{m^{\prime}})\notag\\
=\ 
&\frac{T}{N}
\sum\limits_{\boldk}
\sum\limits_{m}
\sum\limits_{a,b,A,B}
\Bigl\{
\Delta \Lambda_{x;ba}(\boldk,i\epsilon_{m},\boldk,i\epsilon_{m+n})
G_{Bb}(\boldk,i\epsilon_{m})
G_{aA}(\boldk,i\epsilon_{m+n})\notag\\
&\ \ \ \ \ \ \ \ \ \ \ \ \ \ \ \ +
\Lambda_{x;ba}(\boldk,i\epsilon_{m},\boldk,i\epsilon_{m+n})
\sum\limits_{\eta}\dfrac{q_{\eta}}{2}
\Bigl[
G_{Bb}(\boldk,i\epsilon_{m})
\dfrac{\overleftrightarrow{\partial}}{\partial k_{\eta}}
G_{aA}(\boldk,i\epsilon_{m+n})
\Bigr]
\Bigr\}\notag\\
&\ \ \ \ \ \ \ \ \ \ \ \ \ \ \ \ \times 
\dfrac{\partial}{\partial k_{\nu}}
\Bigl[
\Lambda_{y;AB}(\boldk,i\epsilon_{m+n},\boldk,i\epsilon_{m})-(v_{\boldk y})_{AB}
\Bigr]\notag\\
-&
\frac{T}{N}
\sum\limits_{\boldk}
\sum\limits_{m}
\sum\limits_{c,d,C,D}
\Delta \Lambda_{x;DC}(\boldk^{\prime},i\epsilon_{m^{\prime}+n},\boldk^{\prime},i\epsilon_{m^{\prime}})
\dfrac{\partial}{\partial k^{\prime}_{\nu}}
\Bigl[
G_{Dd}(\boldk^{\prime},i\epsilon_{m^{\prime}+n})
\Lambda_{y;dc}(\boldk^{\prime},i\epsilon_{m^{\prime}+n},\boldk^{\prime},i\epsilon_{m^{\prime}})
G_{cC}(\boldk^{\prime},i\epsilon_{m^{\prime}})\Bigr].\label{eq:9-10-terms}
\end{align} 
\end{widetext}
Thus, 
replacing the ninth and tenth terms of Eq. (\ref{eq:phi-third}) 
by the terms of Eq. (\ref{eq:9-10-terms}), 
we can rewrite Eq. (\ref{eq:phi-third}) in a simpler form: 
\begin{widetext}
\begin{align}
K_{xy\nu}(\boldq,i\Omega_{n})
=&
\frac{T}{N}
\sum\limits_{\boldk}
\sum\limits_{m}
\sum\limits_{\{a\}}
\Lambda_{x;ba}(\boldk,i\epsilon_{m},\boldk,i\epsilon_{m+n})
\sum\limits_{\eta}\dfrac{q_{\eta}}{2}
\dfrac{\partial G_{ad}(\boldk,i\epsilon_{m+n})}{\partial k_{\eta}}
\dfrac{\partial G_{cb}(\boldk,i\epsilon_{m})}{\partial k_{\nu}}
\Lambda_{y;dc}(\boldk,i\epsilon_{m+n};\boldk,i\epsilon_{m})\notag\\
&+\frac{T}{N}
\sum\limits_{\boldk}
\sum\limits_{m}
\sum\limits_{\{a\}}
\Lambda_{x;ba}(\boldk,i\epsilon_{m},\boldk,i\epsilon_{m+n})
G_{ad}(\boldk,i\epsilon_{m+n})
\dfrac{\partial G_{cb}(\boldk,i\epsilon_{m})}{\partial k_{\nu}}
\sum\limits_{\eta}\dfrac{q_{\eta}}{2}
\dfrac{\partial \Lambda_{y;dc}(\boldk,i\epsilon_{m+n},\boldk,i\epsilon_{m})}
{\partial k_{\eta}}\notag\\
&-\frac{T}{N}
\sum\limits_{\boldk}
\sum\limits_{m}
\sum\limits_{\{a\}}
\Lambda_{x;ba}(\boldk,i\epsilon_{m},\boldk,i\epsilon_{m+n})
\sum\limits_{\eta}\dfrac{q_{\eta}}{2}
\dfrac{\partial G_{cb}(\boldk,i\epsilon_{m})}{\partial k_{\eta}}
\dfrac{\partial G_{ad}(\boldk,i\epsilon_{m+n})}{\partial k_{\nu}}
\Lambda_{y;dc}(\boldk,i\epsilon_{m+n},\boldk,i\epsilon_{m})\notag\\
&-\frac{T}{N}
\sum\limits_{\boldk}
\sum\limits_{m}
\sum\limits_{\{a\}}
\Lambda_{x;ba}(\boldk,i\epsilon_{m},\boldk,i\epsilon_{m+n})
G_{cb}(\boldk,i\epsilon_{m})
\dfrac{\partial G_{ad}(\boldk,i\epsilon_{m+n})}{\partial k_{\nu}}
\sum\limits_{\eta}\dfrac{q_{\eta}}{2}
\dfrac{\partial \Lambda_{y;dc}(\boldk,i\epsilon_{m+n},\boldk,i\epsilon_{m})}
{\partial k_{\eta}}\notag\\
&+ 
\frac{T}{N}
\sum\limits_{\boldk}
\sum\limits_{m}
\sum\limits_{\{a\}}
\Lambda_{x;ba}(\boldk,i\epsilon_{m},\boldk,i\epsilon_{m+n})
\sum\limits_{\eta}\dfrac{q_{\eta}}{2}
\Bigl[
G_{cb}(\boldk,i\epsilon_{m})
\dfrac{\overleftrightarrow{\partial}}{\partial k_{\eta}}
G_{ad}(\boldk,i\epsilon_{m+n})
\Bigr]\notag\\
&\ \times 
\dfrac{\partial \Lambda_{y;dc}(\boldk,i\epsilon_{m+n},\boldk,i\epsilon_{m})}
{\partial k_{\nu}}
\notag\\
&= 
\frac{T}{N}
\sum\limits_{\boldk}
\sum\limits_{m}
\sum\limits_{\{a\}}
\Lambda_{x;ba}(\boldk,i\epsilon_{m},\boldk,i\epsilon_{m+n})
\Lambda_{y;dc}(\boldk,i\epsilon_{m+n},\boldk,i\epsilon_{m})\notag\\
&\ \ \ \ \times
\sum\limits_{\eta}\dfrac{q_{\eta}}{2}
\Bigl\{
\dfrac{\partial G_{ad}(\boldk,i\epsilon_{m+n})}{\partial k_{\eta}}
\dfrac{\partial G_{cb}(\boldk,i\epsilon_{m})}{\partial k_{\nu}}
-\dfrac{\partial G_{ad}(\boldk,i\epsilon_{m+n})}{\partial k_{\nu}}
\dfrac{\partial G_{cb}(\boldk,i\epsilon_{m})}{\partial k_{\eta}}\Bigr\}\notag\\
&+
\frac{T}{N}
\sum\limits_{\boldk}
\sum\limits_{m}
\sum\limits_{\{a\}}
\Lambda_{x;ba}(\boldk,i\epsilon_{m},\boldk,i\epsilon_{m+n})\notag\\
&\ \ \times 
\sum\limits_{\eta}\dfrac{q_{\eta}}{2}
\Bigl\{
\Bigl[
G_{cb}(\boldk,i\epsilon_{m})
\dfrac{\overleftrightarrow{\partial}}{\partial k_{\eta}}
G_{ad}(\boldk,i\epsilon_{m+n})
\Bigr]
\dfrac{\partial \Lambda_{y;dc}(\boldk,i\epsilon_{m+n},\boldk,i\epsilon_{m})}
{\partial k_{\nu}}\notag\\
&\ \ \ \ \ \ \ \ \ \ \ \ \ 
-
\Bigl[
G_{cb}(\boldk,i\epsilon_{m})
\dfrac{\overleftrightarrow{\partial}}{\partial k_{\nu}}
G_{ad}(\boldk,i\epsilon_{m+n})
\Bigr]
\dfrac{\partial \Lambda_{y;dc}(\boldk,i\epsilon_{m+n},\boldk,i\epsilon_{m})}
{\partial k_{\eta}}
\Bigr\}\notag\\
&=
\frac{1}{2}
(q_{x}\delta_{\nu,y}-q_{y}\delta_{\nu,x})
\frac{T}{N}
\sum\limits_{\boldk}
\sum\limits_{m}
\sum\limits_{\{a\}}
\Lambda_{x;ba}(\boldk,i\epsilon_{m},\boldk,i\epsilon_{m+n})\notag\\
&\ \ \times
\Bigl\{
\Bigl[
G_{cb}(\boldk,i\epsilon_{m})
\dfrac{\overleftrightarrow{\partial}}{\partial k_{x}}
G_{ad}(\boldk,i\epsilon_{m+n})
\Bigr]
\dfrac{\partial \Lambda_{y;dc}(\boldk,i\epsilon_{m+n},\boldk,i\epsilon_{m})}
{\partial k_{y}}\notag\\
&\ \ \ \ \ \
-\Bigl[
G_{cb}(\boldk,i\epsilon_{m})
\dfrac{\overleftrightarrow{\partial}}{\partial k_{y}}
G_{ad}(\boldk,i\epsilon_{m+n})
\Bigr]
\dfrac{\partial \Lambda_{y;dc}(\boldk,i\epsilon_{m+n},\boldk,i\epsilon_{m})}{\partial k_{x}}
\Bigr\}\notag\\
&+\frac{1}{2}
(q_{x}\delta_{\alpha,y}-q_{y}\delta_{\alpha,x})
\frac{T}{N}
\sum\limits_{\boldk}
\sum\limits_{m}
\sum\limits_{\{a\}}
\Lambda_{x;ba}(\boldk,i\epsilon_{m},\boldk,i\epsilon_{m+n})\notag\\
&\ \ \times
\Bigl(
\dfrac{\partial G_{cb}(\boldk,i\epsilon_{m})}{\partial k_{y}}
\dfrac{\partial G_{ad}(\boldk,i\epsilon_{m+n})}{\partial k_{x}}
-\dfrac{\partial G_{cb}(\boldk,i\epsilon_{m})}{\partial k_{x}}
\dfrac{\partial G_{ad}(\boldk,i\epsilon_{m+n})}{\partial k_{y}}
\Bigr)
\Lambda_{y;dc}(\boldk,i\epsilon_{m+n};\boldk,i\epsilon_{m}).\label{eq:phi-fourth}
\end{align}
\end{widetext}
In the above derivation, 
we have used the fact that the surface terms arising from 
the partial integrations about $k_{x}$ or $k_{y}$ become zero 
due to the periodicity of the Brillouin zone, 
while I have not used the replacement used in Refs. \onlinecite{Fukuyama-RH} 
and \onlinecite{Kohno-Yamada}. 

Finally, 
we can rewrite Eq. (\ref{eq:phi-fourth}) as Eq. (\ref{eq:phi-final}) 
by using the equivalence between the $x$ and the $y$ directions. 

\section{Technical details about the numerical calculations 
  of the FLEX approximation} 

In this appendix, 
I remark on several techniques about the numerical calculations 
to solve a set of the equations of the FLEX approximation self-consistently 
by iteration using the fast Fourier transformation (FFT)~\cite{NumericalReceip}. 

To use the FFT for the quantities as a function of fermionic Matsubara frequency, 
we need to use the zero padding~\cite{NumericalReceip}. 
This is because of the antiperiodicity 
in terms of fermionic Matsubara frequency~\cite{Hori-phD}. 
For example, 
when the number of Matsubara frequencies is $2M$, 
the noninteracting single-particle Green's function should satisfy
\begin{align}
& G_{ab}^{0}(\boldk,i\epsilon_{m})
=
\begin{cases} 
\ G_{ab}^{0}(\boldk,i\epsilon_{m}) \ \textrm{for} \ 0\leq m < M\\
\ 0 \ \textrm{for} \ M\leq m < 3M\\
\ G_{ab}^{0}(\boldk,i\epsilon_{m-4M}) \ \textrm{for} \ 3M\leq m < 4M\\
\end{cases}.\label{eq:G0-zeropad}
\end{align}
The similar property is satisfied for 
$G_{ab}(\boldk,i\epsilon_{m})$ and $\Sigma_{ab}(\boldk,i\epsilon_{m})$. 
Furthermore, 
due to this property, 
the quantities as a function of bosonic Matsubara frequency  
such as $\chi_{abcd}(\boldq,i\Omega_{n})$ satisfy the following property: 
\begin{align}
& \chi_{abcd}(\boldq,i\Omega_{n})
=
\begin{cases} 
\ \chi_{abcd}(\boldq,i\Omega_{n}) \ \textrm{for} \ 0\leq n < 2M\\
\ 0 \ \textrm{for} \ n=2M\\
\ \chi_{abcd}(\boldq,i\Omega_{n-4M}) \ \textrm{for} \ 2M< m < 4M\\
\end{cases}.\label{eq:chi-zeropad}
\end{align}

Using the above properties, 
we can utilize the FFT to solve the set of the self-consistent equations 
of the FLEX approximation by iteration. 
First, 
using the input of the self-energy, 
which is zero for the first iteration, 
we determine $G_{ab}(\boldk,i\epsilon_{m})$ from the Dyson equation 
[i.e., Eq. (\ref{eq:FLEX-4})]. 
Second, we calculate the chemical potential from Eq. (\ref{eq:mu-int}) 
by the bisection method; 
in this calculation, 
to reduce the numerical error arising from the cut-off frequency, 
we use Eq. (\ref{eq:mu-int}) instead of the equation 
where only $G_{ab}(\boldk,i\epsilon_{m})$ appears, 
and we set the chemical potentials in $f(\epsilon_{\alpha}(\boldk))$, 
$G_{ab}(\boldk,i\epsilon_{m})$, and $G_{ab}^{0}(\boldk,i\epsilon_{m})$ the same. 
Third, 
we carry out the Fourier transformations of $G_{ab}(\boldk,i\epsilon_{m})$ 
about momentum and frequency: 
\begin{align}
G_{ab}(\boldr,\tau_{l})
=& 
\dfrac{1}{N}
\sum\limits_{\boldk}
e^{i\boldk \cdot \boldr}
e^{\frac{-i\pi l}{4M}}
T
\sum\limits_{m=0}^{4M-1}
e^{-2\pi i\frac{m l}{4M}}
G_{ab}(\boldk,i\epsilon_{m}).\label{eq:G-Fourier}
\end{align}
Fourth, using Eqs. (\ref{eq:G-Fourier}) and (\ref{eq:FLEX-1}), 
we determine $\chi_{abcd}(\boldq,i\Omega_{n})$ from the equation,  
\begin{align}
\chi_{abcd}(\boldq,i\Omega_{n})
=&
-\sum\limits_{\boldr}
e^{-i\boldq \cdot \boldr}
\frac{1}{4MT}
\sum\limits_{l=0}^{4M-1}
e^{2\pi i\frac{n l}{4 M}}\notag\\
&\times 
G_{ac}(\boldr,\tau_{l})
G_{db}(-\boldr,-\tau_{l}).\label{eq:chi-Fourier}
\end{align}
Fifth, 
from Eqs. (\ref{eq:chi-Fourier}), (\ref{eq:FLEX-2}), and (\ref{eq:FLEX-3}), 
we calculate $\chi_{abcd}^{\textrm{S}}(\boldq,i\Omega_{n})$ and 
$\chi_{abcd}^{\textrm{C}}(\boldq,i\Omega_{n})$. 
Sixth, by solving Eq. (\ref{eq:FLEX-6}), 
we obtain $V_{abcd}(\boldq,i\Omega_{n})$. 
Seventh, 
we carry out the Fourier transformations of $V_{abcd}(\boldq,i\Omega_{n})$ 
about momentum and frequency to determine $\Sigma_{ac}(\boldk,i\epsilon_{m})$ 
from the following equation obtained from Eq. (\ref{eq:FLEX-5}): 
\begin{align}
\Sigma_{ac}(\boldk,i\epsilon_{m})
=&
\sum\limits_{\boldr}
e^{-i\boldk \cdot \boldr}
\frac{1}{4MT}
\sum\limits_{l=0}^{4M-1}
e^{2\pi i\frac{m l}{4M}}
e^{\frac{i\pi l}{4M}}\notag\\
&\times
\sum\limits_{b,d}
V_{abcd}(\boldr,\tau_{l})
G_{bd}(\boldr,\tau_{l}).
\end{align}
At this stage, 
we obtain the output of the self-energy. 
If the sum of the difference between the absolute values of 
the output and the input becomes less than $10^{-4}$, 
I assume that 
the solution is obtained; 
otherwise, 
we replace the input of the self-energy by 
the average of the input and the output, 
and solve the above procedures again. 
Note that if we use the relations about the symmetry of the system 
such as time-reversal symmetry and even-parity symmetry 
and utilize the arrays efficiently, 
we can reduce the memory of the arrays and the time of the numerical calculations. 

In the above iterative procedures, 
we should increase the Hubbard interaction terms 
so slowly as to keep the susceptibilities finite~\cite{FLEX2} 
since the calculations are restricted to a PM phase. 
For example, 
if we analyze case at $U=1.8$ eV, $J_{\textrm{H}}=J^{\prime}=\frac{U}{6}$, 
and $U^{\prime}=U-2J_{\textrm{H}}=\frac{2U}{3}$, 
we begin with $U=0.2$ eV and then increase the value of $U$ slowly 
after several iterations. 

\section{Derivation of Eqs. (\ref{eq:Gamma-Matsu-sum}){--}(\ref{eq:AL2-Matsu})}

In this appendix, 
we derive the set of Eqs. (\ref{eq:Gamma-Matsu-sum}){--}(\ref{eq:AL2-Matsu}) 
by adopting Eq. (\ref{eq:Gamma1-consv}) to Eq. (\ref{eq:FLEX-5}). 

Using Eq. (\ref{eq:Gamma1-consv}) in the FLEX approximation, 
we obtain $\Gamma_{abcd}^{(1)}(\boldk,i\epsilon_{m},\boldk^{\prime},i\epsilon_{m^{\prime}}; 
\boldq,i\Omega_{n})$ in the FLEX approximation. 
Setting $\boldq=\boldzero$ and $\Omega_{n}=0$ in Eq. (\ref{eq:Gamma1-consv}) 
and substituting Eq. (\ref{eq:FLEX-5}) into Eq. (\ref{eq:Gamma1-consv}), 
we have 
\begin{align}
&\Gamma_{abcd}^{(1)}(\boldk,i\epsilon_{m},\boldk^{\prime},i\epsilon_{m^{\prime}}; 
\boldzero,0)\notag\\
=&\
\dfrac{T}{N}
\sum\limits_{\boldq^{\prime},n^{\prime}}
\sum\limits_{B,D}
V_{aBbD}(\boldq^{\prime},i\Omega_{n^{\prime}})
\dfrac{\delta G_{BD}(\boldk-\boldq^{\prime},i\epsilon_{m-n^{\prime}})}
{\delta G_{cd}(\boldk^{\prime},i\epsilon_{m^{\prime}})}\notag\\
&+\dfrac{T}{N}
\sum\limits_{\boldq^{\prime},n^{\prime}}
\sum\limits_{B,D}
\dfrac{\delta V_{aBbD}(\boldq^{\prime},i\Omega_{n^{\prime}})}
{\delta G_{cd}(\boldk^{\prime},i\epsilon_{m^{\prime}})}
G_{BD}(\boldk-\boldq^{\prime},i\epsilon_{m-n^{\prime}}).\label{eq:Gamma-Matsu-sum-zero}
\end{align}
The first term of Eq. (\ref{eq:Gamma-Matsu-sum-zero}) gives the MT term, 
\begin{align}
\Gamma_{abcd}^{(1)\textrm{MT}}(\boldk,i\epsilon_{m},\boldk^{\prime},i\epsilon_{m^{\prime}}; 
\boldzero,0)
=& \ V_{acbd}(\boldk-\boldk^{\prime},i\epsilon_{m}-i\epsilon_{m^{\prime}}),\label{eq:MT-Matsu-zero}
\end{align}
and the second term gives the AL1 and the AL2 term, 
\begin{widetext}
\begin{align}
&\Gamma_{abcd}^{(1)\textrm{AL}}(\boldk,i\epsilon_{m},\boldk^{\prime},i\epsilon_{m^{\prime}}; 
\boldzero,0)\notag\\
=&\ 
\dfrac{T}{N}
\sum\limits_{\boldq^{\prime}}
\sum\limits_{n^{\prime}}
\sum\limits_{B,D}
\Bigl[
\dfrac{3}{2}
\sum\limits_{\{a^{\prime}\}}
U_{aBa^{\prime}b^{\prime}}^{\textrm{S}}
\dfrac{\delta \chi_{a^{\prime}b^{\prime}c^{\prime}d^{\prime}}^{\textrm{S}}(\boldq^{\prime},i\Omega_{n^{\prime}})}
{\delta G_{cd}(\boldk^{\prime},i\epsilon_{m^{\prime}})}
U_{c^{\prime}d^{\prime}bD}^{\textrm{S}}
+\dfrac{1}{2}
\sum\limits_{\{a^{\prime}\}}
U_{aBa^{\prime}b^{\prime}}^{\textrm{C}}
\dfrac{\delta \chi_{a^{\prime}b^{\prime}c^{\prime}d^{\prime}}^{\textrm{C}}(\boldq^{\prime},i\Omega_{n^{\prime}})}
{\delta G_{cd}(\boldk^{\prime},i\epsilon_{m^{\prime}})}
U_{c^{\prime}d^{\prime}bD}^{\textrm{C}}\notag\\
&-
\sum\limits_{\{a^{\prime}\}}
U_{aa^{\prime}Bb^{\prime}}^{\uparrow\downarrow}
\dfrac{\delta \chi_{a^{\prime}b^{\prime}c^{\prime}d^{\prime}}(\boldq^{\prime},i\Omega_{n^{\prime}})}
{\delta G_{cd}(\boldk^{\prime},i\epsilon_{m^{\prime}})}
U_{c^{\prime}bd^{\prime}D}^{\uparrow\downarrow}
\Bigr]
G_{BD}(\boldk-\boldq^{\prime},i\epsilon_{m-n^{\prime}})
\notag\\
=&\
\dfrac{T}{N}
\sum\limits_{\boldq^{\prime}}
\sum\limits_{n^{\prime}}
\sum\limits_{B,D}
\Bigl[
\dfrac{3}{2}
\sum\limits_{\{A^{\prime}\}}
\tilde{N}_{aBA^{\prime}B^{\prime}}^{\textrm{S}}(\boldq^{\prime},i\Omega_{n^{\prime}})
\dfrac{\delta \chi_{A^{\prime}B^{\prime}C^{\prime}D^{\prime}}(\boldq^{\prime},i\Omega_{n^{\prime}})}
{\delta G_{cd}(\boldk^{\prime},i\epsilon_{m^{\prime}})}
\tilde{N}_{C^{\prime}D^{\prime}bD}^{\textrm{S}}(\boldq^{\prime},i\Omega_{n^{\prime}})\notag\\
&+\dfrac{1}{2}
\sum\limits_{\{A^{\prime}\}}
\tilde{N}_{aBA^{\prime}B^{\prime}}^{\textrm{C}}(\boldq^{\prime},i\Omega_{n^{\prime}})
\dfrac{\delta \chi_{A^{\prime}B^{\prime}C^{\prime}D^{\prime}}(\boldq^{\prime},i\Omega_{n^{\prime}})}
{\delta G_{cd}(\boldk^{\prime},i\epsilon_{m^{\prime}})}
\tilde{N}_{C^{\prime}D^{\prime}bD}^{\textrm{C}}(\boldq^{\prime},i\Omega_{n^{\prime}})\notag\\
&-
\sum\limits_{\{A^{\prime}\}}
U_{aA^{\prime}BB^{\prime}}^{\uparrow\downarrow}
\dfrac{\delta \chi_{A^{\prime}B^{\prime}C^{\prime}D^{\prime}}(\boldq^{\prime},i\Omega_{n^{\prime}})}
{\delta G_{cd}(\boldk^{\prime},i\epsilon_{m^{\prime}})}
U_{C^{\prime}bD^{\prime}D}^{\uparrow\downarrow}
\Bigr]
G_{BD}(\boldk-\boldq^{\prime},i\epsilon_{m-n^{\prime}})\notag\\
=&
-\dfrac{T}{N}
\sum\limits_{\boldq^{\prime}}
\sum\limits_{n^{\prime}}
\sum\limits_{\{A\}}
\Bigl[
\dfrac{3}{2}
\tilde{N}_{aBcA}^{\textrm{S}}(\boldq^{\prime},i\Omega_{n^{\prime}})
\tilde{N}_{dCbD}^{\textrm{S}}(\boldq^{\prime},i\Omega_{n^{\prime}})
+\dfrac{1}{2}
\tilde{N}_{aBcA}^{\textrm{C}}(\boldq^{\prime},i\Omega_{n^{\prime}})
\tilde{N}_{dCbD}^{\textrm{C}}(\boldq^{\prime},i\Omega_{n^{\prime}})
-
U_{acBA}^{\uparrow\downarrow}U_{dbCD}^{\uparrow\downarrow}
\Bigr]\notag\\
&\ \ \times 
G_{CA}(\boldk^{\prime}-\boldq^{\prime},i\epsilon_{m^{\prime}-n^{\prime}})
G_{BD}(\boldk-\boldq^{\prime},i\epsilon_{m-n^{\prime}})\notag\\
&-\dfrac{T}{N}
\sum\limits_{\boldq^{\prime}}
\sum\limits_{n^{\prime}}
\sum\limits_{\{A\}}
\Bigl[
\dfrac{3}{2}
\tilde{N}_{aBAd}^{\textrm{S}}(\boldq^{\prime},i\Omega_{n^{\prime}})
\tilde{N}_{CcbD}^{\textrm{S}}(\boldq^{\prime},i\Omega_{n^{\prime}})
+\dfrac{1}{2}
\tilde{N}_{aBAd}^{\textrm{C}}(\boldq^{\prime},i\Omega_{n^{\prime}})
\tilde{N}_{CcbD}^{\textrm{C}}(\boldq^{\prime},i\Omega_{n^{\prime}})
-
U_{aABd}^{\uparrow\downarrow}
U_{CbcD}^{\uparrow\downarrow}
\Bigr]\notag\\
&\ \ \times 
G_{AC}(\boldk^{\prime}+\boldq^{\prime},i\epsilon_{m^{\prime}+n^{\prime}})
G_{BD}(\boldk-\boldq^{\prime},i\epsilon_{m-n^{\prime}})\notag\\
=&
-\dfrac{T}{N}
\sum\limits_{\boldq^{\prime}}
\sum\limits_{n^{\prime}}
\sum\limits_{\{A\}}
W^{\textrm{AL}}_{aBcA;dCbD}(\boldq^{\prime},i\Omega_{n^{\prime}};\boldq^{\prime},i\Omega_{n^{\prime}})
G_{CA}(\boldk^{\prime}-\boldq^{\prime},i\epsilon_{m^{\prime}-n^{\prime}})
G_{BD}(\boldk-\boldq^{\prime},i\epsilon_{m-n^{\prime}})\notag\\
&-\dfrac{T}{N}
\sum\limits_{\boldq^{\prime}}
\sum\limits_{n^{\prime}}
\sum\limits_{\{A\}}
W^{\textrm{AL}}_{aBAd;CcbD}(\boldq^{\prime},i\Omega_{n^{\prime}};\boldq^{\prime},i\Omega_{n^{\prime}})
G_{AC}(\boldk^{\prime}+\boldq^{\prime},i\epsilon_{m^{\prime}+n^{\prime}})
G_{BD}(\boldk-\boldq^{\prime},i\epsilon_{m-n^{\prime}})\notag\\
=&\
\Gamma_{abcd}^{(1)\textrm{AL}1}(\boldk,i\epsilon_{m},\boldk^{\prime},i\epsilon_{m^{\prime}}; 
\boldzero,0)
+\Gamma_{abcd}^{(1)\textrm{AL}2}(\boldk,i\epsilon_{m},\boldk^{\prime},i\epsilon_{m^{\prime}}; 
\boldzero,0).\label{eq:AL-Matsu-zero}
\end{align}
\end{widetext}
In the derivation of Eq. (\ref{eq:AL-Matsu-zero}), 
we have used several equations: 
the second line of Eq. (\ref{eq:AL-Matsu-zero}) is obtained by using 
\begin{align}
&U_{aBa^{\prime}b^{\prime}}^{\textrm{S}}
\dfrac{\delta \chi_{a^{\prime}b^{\prime}c^{\prime}d^{\prime}}^{\textrm{S}}(\boldq^{\prime},i\Omega_{n^{\prime}})}
{\delta G_{cd}(\boldk^{\prime},i\epsilon_{m^{\prime}})}
U_{c^{\prime}d^{\prime}bD}^{\textrm{S}}\notag\\
=&\
\sum\limits_{\{A^{\prime}\}}
U_{aBa^{\prime}b^{\prime}}^{\textrm{S}}
(M^{\textrm{S}})^{-1}_{a^{\prime}b^{\prime}A^{\prime}B^{\prime}}(\boldq^{\prime},i\Omega_{n^{\prime}})
\dfrac{\delta \chi_{A^{\prime}B^{\prime}C^{\prime}D^{\prime}}(\boldq^{\prime},i\Omega_{n^{\prime}})}
{\delta G_{cd}(\boldk^{\prime},i\epsilon_{m^{\prime}})}\notag\\
&\times 
N^{\textrm{S}}_{C^{\prime}D^{\prime}c^{\prime}d^{\prime}}(\boldq^{\prime},i\Omega_{n^{\prime}})
U_{c^{\prime}d^{\prime}bD}^{\textrm{S}}\notag\\
=&\ 
\tilde{N}_{aBA^{\prime}B^{\prime}}^{\textrm{S}}(\boldq^{\prime},i\Omega_{n^{\prime}})
\dfrac{\delta \chi_{A^{\prime}B^{\prime}C^{\prime}D^{\prime}}(\boldq^{\prime},i\Omega_{n^{\prime}})}
{\delta G_{cd}(\boldk^{\prime},i\epsilon_{m^{\prime}})}
\tilde{N}_{C^{\prime}D^{\prime}bD}^{\textrm{S}}(\boldq^{\prime},i\Omega_{n^{\prime}}),
\end{align}
and
\begin{align}
&U_{aBa^{\prime}b^{\prime}}^{\textrm{C}}
\dfrac{\delta \chi_{a^{\prime}b^{\prime}c^{\prime}d^{\prime}}^{\textrm{C}}(\boldq^{\prime},i\Omega_{n^{\prime}})}
{\delta G_{cd}(\boldk^{\prime},i\epsilon_{m^{\prime}})}
U_{c^{\prime}d^{\prime}bD}^{\textrm{C}}\notag\\
=&\
\sum\limits_{\{A^{\prime}\}}
U_{aBa^{\prime}b^{\prime}}^{\textrm{C}}
(M^{\textrm{C}})^{-1}_{a^{\prime}b^{\prime}A^{\prime}B^{\prime}}(\boldq^{\prime},i\Omega_{n^{\prime}})
\dfrac{\delta \chi_{A^{\prime}B^{\prime}C^{\prime}D^{\prime}}(\boldq^{\prime},i\Omega_{n^{\prime}})}
{\delta G_{cd}(\boldk^{\prime},i\epsilon_{m^{\prime}})}\notag\\
&\times 
N^{\textrm{C}}_{C^{\prime}D^{\prime}c^{\prime}d^{\prime}}(\boldq^{\prime},i\Omega_{n^{\prime}})
U_{c^{\prime}d^{\prime}bD}^{\textrm{C}}\notag\\
=&\ 
\tilde{N}_{aBA^{\prime}B^{\prime}}^{\textrm{C}}(\boldq^{\prime},i\Omega_{n^{\prime}})
\dfrac{\delta \chi_{A^{\prime}B^{\prime}C^{\prime}D^{\prime}}(\boldq^{\prime},i\Omega_{n^{\prime}})}
{\delta G_{cd}(\boldk^{\prime},i\epsilon_{m^{\prime}})}
\tilde{N}_{C^{\prime}D^{\prime}bD}^{\textrm{C}}(\boldq^{\prime},i\Omega_{n^{\prime}}),
\end{align}
where 
$N_{abcd}^{\textrm{S}}(\boldq^{\prime},i\Omega_{n^{\prime}})$ 
and $N_{abcd}^{\textrm{C}}(\boldq^{\prime},i\Omega_{n^{\prime}})$ are 
\begin{align}
N^{\textrm{S}}_{abcd}(\boldq^{\prime},i\Omega_{n^{\prime}})
=&\ 
\delta_{a,A}\delta_{b,B}
+\sum\limits_{C,D}
U_{abCD}^{\textrm{S}}
\chi_{CDAB}^{\textrm{S}}(\boldq^{\prime},i\Omega_{n^{\prime}}),
\end{align}
and 
\begin{align}
N^{\textrm{C}}_{abcd}(\boldq^{\prime},i\Omega_{n^{\prime}})
=&\ 
\delta_{a,A}\delta_{b,B}
-\sum\limits_{C,D}
U_{abCD}^{\textrm{C}}
\chi_{CDAB}^{\textrm{C}}(\boldq^{\prime},i\Omega_{n^{\prime}}),
\end{align}
respectively, and 
$(M^{\textrm{S}})^{-1}_{abcd}(\boldq^{\prime},i\Omega_{n^{\prime}})$ 
and $(M^{\textrm{C}})^{-1}_{abcd}(\boldq^{\prime},i\Omega_{n^{\prime}})$ 
are the inverse matrices of 
$M^{\textrm{S}}_{abcd}(\boldq^{\prime},i\Omega_{n^{\prime}})$ 
and 
$M^{\textrm{C}}_{abcd}(\boldq^{\prime},i\Omega_{n^{\prime}})$, 
respectively, 
with  
\begin{align}
M^{\textrm{S}}_{abcd}(\boldq^{\prime},i\Omega_{n^{\prime}})
=\delta_{a,c}\delta_{b,d}
-
\sum\limits_{A,B}
\chi_{abAB}(\boldq^{\prime},i\Omega_{n^{\prime}})
U_{ABcd}^{\textrm{S}},
\end{align}
and 
\begin{align}
M^{\textrm{C}}_{abcd}(\boldq^{\prime},i\Omega_{n^{\prime}})
=\delta_{a,c}\delta_{b,d}
+
\sum\limits_{A,B}
\chi_{abAB}(\boldq^{\prime},i\Omega_{n^{\prime}})
U_{ABcd}^{\textrm{C}};
\end{align}  
to obtain the third line of Eq. (\ref{eq:AL-Matsu-zero}), 
we have used 
\begin{align}
\dfrac{\delta \chi_{A^{\prime}B^{\prime}C^{\prime}D^{\prime}}(\boldq^{\prime},i\Omega_{n^{\prime}})}
{\delta G_{cd}(\boldk^{\prime},i\epsilon_{m^{\prime}})}
&=-\delta_{A^{\prime},c}\delta_{C^{\prime},d}
G_{D^{\prime}B^{\prime}}(\boldk^{\prime}-\boldq^{\prime},i\epsilon_{m^{\prime}-n^{\prime}})\notag\\
&-\delta_{D^{\prime},c}\delta_{B^{\prime},d}
G_{A^{\prime}C^{\prime}}(\boldk^{\prime}+\boldq^{\prime},i\epsilon_{m^{\prime}+n^{\prime}}), 
\end{align}
where we have used Eq. (\ref{eq:FLEX-1}); 
we have introduced Eq. (\ref{eq:W-Matsu}) 
at the final line of Eq. (\ref{eq:AL-Matsu-zero}). 

Finally, we can obtain Eqs. (\ref{eq:Gamma-Matsu-sum}){--}(\ref{eq:AL2-Matsu}) 
by labeling $\boldq$ and $i\Omega_{n}$ correctly 
as the electron-hole scattering with 
the momentum transfer $\boldq$ and 
the frequency transfer $i\Omega_{n}$. 
For the correct labeling, 
see Figs. \ref{fig:Fig6}(a){--}\ref{fig:Fig6}(c). 

\section{Analytic continuations 
of Eqs. (\ref{eq:MT-Matsu}){--}(\ref{eq:AL2-Matsu})}

In this appendix, 
we explain the details of the analytic continuations 
of Eqs. (\ref{eq:MT-Matsu}){--}(\ref{eq:AL2-Matsu}) 
in case that 
their frequency variables satisfy the inequalities 
for region 22-II or 22-III or 22-IV of TABLE \ref{tab:4ptVF}. 

First, we can easily carry out the analytic continuation of Eq. (\ref{eq:MT-Matsu}). 
Namely, 
since Im$\epsilon-$Im$\epsilon^{\prime}$ is negative for region 22-II 
and positive for regions 22-III and 22IV (see Table \ref{tab:4ptVF}), 
the MT terms in regions 22-II, 22-III, and 22-IV are 
\begin{align}
\Gamma_{22\textrm{-II};abcd}^{(1)\textrm{MT}}(k,k^{\prime};q)
=& \ \delta_{\boldq,\boldzero}\delta_{\omega,0}
V_{acbd}^{(\textrm{A})}(k-k^{\prime}),\label{eq:MT-analy-22II}\\
\Gamma_{22\textrm{-III};abcd}^{(1)\textrm{MT}}(k,k^{\prime};q)
=& \ \delta_{\boldq,\boldzero}\delta_{\omega,0}
V_{acbd}^{(\textrm{R})}(k-k^{\prime}),\label{eq:MT-analy-22III}
\end{align}
and 
\begin{align}
\Gamma_{22\textrm{-IV};abcd}^{(1)\textrm{MT}}(k,k^{\prime};q)
=& \ \Gamma_{22\textrm{-III};abcd}^{(1)\textrm{MT}}(k,k^{\prime};q),\label{eq:MT-analy-22IV}
\end{align}
respectively. 

Second, 
we can obtain the AL1 terms for regions 22-II, 22-III, and 22-IV  
after several calculations for the analytic continuations 
by using a similar way for the analytic continuations in Sect. II B 1. 
Replacing the sum about bosonic Matsubara frequency by 
the corresponding contour integral~\cite{Takada-text2} 
and using the analytic properties of the single-particle Green's functions 
and $W^{\textrm{AL}}_{abcd;ABCD}(\boldq_{1},i\Omega_{n_{1}};\boldq_{2},i\Omega_{n_{2}})$, 
we obtain the AL1 term in region 22-II: 
\begin{widetext}
\begin{align}
\Gamma_{22\textrm{-II};\{a\}}^{(1)\textrm{AL}1}(k,k^{\prime};q)
=& 
-\frac{1}{N}
\sum\limits_{\boldq^{\prime}}
\sum\limits_{\{A \}}
\int_{\textrm{C}}\frac{d\omega^{\prime}}{4\pi i}
\coth \frac{\omega^{\prime}}{2T}
G_{CA}(\boldk^{\prime}+\boldq^{\prime},i\epsilon_{m^{\prime}}+\omega^{\prime})
G_{BD}(\boldk+\boldq^{\prime},i\epsilon_{m}+\omega^{\prime})\notag\\
&\times 
W^{\textrm{AL}}_{aBcA;dCbD}(\boldq-\boldq^{\prime},i\Omega_{n}-\omega^{\prime};
-\boldq^{\prime},-\omega^{\prime})\notag\\
&-\frac{T}{N}
\sum\limits_{\boldq^{\prime}}
\sum\limits_{\{A \}}
G_{CA}(\boldk^{\prime}+\boldq^{\prime},i\epsilon_{m^{\prime}})
G_{BD}(\boldk+\boldq^{\prime},i\epsilon_{m})
W^{\textrm{AL}}_{aBcA;dCbD}(\boldq-\boldq^{\prime},i\Omega_{n};
-\boldq^{\prime},0)\notag\\
&-\frac{T}{N}
\sum\limits_{\boldq^{\prime}}
\sum\limits_{\{A \}}
G_{CA}(\boldk^{\prime}+\boldq^{\prime},i\epsilon_{m^{\prime}}+i\Omega_{n})
G_{BD}(\boldk+\boldq^{\prime},i\epsilon_{m}+i\Omega_{n})
W^{\textrm{AL}}_{aBcA;dCbD}(\boldq-\boldq^{\prime},0;
-\boldq^{\prime},-i\Omega_{n})\notag\\
\rightarrow 
&
-\frac{1}{N}
\sum\limits_{\boldq^{\prime}}
\sum\limits_{\{A \}}
\int^{\infty}_{-\infty}\frac{d\omega^{\prime}}{2\pi}
\tanh \frac{\omega^{\prime}+\epsilon^{\prime}}{2T}
\textrm{Im}G_{CA}^{(\textrm{R})}(k^{\prime}+q^{\prime})
G_{BD}^{(\textrm{A})}(k+q^{\prime})
W^{\textrm{AL}(\textrm{RA})}_{aBcA;dCbD}(q-q^{\prime};-q^{\prime})\notag\\
&
-\frac{1}{N}
\sum\limits_{\boldq^{\prime}}
\sum\limits_{\{A \}}
\int^{\infty}_{-\infty}\frac{d\omega^{\prime}}{2\pi}
\tanh \frac{\omega^{\prime}+\epsilon}{2T}
G_{CA}^{(\textrm{R})}(k^{\prime}+q^{\prime})
\textrm{Im}G_{BD}^{(\textrm{R})}(k+q^{\prime})
W^{\textrm{AL}(\textrm{RA})}_{aBcA;dCbD}(q-q^{\prime};-q^{\prime})\notag\\
&+(\textrm{Principal integral terms}),\label{eq:AL1-analy-22II-appendix}
\end{align}
\end{widetext}
\begin{figure*}[tb]
\begin{center}
\includegraphics[width=180mm]{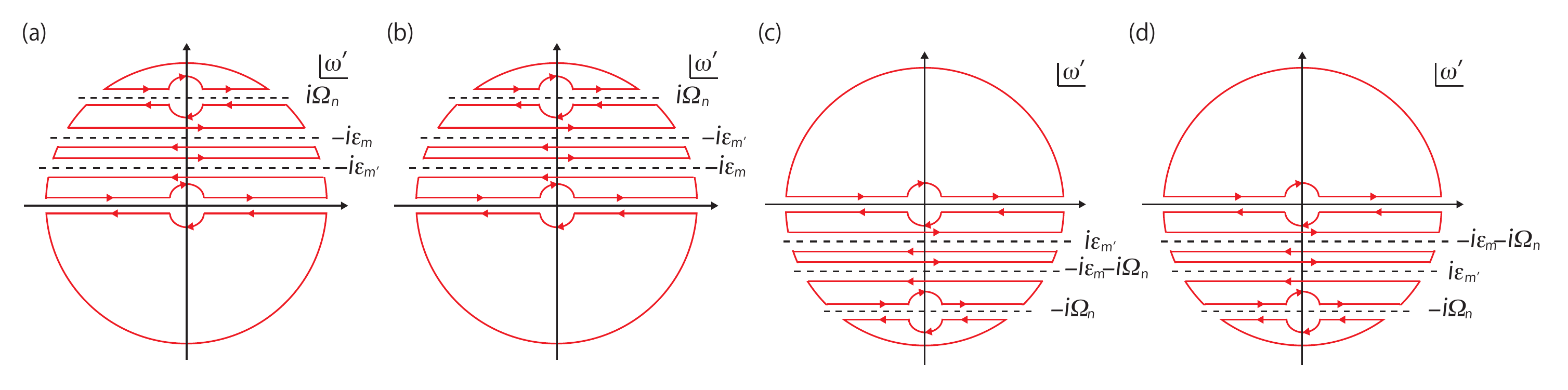}
\end{center}
\vspace{-15pt}
\caption{
Contours used for the analytic continuations of the AL$1$ term 
in (a) region $22$-II and (b) region $22$-III or $22$-IV 
and of the AL$2$ term 
in (c) region $22$-II or $22$-III and (d) region $22$-IV. }
\label{fig:Fig7}
\end{figure*}
where we have used the contour $C$ shown in Fig. \ref{fig:Fig7}(a) and 
introduced $W^{\textrm{AL}(\textrm{RA})}_{abcd;ABCD}(q_{1};q_{2})$,
\begin{align}
W^{\textrm{AL}(\textrm{RA})}_{abcd;ABCD}(q_{1};q_{2})
=&
\dfrac{3}{2}
\tilde{N}_{abcd}^{\textrm{S}(\textrm{R})}(\boldq_{1},\omega_{1})
\tilde{N}_{ABCD}^{\textrm{S}(\textrm{A})}(\boldq_{2},\omega_{2})\notag\\
&
+\dfrac{1}{2}
\tilde{N}_{abcd}^{\textrm{C}(\textrm{R})}(\boldq_{1},\omega_{1})
\tilde{N}_{ABCD}^{\textrm{C}(\textrm{A})}(\boldq_{2},\omega_{2})\notag\\
&-U_{acbd}^{\uparrow\downarrow}U_{ACBD}^{\uparrow\downarrow}.
\end{align}  
Furthermore, 
we have not explicitly written the terms of the principal integral 
since those terms exactly cancel out 
the terms of the principal integral for region 22-III or 22-IV; 
due to this cancellation, 
such terms are unnecessary to calculate the kernel of the CVCs 
since part of the kernel, $\mathcal{J}_{22;cdCD}^{(1)}(k,k^{\prime};0)$, 
is proportional to the difference between the AL1 terms for the different regions 
[see Eq. (\ref{eq:4VC-22})]. 
In deriving Eq. (\ref{eq:AL1-analy-22II-appendix}), 
we have used a relation of the single-particle Green's function 
as a result of the time-reversal and the even-parity symmetry, 
$G_{ab}^{(\textrm{R})}(k)=G_{ab}^{(\textrm{A})}(k)^{\ast}$; 
as described in Sect. II C, 
for more general derivation, 
we should replace $\textrm{Im}G_{ab}^{(\textrm{R})}(k)$ 
by $\frac{1}{2i}[G_{ab}^{(\textrm{R})}(k)-G_{ab}^{(\textrm{A})}(k)]$. 
(The similar remark holding in the other cases of the AL term is not described below.)  
Then, 
in the similar way for region 22-II by using the contour $C$ shown in Fig. \ref{fig:Fig7}(b), 
we obtain the AL1 term in region 22-III or 22-IV: 
\begin{widetext}
\begin{align}
\Gamma_{22\textrm{-III};abcd}^{(1)\textrm{AL}1}(k,k^{\prime};q)
=&
-\frac{1}{N}
\sum\limits_{\boldq^{\prime}}
\sum\limits_{\{A \}}
\int^{\infty}_{-\infty}\frac{d\omega^{\prime}}{2\pi}
\tanh \frac{\omega^{\prime}+\epsilon^{\prime}}{2T}
\textrm{Im}G_{CA}^{(\textrm{R})}(k^{\prime}+q^{\prime})
G_{BD}^{(\textrm{R})}(k+q^{\prime})
W^{\textrm{AL}(\textrm{RA})}_{aBcA;dCbD}(q-q^{\prime};-q^{\prime})\notag\\
&
-\frac{1}{N}
\sum\limits_{\boldq^{\prime}}
\sum\limits_{\{A \}}
\int^{\infty}_{-\infty}\frac{d\omega^{\prime}}{2\pi}
\tanh \frac{\omega^{\prime}+\epsilon}{2T}
G_{CA}^{(\textrm{A})}(k^{\prime}+q^{\prime})
\textrm{Im}G_{BD}^{(\textrm{R})}(k+q^{\prime})
W^{\textrm{AL}(\textrm{RA})}_{aBcA;dCbD}(q-q^{\prime};-q^{\prime})\notag\\
&+(\textrm{Principal integral terms}).\label{eq:AL1-analy-22III}
\end{align}
\end{widetext}
and 
\begin{align}
\Gamma_{22\textrm{-IV};abcd}^{(1)\textrm{AL}1}(k,k^{\prime};q)
=\Gamma_{22\textrm{-III};abcd}^{(1)\textrm{AL}1}(k,k^{\prime};q).\label{eq:AL1-analy-22IV}
\end{align}

Third, 
carrying out the analytic continuation of the AL2 term 
in the similar way for the AL1 term 
by using the contour $C$ shown in Fig. \ref{fig:Fig7}(c) or \ref{fig:Fig7}(d), 
we obtain the AL2 term 
for region 22-II or 22-III or 22-IV. 
Namely, 
the AL2 terms in regions 22-II, 22-III, and 22-IV are given by, respectively,
\begin{widetext}
\begin{align}
\Gamma_{22\textrm{-II};abcd}^{(1)\textrm{AL}2}(k,k^{\prime};q)
=& 
-\frac{1}{N}
\sum\limits_{\boldq^{\prime}}
\sum\limits_{\{A \}}
\int^{\infty}_{-\infty}\frac{d\omega^{\prime}}{2\pi}
\tanh \frac{\omega^{\prime}-\epsilon^{\prime}}{2T}
[-\textrm{Im}G_{AC}^{(\textrm{R})}(k^{\prime}-q^{\prime})]
G_{BD}^{(\textrm{R})}(k+q+q^{\prime})
W^{\textrm{AL}(\textrm{RA})}_{aBAd;CcbD}(-q^{\prime};-q-q^{\prime})\notag\\
&
-\frac{1}{N}
\sum\limits_{\boldq^{\prime}}
\sum\limits_{\{A \}}
\int^{\infty}_{-\infty}\frac{d\omega^{\prime}}{2\pi}
\tanh \frac{\omega^{\prime}+\epsilon+\omega}{2T}
G_{AC}^{(\textrm{R})}(k^{\prime}-q^{\prime})
\textrm{Im}G_{BD}^{(\textrm{R})}(k+q+q^{\prime})
W^{\textrm{AL}(\textrm{RA})}_{aBAd;CcbD}(-q^{\prime};-q-q^{\prime})\notag\\
&+(\textrm{Principal integral terms}),\label{eq:AL2-analy-22II}\\
\Gamma_{22\textrm{-III};abcd}^{(1)\textrm{AL}2}(k,k^{\prime};q)
=& \Gamma_{22\textrm{-II};abcd}^{(1)\textrm{AL}2}(k,k^{\prime};q),\label{eq:AL2-analy-22III}
\end{align}
and 
\begin{align}
\Gamma_{22\textrm{-IV};abcd}^{(1)\textrm{AL}2}(k,k^{\prime};q)
=
&
-\frac{1}{N}
\sum\limits_{\boldq^{\prime}}
\sum\limits_{\{A \}}
\int^{\infty}_{-\infty}\frac{d\omega^{\prime}}{2\pi}
\tanh \frac{\omega^{\prime}-\epsilon^{\prime}}{2T}
[-\textrm{Im}G_{AC}^{(\textrm{R})}(k^{\prime}-q^{\prime})]
G_{BD}^{(\textrm{A})}(k+q+q^{\prime})
W^{\textrm{AL}(\textrm{RA})}_{aBAd;CcbD}(-q^{\prime};-q-q^{\prime})\notag\\
&
-\frac{1}{N}
\sum\limits_{\boldq^{\prime}}
\sum\limits_{\{A \}}
\int^{\infty}_{-\infty}\frac{d\omega^{\prime}}{2\pi}
\tanh \frac{\omega^{\prime}+\epsilon+\omega}{2T}
G_{AC}^{(\textrm{A})}(k^{\prime}-q^{\prime})
\textrm{Im}G_{BD}^{(\textrm{R})}(k+q+q^{\prime})
W^{\textrm{AL}(\textrm{RA})}_{aBAd;CcbD}(-q^{\prime};-q-q^{\prime})\notag\\
&+(\textrm{Principal integral terms}).\label{eq:AL2-analy-22IV}
\end{align} 
\end{widetext}

\section{Technical details about numerically solving 
  the Bethe-Salpeter equation for the current}

In this appendix, 
I give several technical remarks about the numerical calculations 
to self-consistently determine the current from the Bethe-Salpeter equation. 

Before the remarks about the iterative self-consistent procedures of 
the Bethe-Salpeter equation, 
I remark on how to numerically calculate the frequency integral. 
We calculate the frequency integral by 
discretizing it with finite interval such as $\Delta \epsilon_{j}$ 
and approximating the upper and the lower value of the integral 
by the finite cutoff values 
such as $\epsilon_{\textrm{c}}$ and $-\epsilon_{\textrm{c}}$, respectively; 
those values can be appropriately chosen within the numerical accuracy. 
(In the similar way, 
we numerically calculate the frequency integrals of the conductivities.)

We can self-consistently solve the Bethe-Salpeter equation 
for the current including the $\Sigma$ CVC, the MT CVC, and the AL CVC by iteration 
in the following procedures.

First, 
we determine the MT CVC in the presence of the $\Sigma$ CVC. 
Using the input of the current,  
which is $\Lambda_{\nu;2;AB}^{(0)}(\boldk,\epsilon_{j};0)$ for the first iteration, 
we determine $\tilde{\Lambda}_{\nu;2;ab}(\boldk,\epsilon_{j};0)$ from Eq. (\ref{eq:tild-Lamb}). 
After carrying out the Fourier transformations of 
$\tilde{\Lambda}_{\nu;2;CD}(\boldk,\epsilon_{j};0)$ 
and $\textrm{Im}V_{cCdD}^{(\textrm{R})}(\boldk-\boldk^{\prime},\epsilon_{j}-\epsilon^{\prime}_{j^{\prime}})$ 
about momentum, 
we calculate $\Delta \Lambda_{\nu;2;cd}^{\textrm{MT}}(\boldr,\epsilon_{j};0)$ as follows: 
\begin{align}
\Delta \Lambda_{\nu;2;cd}^{\textrm{MT}}(\boldr,\epsilon_{j};0)
=&
\sum\limits_{C,D}
\sum\limits_{\epsilon_{j^{\prime}}^{\prime}}\dfrac{\Delta \epsilon_{j^{\prime}}^{\prime}}{2\pi}
F_{ct}^{\textrm{MT}}(\epsilon_{j},\epsilon^{\prime}_{j^{\prime}};T)\notag\\
&\times \textrm{Im}V_{cCdD}^{(\textrm{R})}(\boldr,\epsilon_{j}-\epsilon^{\prime}_{j^{\prime}})\notag\\
&\times 
\tilde{\Lambda}_{\nu;2;CD}(\boldr,\epsilon_{j^{\prime}}^{\prime};0).
\end{align}
Since $F_{ct}^{\textrm{MT}}(\epsilon_{j},\epsilon^{\prime}_{j^{\prime}};T)$ includes 
the hyperbolic cotangent [see Eq. (\ref{eq:FctMT})] 
and its principal integral has the $0$/$0$ structure due to 
$\textrm{Im}V_{cCdD}^{(\textrm{R})}(\boldr,0)=0$, 
the principal integral can be calculated as follows~\cite{NA-review}: 
\begin{align}
&\sum\limits_{\epsilon_{j^{\prime}}^{\prime}}
\dfrac{\Delta \epsilon_{j^{\prime}}^{\prime}}{2\pi}
\coth \frac{\epsilon_{j}-\epsilon_{j^{\prime}}^{\prime}}{2T}
\textrm{Im}V_{cCdD}^{(\textrm{R})}(\boldr,\epsilon_{j}-\epsilon^{\prime}_{j^{\prime}})\notag\\
&\times 
\tilde{\Lambda}_{\nu;2;CD}(\boldr,\epsilon^{\prime}_{j^{\prime}};0)\notag\\
=& 
\sum\limits_{\epsilon_{j^{\prime}}^{\prime}\neq \epsilon_{j}}
\dfrac{\Delta \epsilon_{j^{\prime}}^{\prime}}{2\pi}
\coth \frac{\epsilon_{j}-\epsilon_{j^{\prime}}^{\prime}}{2T}
\textrm{Im}V_{cCdD}^{(\textrm{R})}(\boldr,\epsilon_{j}-\epsilon^{\prime}_{j^{\prime}})\notag\\
&\times 
\tilde{\Lambda}_{\nu;2;CD}(\boldr,\epsilon^{\prime}_{j^{\prime}};0)\notag\\
&-\dfrac{\Delta \epsilon_{j^{\prime}}^{\prime}}{2\pi}
T\dfrac{\partial}
{\partial \epsilon^{\prime}_{j^{\prime}}}
\bigl[
(e^{\frac{\epsilon^{\prime}_{j^{\prime}}-\epsilon_{j}}{T}}+1)
\textrm{Im}V_{cCdD}^{(\textrm{R})}(\boldr,\epsilon_{j}-\epsilon^{\prime}_{j^{\prime}})\notag\\
&\ \ \ \ \ \ \ \ \ \ \ \ \ \ \ \ \ \ \ \ \ \ \ \ \times 
\tilde{\Lambda}_{\nu;2;CD}(\boldr,\epsilon^{\prime}_{j^{\prime}};0)\bigr]
\Bigl|_{\epsilon^{\prime}_{j^{\prime}}=\epsilon_{j}}.
\end{align}
After the Fourier transformation of 
$\Delta \Lambda_{\nu;2;cd}^{\textrm{MT}}(\boldr,\epsilon_{j};0)$ about $\boldr$, 
we obtain the MT CVC, $\Delta \Lambda_{\nu;2;cd}^{\textrm{MT}}(\boldk,\epsilon_{j};0)$, 
and then add this to the input of the current. 
(If we consider only the MT CVC, i.e. neglect the AL CVC, 
we skip the following second and third steps, 
and the sum of the input and the MT CVC becomes the output.)

Second, 
we turn to the calculation of the AL1 CVC. 
Carrying out the Fourier transformation of  
$\textrm{Im}G_{cc}^{(\textrm{R})}(\boldk,\epsilon_{j^{\prime}}^{\prime})$ about $\boldk$ 
and using $\tilde{\Lambda}_{\nu;2;cc}(\boldr,\epsilon_{j};0)$, 
which is the same for the calculation of the MT CVC, 
we calculate $\bigl[\textrm{Im}G\tilde{\Lambda}\bigr]_{c}(-\boldk;
\epsilon_{j^{\prime}}^{\prime},\omega^{\prime}_{j^{\prime\prime}})$ from the equation,  
\begin{align}
\bigl[\textrm{Im}G\tilde{\Lambda}\bigr]_{c}(-\boldk;
\epsilon_{j^{\prime}}^{\prime},\omega^{\prime}_{j^{\prime\prime}})
=&
\sum\limits_{\boldr}e^{i\boldk \cdot \boldr}
\textrm{Im}G_{cc}^{(\textrm{R})}(\boldr,\epsilon^{\prime}_{j^{\prime}}+\omega^{\prime}_{j^{\prime\prime}})\notag\\
&\times 
\tilde{\Lambda}_{\nu;2;cc}(-\boldr,\epsilon_{j^{\prime}}^{\prime};0).
\end{align}
After calculating $\tilde{N}_{cccc}^{\textrm{S}(\textrm{R})}(\boldq,\omega_{j})$ 
and $\tilde{N}_{cccc}^{\textrm{C}(\textrm{R})}(\boldq,\omega_{j})$ from 
Eqs. (\ref{eq:NS-AL-simple}) and (\ref{eq:NC-AL-simple}), respectively, 
and using Eq. (\ref{eq:W-AL-simple}), 
we determine $X_{c}^{\textrm{AL}1}(\boldk;\epsilon_{j},\omega_{j^{\prime\prime}}^{\prime})$ 
from the following equation: 
\begin{align}
X_{c}^{\textrm{AL}1}(\boldk;\epsilon_{j},\omega_{j^{\prime\prime}}^{\prime})
=&
\sum\limits_{\epsilon_{j^{\prime}}^{\prime}}\Delta \epsilon_{j^{\prime}}^{\prime}
F_{\textrm{tt}}^{\textrm{AL}1}(\epsilon_{j},\epsilon^{\prime}_{j^{\prime}},\omega^{\prime}_{j^{\prime\prime}};T)\notag\\
&\times
F_{\textrm{ct}}^{\textrm{AL}1}(\epsilon_{j},\epsilon^{\prime}_{j^{\prime}};T)
W_{c}^{\textrm{AL}(\textrm{RA})}(\boldk,-\omega_{j^{\prime\prime}}^{\prime})\notag\\
&\times 
\bigl[\textrm{Im}G\tilde{\Lambda}\bigr]_{c}(-\boldk;
\epsilon_{j^{\prime}}^{\prime},\omega^{\prime}_{j^{\prime\prime}}).\label{eq:AL1-tildeW}
\end{align}
In the above equation, 
we can calculate the principal integral of the hyperbolic cotangent of 
$F_{\textrm{ct}}^{\textrm{AL}1}(\epsilon_{j},\epsilon^{\prime}_{j^{\prime}};T)$ 
in the similar way for the MT CVC 
since that principal integral also has the $0$/$0$ structure 
due to $F_{\textrm{tt}}^{\textrm{AL}1}(\epsilon_{j},\epsilon_{j},\omega^{\prime}_{j^{\prime\prime}};T)=0$ 
[see Eq. (\ref{eq:FttAL1})]. 
Carrying out the Fourier transformation of 
$X_{c}^{\textrm{AL}1}(\boldk;\epsilon_{j},\omega_{j^{\prime\prime}}^{\prime})$ 
about $\boldk$ 
and using the equation, 
\begin{align}
\Delta \Lambda_{\nu;2;cc}^{\textrm{AL}1}(\boldk,\epsilon_{j};0)
=&
-\frac{1}{4\pi^{2}}
\sum\limits_{\boldr}e^{-i\boldk \cdot \boldr}
\sum\limits_{\omega_{j^{\prime\prime}}^{\prime}}
\Delta \omega_{j^{\prime\prime}}^{\prime}\notag\\
&\times 
\textrm{Im}G_{cc}^{(\textrm{R})}(\boldr,\epsilon_{j}+\omega^{\prime}_{j^{\prime\prime}})\notag\\
&\times 
X_{c}^{\textrm{AL}1}(\boldr;\epsilon_{j},\omega_{j^{\prime\prime}}^{\prime}),
\end{align}
we obtain the AL1 CVC, $\Delta \Lambda_{\nu;2;cc}^{\textrm{AL}1}(\boldk,\epsilon_{j};0)$. 
At this point, 
we add the AL1 CVC to the sum of the input of the current and the MT CVC. 

Third, 
we calculate the AL2 CVC. 
Using $\bigl[\textrm{Im}G\tilde{\Lambda}\bigr]_{c}(\boldk;
\epsilon_{j^{\prime}}^{\prime},-\omega^{\prime}_{j^{\prime\prime}})$ 
and $W_{c}^{\textrm{AL}(\textrm{RA})}(\boldk,-\omega_{j^{\prime\prime}}^{\prime})$ 
(which have been determined in the above second step), 
we calculate $X_{c}^{\textrm{AL}2}(\boldk;\epsilon_{j},\omega_{j^{\prime\prime}}^{\prime})$ 
from the following equation: 
\begin{align}
X_{c}^{\textrm{AL}2}(\boldk;\epsilon_{j},\omega_{j^{\prime\prime}}^{\prime})
=&
\sum\limits_{\epsilon_{j^{\prime}}^{\prime}}\Delta \epsilon_{j^{\prime}}^{\prime}
F_{\textrm{tt}}^{\textrm{AL}2}(\epsilon_{j},\epsilon^{\prime}_{j^{\prime}},\omega^{\prime}_{j^{\prime\prime}};T)\notag\\
&\times
F_{\textrm{ct}}^{\textrm{AL}2}(\epsilon_{j},\epsilon^{\prime}_{j^{\prime}};T)
W_{c}^{\textrm{AL}(\textrm{RA})}(\boldk,-\omega_{j^{\prime\prime}}^{\prime})\notag\\
&\times 
\bigl[\textrm{Im}G\tilde{\Lambda}\bigr]_{c}(\boldk;
\epsilon_{j^{\prime}}^{\prime},-\omega^{\prime}_{j^{\prime\prime}}).\label{eq:AL2-tildeW}
\end{align}
Since 
$F_{\textrm{tt}}^{\textrm{AL}2}(\epsilon_{j},-\epsilon_{j},\omega^{\prime}_{j^{\prime\prime}};T)=0$ 
is satisfied in the above equation [see Eq. (\ref{eq:FttAL2})], 
we can calculate the principal integral of the hyperbolic cotangent of 
$F_{\textrm{ct}}^{\textrm{AL}2}(\epsilon_{j},\epsilon^{\prime}_{j^{\prime}};T)$ 
in the similar way for the MT CVC. 
After the Fourier transformation of 
$X_{c}^{\textrm{AL}2}(\boldk;\epsilon_{j},\omega_{j^{\prime\prime}}^{\prime})$ about $\boldk$, 
we obtain the AL2 CVC, 
$\Delta \Lambda_{\nu;2;cc}^{\textrm{AL}2}(\boldk,\epsilon_{j};0)$: 
\begin{align}
\Delta \Lambda_{\nu;2;cc}^{\textrm{AL}2}(\boldk,\epsilon_{j};0)
=&
-\frac{1}{4\pi^{2}}
\sum\limits_{\boldr}e^{-i\boldk \cdot \boldr}
\sum\limits_{\omega_{j^{\prime\prime}}^{\prime}}
\Delta \omega_{j^{\prime\prime}}^{\prime}\notag\\
&\times 
\textrm{Im}G_{cc}^{(\textrm{R})}(\boldr,\epsilon_{j}+\omega_{j^{\prime\prime}}^{\prime})\notag\\
&\times X_{c}^{\textrm{AL}2}(\boldr;\epsilon_{j},\omega_{j^{\prime\prime}}^{\prime}).
\end{align}
Adding the AL2 CVC to the sum of the input of the current, 
the MT CVC, and the AL1 CVC, 
we obtain the output of the current for the Bethe-Salpeter equation 
with the $\Sigma$ CVC, the MT CVC, and the AL CVC. 

Finally, 
estimating the sum of the difference between 
the absolute values of the output and the input of the current, 
we judge whether 
the output can be regarded as the solution of the Bethe-Salpeter equation 
within the numerical accuracy: 
if the difference becomes less than $10^{-4}$, 
the solution is assumed to be obtained; 
otherwise, 
after replacing the input by the output, 
we solve the above procedures again. 

\section{Method to determine the dominant fluctuations 
  in a multiorbital Hubbard model}

In this appendix, 
I explain how to determine the dominant fluctuations 
in a multiorbital Hubbard model 
among the four kinds of fluctuations. 

To discuss the dominant fluctuations in a multiorbital Hubbard model, 
we need to consider 
charge fluctuations, spin fluctuations, 
orbital fluctuations, and spin-orbital-combined fluctuations~\cite{Ueda-paramag}. 
This is because 
the Hubbard interaction terms can be expressed 
in terms of their operators~\cite{Ueda-paramag}. 
Then, 
spin and spin-orbital-combined fluctuations 
are described by $\chi_{abcd}^{\textrm{S}}(\boldq,i\Omega_{n})$, 
and charge and orbital fluctuations 
are described by $\chi_{abcd}^{\textrm{C}}(\boldq,i\Omega_{n})$~\cite{Ueda-paramag,NA-paramag}. 
For example, 
spin fluctuations are characterized by 
$\chi^{\textrm{S}}(\boldq,i\Omega_{n})
=\sum_{a,b}\chi^{\textrm{S}}_{aabb}(\boldq,i\Omega_{n})$~\cite{Ueda-paramag,NA-paramag}; 
spin-orbital-combined fluctuations are characterized by 
the correlation function between the products of the spin and the orbital operator, 
e.g. $\sum_{a}[\chi^{\textrm{S}}_{aa32}(\boldq,i\Omega_{n})
+\chi^{\textrm{S}}_{aa23}(\boldq,i\Omega_{n})]$~\cite{Ueda-paramag,NA-paramag}. 

Since those four kinds of fluctuations have 
the different dependence on $J_{\textrm{H}}$ or $U$, 
the dominant fluctuations can be determined 
by analyzing $J_{\textrm{H}}$ and $U$ dependences 
of $\lambda^{\textrm{S}}_{\textrm{max}}(\boldq)^{-1}$ 
and $\lambda^{\textrm{C}}_{\textrm{max}}(\boldq)^{-1}$. 
The difference in the $J_{\textrm{H}}$ or $U$ dependence 
arises from the difference in the dependence of 
the bare four-point vertex function characterizing the fluctuation. 
For example, 
in case~\cite{NA-paramag} of a $t_{2g}$-orbital Hubbard model 
with $J^{\prime}=J_{\textrm{H}}$ and $U^{\prime}=U-2J_{\textrm{H}}$, 
the bare four-point vertex function for charge fluctuations 
is $-U-4U^{\prime}+2J_{\textrm{H}}=-5U+10J_{\textrm{H}}$; 
that for spin fluctuations is $U+2J_{\textrm{H}}$; 
that for orbital fluctuations is 
$U^{\prime}-2J_{\textrm{H}}-J^{\prime}=U-5J_{\textrm{H}}$ 
or $U^{\prime}-2J_{\textrm{H}}+J^{\prime}=U-3J_{\textrm{H}}$ 
or $-U+2U^{\prime}-J_{\textrm{H}}=U-5J_{\textrm{H}}$; 
that for orbital-spin-combined fluctuations is 
$U^{\prime}+J^{\prime}=U-J_{\textrm{H}}$ or $U^{\prime}-J^{\prime}=U-3J_{\textrm{H}}$ or $U-J_{\textrm{H}}$.
Thus, if orbital or spin-orbital-combined fluctuations are dominant, 
we obtain non-monotonic $J_{\textrm{H}}$ dependence of 
$\lambda^{\textrm{C}}_{\textrm{max}}(\boldq)^{-1}$ 
or $\lambda^{\textrm{S}}_{\textrm{max}}(\boldq)^{-1}$, respectively, 
with increasing $J_{\textrm{H}}$ 
since the bare four-point vertex function characterizing those fluctuations 
changes from repulsive to attractive at a critical value of $J_{\textrm{H}}$ 
as a function of $U$; 
if spin fluctuations are dominant, 
we obtain monotonic $J_{\textrm{H}}$ dependence of $\lambda^{\textrm{S}}_{\textrm{max}}(\boldq)^{-1}$ 
with increasing $J_{\textrm{H}}$ 
due to the always repulsive bare four-point vertex function 
characterizing those. 
Note that 
charge fluctuations are always suppressed 
in a realistic set of the Hubbard interaction terms.


\begin{acknowledgments}
I thank 
Y. Yanase and T. Kariyado for several useful comments about 
the numerical calculations of the FLEX and the Pad\'{e} approximation 
and H. Kontani for a useful comment 
about the numerical calculations of the Pad\'{e} approximation. 
Almost all the numerical calculations were performed 
using the facilities of the Super Computer Center, 
the Institute for Solid State Physics, the University of Tokyo. 
\end{acknowledgments}


\begin{thebibliography}{99}
\bibitem{Fazekas}
P. Fazekas, Lecture Notes on Electron Correlation and
Magnetism (World Scientific, Singapore, 1999).

\bibitem{Moriya-review}
T. Moriya, 
J. Magn. Magn. Mater. \textbf{14}, 1, (1979).

\bibitem{Kon-review} 
H. Kontani, 
Rep. Prog. Phys. {\bf 71}, 026501, (2008). 

\bibitem{MIT-review}
M. Imada, A. Fujimori, and Y. Tokura, 
Rev. Mod. Phys. {\bf 70}, 1039, (1998).

\bibitem{Ashcroft-Mermin}
N. W. Ashcroft and N. D. Mermin, 
Solid State Physics (Thomson Learning, Ithaca, 1976).

\bibitem{Ziman}
J. M. Ziman, 
Principles of the Theory of Solids 
(Cambridge University Press, Cambridge, 1979). 

\bibitem{Landau}
L. D. Landau,  
ZhETF {\bf 30}, 1058 (1956)
[Sov. Phys. JETP {\bf 3}, 920 (1957)]. 

\bibitem{Nozieres}
P. Nozi$\grave{\textrm{e}}$res, 
Theory of Interacting Fermi Systems 
(Westview Press, Colorado, 1997). 

\bibitem{AGD}
A. A. Abrikosov, L. P. Gor'kov, I. E. Dyaloshinski, 
Methods of Quantum Field Theory in Statistical Physics
(Dover Publications, Mineola, 1963).

\bibitem{NA-review}
N. Arakawa, 
Mod. Phys. Lett. B {\bf 29}, 1530005, (2015).

\bibitem{Maeno-RW}
Y. Maeno, K. Yoshida, H. Hashimoto, S. Nishizaki, S. Ikeda, 
M. Nohara, T. Fujita, A. P. Mackenzie, N. E. Hussey, 
J. G. Bednorz, and F. Lichtenberg, 
J. Phys. Soc. Jpn. \textbf{66}, 1405 (1997).

\bibitem{resistivity-x2}
N. E. Hussey, A. P. Mackenzie, J. R. Cooper, 
Y. Maeno, S. Nishizaki, and T. Fujita, 
Phys. Rev. B \textbf{57}, 5505 (1998).

\bibitem{dHvA-x2}
A. P. Mackenzie, S. R. Julian, A. J. Diver, G. J. McMullan, M. P. Ray,
G. G. Lonzarich, Y. Maeno, S. Nishizaki, and T. Fujita, 
Phys. Rev. Lett. \textbf{76}, 3786 (1996).

\bibitem{Mackenzie-review}
A. P. Mackenzie and Y. Maeno, 
Rev. Mod. Phys. {\bf 75}, 657 (2003).

\bibitem{Yamada-Yosida}
K. Yamada and Y. Yosida,
Prog. Theor. Phys. {\bf 76}, 621 (1986).

\bibitem{Ti214-nFL1} 
M. Minakata and Y. Maeno, 
Phys. Rev. B {\bf 63}, 180504(R) (2001).

\bibitem{Ti214-nFL2} 
N. Kikugawa and Y. Maeno, 
Phys. Rev. Lett. {\bf 89}, 117001 (2002).

\bibitem{CSRO-nFL1}
S. Nakatsuji and Y. Maeno, 
Phys. Rev. Lett. {\bf 84}, 2666 (2000). 

\bibitem{CSRO-nFL2}
L. M. Galvin, R. S. Perry, A. W. Tyler, A. P. Mackenzie, 
S. Nakatsuji, and Y. Maeno, 
Phys. Rev. B \textbf{63}, 161102(R) (2001).

\bibitem{Neutron-Ti214}
M. Braden, O. Friedt, Y. Sidis, P. Bourges, M. Minakata, and Y. Maeno, 
Phys. Rev. Lett. {\bf 88}, 197002 (2002).

\bibitem{Neutron-x2}
Y. Sidis, M. Braden, P. Bourges, B. Hennion, S. NishiZaki, 
Y. Maeno, and Y. Mori, 
Phys. Rev. Lett. \textbf{83}, 3320 (1999).

\bibitem{xray-CSRO}
O. Friedt, M. Braden, G. Andr$\acute{\textrm{e}}$, P. Adelmann, 
S. Nakatsuji, and Y. Maeno, 
Phys. Rev. B \textbf{63}, 174432 (2001).

\bibitem{Terakura}
Z. Fang and K. Terakura, 
Phys. Rev. B \textbf{64}, 020509(R) (2001). 

\bibitem{NA-GA}
N. Arakawa and M. Ogata, 
Phys. Rev. B \textbf{86}, 125126 (2012).

\bibitem{X-ray10Dq}
H.-J. Noh, S.-J. Oh, B.-G. Park, J.-H. Park, J.-Y. Kim, H.-D. Kim, 
T. Mizokawa, L. H. Tjeng, H.-J. Lin, C. T. Chen, S. Schuppler, 
S. Nakatsuji, H. Fukazawa, and Y. Maeno, 
Phys. Rev. B \textbf{72}, 052411 (2005).

\bibitem{Oguchi}
T. Oguchi, 
Phys. Rev. B \textbf{51}, 1385 (1995).

\bibitem{Mazin-LDA} 
I. I. Mazin and D. J. Singh, 
Phys. Rev. Lett. \textbf{79}, 733 (1997). 

\bibitem{ARPES-x2}
A. Damascelli, D. H. Lu, K. M. Shen, N. P. Armitage, 
F. Ronning, D. L. Feng, C. Kim, Z.-X. Shen, 
T. Kimura, Y. Tokura, Z. Q. Mao, and Y. Maeno, 
Phys. Rev. Lett. \textbf{85}, 5194 (2000).

\bibitem{Yanase-review}
Y. Yanase, T. Jujo, T. Nomura, H. Ikeda, T. Hotta, and K. Yamada,  
Physics Reports {\bf 387}, 1 (2003).

\bibitem{FLEX1}
N. E. Bickers, D. J. Scalapino, and S. R. White, 
Phys. Rev. Lett. {\bf 62}, 961 (1989). 

\bibitem{FLEX2} 
N. E. Bickers and D. J. Scalapino, 
Annals of Physics {\bf 193}, 206 (1989).

\bibitem{FLEX3} 
N. E. Bickers and S. R. White, 
Phys. Rev. B {\bf 43}, 8044 (1991).

\bibitem{multi-FLEX1}
T. Takimoto, T. Hotta, and K. Ueda, 
Phys. Rev. B \textbf{69}, 104504 (2004).

\bibitem{multi-FLEX2}
H. Ikeda, R. Arita, and J. Kune\ifmmode \check{s}\else \v{s}\fi{}, 
Phys. Rev. B {\bf 81}, 054502 (2010).

\bibitem{MT1}
K. Maki, 
Prog. Theor. Phys. \textbf{40}, 193 (1968).

\bibitem{MT2}
R. S. Thompson, 
Phys. Rev. B \textbf{1}, 327 (1970).

\bibitem{Kon-CVC} 
H. Kontani, K. Kanki, and K. Ueda, 
Phys. Rev. B {\bf 59}, 14723 (1999).

\bibitem{NA-CVC}
N. Arakawa, 
Phys. Rev. B {\bf 90}, 245103 (2014).

\bibitem{cuprate-chiS}
S. Ohsugi, Y. Kitaoka, K. Ishida, and K. Asayama, 
J. Phys. Soc. Jpn. \textbf{60}, 2351 (1991).

\bibitem{cuprate-rho}
S. W. Tozer, A. W. Kleinsasser, T. Penney, D. Kaiser, and F. Holtzberg, 
Phys. Rev. Lett. \textbf{59}, 1768 (1987). 

\bibitem{cuprate-RH}
T. Penney, S. von Moln\'ar, D. Kaiser, F. Holtzberg, and A. W. Kleinsasser, 
Phys. Rev. B \textbf{38}, 2918 (1988).

\bibitem{Hall-x2}
A. P. Mackenzie, N. E. Hussey, A. J. Diver, 
S. R. Julian, Y. Maeno, S. Nishizaki, and T. Fujita, 
Phys. Rev. B \textbf{54}, 7425 (1996).

\bibitem{Ti214-RH}
N. Kikugawa, A. P. Mackenzie, C. Bergemann, and Y. Maeno,  
Phys. Rev. B \textbf{70}, 174501 (2004).

\bibitem{AL}
L. G. Aslamasov and A. I. Larkin, 
Fizika tvervd. tela, \textbf{10 (4)}, 1104 (1968)
[Sov. Phys. Solid State \textbf{10}, 875 (1968)].

\bibitem{Tewordt-AL}
S. Wermbter and L. Tewordt, 
Physica C \textbf{199}, 375 (1992).

\bibitem{Eliashberg-theory} 
G. M. $\acute{\textrm{E}}$liashberg, 
ZhETF \textbf{41}, 1241 (1962) 
[Sov. Phys. JETP \textbf{14}, 886 (1962)]. 

\bibitem{Fukuyama-RH} 
H. Fukuyama, H. Ebisawa, and Y. Wada, 
Prog. Theor. Phys. \textbf{42}, 494 (1969).

\bibitem{Kohno-Yamada} 
H. Kohno and K. Yamada, 
Prog. Theor. Phys. \textbf{80}, 623 (1988).


\bibitem{Oguchi-LS}
T. Oguchi, 
J. Phys. Soc. Jpn. \textbf{78}, 044702 (2009).

\bibitem{Yamada-text}
K. Yamada, 
Electron Correlation in Metals 
(Cambridge University Press, Cambridge, 2004).

\bibitem{Kubo-formula}
R. Kubo, 
J. Phys. Soc. Jpn. \textbf{12}, 570 (1957).

\bibitem{Onsager-thm1}
L. Onsager,  
Phys. Rev. {\bf 37}, 405 (1931).

\bibitem{Onsager-thm2}
L. Onsager,  
Phys. Rev. {\bf 38}, 2265 (1931).

\bibitem{Konno}
T. Konno, Busshitsu no taishousei to gunron [in Japanese] 
(Kyoritsu shuppan, Tokyo, 2001).

\bibitem{Luttinger-Ward}
J. M. Luttinger and J. C. Ward, 
Phys. Rev. {\bf 118}, 1417 (1960).

\bibitem{Baym-Kadanoff}
G. Baym and L. P. Kadanoff, 
Phys. Rev. {\bf 124}, 287 (1961).

\bibitem{DMFT-trans-QP}
W. Xu, K. Haule, and G. Kotliar,  
Phys. Rev. Lett. {\bf 111}, 036401 (2013).

\bibitem{Georges-diag}
E. Kozik, M. Ferrero, and A. Georges, 
Phys. Rev. Lett. {\bf 114}, 156402 (2015).

\bibitem{Eder}
R. Eder, 
arXiv:1407.6599. 

\bibitem{Hori-phD}
C. Hori, 
Ph.D. thesis [in Japanese], The University of Tokyo (2011).

\bibitem{Pade-approx}
H. J. Vildberg and J. W. Serene, 
J. Low Temp. Phys. {\bf 29}, 179 (1977).

\bibitem{Kusunose-Pade}
H. Kusunose, unpublished. 

\bibitem{Tsunetsugu-RPA}
H. Tsunetsugu, 
J. Phys. Soc. Jpn. \textbf{71}, 1844 (2002).

\bibitem{NA-RPA}
N. Arakawa and M. Ogata, 
Phys. Rev. B \textbf{87}, 195110 (2013). 

\bibitem{Ueda-paramag}
Y. Yamashita and K. Ueda, 
Phys. Rev. B \textbf{67}, 195107 (2003).

\bibitem{NA-paramag}
N. Arakawa, 
unpublished; 
I adopted the method of Ref. \onlinecite{Ueda-paramag} 
to a $t_{2g}$-orbital Hubbard model including not only the $U$, $U^{\prime}$, 
and $J_{\textrm{H}}$ terms but also the $J^{\prime}$ term, 
which was neglected in Ref. \onlinecite{Ueda-paramag}. 

\bibitem{vHs-cuprate}
S. Koikegami, S. Fujimoto, and K. Yamada, 
J. Phys. Soc. Jpn. \textbf{66}, 1438 (1997).

\bibitem{Hlubina-Rice}
R. Hlubina and T. M. Rice, 
Phys. Rev. B {\bf 51}, 9253 (1995).

\bibitem{Ishida-NMR-x2}
K. Ishida, H. Mukuda, Y. Minami, Y. Kitaoka, Z. Q. Mao, 
H. Fukazawa, and Y. Maeno, 
Phys. Rev. B \textbf{64}, 100501(R) (2001). 

\bibitem{Ishida-NMR-x05}
K. Ishida, Y. Minami, Y. Kitaoka, S. Nakatsuji, N. Kikugawa, 
and Y. Maeno, 
Phys. Rev. B \textbf{67}, 214412 (2003). 

\bibitem{Yanase-Ru}
Y. Yanase and M. Ogata, 
J. Phys. Soc. Jpn. \textbf{72}, 673 (2003).

\bibitem{Ikeda-FLEX}
H. Ikeda, 
J. Phys. Soc. Jpn. \textbf{77}, 123707 (2008).

\bibitem{Kon-AHE}
H. Kontani and K. Yamada, 
J. Phys. Soc. Jpn. \textbf{63}, 2627 (1994).

\bibitem{Haule-DMFT}
J. Mravlje, M. Aichhorn, T. Miyake, K. Haule, G. Kotliar, and A. Georges, 
Phys. Rev. Lett. \textbf{106}, 096401 (2011). 

\bibitem{Georges-DMFT-Ru}
J. Mravlje and A. Georges, 
arXiv:1504.03860. 

\bibitem{Hall-theory-x2}
C. Noce and M. Cuoco, 
Phys. Rev. B \textbf{62}, 9884 (2000).

\bibitem{Metzner} 
W. Metzner and D. Vollhardt, 
Phys. Rev. Lett. {\bf 62}, 324 (1989). 

\bibitem{Georges-review}
A. Georges, G. Kotliar, W. Krauth, and M. J. Rozenberg, 
Rev. Mod. Phys. {\bf 68}, 13 (1996).

\bibitem{Kotliar-review}
G. Kotliar, S. Y. Savrasov, K. Haule, V. S. Oudovenko, O. Parcollet, 
and C. A. Marianetti, 
Rev. Mod. Phys. {\bf 78}, 865 (2006).

\bibitem{cDMFT-Kotliar} 
O. Parcollet, G. Biroli, and G. Kotliar, 
Phys. Rev. Lett. \textbf{92}, 226402 (2004).

\bibitem{cDMFT-Arita}
Y. Nomura, S. Sakai, and R. Arita, 
Phys. Rev. B \textbf{89}, 195146 (2014).

\bibitem{DMFT-CVC1}
A. Khurana, 
Phys. Rev. Lett. \textbf{64}, 1990 (1990).

\bibitem{DMFT-CVC2} 
Th. Pruschke, D. L. Cox, and M. Jarrell, 
Phys. Rev. B \textbf{47}, 3553 (1993).

\bibitem{Morel-Nozieres}
P. Morel and P. Nozi$\grave{\textrm{e}}$res, 
Phys. Rev. {\bf 126}, 1909 (1962). 

\bibitem{Ru113}
L. Klein, L. Antognazza, T. H. Geballe, M. R. Beasley, 
and A. Kapitulnik,  
Phys. Rev. B {\bf 60}, 1448 (1999).

\bibitem{SC-trans1}
K. Izawa, H. Takahashi, H. Yamaguchi, Y. Matsuda, M. Suzuki, 
T. Sasaki, T. Fukase, Y. Yoshida, R. Settai, and Y. Onuki,  
Phys. Rev. Lett. \textbf{86}, 2653 (2001).

\bibitem{SC-trans2}
J. Xia, Y. Maeno, P. T. Beyersdorf, M. M. Fejer, and A. Kapitulnik, 
Phys. Rev. Lett. \textbf{97}, 167002 (2006).

\bibitem{Ce115} 
Y. Nakajima, H. Shishido, H. Nakai, T. Shibauchi, K. Behnia, 
K. Izawa, M. Hedo, Y. Uwatoko, T. Matsumoto, R. Settai, 
Y. \={O}nuki, H. Kontani, and Y. Matsuda, 
J. Phys. Soc. Jpn. {\bf 76}, 024703 (2007).

\bibitem{UPt3}
G. R. Stewart, Z. Fisk, J. O. Willis, and J. L. Smith,  
Phys. Rev. Lett. {\bf 52}, 679 (1984).

\bibitem{organic-review} 
M. Dressel, 
J. Phys.: Condens. Matter {\bf 23}, 293201 (2011). 

\bibitem{Takada-text}
Y. Takada, Tataimondai tokuron [in Japanese] 
(Asakura shoten, Tokyo, 2009).

\bibitem{NumericalReceip}
W. H. Press, S. A. Teukolsky, W. T. Vetterling, and B. P. Flannery, 
NUMERICAL RECIPES in C (Cambridge University Press, Cambridge, 1988). 

\bibitem{Takada-text2}
Y. Takada, Tataimondai [in Japanese] 
(Asakura shoten, Tokyo, 1999).

\end{thebibliography}

\end{document}